# MAPPING RESEARCH PRODUCTIVITY OF BRICS COUNTRIES WITH SPECIAL REFERENCE TO CORONARY ARTERY DISEASE (CAD): A SCIENTOMETRIC STUDY

**A Thesis**

**Submitted by**

**MUNEER AHMAD**

**(Enrollment No. 1710060003)**

**In partial fulfillment for the requirement of the award of the degree of Doctor of Philosophy in Library and Information Science**

**DEPARTMENT OF LIBRARY & INFORMATION SCIENCE**

**FACULTY OF ARTS**

**ANNAMALAI UNIVERSITY**

**ANNAMALAI NAGAR – 608 002**

**TAMIL NADU, INDIA**

**DECEMBER – 2020**

# ANNAMALAI UNIVERSITY

**Dr. M. SADIK BATCHA**
**Professor & University Librarian**
**Department of Library and**
**Information Science, Annamalai University**
**Annamalainagar-608 002**
**Tamil Nadu, India**

## CERTIFICATE

This is to certify that the thesis entitled **"MAPPING RESEARCH PRODUCTIVITY OF BRICS COUNTRIES WITH SPECIAL REFERENCE TO CORONARY ARTERY DISEASE (CAD): A SCIENTOMETRIC STUDY"** is a bonafide work of **Mr. Muneer Ahmad (Roll No. 1710060003,** Ph. D Research Scholar, Department of Library and Information Science, Annamalai University who carried out research under my supervision. Certified further, that to the best of my knowledge this thesis has not previously formed the basis for the award of any Degree, Diploma, Associateship, Fellowship or other similar title to any candidate.

**Station:** Annamalainagar **Signature of the Research Supervisor**

**Date:**

**ANNAMALAI UNIVERSITY**

**Annamalainagar**

# DECLARATION

I, **Muneer Ahmad (Roll No. 1710060003)** Research Scholar in the Department of Library and Information Science declare that the work embodied in this Ph. D. thesis entitled is a result of my own bonafide work carried out with my personal effort and submitted by me under the supervision of **Dr. M. Sadik Batcha,** Professor & University Librarian, at Annamalai University, Annamalainagar. The contents of this thesis have not formed the basis for the award of any Degree/ Diploma/ Fellowship/ Titles in this University or any other University or similar Institutions of Higher Learning.

I declare that I have faithfully acknowledged and given credit and referred to the researchers wherever their works have been cited in the body of the thesis. I further declare that I have not willfully copied some other's data/ work/ results etc. reported in the journals, magazines, books, reports, dissertations, theses, internet etc. and claimed as my own work.

**Station:** Annamalainagar **Signature of the Research Scholar**

**Date:**

## ACKNOWLEDGEMENT

In the name of Allah, the most Gracious and the most Merciful, I am humbled to Almighty, the elevated, for the achievement of this research work. The completion of this dissertation has been a joint effort in the sense that I have received abundant inspiration, encouragement and assistance from several colleagues, friends and respondents.

I place on record my deepest sense of gratitude to my esteemed guide, supervisor and my teacher **Dr. M. Sadik Batcha,** Professor & University Librarian, Annamalai University, for his valuable guidance, scholastic criticism and affectionate encouragement for the completion of this dissertation. I am feeling short of words to thank him for all the guidance, support and time in spite of having a very busy schedule.

I am indeed grateful and extend my worm thanks to the Vice-Chancellor, the Registrar, the Controller of Examinations, Dean, Faculty of Arts, for giving me an opportunity to complete this research work and for their support. I thank all the University authorities to have provided me all facilities and permission to complete my work.

I would like to thank Dr. P. Ravichandran, Professor and Head, Department of Library and Information Science, Annamalai University for providing me the necessary facilities in my study. I am also thankful to all other faculty members of the Department of Library and Information Science for their words of encouragement.

I reserve a special word of thanks and am grateful to the unforgettable friendship, encouragement of Auwalu Abdullahi Umar, MLIS Student, Department of Library and Information Science for his invaluable help.

I acknowledge with heartful thanks to Ph. D Research Scholars S. Roselin Jahina, Gana. G. S, R. Mohanadevi, R. Sonia, K. Thirumal, P. Balachandar, and M. Phil Scholars Mohammad Amin Dar and Saddam Hossain for their continuous help and for their inspiration in completing my research work.

I am extremely grateful to my father Mr. Ghulam Qadir Bangroo and my mother Mrs. Amina Bano, my Sisters Haseena Bano, Asiya Amin & Nazima Qadir, my Brother Mohammad Abdullah Bangroo and my nephews Taha Mukhtar Bhat, Soliha Lateef & Mohammad Azam for their constant encouragement, inspirations, love and affection which acted as a premium to achieve my goal.

I am also thankful to my friend Younis Rashid Dar, Ph. D Research Fellow, University of Kashmir for his continuous help and the inspiration in completing my research work.

I also express my thanks to the non-teaching staff Anand, Arumugam, Murali and Sreenivasan for their cooperation and constant help.

Finally, my sincere thanks are for those who have helped me directly or indirectly in the completion of this research work.

**Muneer Ahmad**

# TABLE OF CONTENTS

# LIST OF TABLES

## LIST OF FIGURES

# ABBREVIATIONS

| | | |
|---|---|---|
| AAPP | : | Average Number of Authors per Paper |
| CAD | : | Coronary Artery Disease |
| CC | : | Collaborative Coefficient |
| CI | : | Collaborative Index |
| CHD | : | Coronary Heart Disease |
| DT | : | Doubling Time |
| ISI | : | Institute for Scientific Information |
| JIF | : | Journal Impact Factor |
| MeSH | : | Medical Subject Headings |
| MCC | : | Modified Collaborative Coefficient |
| NA | : | Number of Authors |
| RGR | : | Relative Growth Rate |
| SCI | : | Science Citation Index |
| SSCI | : | Social Science Citation Index |
| WoS | : | Web of Science |
| % | : | Percentage |

# CHAPTER I

# INTRODUCTION

A scientist is not merely a gifted person involved in devising new information claims through scientific publications. More prosaically, he or she is an individual arising out of a distinct, considerably irreproducible string of biological, biographical, and factual situations. Therefore, it might be plausibly contested that a comprehensive scope of science should be implemented at varying layers, practising mathematical means not only to the concluding output, the stylish and impeccable book or journal paper but to any type of quantitative data moderately referable to scientific accomplishments. Furthermore, such a case is indeed more rational because an extra-bibliographic concern with science standards emerged in the past long before the bibliometric zooming in on publications and citations, signifying the deep purpose of many scholars to apprehend in precise, mathematical expressions the material stipulations for the phenomenon of inspiration and creativity given their artificial generation for the purpose of advancement.

Eighteenth-century scientists' attention with the statistical dispersion of scientific character, is fundamentally motivated by the exploration for the true, corporeality causes (physiological, emotional) of its exhibitions, inclined on a conventional interpretation of scientific value that twirled throughout the sealing of individual perfection by past performances, such as the affiliation with a prestigious association, the insertion in a vocabulary, or the idea of qualified companions. Bibliometrics, in contrast, developed from the investigation of quantitative patterns about the arrangement of scientific papers generated by the specialists themselves. When it scrutinised specifications, it did not urge external agents or material causes but to Lotka, Bradford, and Zipf's empirical laws. Moreover, when it adhered to citation indexes, its strength to support or compete with peer evaluations for the assessment of scientific value initiated an entirely new set of possibilities.

The fundamental purpose of science is to create and transfer scientific information. As Merton affirmed: 'for science to be exceptional, it is not sufficient that profitable concepts are introduced, or new operations evolved or new problems formulated or new systems initiated. The innovations must be efficiently communicated to others. That, after all, is something we indicate by an augmentation to science – something is given to the prevalent stock of education. In the end, then, science is a humanly bestowed and socially authorised body of knowledge. For the expansion of science, only work effectively perceived and utilised by other scientists, then and there, matters.'

Scientific research is an information-producing activity (Nalimov & Mul'čenko, 1969), the essence of communication (Garvey, 1979). The representatives are operating in scientific communication form a highly heterogeneous method.

### 1.1 Coronary Artery Disease

*Coronary artery disease (CAD)*-often described as coronary heart disease or CHD, is usually related to indicate to the pathologic means concerning the coronary arteries (regularly atherosclerosis). CAD incorporates the examinations of angina pectoris, myocardial infarction (MI), silent myocardial ischemia, including CAD fatality that occurs from CAD. Hard CAD endpoints usually incorporate MI and CAD death. The title CHD is frequently employed conversely with CAD. *CAD death*—Involves unexpected cardiac death (SCD) for incidents when the death has befallen within 24 hours of the unexpected encounter of indications, and the word non-SCD utilises when the time passage from the clinical exhibition unto the time of death overtakes 24 hours or has not been explicitly distinguished. *Atherosclerotic cardiovascular disease (ASCVD, oftentimes shortened to CVD)*-the pathologic means concerning the complete arterial circulation, not merely the coronary arteries. Stroke, transient ischemic attacks, angina, MI, CAD death, claudication, and severe limb ischemia manifest as ASCVD (Lemos & Omland, 2018).

Cardiovascular disease is established as the preeminent cause of death worldwide. According to the World Heart Federation, cardiovascular disease is responsible for 17.1 million deaths globally each year. Surprisingly, 82% of these deaths occur in the developing world. Such figures are oftentimes hard to perceive. The situation's gravity is magnified when depicted as heart disease kills one person every 34 seconds in the USA alone. Thirty-five people under 65 expire precipitately in the UK every day due to cardiovascular disease (12,500 deaths per annum). Despite the leading destroyer, the occurrence of cardiovascular disease has dwindled in recent years due to a higher immeasurable knowledge of pathology, implementation of lipid-lowering treatment unique medication regimens including moderate molecular mass heparin and antiplatelet medications such as glycoprotein receptor inhibitors and stringent operational interference (Beltrame, 2012).

### 1.1.1 Ageing

Ageing is an unmodifiable danger constituent for CAD, with males clinically exhibiting this situation at 50-65 years of age and females about ten years later, following menopause (Lerner & Kannel, 1986). The WHO describes that the leading reason for people's death over 65 years is CAD, and as age advances, an abundant proportion of deaths are amongst females. In several advanced countries, the amount and proportion of older people (i.e., over 65 years) are progressing, manifested chiefly by potency and mortality deterioration. The ageing residents of various countries have expedited the augmentation of CAD to the cumulative disease burden. It is prognosticated that the global ageing community will sustain CAD as a surpassing purpose of death worldwide (Mensah, 2010).

Amongst countries besides high but waning CAD fatality, it is proposed that certain inclinations are developing concerning more growing generation subgroups (O'Flaherty et al., 2009). A slowing or levelling of the drop in CAD mortality in growing adults has now been proclaimed in England and Wales, the US, France, Australia, and New Zealand. These conclusions prompt attention, indicating that decades of advancement in lessening losses from CAD resemble stalling. Alterations in lifestyle constituents in the young (progressing obesity and sedentary lifestyles) may subdue development.

### 1.1.2 Coronary Artery Disease at Developing Nations

CAD is related to the world's leading agent of mortality for men and women, responsible for more than 7 million deaths every year. Although CAD is the most common reason for death in advanced nations, globally, covering 60% of deaths promptly befall in developing nations. It is apparent that an extensive spectrum in the predominance of CAD mortality survives. Notwithstanding many efforts to enhance the disproportional mortality rates, a human inclination in CAD still resides. This is visible by the higher CAD death rates in lower SES areas within regions and too within nations and apparent gender segregation, curiously amongst younger women.

With a slowing down of age-adjusted dying, social differences will likely increase. By 2030, it is calculated that the amount of CAD deaths will surge by up to 137% in developing countries and by up to 48% in regions wherever CAD is in decline; as such, CAD will prevail the principal element of death worldwide (Beltrame, 2012).

### 1.1.3 Mortality Rate of Coronary Artery Disease

There transpired a peak in CAD mortality in advanced countries in the 1950s, with a consecutive decline since the 1960s. The WHO Multinational MONItoring of trends and determinants in CArdiovascular disease (MONICA) scheme recognised an annually 4% drop in CAD mortality rate drifts over ten years from the 1980s crosswise 21 countries (Tunstall-Pedoe et al., 2000).

Consequently, while many Western European nations have registered visible advancements in CAD mortality, Eastern European nations (such as Hungary) frequently revealed less recovery. Certain trends typically correspond to socioeconomic inequalities, with the decline in CAD mortality remaining visible in nations beside a more favoured socioeconomic situation.

In contrast, some developing nations beget an accelerating pace of CAD mortality. Admittedly, the WHO predicts that 60% of the global load of CAD happens in advancing countries. Notwithstanding, mortality assessments are trying to achieve in some countries; broad estimations of overall CVD epidemiology report growing CVD mortality in metropolitan China, Taiwan, Korea, and Malaysia. In China, CVD mortality progressed as a dimension of cumulative deaths from 12.8% in 1957 to 35.8% in 1990 (Khor, 2001). Similar to many developing nations, it has undergone accelerated urbanisation, socioeconomic, and well-being changes, contemporaneously including an increase in life prospects - features consistent with stage 2 of the epidemiologic transformation.

### 1.1.4 Coronary Artery Disease in BRICS Nations

Cardiovascular diseases continue the number one determinant of death globally and in Brazil, resembling one-third of total deaths. The CVDs and their complexities produce a strong influence on the loss of productivity in the workplace and the household earnings, ending in a US$ 4.18 billion debt in the Brazilian economy from 2006 to 2015. Researchers conducted in various nations have

conferred a decline in the proportion of CVDs and in CVD mortality since the 1960s. In Brazil, that decrease transpired later, in the 1990s. Though other investigations have related worsening of health indicators in Brazil, which must be associated to the economic crisis, the growth in poverty, and the reductions in health and social policies emanating after the Constitutional Amendment 95/2016 and the restriction on common investments, health covered, for 20 years (Malta, Teixeira, Oliveira, & Ribeiro, 2020).

Cardiovascular disease (CVD) persists the preeminent cause of death and disability in most countries of the world, including Russia. Mortality from CVD is particularly high in the Russian Federation compared with the average in Europe (55.7% vs 46%). In Russia, 29.4% of deaths occur from coronary heart disease (CHD) and 17.6% from cerebrovascular disease, mostly strokes. CVD kills more women than men: 51 and 42% respectively in Europe, 60.2 and 47.2% in Russia, with women dying at older ages ("Positive trends in cardiovascular mortality in Russia and Moscow: potential confounders," 2016).

There has been a frightening increase over the past two decades in the predominance of CHD and cardiovascular deaths in India and other South Asian nations. India is transpiring through an epidemiologic transformation whereby the burden of transmittable infections has diminished moderately, but that of non-communicable diseases (NCD) has surged immediately, thus leading to multiplying trouble. There has been a 4-fold acceleration of CHD pervasiveness in India throughout the past 40 years. Contemporary estimations from epidemiologic investigations from several parts of the country intimate a predominance of CHD to be between 7% and 13% in urban and 2% and 7% in rural residents. Epidemiologic researches have revealed that there are at present over 30 million cases of CHD in India. The Global Burden of Diseases Study proclaimed that the disability-adjusted life years dissipated by CHD in India throughout 1990 was 5.6 million in men and 4.5 million in women; the calculated numbers for 2020 were 14.4 million and 7.7 million in men and women sequentially (Krishnan, 2012).

Coronary heart disease (CHD) is the second foremost cause of cardiovascular death in the Chinese population. It estimates 22% of cardiovascular deaths in metropolitan regions and 13% in provincial areas. Although the mortality from CHD in China is comparatively low compared with Western levels, the burden of CHD has been progressing. This is notably because of a deteriorating profile of risk determinants, such as an extended pervasiveness of hypertension, hyperlipidaemia, overweight/ obesity, diabetes, etc. and notably because of an advance in the aged population. The large predominance of overweight (25.7% in urban vs 19.3% in rural areas) and still expanding trends in the Chinese population make it the third most important contributor to the occurrence of CHD in the Chinese adult community. Overweight records for 25.7% of CHD incidence in metropolitan and 20.5% in provincial areas for adults aged >18 years (Zhang, Lu, & Liu, 2008).

Cardiovascular diseases (CVDs) were accountable for nearly half of total deaths worldwide in 2008. The majority of those deaths happened in low-to-middle income nations, with >50% occurring in those aged <70 years. Africa is a continent to >1 billion people and is a significant contributor to the global burden of CVD. In 2013, an assessed 1 million deaths occurred attributable to CVD in sub-Saharan Africa only, which aggregated 5.5% of all global CVD-related deaths and 11.3% of all mortality in Africa. CVD-related deaths contributed to 38% of all non-communicable disease-related deaths in Africa, reflecting the expanding threat of both non-communicable disease and CVD. An almost twofold increase in the overall amount of CVD-related deaths since 1990 has remained reported, with a >10% variance in mortality amongst women associated with men (Keates, Mocumbi, Ntsekhe, Sliwa, & Stewart, 2017).

### 1.2 Bibliometrics: A Science of Science

The study of scientific literature has a long history dating back to the past century's early decades. However, despite the amount of research in this area, it was not until 1969 that the term *bibliometrics* first appeared in print (Groos & Pritchard, 1969). It was defined as the 'application of mathematical and statistical methods to books and other media of communication,' and the term was quickly adopted and used, particularly in North America (Wilson, 1999). At almost the same time, (Nalimov & Mul'čenko, 1969) coined the term *scientometrics* to refer to 'the application of quantitative methods which are dealing with the analysis of science viewed as an information process.' In contrast, this term was widely used in Europe (Wolfram, 2003). Initially, therefore, scientometrics was restricted to the measurement of science communication, whereas bibliometrics was designed to deal with more general information processes. At present, however, bibliometrics and scientometrics are used as synonyms (Glänzel, 2003).

## 1.3 Scientometrics

The definition of scientometrics focused on the study of scientific information is given by (Braun, Bujdos, & Schubert, 1987): 'Scientometrics analyses the quantitative aspects of the generation, propagation, and utilisation of scientific information in order to arrive at a better understanding of the mechanism of scientific research activities.'

Scientometrics is a science field dealing with the quantitative aspects of people or groups of people, matters and phenomena in science, and their relationships, but which do not primarily belong within the scope of a particular scientific discipline (Péter Vinkler, 2001). Scientometrics aims to reveal characteristics of scientometric phenomena and scientific research processes for more efficient management of science.

### 1.3.1 Scientometrics: Its Origin and Development

Scientometrics may belong to the discipline of 'the science of science' (Bernal, 1939; Merton & Storer, 1973; Price & Tukey, 1963). However, the term 'the science of science' may be understood as indicating a discipline that is superior to others. In this respect, the relationships between scientometrics and other disciplines would be similar to philosophy, as had been assumed earlier. However, scientometrics should not be regarded as a field 'above' other scientific fields: scientometrics is not the science of sciences but science for science.

As with all scientific disciplines, scientometrics involves two main approaches: theoretical and empirical. Both theoretical and empirical studies are concerned primarily with the impact of scientific information.

An important step on the road to the development of evaluative scientometrics was made by Martin and Irvine, who applied several input and output indicators and developed the method of converging partial indicators for evaluating research performance of large research institutes (Irvine & Martin, 1984; Martin & Irvine, 1983; Martin, 1996; Martin & Irvine, 1984). With the conclusion drawn by Martin: "…all quantitative measures of research are, at best, only partial indicators – indicators influenced partly by the magnitude of the contribution to scientific progress and partly by other factors. Nevertheless, selective and careful use of such indicators is surely better than none at all. Furthermore, the most fruitful approach is likely to involve the combined use of multiple indicators" (Martin, 1996).

(Braun, Glänzel, & Grupp, 1995) introduced several sophisticated indicators for studying publications of particular countries. (Moed, Burger, Frankfort, & Van Raan, 1985a, 1985b) provided a standardised method for evaluating publications of research teams at universities. Furthermore, they have developed several indicators and methods for assessing research institutes and teams (Vinkler, 2000).

According to (Kostoff, 1995) 'the bibliometric assessment of research performance is based on one central assumption: scientists who have to say something important do publish their findings vigorously in the open international journal ("serial") literature.' In his opinion: 'Peer review undoubtedly is and has to remain the principal procedure of quality judgment.' This may be true, but we can easily prove that most evaluative scientometrics indicators are based directly or indirectly on detailed expert reviews (e.g., acceptance or rejection of manuscripts, referencing or neglecting publications).

Scientific information may be regarded as goods (Koenig, 1995) with features characteristic of goods, namely *value* and *use-value*. Here, 'value' may be assumed as scientific value referring to information's innate characteristics (i.e., originality, validity, brightness, generality, coherence, etc.). 'Use the value' refers to the applicability of the information in generating new information or to its immediate application in practice. References may be considered as manifested signs of use-value of information.

### 1.3.2 Scientometric Indicators

Scientometric indicators can be classified according to *the number of scientometric sets* they represent and the *application of reference standard(s)* (Vinkler, 1988; Vinkler, 2001). Scientometric indicators referring to the measure of a single scientometric aspect of scientometric systems represented by a single scientometric set with a single hierarchical level are termed *gross* indicators. Indicators referring to two or more sets or a single set with more than a single hierarchical level are *complex* indicators. Those indicators, which consist of several gross or complex indicators, preferably with weighting factors, and each representing an unusual aspect of a given scientometric system, are *composite* (or compound) indexes. Complex indicators may characterise a particular scientometric aspect of a system, and as such, they have a well-defined physical meaning (in contrast to composite indicators). Complex indexes may incorporate reference standards; gross indicators do not.

The fundamental criterion of scientometrics is that science or scientific research as a system has *quantitative perspectives* that *can be described* by mathematical (mainly statistical) techniques. According to (Holton, 1978) "… nothing is more reasonable than to produce indicators about science that themselves consist of quantifiable measures". Scientometrics is concerned primarily with the description of regularities in the production, flow, and application of the information in science. In order to characterise information phenomena quantitatively, *reliable data* must be obtained, *appropriate methods* and *relevant indicators* must be constructed and applied (Moravcsik, 1988).

According to (Braun, Glänzel, & Schubert, 1985): 'Statistical indicators are selected or constructed from empirical statistical data, in a way to form a coherent system based explicitly or implicitly on some theoretical model of the phenomenon under study.'

### 1.4 Mapping Science Using the Web of Science

Since Derek de Solla Price first proposed turning the tools of science on science itself (Price & Tukey, 1963), measuring and mapping the scientific enterprise using the scholarly literature in the Web of Science has been a desire of policymakers, researchers, and Scientometricians. Beyond only counting the papers published in specific journals or subject categories of Web of Science, the citation relationships that have been comprehensively indexed for decades allows for clustering of papers to represent the real structure and dynamics of speciality areas and, when aggregated, domains of investigation. Often analysts will follow direct citation paths through subsequent generations of papers to map the evolution of our understanding of, say, a given disease or physical phenomenon. However, Scientometricians can also use bibliographic coupling and co-citation methods to reveal more about how scholarly research forms, ebbs, and flows, grows or dies (Cantu-Ortiz, 2017).

For close to half a century, the Web of Science has been a critical resource for discovering important literature, evaluating the impact of journals, and the productivity of research organisations and scientists, often within a subject area. In more recent years, the Web of Science has also become an analytic resource for researchers interested in using the citation histories contained within it as a proxy for large-scale analysis of the knowledge flows in the scientific enterprise, especially in networks or graph theory.

### 1.5 Citation Analysis

The Web of Science is made up of citation indexes, and the editors also leverage the citation data within the various databases to determine the influence and impact of journals being evaluated for coverage. Two primary indicators are used: Total Citation (TC) counts and the Journal Impact Factor (JIF). These two indicators provide the editors with a better understanding of how a journal contributes to the overall historical citation activity within the scholarly community in a size-dependent way (TC) and the recent average impact of the journal in a size-independent way (JIF). Citation data are considered within the journal's overall editorial context, including looking at the citation data of editors and authors and the general citation dynamics within the journal's subject area. The self-citation phenomenon is normal; however, the *Journal Citation Report* (JCR) editors monitor the level of self-citation that occurs within a journal's overall citation activity. Excessive self-citation rates are examined further, with possible suppression or deselection of the journal if citation patterns are found to be anomalous (Cantu-Ortiz, 2017).

### 1.6 Bibliometric Databases

A study conducted by Lutz Bornmann and Rüdiger Mutz revealed that the number of papers stored in the Clarivate Analytics' Web of Science (WoS) Bibliometric database grew exponentially between 1980 and 2010, moving from around 700,000 documents in 1980 to nearly 2 million documents in 2010. Additionally, they discovered that the number of references and citations increased exponentially from 1650 to 2000 (Bornmann, Stefaner, de Moya Anegon, & Mutz, 2016). Based on these findings, it has been observed that the amount of scientific knowledge measured in the number of published documents in Bibliometric databases doubles every 9 to 15 years, and this is a trend that will remain for the next few years. This sprouting expansion poses a challenge for most institutions that lack the means or formal mechanisms to adequately document the inception of intellectual contributions taking place within themselves (Cantu-Ortiz, 2017).

Bibliometric databases are of various types. They may be proprietary products, like Clarivate Analytics's WoS (formerly owned by Thomson Reuters) and Elsevier's Scopus, or they can be public products, such as Google Scholar (GS), Research Gate, and others entering into the Open Access movement. Other databases that hold institutional repositories or current-research information systems (CRIS) are reservoirs of publications used by a growing number of organisations. To give a sense of scale, as of September 3, 2014, WoS held around 50,000 scholarly books, 12,000 journals, 160,000 conference proceedings with 90 million total records, and 1 billion total cited references. Also, 65 million new records are added per year.

Elsevier, an extensive research publisher and digital information provider publishing over 2,500 scientific journals, launched Scopus in 2004. As of 2017, Scopus covers 67 million items drawn from more than 22,500 serial titles, 96,000 conferences, and 136,000 books from over 7,500 different publishers worldwide (Cantu-Ortiz, 2017).

### 1.7 Statement of the Problem

The present study analyses Coronary Artery Disease research from BRICS countries through three decades (1990-2019). Coronary heart disease (CHD) is the single largest cause of death in developed countries and is one of the principal causes of disease burden in developing countries. In 2017, 17.8 million (17.5-18.0 million) deaths were attributed to CAD globally, which amounted to an increase of 21.1% (19.7%–22.6%) from 2007. Overall, the crude prevalence of CVD was 485.6 million cases (468.0–505.0 million) in 2017, an increase of 28.5% (27.7%–29.4%) compared with 2007. On the other side, the publications on CAD in 2007 were 14842 and in 2017 publications grow 22454. The growth is not at par with the growth of disease and thus needs to be an enormous amount of research in the field. Hence the problem with this study attempts to convert the publications in the exclusive database. This database covers information, titles, publications, authors, address, and authorship pattern adopted. This study aims to arrive in the future course of projection in authorship pattern and publications in the study period 1990-2019. The research is mostly examining in nature in recognising the research output of scientists in coronary artery disease research and is also systematic in nature with appropriate statistical tools application in strengthening the empirical validity. Assessment studies on coronary artery disease are to know the trends and opportunities of academic and research practices and to identify the areas in which such assessment studies have not been carried out as there are no major scientometric and bibliometric studies on analysing the publication outcome in the field of coronary artery disease, which is being one of the prominent areas of research. There is a requirement to appraise the achievement of scientists and researchers of countries and to estimate their scientific activities by scientists and researchers to divert the innovative activities where the need is greater. Doing this requires an evaluation mechanism. Scientific output, particularly literature, are good indicators of innovative activity. Therefore, it is necessary to evaluate the performance of coronary artery disease research and development activities for accessing the quality of technical knowledge and to incite the learning experience so as to benefit researchers engaged in research and development activities and to assess the quality of technical knowledge as well as the intensity of specialisation in high technology in a country.

### 1.8 Objectives of the Study

The primary objectives are framed with the particular theory of the present study as mentioned below;

1. To analyse the growth of literature in Coronary Artery Disease research output during the years 1990-2019.
2. To examine Relative growth rate and Doubling time of Coronary Artery Disease research.
3. To compare and measure the analysis of country-wise Coronary Artery Disease research output performance.
4. To measure research productivity through the Activity index concerning countries during the study period.
5. To determine the document wise research concentrations in Coronary Artery Disease research in the study period.

6. To evaluate the language-wise distribution in Coronary Artery Disease research.
7. To assess the nature of the authorship pattern and find out the degree of collaboration.
8. To identify the most prolific authors and productive sources on Coronary Artery Disease research in BRICS countries.
9. To test the applicability of Lotka's law of author production in Coronary Artery Disease research.
10. To test the applicability of Bradford's law of scattering in Coronary Artery Disease research.
11. To test the applicability of Zipf's law of word frequencies on the Coronary Artery Disease literature.
12. To evaluate the various indices in authors research output performance.

### 1.9 Hypotheses

In the present study, the hypothesis with a vision to investigate the exact legitimacy of kept targets of the present assessment. Therefore, the hypotheses are formulated based on framed objectives.

1. There is an expanding trend in the relative growth rate and correspondingly a decreasing pattern in the doubling time in Coronary Artery Disease research.
2. The Journal source of distribution of Coronary Artery Disease output involves a predominant spot compared to other sources of productions.
3. There has been an increasing trend in collaborative research in Coronary Artery Disease examined in recent years.
4. There has been a logical efficiency of the authors contributing to the Coronary Artery Disease in conformity to the Lotka's Law.
5. There has been a delivery of Coronary Artery Disease research productivity in journals and articles that comply with the implication of Bradford's law.
6. The distribution of Coronary Artery Disease literature on output articles relatively confirms the implications of Zipf's law.

### 1.10 Chapterisation

The present study is structured into five parts.

- **Chapter I** introduces the thesis and its background provide the context of the growth and development of scientometrics. The chapter also provides a general introduction to coronary artery disease and the metric areas such as bibliometrics, scientometrics, its origin, scientometric indicators, and mapping science using the web of science. The chapter also consists of the objectives of the research, hypotheses, and statement of the problem.

- **Chapter II** focuses on reviewing the available literature in the context of the study and covers critical sources. The literature review focuses on scientometric studies and has been arranged according to various subheadings and chronological order from 1990 to 2020 and consists of various studies like individual scientists works, scientific productivity of individual institutions, scientific productivity of individual journals, studies based on country's literary output, studies based on scientometric analysis of various diseases, and other scientometric works. The chapter also provides a quick recap of the various ways the scientometric concepts are being applied and how they were influential to the research here. The section is concluded with a summary of how the findings have helped the researcher find the gaps that have formed a basis for the research.

- **Chapter III** focuses on the research design and the methodology applied in the investigation. It also discussed data collection and limitations and various statistical indicators and statistical tools used in the study.

- **Chapter IV** illustrates the study results are published and discussed, and details of all the indicators and the outcome are presented.

- **Chapter V** sums up findings based on the objectives set. Its limitations and contributions are highlighted. The directions for possible avenues for future work are given before the conclusion.

- The bibliography is provided at the end. The bibliography and the in-text citations are given in APA citation style, which is popular in the scholarly communication literature.

# CHAPTER II

# REVIEW OF LITERATURE

This chapter examines the review of works related to various aspects of Scientometric studies. It has been observed that various research studies are highlighting the importance of 'Scientometric Analysis' and its applications in measuring the quality of research through various indicators. This type of analysis enables the researcher to identify the research gaps in the previous studies. Review of related studies further avoids the duplication of work that has already been done in that area. Moreover, it helps the researcher in studying the various aspects of a problem which further allows him to identify and create new grounds for the research. Research is a journey to discover the unknown. A vital component of the research process is the literature review. This chapter reviews the relevant literature on many areas within scientometrics. It also examines the outcome of other research findings, which fueled the whole output of this investigation. A researcher is always measured by the quality of his/her literature review, which provides an understanding of the whole subject. The literature review helps the researcher frame the research study on the chosen topic by providing new ideas, concepts, methods, techniques, and approaches. The literature review process's eventual goal is to identify published information in the area, analyse a part of the published knowledge (scholarly articles, books, dissertations, etc.) critically through summary, classification, and comparison of theoretical articles area of the present study. The significance and advantage of a well-conducted and thorough literature review cannot be emphasised in the context of designing a research study. The principal objective of a literature review is to assist researchers in becoming accustomed to the work that has previously been conducted in their decided topic areas. It also helps to find out the different aspects of the problem. It enables us to discover unexplored or new areas to create new growth for research.

Usually, the outcomes of a well-conducted research review will reveal that the investigation is intended, in fact, previously been conducted. This would unquestionably be important to know during the outlining stage of a study, and it would definitely be advantageous to be informed of this fact sooner rather than later.

Other times, researchers may change their studies' focus or methodology based on the varieties of inquiries that have previously been administered. Literature reviews can oftentimes be intimidating for novice researchers, but like most other concerns associating to research, they become easier as one gains experience.

Acknowledging the indispensable importance of pursuing sustainable competitive advantage and long-term success for enterprises, the sustainable operation has stimulated extensive interest in academia and industry. Understanding the course can fill the research gap, which may be there between theory and practice. It is significant not only to implement relevant research to help organisations obtain sustainability but also to find what has been studied till now and what needs further exploration shortly.

In view of the huge amount of literature available in this field, in this chapter, an attempt has been made to review only the significant and the recent literature on the various aspects of scientometric research under the following subheadings:

## 2.1 Studies Based on the Works of Individual Scientists

Klaić (1990) examined the research activity of chemists from the "Rugjer Bogkovid" Institute (RBI, Zagreb, Yugoslavia) for the period 1976-1985, comprising 2018 research years of systematic work, and 1149 SCI recorded papers (0.57 publications per research year). On average, one paper was written by 3.05 scientists. The articles were published in 235 different journals, most usually in the national *Croatica Chemica Acta* (171 papers). The publications were classified into two groups: for the periods 1976-1980 and 1981-1985, and for each paper, citations were received in the corresponding time period. An average publication produced 2.58 citations. Chemical papers from the second period got 2.73 citations per paper, which is 85% of the anticipated value, which remained considerably more than for Yugoslav papers (66%) in general. *The* distribution of citations was also examined from 1975 through 1985; the RBI chemists published more than 30% of the Yugoslav chemical output. Papers were published by, on average, nearly three authors. The study of Klaic indicates that an average publication produced 2.58 citations, whereas in the present study the overall citation per paper is 15.16. Hence the study of researcher and review taken differs in terms of citations per paper.

Kalyane & Munnolli (1995) carried out the scientometric study of T. S. West, the globally well-known analytical chemist who has been internationally known as a very thriving scientist. His research productivity and collaboration pattern were interpreted by years, papers, authorships, and author wise productivity. He has published 410 research papers. The years 1969-70, when he was 42-43 years ago, was most fruitful with 41 research papers in 1969 and seven single authorship papers in 1970. Quinquennial collaboration coefficients fluctuated between 0.57 to 1.00, obviously intimating distinguished collaboration team spirit in his research association. His productivity coefficient was 0.45, showing rapid publication activity throughout the early period of his research career. His most conspicuous collaborators in different papers were: R. M. Dagnall (92), G. F. Kirkbright (77), R. Belcher (56), K. C. Thompson (19), J. D. Norris (13), and J. F. Alder (11). Topmost ranking journals, with papers, to which he had contributed were: *Anal Chim. Acta* (106), *Talanta (84), The Analyst* (49), *Anal Chem.* (23), and J. *Chem. Soc.* (20). Publication density was 8.54, publication consistency was 6.25, and the average Bradford multiplier was 3.9. The study of Kalyan and Munnolli reveals that the average Bradford multiplier was 3.9, whereas in present study the average Bradford multiplier value is 8.7. Hence the study of researcher and the review taken differs.

Kademani & Kalyane (1996) evaluated and studied 164 papers by R. Chidambaram, a nuclear physicist from India, published during 1958-93, and recognises highly cited papers as per Science Citation Index. The study's remarkable finding is that out of thirteen papers deemed by the scientist as most notable; four are outstandingly cited, four are remarkably cited, one is fairly cited, and one which was published in 1990 received two citations till 1992, and two papers did not get any citation. The study of Kademani and Kalyane detected that 2 (1.22%) articles out of 164 publications were uncited, whereas in the present study 11714 (23.41%) of research articles are uncited. Hence the study of researcher and the review taken differs. The citation analysis of papers published in 1993 was not brought out. The well-cited papers of the scientist did not find any place in the aforementioned list of thirteen. The conclusion indicates towards a probability that a scientist's self-assessment about the importance of his papers may not always correspond with the world opinion.

Kademani, Kalyane, & Kumar (2001) registered Scientometric analysis of 246 papers by Ahmed Hassan Zewail, the Nobel laureate in chemistry (1999), written between 1976 and 1994 in distinct fields: femtochemistry (62), reaction rates and IVR (56), extensive reviews (49), coherence and optical dephasing phenomena (27), solids: magnetic resonance and optical studies (13), liquids and biological systems(9), local modes in large molecules (9), molecular structure from rotational coherence (8), solar energy concentrators(7), and different subjects(6). Data were examined for authorship patterns with his 103 associates. The highest collaborations were with P. M. Felker (39), M. Dantus (19), and L.R. Khundkar (16). The highest number of collaborators (38) was in 1986 – 90, followed by 30 throughout 1981 – 1985. His productivity coefficient was 0.52, which is clear evidence of constant publication productivity behaviour throughout his 19 years of research. The study of Kademani, Kalyane and Kumar exposed that collaborative coefficient of their study is 0.52, whereas in the present study, collaborative coefficient value is 0.79. Hence the study of researcher and review taken differs.

Kalyane & Sen (2003) examined and undertook the analysis work of an analytical chemist, "Tibor Braun," who has an outstanding record of accomplishments. Quantitative documentation about Tibor Braun comprises his papers (single-authored 40; and multi-authored 140) during 1954-1995. The productivity coefficient is 0.78. Tibor Braun had 80 collaborators, of which Schubert, Glanzel, Zsindely, including Farag, were the most productive. Author productivity in the research group of Tibor Braun supports the trend of Lotka's Law. The study of Kalyane and Sen explored that Lotka's law acknowledges the authorship distribution and in the same way the present study on CAD fits Lotka's law. Hence, the study of researcher and the review taken correlates. He had practised 49 channels of communication to propagate his research results, of which Scientometrics (33 papers) surpasses the list, succeeded by Anal Chim Acta (21 articles). The publication concentration is 10.2, and publication density 3.7.

Meyer et al. (2004) conducted the bibliometric analysis of Keith Pavitt's performance and the influence that he has had. Keith Pavitt has executed pioneering enrichment to the knowledge of science, technology, and innovation. First, the paper observes how Pavitt's publication profile evolves over time. Then it pursues his most cited works and investigates the collections of references in his papers. Author and journal co-citation maps demonstrate Pavitt's rational setting and the central role Research Policy performed in this context. An analysis of the most commonly cited authors in Research Policy and Scientometrics emphasises Keith Pavitt's position as both a shaper of and a link between science and technology policy and bibliometric examination.

### 2.2 Scientific Productivity Study on Individual Institutions

Beck & Gáspár (1991) assessed a scientometric evaluation of the dissemination activities of various departments of the Faculty of Natural Sciences of Kossuth Lajos University. The essence of the strategy is the deliberation of the abundance and variety of the papers written. For a measure of this quality, the author considered the journal's impact factor, in which a paper was distributed. The somewhat different range of the impact factors of different areas was taken into account throughout the evaluation. As a whole, no noteworthy discrepancy was found between the writing activity (impact per number of researchers) of the research institutes of the Hungarian Academy of Sciences and their Faculty's corresponding departments. However, notable divergences occur in distinct disciplines. Based on this study, differences in the publication policies of the different departments were prescribed.

Braun, Glänzel, & Grupp (1995) attempted some new approaches to the presentation of bibliometric macro-level indicators. All results are based on rough bibliographic data extracted from the 1989-1993 annual cumulations of the Science Citation Index (SCI) of the Institute for Scientific Information (ISI), cleaned up and processed to indicators according to the rules ISSRU's Scientometric Indicators Data files. The country assignment criterion was the first address; that is, each paper has been assigned to the country indicated in its first author's corporate address. All papers of the *article, letter, note,* and *review* type recorded in the 1989-1993 volumes of the SCI have been considered. Only countries that have produced at least 1000 papers in all science fields in the given five-year period have been considered. This criterion was satisfied with 50 countries. USA contributed the highest number of publications 878866 (33.78%) in all fields combined from 1989-1993. The study of Braun, Glanzel and Grupp examined that USA produced highest number of publication 878866 (33.78%) and in present study China contributed the most number of publication 32770 (65.49%). Hence the study of researcher and the review taken differs in terms of most productive countries.

Fernández-Cano & Bueno (1999) studied Spanish educational research systems using scientometric tools. The shreds of evidence presented here endeavour to apprehend an illuminative background of the Spanish educational research system, applying scientometric models that mix a time perspective with quantitative tools. Based on the results obtained, a general impression is that the scientific study of the Spanish bibliometric production in the field of educational sciences has been growing in the way of concentrical circles in the first of the three clusters commented, beginning from the pro-quantified review, going towards studies about diachronic production and expanding into institutional studies. The general finding inferred from the available studies here shows a system not yet firmly established. The soft social sciences are still far from the scientometric patterns of highly consolidated disciplines and national systems. The degree of collaboration is so low that none of the studies considered goes above two authors on the average. The study of Fernandez-Cano and Bueno figured out that none of the studies considered goes above two authors on the average, whereas in the present study, 97.91 % of research output published collaboratively. Hence the study of researcher and the review taken contradicts.

Lee (2003) describes the results of a scientometric study of the Institute of Molecular and Cell Biology (IMCB). The study's purpose is to appraise the research accomplishment of IMCB in the first ten years since its establishment. Research inputs and three research outputs: publications, graduate students, and patents registered, are reviewed. The findings indicate that IMCB yielded 395 research papers, 33 book chapters, 24 conference papers, and 4 monographs, graduated 46 PhDs and 14 MScs, and filed ten patents in the ten years. The number of patents filed (ten in ten years) was smaller than expected in a field where technology is very close to science. In its quest to become world-class, IMCB researchers have remained extremely particular in where they publish 95.6% of the articles were published in ISI journals. The articles gained an average of 25 to 35 citations per article, and the rate of uncited articles is 11.6%. Four articles received more than 200 citations, and 18 received between 100 to 200 citations. The study of Lee exposed that 11.6% articles are uncited whereas the present study gives the results of 23.41% of articles are uncited. Hence the study of researcher and the review taken differs.

### 2.3 Scientific Productivity of Individual Journals

Schubert & Maczelka (1993) explored the "Scientometric" journal, and this proposition has been tested and supported by investigating the references of the research articles published in the journal in the periods 1980-81 and 1990-91, sequentially. Using the analysis of references, the proof was collected to establish the opinion that the research field of scientometrics - as exhibited in the journal of that name - has experienced a crystallisation process and migrated from the 'soft' towards the 'harder' sciences between the periods 1980-81 and 1990-91. This sign consists of the establishment of a 'standard' arrangement of the number of references per paper; an increase of Price's index by about 20%; an expansion of 'first-paged' references (a surrogate for journal references) by more than 20%; a decrease of the percentage of journals cited only once by one third.

Courtial (1994) studied the field through the problematical network constituted by scientific articles, using actor-network theory (and consequently co-word analysis) as a scientific knowledge model (regarded as a social process) growth. Scientometrics is a hybrid field made of invisible colleges and many users, thus established by scientific research and final uses. Co-word analysis gives the same weight to all articles, cited or not, and consequently estimates the interaction network within all sorts of authors. According to the previously specified network characteristics of scientific communication, the co-word analysis illustrates the field's fervent following what has been discerned and recommends determining for the future.

Wouters & Leydesdorff (1994) exhibited a combined bibliometric and social network analysis of "Scientometrics" journal at the moment of the achievement of the 25$^{th}$ volume. In more than one respect, *Scientometrics* demonstrates the qualities of a social science journal. Its Price Index amounts to 43.0 per cent and is exceptionally enduring over time. A single author has written the bulk of the published items in *Scientometrics*. The study of Wouters and Leydesdorff investigated that single author has written the bulk of published items, where as in the present study single author has written only 2.09% of publications while major portion of publications (97.91%) is collaborative in nature. Hence the study of researcher and the review taken contradicts. Furthermore, most authors collaborate with no more than one or two collaborators because the co-authorships arrangement is profoundly fragmented.

Garg & Padhi (1998) endeavoured the examination of the patents deposited and scientific papers published and extracted in the *Journal of Current Laser Abstracts* (JCLA) for the years 1967-95 registers that innovative venture in laser science and technology was at its zenith in the early 70s. Nevertheless, scientific activity transcended innovative pursuit in the early 80s. There was a constant alteration in emphasis from "applications of lasers" to "experimental laser research" and to "theoretical laser research." Additional analysis of the 1840 patents registered in 1970-71, 1975-76, and 1980-85 intimates that most of the firms registering patents were located in the USA. Consequently, the USA is the preeminent country filing patents in this area, succeeded by Japan. "Spectroscopy of laser output" followed by "Communication applications of laser" reached the maximum importance. The study of Garg and Padhi attempted quantitative analysis of patents but in the present study, analysis of patents haven't been under consideration but other document types like article, review, meeting abstract, etc have been taken for study. Hence the study of researcher and review taken differs.

Uzun (1998) studied the social sciences journal literature for the decade period 1987-1996 regarding papers beside authors, or at least one co-author providing an address of an organisation in Turkey. The number of such papers contributed approximately tripled from 1987 to 1996. It was observed that the papers are diffused into 341 journals, and roughly one-third of all research papers published in nine journals, each of which included an average of a least one Turkish paper per year. Only two of these research papers, on archaeology and anthropology, appeared to be of high citation impact. Psychology and psychiatry, connected with business and economics, are the most prolific subjects estimating for about half of the research output. A vast majority of the research papers were articles in English (96%), and an average article comprised around 24 bibliographic references. The study of Uzun explored that English has been preferred almost 96% of researchers to communicate their scholarly output, similarly in the present study English has also been chosen by 95 % of scientists to communicate their research output. Hence the study of researcher and the review correlates.

Arkhipov (1999) carried out the scientometric analysis of Nature journal and evaluated 300000 records throughout the 1869-1998 period. The combination of articles by subfields was ascertained. Supplementary sources of information obtained several journals on analytical chemistry and papers at the Pittsburg conference category during 1950-1999. The methodology adopted was based on the review of the average age of operating instruments. The transaction between scientometric data from multiple sources of information depends on the developing stage of science. Estimated and measured scientometric curves were correlated. One of the key trends in the growth of basic sciences, precisely, the expansion of articles dealing with instrumental analytical chemistry, in *Nature* were explained.

Sin & Sen (2002) highlighted the acknowledgements included in the research articles and short communications published in Journal of Natural Rubber Research (1986-1997) concerning types, frequency of appearance; individuals acknowledged, etc. Results symbolise that the usage of acknowledgement in natural rubber research information is found to be quite common, considering that 74% of communications included acknowledgements. The average acknowledgement per research communication is 2.2, which intimates the composite nature of the acknowledgements. The number of PIC acknowledgements deems for 44% of the total acknowledgements, which is more or less at par with those found in LIS journals, where PIC acknowledgements vary from 42.6% to 56.5%. Though, it is low compared to those witnessed in humanities and social science journals, where PIC acknowledgements extend from 78.1% to 95.5%. The study of Sin and Sen evaluated acknowledgements published in Journal of Natural Rubber Research, whereas in the present study acknowledgements haven’t been evaluated but in contrast references have been quantified. Hence the study of researcher and the review taken contradicts.

Schloegl & Stock (2004) investigated international and regional (i.e., German-language) publications in library and information science (LIS). This was done contracting a citation analysis and a user survey. For the citation analysis, impact factor, citing half-life, amount of references per article, and the rate of self-references of a periodical were employed as indicators. Furthermore, the preeminent LIS periodicals were outlined. For the 40 international periodicals, data were obtained from ISI's *Social Sciences Citation Index Journal Citation Reports* (JCR); the 10 German-language journals' citations were included manually (overall 1,494 source articles with 10,520 citations). The study of Schloegl and Stock exposed that ISI's Social Science Citation Index is used to acquire bibliographical data for analysis, whereas in the present study, Science Citation Index is used to obtain bibliographical data for analysis instead of Social Science Citation Index. Hence the study of researcher and review taken contradicts in choosing the database for retrieving of bibliographical data. Collectively, the citation analysis's observational base consisted of approximately 90,000 citations in 6,203 source articles published within 1997 and 2000.

Ahmad & Batcha (2019) assessed research productivity in Journal of Documentation (JDoc) for a period of 30 years between 1989 and 2018. Web of Science, a service of Clarivate Analytics, has been conferred to retrieve bibliographical details, and it has been investigated within Bibexcel and Histcite tools to deliver the datasets. The study of Ahmad and Batcha explored Bibexcel and Histcite tools for analysis of bibliographical data, likewise in the present study Bibexcel and Histcite tools have also been used to analyse the bibliographical data retrieved from web of science. Thus the study is correlating in terms of tools used for analysis of researcher and review taken. The examination part deals with local and global citation level impact, profoundly prolific authors and their research output, and ranking of notable organisations and nations. In addition to this, the scientographical mapping of bibliographical data is obtained through VOSviewer, which is open source mapping software.

## 2.4 Studies Based on the Literary Output of Country

Haiqi & Yuhua (1997) administered quantitative examination on China's research achievement, which has developed appreciably throughout the past few years, both in the comparative output of publications and their impact on international research productivity. The objective of the investigation, based on the data registered in the Science Citation Index (SCI) database between 1987 and 1993, was to study the research accomplishment in the People's Republic of China. The 35,087 papers written in domestic or foreign periodicals were chosen to investigate and assess the dissemination of publications and citations for China's research performance's numerical characterisation. The conclusions register that 17,687 papers reported by the Source Indexes of the SCI in 1990-1992 received 7944 citations in 1993 and that the mean citation rate is 0.45. The number of cited papers is 4491, and the relationship of cited papers to the sum is 0.25. The study of Haiqi and Yuhua examined that China dominates in terms of publication productivity; similarly in the present study China has contributed remarkably 65.49% of publications on coronary artery disease. Hence the study of researcher and review taken correlates in terms of research productivity of China.

Gupta et al. (2002) studied the scientific collaboration between India and Australia has been considered based on the number of Indian and Australian scientists' number of joint publications, as revealed through co-authored papers during the period 1995-1999. The study reveals the extent, mode, and direction of collaborative research between the two nations. It also has been endeavoured to crystallise and distinguish the priority S&T areas for collaborative research between the two countries. The impact of such collaborative research has been studied by analysing the impact factor of publications where this joint research is published. It has been explored that the average value of impact factor per paper for multilateral papers is far above the average impact factor of all papers. In about 38% of the India-Australian collaborated papers, the other countries were also incorporated, including the USA, the UK, Russia, New Zealand, France, Switzerland, and even Italy, Japan, and China. The study of Gupta et al. scrutinized that USA, UK, New Zealand, and France collaborated maximum number of publications, likewise the present study reveals that USA, UK, New Zealand, and France have also collaborated maximum number of publications throughout the study with BRICS nations. Hence the study of researcher and review taken correlates in terms of collaborating countries.

Kumar & Garg (2005) analysed 2058 papers published by Chinese authors and 2678 papers published by Indian authors in computer science during 1971-2000 indicates that India's output is significantly higher than the Chinese output. A major portion of the research results is published in journals. Chinese researchers favour distributing their research outcomes in national journals, while Indian researchers favour writing their research results in journals proclaimed in the west. China publishes many domestic journals in computer science, while the number of domestic journals for India was much smaller. Indian research output looks better pertinent to mainstream research associated with China. Both countries have indicated similar sub-fields. Emphasis on computational mathematics has declined during 1986-2000 as compared to 1971-1985 for both countries.

Garg, Kumar, & Lal (2006) studied the pattern of growth of agricultural research output, its existence in different sub-fields, the output of different agencies, expression pattern of Indian agricultural scientists, citation pattern of the research output, identified highly productive institutions, studied their activity profile and the impact of their research output as seen through citations, and identified prolific authors and highly cited papers. The study indicates that, like the decline in Indian scientific output during the last two decades, depicts decline in the agricultural research product during the subsequent period, *i.e.*, after 1997. The study of Garg, Kumar and Lal inquired that scientific output on agriculture of Indian researchers is declining for the last two decades, whereas in the present study, Indian scientists have contributed large amount of research publications on CAD and there is growth of publications from year 1997 onwards. Hence the study of researcher and the review taken contradicts. 'Dairy and animal sciences' followed by 'veterinary sciences' establish the largest Indian agricultural research output component. Agricultural universities and institutes under the administration of ICAR are the major contributors to the study output.

Jacobs (2006) proclaims the preliminary findings on South Africa's most productive authors, journals, and research universities. The paper employs

Scientometric procedures to evaluate the quantity and quality of scientific research papers published by researchers in numerous journals. The results explain that four of the most cited authors reproduce 40.80% of the total number. The study of Jacobs implored that four of the most cited authors from South Africa contributed 40.80% of publications, whereas in the present study the four most cited authors produced 285 articles comprising of 22.20 % of total contributions from South Africa, which is comparatively less than revealed in the study of researcher and thus differs from the review. The citations per paper for specific authors are Bilic N (16.40), Michael JP (6.36), Sacht C (6.00), and Marques HM (4.60). The preponderance of citations are detected in Chemistry (37.0%), followed by Physics (26.0%), Medicine (7.40%), and Biology (7.40%).

Armenta et al. (2007) scrutinised the evolution of spectroscopy in Morocco during 1984–2006, treating research only in the journal articles indexed in the SCI database of ISI. The most productive cities based on the total number of publications were Rabat, Marrakech, Kenitra, and Oujda. The research venture by Moroccan authors in spectroscopy is chiefly concentrated on qualitative studies of new materials' characterisation. It is concentrated in a small number of fields; physics and physical chemistry, and materials science. The author intimated that political actions need to be practised to create reference centres to encourage the research teams' activity and satisfy the lacked tools to characterise synthesised products accurately. The study of Armenta et al revealed that only "article" type of document have been retrieved from Science Citation Index and considered for study, whereas the present study involved all the document types like article, review, meeting abstract, letter, etc for study. Hence the study of researcher and the review taken differs.

Walke, Dhawan, & Gupta (2007) studied the publication output by Indian scientists in Material Science research in India during 1993-2001, with metrics on its publication size and growth rate, and reviewing its media of communication, strength, and vulnerability in the fields of research, quality of research output, nature of collaboration, and institutional productivity. The study finds that India's publications output in Materials Science has been growing steadily at about 7% per year. The study of Walke, Dhawan and Gupta exposed that India's research output in Material

Science has been growing at about 7% per year, whereas in the present study, Indian output on CAD is also persistently increasing at about 11.47 % growth rate per annum showing a little bit difference of 4.47 %. Hence the study of researcher and the review taken differs. India's publications output in Materials Science during 1993-2001 was published in 108 journals (both Indian and foreign), but a larger share of the country output (58.79%) was published in top 10 journals. The top 40 journals accounted for 90.94% of papers of the total output in Materials Science. The collaborative research in Materials Science grew faster (368.2%) than the research conducted indigenously in Materials Science (7.09%).

Fritzsche et al. (2008) analysed the European Union's 15 primary member states' contributions and chosen non-European countries to pathological research within 2000 and 2006. Pathological journals were determined utilising the ISI Web of the Knowledge database. The number of publications and relevant impact factors was confined to each country. Relevant socioeconomic indicators were recounted to the scientific output. Consequently, results were correlated to publications in 10 of the leading biomedical journals. The research output usually remained constant. In Europe, the UK, Germany, France, Italy, and Spain placed top concerning contributions to publications and impact factors in the pathological and leading general biomedical journals. Results imitate the USA's preeminent role in pathology research and confirm the significance of European scientists. The study of Fritzsche et al evaluated that USA have produced the maximum number of publications on Pathology research, whereas the present study reveals that China's contribution is outstanding and dominates on CAD research publications. Hence the study of researcher and the review contradicts.

Buylova & Osipov (2009) evaluated scientometric data on the participants, their origin by region and research centre, and the analysis of improvements and problems of Russian studies on nanotechnology. Analysis of the papers revealed that although the highest number of authors live in Moscow and St. Petersburg, the number of authors working outside the capital cities (57%) has become significant. The co-authors of papers presented at the forum include people from more than twenty countries. The papers co-authored by foreign scientists testify to an expansion of scientists' international cooperation, which is particularly important in Russia, increasing its development speed in nanotechnologies. Papers relating to the acumen of nanotechnology's ideas, methods, and achievements in biology, medicine, biomechanics, manufacturing, etc., remain outside this review and need appropriate analysis.

Geracitano, Chaves, & Monserrat (2009) manifested the analysis of scientific result and is scrutinised by scientometric systems that measure the improvement and advancement in science by investigating the productivity and impact of the scientific production in different universities, countries, research groups, etc. In this connection, the study intended to examine scientific literature in environmental studies that implement biomarkers in Latin America. The chosen period of analysis was from 1999 to 2008. Brazil was the country that exhibited the highest number of published articles (872), followed by Mexico (559), Argentina (368), and Chile (232). The h index analysis revealed that the four Latin American countries with tremendous scientific productivity displayed lower values than countries outside this region, meaning that the establishment of collaborative studies could be one of the policies to enhance Latin American distinctness in environmental studies.

Kumaran & Manoharan (2016) attempted an analysis of 2676 publications on Artificial Intelligence published by Indian scientists during 1986-2015 and indexed by Scopus online database. The year-wise distribution of research output on Artificial Intelligence reveals that the highest number of publications is 907 in 2015, followed by 467 papers in 2014 and 263 papers in 2013. From 1986 to 2009, the number of publications is less than 100. The finding of the authors' ranking based on their publications reveals that Pal S K has published 16 papers with 746 Citation Scores (h-index 11) ranked first based on the number of publications in the field of Artificial Intelligence. To conclude, the sum of citations of the Artificial Intelligence research publications and the h index scored is reasonable.

Farooq et al. (2018) outlined the contribution of researchers in the energy output of Pakistan in the years 1990–2016. A scientometric procedure was implemented to examine the scientific publications in the area utilising the Scopus Elsevier database. Diverse features of the publications were interpreted, such as publication type, influential research areas, journals, citations, authorship pattern, affiliations, and the keyword occurrence frequency. Pakistan has recognised a notable increase in research publications in the energy sector in recent years. The scientometric analysis reveals that 2139 authors have published 991 research papers from 213 institutes from 1990 to 2016. The impact factor and cites per paper correspond to 2.32 and 10, respectively. The most productive journal, authors, and institutes are Renewable and Sustainable Energy Reviews, Shahbaz M., and COMSATS, respectively.

### 2.5 Studies Based on Scientometric Analysis of Various Diseases

Chen, Chiu, & Ho (2005) evaluated the publication output connected with research on asthma in children. The data contained the period from 1991 to 2002 and were derived from the Science Citation Index online version. Selected documents included 'asthmatic children' and 'asthma children' as a part of its title, abstract, or keyword. Parameters investigated involved language, type of document, page count, publication output, country of publication, authorship, publication pattern, and the most regularly cited paper. The results show that the annual publications have progressed from 1991 to 2002. The seven industrialised countries produce high productivity in this research domain. English was the predominant language, and four or five authors were the most common number of co-author. The US was the world leader and managed most of the publications, followed by the UK. The study of Chen, Chiu and Ho revealed that 1617 (94.9%) of publications published in English, similarly in the present study, English also remains the main language for communication of research articles contributing 46660 (93.25%) publication on CAD. Hence the study of researcher and the review correlates.

Bolaños-Pizarro, Thijs, & Glänzel (2010) presented a bibliometric analysis of Spanish cardiovascular research. The research emphases on the productivity, visibility, and citation impact in an international, notably European context. This study has confirmed and deepened the results of earlier related studies. In particular, Spanish cardiovascular research showed increasing international visibility as reflected by the expanding number of publications recorded in the Web of Science database. The study of Bolanos-Pizarro, Thijs and Glazel studied that web of science has been employed for fetching data for analysis, likewise the present study also uses web of science for retrieval of bibliographical data. Hence the study of researcher correlates with the review. Spain holds a constant leading spot in the world ranking of most productive countries in the field. Strengthening collaboration has certainly contributed to increasing visibility.

Gupta et al. (2011) conducted analyses on India's research output in typhoid during 2000-2009, its growth, rank and global publications share, citation impact, the share of international collaborative papers, the contribution of major collaborative partner countries, the contribution of various subject fields, and patterns of research communication in most productive journals. Indian scientists have published 940 papers in typhoid research in 2000-2009, compared to 322 papers each by China and Brazil during the same period. The average number of citations per paper registered by India's publications in Typhoid research in 2000-20009 was 2.36. The study of Gupta et al exposed that citation per paper registered by India's publications in Typhoid research was 2.36, whereas in the present study citation per paper for Indian researchers on CAD is 24.82. Hence the study of researcher and the review taken differs in terms of CPP. India ranks at third position among the top 21 countries in typhoid research, with its global publication share of 5.61% in 2000-2009. The average number of citations per paper registered by India's publications in typhoid research in 2000-2009 was 2.36, which is lower than China (3.7) and Brazil (3.47).

Gupta & Bala (2011) studied and analyses the research output of India in asthma during the period from 1999 till 2008. It analyses the growth, rank, and global publications share, citation impact, the share of international collaborative papers, major collaborative partner countries, and contribution of various subject fields. Scopus database has been utilised to reclaim the data on publication output in asthma research. Indian scientists had published 862 papers in asthma research during 1999-2008 and registered an average citation per paper of 3.43. India ranks 15th position among the top 23 countries in asthma research and achieved h-index as 33, with its global publication share of 1.27% and international collaborative publications share of 10.09% during 1999-2008. Among India's major collaborative partners during 1999-2008, the USA contributed the largest publications share of 51.72%. The study of Gupta and Bala explored that Indian publications on asthma achieve h-index as 33, whereas in the present study Indian publications attained 112 h-index on CAD. Hence the study of researcher and the review taken differs.

Gupta, Kaur, & Kshitig (2012) analysed the dementia research output from India during 2002-11 on different parameters including the growth; global publications share citation impact, the share of international collaborative papers, the contribution of major collaborative partner countries, the contribution of various subject fields and by type of dementia, productivity and impact of most productive institutions and authors and patterns of research communication in most productive journals. Among the top 20 most prolific nations in dementia research, India ranks 16th (with 1109 papers) with a global publication share of 1.24% and an annual average publication growth rate of 25.58% during 2002-11. Its global publication part has grown over the years, growing from 0.54% in 2002 to 2.20% during 2011. Its international collaborative publications part was 24.54% during 2002-11, which declined from 28.57% during 2002-06 to 23.07% during 2007-11.

Gupta et al. (2014) investigated 1832 papers in Indian mouth cancer, as comprised in the Scopus database throughout 2003-2012, undergoing a yearly average growth rate of 14.37% and citation impact of 4.51. The world mouth cancer output (37,049 papers) evolved from several countries, of which the top 10 (United States, Japan, UK, Germany, India, China, etc.) estimates for 75.59% share of the global output during 2003- 2012. India's global publication share was 4.94% and held the seventh rank in global publication output during 2003-2012. The Indian mouth cancer output came from several organisations and authors, of which the top 15 contributed 43.39% and 21.89% share, respectively during 2003-2012. The medical colleges added the highest publications share (36.68%) to Indian publications in mouth cancer during 2003-2012, followed by hospitals (19.81%), universities (18.45%), research institutes (12.66%), institutes of national importance (11.74%, industrial units (0.49%), etc., through 2003-2012.

Brás et al. (2017) carried out the developmental dynamics of oncology research in Portugal through the second half of the twentieth century and early twenty-first century, concentrating on certain characters that can be determined from studying publication patterns and the network analysis of institutional collaborations and thematic realms. The examination has revealed that the dynamics of Portuguese oncology research was strictly linked to the developments in the policies for science and technology in the country. The collaboration networks exhibited a strong association between the laboratory and the clinic with an equitable quantity of non-clinical institutions (universities, research centres, and associate laboratories) and clinical institutions managing research and consistently operating together.

Ahmad & Batcha (2020) studied and examined 4698 Indian Coronary Artery Disease research publications, as listed in Web of Science database through 1990-2019, reserving to experience their growth rate, global share, citation impact, international collaborative papers, distribution of publications by broad subjects, productivity and citation profile of top institutions and authors, and selected media of communication.

Ahmad & Batcha (2020) explored and studied the trend of world literature on "Coronavirus Disease" in terms of the output of research publications as recorded in (SCI-E) of Web of Science during the period from 2011 to 2020. The study affirmed that 6071 research documents had been published on Coronavirus Disease. The multiple scientometric components of the research records published in the study period were analysed. The study exhibits the several features of Coronavirus Disease research publications such as year wise contribution, relative growth rate, doubling time trend, country wise production, organisation wise, language-wise, form-wise, most prolific authors and source wise.

Ahmad & Batcha (2020) examined Brazil's research production on Coronary Artery Disease as considered in indexed publications in Web of Science to know the concentration of research output, top journals for publications, most prolific authors, authorship pattern, and citations design on CAD. The conclusions designate that the highest growth rate of publications happened between the years 1995-1999. University Sao Paulo topped the scene among all institutes. The study of Ahmad and Batcha investigated that Ramires JAF contributed maximum number of publication on CAD 231 (3.72%) from Brazil and similarly in the present study Ramires JAF also contributed abundantly on CAD from Brazil. Hence the study of researcher correlates with the review. The leading publications were more than ten authored publications. Ramires JAF and Santos RD were observed to be the most productive authors. It is further depicted that most of the prolific authors (by several publications) do not emerge in highly cited publications' lists. CAD researchers often favoured employing article publications to publish their findings.

**2.6 Studies Based on Heart Disease**

Batcha & Baskaran (2007) analysed the publication activity of G8 Countries on Cardiology output of USA, UK, Japan, Italy, Germany, France, Canada, and Russia. Most of the prolific institutions are located in G8 Countries and produced 13028 records in the period from 1964 to 2006. The research based on MEDLARS database which has been published by the National Library of Medicine. The publication of journals scattering among G8 countries, first placed of Journal of American College of Cardiology produced the highest output (8%) followed by Circulation (6.5%). The study of Batcha and Baskaran exposed that journal of American College of Cardiology contributed the highest output (8%), whereas in the present study Journal of American College of Cardiology produced 201 (0.40%) of publication on CAD and ranked 28$^{th}$ in terms of total publications contributed. Hence the study of researcher and the review taken differs. The leading Institutions contributing publication in G8 countries are Massachusetts Medical School, Worcester, USA (220), University College London, Grafton, UK (196), Justus-Liebig University of Gissen, Germany (167), Institute of Clinical Physiology Pisa, Italy (138) University Hospital of Anger, France (124), University of Toronto, Canada (112), Jubetendo University School Medicine Tokyo, Japan (72) and Russian Academy of Medical Science Moscow, Russia (56).

Chuang, Huang, & Ho (2007) studied stroke-related research articles published by Taiwan researchers which were indexed in the SCI from 1991 to 2005. The study uncovered that the quantity of publications has increased at a more expeditious pace than the worldwide trend. Over the years, there has been a growth in international collaboration, mainly with researchers in the USA Article visibility, measured as the frequency of being cited, also increased during the period. The study of Chuang, Huang and Ho explored that USA remains the top collaborator for research publications, and in the same way in the present study USA is also the top country which collaborated with BRICS countries. Thus the study of researcher and the review correlates. It develops that stroke research in Taiwan has become more globally equated and has also enhanced in quality. The publication output was concentrated in several institutes, but there was a wide divergence among these institutes in the ability to conduct research autonomously.

Baskaran & Batcha (2012) studied and presented the field of Cardiology literature records retrieved from MEDLINE database for the period 1991-2010. The research shows that the maximum number of records, i.e. 829 was throughout 2000, followed by 826 in 2003 and 789 in 2002. Relative Growth Rate (RGR) and Doubling Time (Dt) observed to be an increasing and decreasing trend shown during the period of study. The study of Baskaran and Batcha implored that relative growth rate and doubling time shows an increasing and decreasing trend respectively, likewise in the present study RGR is increasing from 0.4662 in the year 1990 to 2.1721 in the year 2019 and doubling time is decreasing from 1.4864 to 0.3190. Hence the study of researcher and the review correlates. The paper explains a study of the authorship pattern and collaborative research in the field of Cardiology. The study measures the performance based on numerous parameters, country annual growth rate and collaborative index. The degree of collaboration mean score is 0.70, and the highest score is 0.88 in 1991 exhibits during the period of study.

Yu, Shao, & Duan (2013) revealed the status of the collaboration activities in Chinese Cardiology and Cardiovasology field. Articles published in 5 journals related to C&C from 2000 to 2010 were retrieved from China National Knowledge Infrastructure (CNKI) and VIP Journal Integration Platform (VJIP). Methods such as co-authorship, co-word analysis, centrality, k-core, m-slice were used in this study. Although the percentage of co-authored papers and the average number of authors per paper in Chinese C&C field were generally increasing, the geographic distribution of the research collaboration activities was extremely uneven. There were 87 authors and 5 institutions ranking in the top 1% of all the three centralities, but 92.8% of authors belonged to 10-Core and below 90.93% of authors are among 1-slice, 2-slice and 3-slice. The study found 63 cohesive research groups in the focuses of research collaboration for Coronary artery disease and myocardial infarction, etc.

Kovlen & Ponomarenko (2015) summarises the results of the recent scientometric analysis of evidence-based investigation and describes the possibilities for their assiduousness to the expansion of the strategy of physiotherapy of coronary heart disease. The intention of the research work was the scientometric analysis of evidence-based research concerning the application of physical therapeutic factors for the treatment of coronary heart disease. The authors present a detailed analysis of the clinical effects and inherent mechanisms of action of the physical therapeutic factors that find utilisation in the treatment of the patients exhibiting with coronary heart disease. Special attention is given to evidence-based research involved with the application of dosed physical exercises, health-promoting gymnastics, and instrumental methods of physiotherapy for the execution of the patients with diverse kinds of coronary heart disease (CHD).

Okhovati, Zare, & Bazrafshan (2015) endeavoured to explain the global distribution of IHD research activities by studying at the countries' burden of disease, income and development data. As a scientometric study, Scopus database was explored for research publications indexed under the medical subject heading (MeSH) 'myocardial ischemia' including the following terms: coronary artery disease, coronary heart disease, and ischemic heart disease. The study of Okhovati, Zare and Bazrafshan revealed that relevant medical subject heading (MeSH) have been employed like 'coronary artery disease', 'coronary heart disease', etc similarly the present study used various medical subject headings like 'ischemic', 'arterioscleroses' 'coronary heart disease', etc. Hence the study of researcher and the review correlates.

Liao et al. (2016) performed a comprehensive analysis of the 100 most cited articles concentrated on CHD in recent decades, which contributes insights into the features and courses in anticipating and managing CHD. Research on coronary heart disease (CHD) persists one of the major concerns in the medical and health areas in recent decades, yet data on the circumstances of CHD are unsatisfying. The investigation intended to assess the conditions and trends of the most cited articles in CHD via bibliometric procedures. The WoS database was utilised to recognise the 100 most cited articles involving CHD. General and bibliometric information was consolidated and interpreted. The total citations extended from 7829 to 1157.

Saquib et al. (2017) investigated the research productivity trends and distinguished the varieties and focus of all CVD research investigations from Saudi Arabia. Data were obtained from studies published up until December 2015 and recorded in the PubMed database. The study of Saquib et al implored that PubMed database has been chosen for obtaining data for analysis, whereas in present study web of science has been used to retrieve bibliographical data. Therefore the study of researcher and review taken contradicts. Examination acceptability standards covered: sample chosen within Saudi Arabia and CVD or a risk agent for CVD as an outcome, or subjects with CVD as study members. Bibliometric data and subject characteristics were deduced from each study; illustrations involve authorship (number, gender, affiliation), journal, publication year, study location, research design, sample size, sample type (general or patient), sample structure (male or female), and sampling procedure (random or non-random).

Batcha (2018) examined the data on cardiovascular disease, which amounts to about 24.8% of deaths in the SAARC nations. The research explores the research trend, authorship, collaborative pattern and activity index of five SAARC countries. The outcomes of the research demonstrate that India is a preeminent country amongst SAARC nations with significant research output followed by Pakistan in cardiovascular disease research. The international collaboration results that USA, England and Australia are the top collaboration countries for SAARC nations. The study of Batcha exposed that USA, UK and Australia are the top collaborating countries for SAARC nations, in the same way in the present study results depicts that USA, UK and Australia also is the top collaborating country with BRICS nations. Hence the study of researcher and the review correlates. India is competing with other developed countries and shows higher activity within the context of their individual productivity.

Ullah et al. (2019) quantitatively studied and applied bibliometric methods to analyse original articles, authorship pattern, citations, contributions from different regions and other relevant parameters of Pakistan Heart Journal covering the period from January 2005 to December 2018. The data was collected from Postgraduate Medical Institute (PGMI) Library Hayatabad Peshawar and official website of PHJ. The study acknowledges that the number of articles published in issues of the journal per year extended from 09 to 57. Variation was found in the number of references cited in each article (40.05%) (153) articles had 11-20 references. Article's length was analysed, and it was reported that a majority (30.22%) of articles contained five pages. The study of Ullah et al, investigated that a majority (30.22%) of articles contained five pages, whereas in the present study majority of research articles used greater than 10 pages (45.07%). Hence the study of researcher and the review taken differs.

**2.7 Studies Related to New Trends and Technology**

Garg & Sharma (1991) analysed the output of the literature scanned in Engineering Index during 1970-84 on solar power research indicates that the literature's growth had been strenuous after the energy crisis in 1973 till 1982. The research in solar power appears to be deferred till 1973 when the energy crisis took place. Though, after 1982, the number of publications has started decreasing, implying that the study's urgency has dwindled. New frontiers in solar power research, like solar power plants, emerged after the energy crisis. The USA is the major producer of scientific output in this field, and the distribution of output follows the world trend in basic sciences. The present study of Garg and Sharma examines that USA is the major contributor on solar power publications, while in the present study; China (65.49%) has contributed remarkable publications on CAD. Hence the study of researcher and review taken differs.

Jain & Garg (1992) examined 785 papers, books, and reports in laser, published from India during 1967-84, intimates that Indian output constitutes approximately 1% of the global output. The total output evolved from 77 educational and research organisations, out of which ten organisations shared around 23%. A significant share of these publications emerged in foreign journals of repute, as reflected by their impact factors. Emphasis has remained on the theoretical aspects of laser research. The laser research conducted in India seems to be a component of mainstream science, as confirmed by the pattern of papers and citations. The investigation also registers that Indian scientists have few global collaborations in this domain. The study of Jain and Garg exposed that Indian scientists have few global collaborations on laser publications but the present study reveals that India collaborated 1424 (30.26%) publications with other countries. Hence the study of researcher and review taken contradicts.

Coursaris & Van Osch (2014) demonstrated the conclusions of a scientometric examination of the corresponding literature to explain the immediately expanding social media research field. The study carried a research productivity study and citation analysis of individuals, institutions, and countries based on 610 peer-reviewed social media articles published in journals and conference proceedings between October 2004 and December 2011. Conclusions show that research productivity is splitting and that numerous leading authors, institutions, countries, and a minute set of foundational papers have appeared. Based on the results-indicating that the social media area represents the restricted variety and is still heavily inspired by practitioners, the paper suggests two primary challenges facing the social media domain and its future advancement: the lack of scholarly maturity and the Matthew Effect.

Karpagam (2014) evaluated nanobiotechnology research output through an efficient scientometric examination based on the Scopus database from a distinct aspect for 2003–2012. The existing study outlines nanobiotechnology research output during 2003–2012 on diverse parameters, including the growth, global papers share and citation impact, the share of international collaborative papers, and major collaborative partner countries' contributions. During the course of ten years 114,684 research papers were published and received 2,503,795 total citations with average citation of 21.83. The study of Karpagam investigated that nanotechnology publications from India received 21.83 average citations per paper, whereas in the present study, Indian scientists received 15.16 citations per paper. Hence the study of researcher and review taken differs. It has been recognised that during 2003–2012, the USA secured the first position by the number of research publications (34,736), h-index (349), g-index (541), hg-index (434.52), and p-index (326.47).

Santha Kumar & Kaliyaperumal (2015) focused and analysed the number of contributions performed by the researchers in mobile technology published on the Web of Science database during 2000–2013. The analysis revealed that 10,638 publications were published in the area of mobile technology. The single most prevailing form of communication is the Journal articles, in which 79.66 % of the total literature is published. This determines that mobile technology researcher's favoured medium of communication is journal articles. The study undertaken by Santha Kumar and Kaliyaperumal revealed that mobile technology researcher's favoured medium of communication is journal articles, similarly in the present study researchers preferred English (79.689%) as the medium of communicating their research publications. Hence the study of researcher and the review taken correlates. The majority of research publications written were observed in the English language. The author's affiliations prove that countries like the USA, UK, China, and Korea are actively engaged in research in the field.

Fu, Niu, & Yeh (2016) employed an automatic content analysis approach from scientometrics to distinguish the trend of researches on a sustainable operation. The data originated from the Web of Science during 1988–2015. Precisely, a multi-stage clustering procedure based on bibliographic coupling has also been included to explore which themes, what the research trend, and which new ideas contribute to sustainable operation's scientific journal fields. The results recognised that energy-related journals were the classic publications in the sustainable operation area, and energy technology held the top topic. USA, UK, and Germany given the most journal articles in this field. Including the accelerated augmentation of Asia, Asian scientists, like South Korea and Singapore, also published numerous sustainable operation papers.

Ahmad, Batcha, & Jahina (2019) quantitatively estimated the research productivity in artificial intelligence at a global level covering the study period of ten years (2008-2017). The investigation recognised the trends and features of growth and collaboration pattern of artificial intelligence research output. The average growth rate of artificial intelligence per year progresses at a rate of 0.862. The study's multi-authorship pattern is high, and the average number of authors per paper is 3.31. The year 2014 is observed to be having the highest Collaborative Index with 3.50. Mean CI during the period of investigation is 3.24. This is also approved by the mean degree of collaboration at a percentage of 0.83. The mean CC perceived is 0.4635. Regarding the applicability of Lotka's Law of authorship productivity in artificial intelligence research, it confirmed to be a fit for the study. The present study of Ahmad, Batch and Jahina employed that Lotka's law fits for the publications on artificial intelligence; similarly in the present study Lotka's law is confirmed for the CAD publications. Hence the study of researcher and the review taken correlates.

### 2.8 Summary of Reviews

The present study summarises 58 reviews about scientometrics. Among them six focuses on the studies based on the works of individual scientists, four focuses on scientific productivity on individual institutions, nine focuses on scientific productivity of individual journals, twelve focuses on studies based on the literary output of countries, ten focuses on studies based on scientometric analysis on various diseases, ten focuses on studies based on heart disease, and seven studies related to new trends and technology. In summary, this section provides the various kinds of scientometric techniques that are directly or indirectly related to the research work taken here. It has discussed the important literature about the present study.

### 2.9 Inferences from Reviews

The researcher found very few studies on coronary artery disease, and no study has been conducted to measure the research performance of BRICS scientists on CAD. The review was carried out to identify various techniques the researchers adopt towards achieving the objectives. The review has provided a solid foundation for laying out the objectives. The section is also revisited while finalising the analysis section to include some of the recent work. The reviews' analyses further reflect that the applications of statistical techniques and tools are using varieties of formulas and equations that facilitate future research to test. There is a research gap found in the field of coronary artery disease literature, and that too from the BRICS countries, which needs a quantitative analysis for measuring the research performance of BRICS scientists. Hence the present study will bridge a gap and will result for futuristic analysis.

# CHAPTER III

# RESEARCH DESIGN

Once the researcher has resolved the particular question to be answered and has operationalised the variables and research question into a transparent, proscribed hypothesis, it is time to contemplate a fitting research design. The investigation has employed a descriptive and analytical research design for directing the research. In this chapter, an exploration about methodology, Data collection and limitations, the scientometric indicators and statistical tools, bibliometric laws, various indices, and mapping tools have been summarised.

## 3.1 Methodology

For the present study, the publication data was retrieved and downloaded from the Web of Science (http://apps.webofknowledge.com) from the Web of Science core collection database on coronary artery disease research during 1990-2019. The advanced search strategy for BRICS countries output was formulated; the search string used for data extraction was:

"TS=(Artery Disease, Coronary OR Artery Diseases, Coronary OR Coronary Artery Diseases OR Disease, Coronary Artery OR Diseases, Coronary Artery OR Coronary Arteriosclerosis OR Arterioscleroses, Coronary OR Coronary Arterioscleroses OR Atherosclerosis, Coronary OR Atheroscleroses, Coronary OR Coronary Atheroscleroses OR Coronary Atherosclerosis OR Arteriosclerosis, Coronary OR Ischaemic OR Ischemic OR hardening of the Arteries OR Induration of the Arteries OR Arterial Sclerosis ) AND CU=(Brazil OR Brasil OR Federative Republic of Brazil OR Russia OR USSR OR Russian Federation OR Union of Soviet Socialist Republics OR India OR China OR People's Republic of China OR South Africa OR Republic of South Africa)."

Further, the search was done in "all languages" and "all document types" tags, and then, this search has been refined to limit the period from 1990 to 2019 within the "Timespan" tag. Data filtering has been performed manually to remove irrelevant and duplicate record entries. And finally, the search strategy generated for 50036 publications on coronary artery disease from the Web of Science database. The detailed analysis was carried out using Bibexcel, Histcite, and Bibliometrix Package in RStudio tools to get the required number of tabular data as per the study's objectives. The data was analysed by subject, collaborating countries, author-wise, organization-wise, and journal-wise. Further, mapping tools such as VOSviewer and Pajek were used to study the collaboration behaviour and citation network.

### 3.2 Data Collection and Limitations

Several sources contribute to the research output in the field of Coronary Artery Disease through BRICS countries scientists. For the present study, the secondary sources are taken for analysis. The required data were retrieved from the Web of Science. Web of Science is a compilation of databases that record the world's leading scholarly research in the sciences, social sciences, arts, and humanities, as published in journals, conference proceedings, symposia, seminars, colloquia, workshops, and convocations over the globe. The Web of Science (WoS) abstract database is one of the world's most extensive resources for citation, indexing, and citation analysis of a wide variety of scientific works in all possible scientific fields. This database, created by Thomson Reuters, now owned by Clarivate Analytics, regularly indexes thousands of various scientific journals and periodicals, which is why many experts and researchers prefer it to prepare new materials or to improve their qualifications.

The study used a search string in the advanced search field selected period from 1990 to 2019 and chooses the Science Citation Index Expanded (SCI-EXPANDED) from the Web of Science Core Collection database for the present study. The data was downloaded on 7th March 2020. Applying the search string, a total of 50036 records were downloaded and analysed by Bibexcel, Histcite, Bibliometrix Package in RStudio, and Microsoft Office 2019 as per the objectives of the study. VOSviewer and Pajek application tools were used for mapping. This research selected and downloaded only thirty years (1990-2019) on Coronary Artery Disease research records in Web of Science database of BRICS countries scientists' publications.

In this study, the researcher performed a scientometric analysis on Coronary Artery Disease research of the BRICS countries using data from the Web of Science indexing and abstracting database for 1990-2019. The data depicted in Scopus,

PubMed and Google Scholar are not taken for present study due to passivity of time and other non-standardised factors. The databases such as Scopus, PubMed, or Google Scholar may produce a different set of publication records using similar search criteria. Although this comparison was beyond this study's scope, future work may attempt to verify this study's findings using data from these alternate sources. The bibliographical details of publications covered in Web of Science from 1990 to 2019 alone taken for analysis.

### 3.3 Scientometric Indicators and Statistical Tools

In the study, the following Scientometric/Bibliometric indicators and statistical tools were applied while analysing the data on Coronary Artery Disease research output, which has been retrieved from the Web of Science database.

- Exponential Growth Rate
- Relative Growth Rate & Doubling Time
- Activity Index
- The ratio of Growth Rate
- Degree of Collaboration
- Collaborative Index
- Collaborative Co-efficient
- Modified Collaborative Co-efficient
- Co-authorship Index
- Citation per Paper
- Lotka's Law of Authorship Productivity
- K-S Test
- Price Square Root Law
- Bradford's Law of Scattering
- Zipf's Law of Word Occurrence
- H-index, e-index, p-index, m-index, a-index, R-index, AR-index, $h_{nom}$ index, and $Q^2$-index

#### 3.3.1 Exponential Growth Rate

The Exponential Growth Rate computes the pace of populace development and the distinctions to figure exponential growth rate. There are two sorts of the development rate: exponential growth rate and direct development rate. The exponential growth rate gives the populace's relative growth rate as it relies on the present populace. Then again, the direct growth rate does not rely on the present development rate, and subsequently, the exponential development rate is achievable. The Exponential growth rate can be utilised to foresee the future populace of any creature. It is utilised all-inclusive to anticipate human populace. With the occasional rate information, i.e., the number of years through which the development rate is to be determined for the first populace, figuring the exponential growth rate should be possible effortlessly. The equation for figuring exponential growth is given as:

$$N_{(t)} = N_{(0)}e^{rt}$$

Where,

N (t) is the population when the time elapsed is "t" years

N (0) is the initial population

r - Growth rate

t - Number of years

e - Natural base of logarithms whose value is 2.711828

### 3.3.2 Relative Growth Rate

Relative Growth Rate (RGR) means the increase in the number of articles per unit of time. The mean RGR of articles over the specific period of the interval is mathematically given by:

Rt (P) = [logP (t)-logP (0)]

Rt = Relative growth rate of articles over a specific period of time.

LogP (0) = Logarithm of initial number of articles logP (t)

= Logarithm of the final number of articles.

### 3.3.3 Doubling Time

Doubling time is determined as the time expected for the articles to become double the existing amount. It has been measured employing the following formula:

Dt is given by (t) =0.693/R

Where R is the relative growth rate of articles

Dt = It is directly related to RGR.

### 3.3.4 Activity Index

Activity index characterises the relative research effort of a country into a given field, and it is explained as:

"Activity Index suggested by (Price, 1981) and elaborated by (Karki & Garg, 1997) has been used to measure the relative research effort of a country in a given field".

Mathematically:

$$AI = \left[ \frac{\left( \frac{C_i}{C_o} \right)}{\left( \frac{W_i}{W_o} \right)} \right] X100$$

Whereas,

$C_i$ = individual Country output in the year i

Co = Total of Individual Country output

Wi = World output in the year i

Wo = Total output

### 3.3.5 Co - Authorship Pattern

To study the shift in the co-authorship pattern during 1990-2019, CAI suggested by (Garg & Padhi, 2001) was used.

CAI is computed as follows:

$$CAI = \left[ \frac{\left( \frac{N_{ij}}{N_{io}} \right)}{\left( \frac{N_{oj}}{N_{oo}} \right)} \right] \times 100$$

Where $N_{ij}$: number of papers having j authors in year/block;

$N_{io}$ : total output of year block i;

$N_{oj}$ : number of papers having j authors for all years/blocks;

$N_{oo}$ : total number of papers for all authors and all years/blocks.

J = 2, (3 or 4), >= 5

### 3.3.6 Degree of Collaboration

Subramanyam propounded the DC, a measure to calculate the proportion of single and multi-author papers and interpret it as a degree. According to (Subramanyam, 1983):

$$DC = \frac{Nm}{Ns+Nm} \qquad \frac{\text{No of Muti−authored papers}}{\text{No of Single+No of Multi−authored Papers}}$$

Where,

C = Degree of collaboration in a discipline

$N_m$ = Number of multi-authored papers

$N_s$ = Number of single-authored papers

Applying the formula, the degree of collaboration in the Coronary Artery Disease research amongst BRICS nations is 0.98 throughout the investigation period.

### 3.3.7 Collaborative Index

(Lawani, 1986) proposed and coined the term Collaborative Index to describe the average number of authors per paper for a given set of papers and used it as a quantitative measure of research collaboration. It can be calculated easily, but it cannot be interpreted as a degree because it has no upper- value limit. The formula denotes it:

$$CI = \frac{\text{Total Number of Authors}}{\text{Total Number of Papers}}$$

Where,

CI = the number of authors per paper

### 3.3.8 Collaborative Coefficient

(Ajiferuke, Burell, & Tague, 1988) recommended a different standard to measure collaborative research and termed it as collaborative co-efficient. The method is based on fractional productivity established by Price and Beaver. The following formula expresses CC. The symbols employed have been described as under:

$$CC = 1 - \frac{\sum_j^k (1/j) f_j}{N}$$

Where $f_j$ is the number of j authored papers, N is the total number of research papers published, and k is the highest number of authors per paper according to Ajiferuke, CC conduces to zero as single-authored papers dominate and to 1-1/j as j-authored papers dominate. This indicates that the greater the value of CC, the larger the probability of papers with multi or mega authors.

### 3.3.9 Modified Collaborative Coefficient (MCC)

It is tenderly changed that the new measure is nearly equivalent to that of CC, as given in (Ajiferuke et al., 1988). Think about that each paper takes with it a solitary "credit," and this acknowledgement is being shared for the worked together authors. In this manner, if a paper has solo authors, the author gets one credit; so also with two authors, each author gets 1/2 credits and, when all is said in done, if a distribution has X authors, each gets 1=X credits (it was equivalent to that of the possibility of partial efficiency characterised by Price and Beaver as the score of authors when he is doled out 1=n of a unit for one thing for which n authors have been credited.) Henceforth, the standard credit granted to each author of a random paper is E [1=X], a worth lying somewhere in the range of 0 and 1. Since worth 0 is comparing to single authorship, it very well may be characterised as the Modified Collective Coefficient (MCC).
(Savanur & Srikanth, 2010) modified the CC and derived MCC as follows:

$$\text{MCC} = \frac{A}{A-1}\left\{1 - \frac{\sum_{jn1}^{A}(1/j)f_j}{N}\right\}$$

Whereas

A = Total number of papers of the specific year

N = All total number of authors in the collection

J = the collaboration of a number of authors like two, three, four, etc.

fj = every one of the authors in the collaboration

### 3.3.10 Citation per Paper

Given the distribution output and the number of citations gotten by these papers, citation per paper (CPP) for a year or various nations and various organisations has been determined. Citation per paper has been determined by utilising the accompanying formula:

$$\text{CPP} = \frac{\text{Total number of citations for a year /Country}}{\text{Total number of cited publications of that year /country}} \times 100$$

## 3.4 Bibliometric Laws

### 3.4.1 Bradford's Law of Scattering

Bradford's law was formulated in 1934 by Samuel C. Bradford to scrutinise the dissemination of scientific literature (Bradford, 1934). His work was developed in the area of geophysics between 1931 and 1933, during which time he deduced all the articles he could find relevant to this area. Upon examining the journals in which these articles were published, he found consistency, specifically an inverse relationship between the number of articles published in a subject area and the number of journals in which the articles appear. This means that, in a given subject area, a small number of journals account for a sizeable portion of the total publications in that area. In contrast, progressing numbers of journals publish fewer articles in that area.

The foundation for ascertaining these zones is that the number of articles in each zone must be the same. However, the number of journals distributing these articles will not be the same in each zone, as some journals will be more productive than others.

Given that Bradford's law ranks journals according to their productivity, a small group of articles will be located in the first central zone. In contrast, an increasing number of journals will be found in each subsequent zone. The first group comprises the *core* journals and will contain a given number of articles. While the number of articles will remain constant in all zones, the number of journals will increase across the zones. The ratio between the number of journals in subsequent zones has been observed to be approximately $1{:}n{:}n^2$

### 3.4.2 Lotka's Law of Author Productivity

(Lotka, 1926) is a conventional procedure used to test the consistency in the publication activity of authors of scientific research. It portrays the repeat of preparations by authors in a given field. It communicates that the amount of contributors making n contributions is around $1/n^2$ of those creation ones. The degree of all supporters that make a single contribution is in the zone of 60 per cent. This suggests out of the significant number of contributors in a given field, 60 per cent will have just a single appropriation; 15 per cent will have two preparations ($1/2^2$ times 60); 7 per cent will have three publications ($1/3^2$ times 60), hence on10-13. This law can be conveyed as:

$$Y = C \times N^2$$

where x is the number of publications of attention (1, 2, etc.); n is a sort that is constant for a given plan of data; y is the expected percentage of scholars with repeat x of productivity, and C is an unfaltering. The effectiveness relates not to the number of articles disseminated by an author yet to its logarithm; a multiplicative, instead of just included substance, model gives a better fit than this measure or counting system. The sort n is normally fixed at 2, in which case the law is known as the inverse square law of sensible productivity. In any case, given that the model n predicts the general number of authors at each benefit level, it would seem, by all accounts, to be useful to figure it. In the present examination, the least-square procedure has been used. It might be conveyed as:

$$n = \frac{N\sum XY - \sum X \sum Y}{N\sum X^2 - \left(\sum X\right)^2}$$

Where N is the number of data pairs considered

X is the logarithm of x (x=number of articles) and

Y is the logarithm of y (y=number of authors)

The constant C is calculated using the formula:

$$c = \frac{1}{[\sum_{1}^{p-1} \frac{1}{x^n} + \frac{1}{(n-1)(p^{n-1})} + \frac{1}{2p^n} + \frac{n}{24(p-1)^{n+1}}]}$$

**3.4.3 Zipf's Law**

(Zipf, 1949) made and expanded a careful law, as observed by directing an association between the rank of a word and the repeat of its appearance in a long message. Zipf proposed in his book Human Behaviour and the standard of the Least Effort from 1949 empirical law on word frequencies in ordinary language talk and texts. Zipf's was subtleties that, while only several words are used repeatedly, various or by and large are used seldom.

Zipf's law can be calculated as follows:-

Rf=c

Where,

'r' is the rank of a word,

'f' is its frequency and

'c' is the constant.

In the long literary issue, if words are masterminded in their decreasing request of frequency, the position of some random expression of the content will be conversely corresponding to the recurrence of the word's event.

### 3.5 Price Square Root Law

This law expresses that "half of the logical articles are given by the square root of the total number of logical writers. As it were, the wellsprings of $N^{1/2}$ produce a small amount of A of Articles. This law is generally called "Rousseau's Law" from Jean Jacques Rousseau; he unmistakably referenced something very similar in his "social contract" about the size of the first class, that is, the individuals who contribute to the administration.

### 3.6 Various Indices

Standard bibliometric indicators, for example, the number of productions (P) during the analysing time frame, number of citations (C) during the study time frame, and the average citation per paper (CPP) have various boundaries. The h-index should quantify the expansive effect of an individual researcher and stay away from every one of the impediments. Additionally, the online database, for example, Web of Science, Scopus, and Google Scholar, gives the h index. Different records, for example, the a-index and m-index, portray the effect of the papers in the centre.

The h-index, otherwise called the Hirsch file, was presented (Hirsch, 2005) as an indicator for lifetime accomplishment. Thinking about a researcher's list of productions, positioned by the number of citations gotten, the h-index is the most crucial position. The end goal is that the primary h distributions got each in any event h reference.

The g-index (Egghe, 2006) is an h-index record for measuring the scientists' profitability of physicists and different researchers dependent on their production records. Egghe's g-index is relatively unique concerning both h and h2 in 78. It changes consideration from the quantity of most productive papers to the genuine number of references pulled in by these most beneficial papers.

A-index (Jin, 2006) accomplishes a similar objective as the g-index to explicitly remedy how the first h-index does not take the careful number of citations of articles incorporated into the h-core into account. This index is basically characterised as the average number of citations received by the Hirsch core distributions. This index's name is determined from the way that it is only an average (A).

$$A = \frac{1}{h}\sum_{j=1}^{h} cit_j$$

Since researchers do not distribute a similar number of articles (Sidiropoulos, Katsaros, & Manolopoulos, 2015), the first h-index is undoubtedly not a reasonable enough measurement. In this way, they characterised the Normalised h-index (hnom).

$$h^M = \frac{h}{N_F}$$

R-index (Jin, Liang, Rousseau, & Egghe, 2007) is determined as R= Axh. When all is said in done one way, compose R (X, Y), where X indicates a specific researcher and Y the year for which the R-index has been determined. As this is of no significance in our examinations, we discard the symbols X and Y. Obviously, h = R as each *citj* is at any rate equivalent to h. In the exceptional situation where each *citj* is equivalent to h, R= h. This outcome is another preferred position of utilising the entirety's square base, not merely the sum.

$$R = \sqrt{\sum_{j=1}^{h} cit_j}$$

(Jin et al., 2007) proposed a subordinate age indicator: The AR-index is characterised as though aj signifies the period of article j; we represent the age-dependent R-index, indicated by AR, by the following equation. If there are a few distributions with precisely h citations, then we incorporate the latest ones in the h-core were incorporated.

$$AR = \sqrt{\sum_{j=1}^{h}} \frac{cit_j}{a_j}$$

The e-index is a fundamental h-index supplement, particularly for assessing exceptionally referred to researchers or for exactly looking at the scientific output of a gathering of researchers having an indistinguishable h-index. The e-index is characterised as the square root of the excess references over those utilised for ascertaining the h-index (Zhang, 2009). That is, e2 = S(h) - h2, where S(h) is the complete citations got by the h papers for a scientist if their h-index is h.

$$e = \sqrt{\sum_{j=1}^{h} cit_j} - h^2$$

(Alonso, Cabrerizo, Herrera-Viedma, & Herrera, 2010) displayed another index, called hg-index, which depends on both h-index and g-index that attempt to balance the advantages of interest of the two measures to limit the disadvantages. Hg-index depends on a blend of h-index and g-index. The hg-index (Alonso et al. 2010) was proposed as the geometric mean of the h-index and the g-index. An analyst's hg-index is computed as the geometric mean of his h and g-index, which is hg $= \sqrt{h\ xg}$ .

$$hg = \sqrt{h\ xg}$$

(Prathap, 2011) proposed a file called p-index (a composite execution index that can adequately join the size and nature of scientific papers) can be expanded out for scientometrics research into appraisal in situations where numerous creations are considered. The p-index strikes the best balance between movement (all-out citations C) and greatness (mean citation rate C/P). The p-index gives the best balance between quality (C/P) and quantity (C).

$$P = h_m = \left(\frac{C^2}{P}\right)^{1/3}$$

**3.7 Bibexcel Tool**

Bibexcel is a handy bibliometric toolbox developed by Olle Persson. In Bibexcel, it is expedient to do most bibliometric analysis types, and Bibexcel enables easy interplay with other software, e.g., Pajek, Excel, SPSS, etc. The application allows the user a high degree of versatility in both data superintendence and interpretation, and this adaptability is one of the program's actual strengths. For example, it is possible to practice other data references than Web of Science, and Bibexcel can administer with data other than bibliographic records. If the user simply receives the necessary file constructions that Bibexcel needs, it is possible to carry many different data types. However, affability has its value, and the flexibility may initially cause new users to observe it as stimulating to use. This product is estimated to help a client investigate bibliographic knowledge or any printed nature information designed along these lines. This toolbox consolidates numerous devices, some outstanding in the window and others take cover behind the menu. Vast numbers of the appliances can be appropriated in a mix to perform the excellent results. Bibexcel authorises presenting a few maps employing multi-dimensional scaling methods. A guide is made by first comprehending the occasion sets of units, such as originators, co-occur in the report record. At that point, coming about co-occurrence framework is exerted as an offering to a multi-dimensional scaling program that finds the best fitting two-dimensional depiction of the info esteems. The division between the guide units is conversely applicable to the number of co-occurrences, which indicates that the more two units co-occur, the closer they will be established on the guide. The co-citation maps and coordinated effort among creators of various foundations in examining performance can be addressed by the utilisation of this product program (Persson, Danell, & Schneider, 2009)

### 3.8 Mapping Tools

#### 3.8.1 VOSviewer

VOSviewer is a program that has been developed for forming and comprehending bibliometric maps. The program is easily obtainable to the bibliometric research community. VOSviewer package, for example, is used to create graphs of authors or journals based on co-citation data or to create maps of keywords based on co-occurrence data. The program allows a viewer that provides bibliometric maps to be explored in adequate detail. VOSviewer can represent a map in numerous styles, each highlighting a different perspective of the map. It has functionality for zooming, scrolling, and examining, promoting the comprehensive analysis of a map. The viewing capacities of VOSviewer are profitable for maps comprising at least a reasonably large number of objects (e.g., at least 100 items). Most computer applications that are practised for bibliometric mapping do not perform such maps in a thoughtful way (Van Eck & Waltman, 2010).

#### 3.8.2 Pajek

Pajek is a generic, more than 20 years old, Microsoft Windows-based interface visualisation tool, originally executed for social network investigation, yet a compelling application for analysis and visualisation of extensive networks. Pajek can readily envision a million nodes with billion attachments in an average computer by outperforming any other convenient tool in the field. Pajek’s user interface is simple, easy to get familiar with, and very receptive to the analysis of massive networks. It was never expected to be the most advanced visualise. However, it offers enormous graph analysis methodologies, delivering it a great applicant for analysis of massive networks and a great correlative to the current tools (Pavlopoulos, Paez-Espino, Kyrpides, & Iliopoulos, 2017).

### 3.8.3 MS Excel

MS Excel is a generally utilised Microsoft Office application. It is a spreadsheet application that is practised to save and investigate statistical data. It emphasises computation, graphing devices, pivot tables, and a macro programming language called Visual Basic for Applications.

### 3.8.4 RStudio

RStudio is a combined advancement environment for R, a programming language for arithmetical computing and graphics. It is prepared in two formats: RStudio Desktop is a conventional desktop application, while RStudio Server operates on a remote server and provides obtaining RStudio working a web browser.

# CHAPTER IV

# ANALYSIS AND INTERPRETATION OF DATA

The focus of this study is to access the research output of BRICS countries through scientometric analysis for the framed hypothesis with bibliographical data taken from the Web of Science database, a comprehensive and in-depth database containing almost all subjects of science, social science, arts, and technology. Its coverage in the medical science field is quite comprehensive and well-acknowledged as it contains one of the databases known as the Science Citation Index (SCI). The database was searched for collecting documents pertaining to the areas related to Coronary Artery Disease published between 1990 and 2019 pertaining to BRICS countries, i.e., Brazil, Russia, India, China, and South Africa in various parameters have been downloaded.

In Chapter I, an overview of Coronary Artery Disease has been presented, projecting the subject's dimensions. It has been noticed that the literature on Coronary Artery Disease is being published in multi-channels of communication, and the same is covered in secondary sources. In this chapter, the published literature on Coronary Artery Disease has been analyzed quantitatively using various scientometric indicators and statistical techniques.

The purpose of the study is based on the scientific literature productivity on Coronary Artery Disease in BRICS countries reflects on observing the enactment at complete and narrow perception. The analytical part of the thesis deals with the source database of Web of Science, i.e., Science Citation Index (SCI). The applied analytical tools are Exponential Growth Rate, Relative Growth Rate (RGR), and Doubling Time for research output in Coronary Artery Disease, Authorship Productivity, Authorship pattern, Collaborative Index, Lotka's Law, Bradford's Law of Scattering, Zipf's Law, Price Square Law, Pareto Principle (80 X 20). Besides, some other investigation has also been carried out to identify the research output on Coronary Artery Disease in BRICS countries.

This chapter presents the analysis and interpretation of data collected for the identified period from the Web of Science database from 1990 to 2019. The investigation was done using scientometric techniques for further analysis of the research output on Coronary Artery Disease output in BRICS countries. Total records were 50036, obtained from Web of Science for 30 years' time, which covers the period from 1990 to 2019.

### 4.1 Analysis of Research Literature Growth Study

**Table 1: Year Wise Distribution of Publications**

| S.No. | Year | Publication | Cum. Publication | Percentage of Publication | Percentage of Cum. | Exponential Growth Rate |
|---|---|---|---|---|---|---|
| 1 | 1990 | 158 | 158 | 0.32 | 0.32 | -- |
| 2 | 1991 | 266 | 424 | 0.53 | 0.85 | 0.017515 |
| 3 | 1992 | 234 | 658 | 0.47 | 1.32 | -0.004263 |
| 4 | 1993 | 182 | 840 | 0.36 | 1.68 | -0.008342 |
| 5 | 1994 | 196 | 1036 | 0.39 | 2.07 | 0.002473 |
| 6 | 1995 | 197 | 1233 | 0.39 | 2.46 | 0.000170 |
| 7 | 1996 | 291 | 1524 | 0.58 | 3.05 | 0.013089 |
| 8 | 1997 | 297 | 1821 | 0.59 | 3.64 | 0.000681 |
| 9 | 1998 | 311 | 2132 | 0.62 | 4.26 | 0.001537 |
| 10 | 1999 | 343 | 2475 | 0.69 | 4.95 | 0.003270 |
| 11 | 2000 | 399 | 2874 | 0.80 | 5.74 | 0.005054 |
| 12 | 2001 | 360 | 3234 | 0.72 | 6.46 | -0.003423 |
| 13 | 2002 | 490 | 3724 | 0.98 | 7.44 | 0.010330 |
| 14 | 2003 | 559 | 4283 | 1.12 | 8.56 | 0.004401 |
| 15 | 2004 | 737 | 5020 | 1.47 | 10.03 | 0.009257 |
| 16 | 2005 | 794 | 5814 | 1.59 | 11.62 | 0.002486 |
| 17 | 2006 | 994 | 6808 | 1.99 | 13.61 | 0.007517 |
| 18 | 2007 | 1224 | 8032 | 2.45 | 16.05 | 0.006962 |
| 19 | 2008 | 1486 | 9518 | 2.97 | 19.02 | 0.006486 |
| 20 | 2009 | 1799 | 11317 | 3.60 | 22.62 | 0.006392 |
| 21 | 2010 | 2149 | 13466 | 4.29 | 26.91 | 0.005943 |
| 22 | 2011 | 2400 | 15866 | 4.80 | 31.71 | 0.003689 |
| 23 | 2012 | 2727 | 18593 | 5.45 | 37.16 | 0.004267 |
| 24 | 2013 | 3336 | 21929 | 6.67 | 43.83 | 0.006742 |
| 25 | 2014 | 3685 | 25614 | 7.36 | 51.19 | 0.003322 |
| 26 | 2015 | 4130 | 29744 | 8.25 | 59.45 | 0.003807 |
| 27 | 2016 | 4577 | 34321 | 9.15 | 68.59 | 0.003431 |
| 28 | 2017 | 4822 | 39143 | 9.64 | 78.23 | 0.001740 |
| 29 | 2018 | 5192 | 44335 | 10.38 | 88.61 | 0.002467 |
| 30 | 2019 | 5701 | 50036 | 11.39 | 100.00 | 0.003122 |
| Total | | 50036 | | | | |

To achieve the first objective of the growth of literature in Coronary Artery Disease research output during the years 1990-2019, the analysis of growth of literature is carried out and it is explained in tables 1, 1A, and 2.

The researcher has chosen the data for analysis from 1990 to 2019 (three decades) periods. The research output cumulated to 50036 records downloaded from the Web of Science database to analyze the subject of coronary artery disease research productivity in BRICS countries. The table value reveals that the year wise growth trend is gradually increasing. It consists of 50036 records, of which total publications less than 50% were published between 1990 to 2013. The literature output on coronary artery disease in BRICS takes a big heap in 2000 and above publications from 2010. Overall publications, the output is steadily increasing.

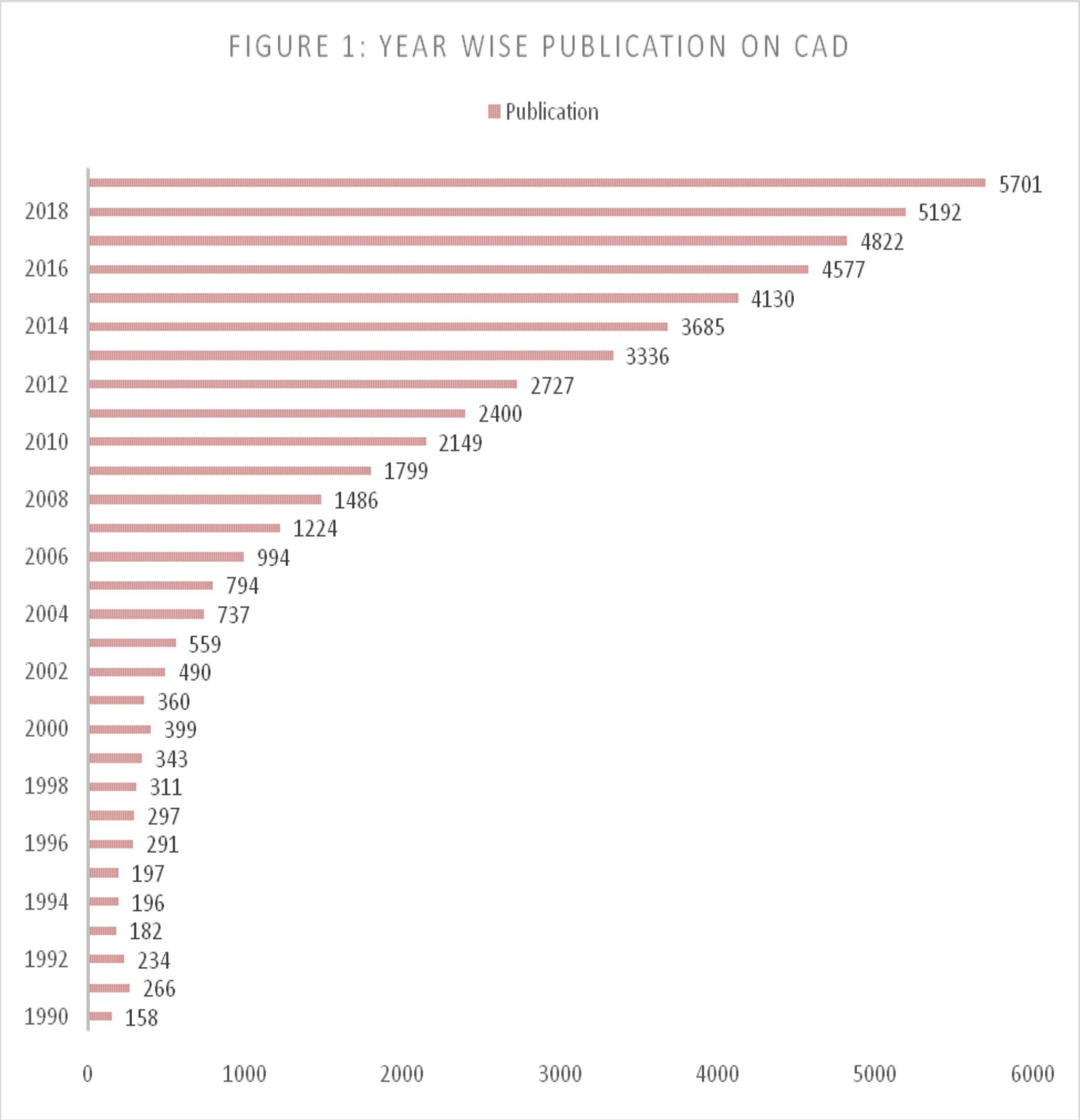

It is observed from the table and figure that during 30 years i.e., from 1990 to 2019, the year 2019 occupies the first place with 5701 (11.39%) publications, 2018 records second place with 5192 (10.38%) publications, 2017 settled third place with 4822 (9.64%) publications, 2016 got fourth place with 4577 (9.15%) publications, 2015 obtained fifth place with 4130 (8.25%) publications followed by 2014, 2013, 2012, 2011, 2010, 2009, 2008, 2007, 2006, 2005, 2004, 2003, 2002, 2000, 2001, 1999, 1998, 1997, 1996, 1991, 1992, 1995, 1994, 1993, and 1990 with 3685 (7.36%), 3336 (6.67%), 2727 (5.45%), 2400 (4.80%), 2149 (4.29%), 1799 (3.60%), 1486 (2.97%), 1224 (2.45%), 994 (1.99%), 794 (1.59%), 737 (1.47%), 559 (1.12%), 490 (0.98%), 399 (0.80%), 360 (0.72%), 343 (0.69%), 311 (0.62%), 297 (0.59%), 291 (0.58%), 266 (0.53%), 234 (0.47%), 197 (0.39%), 196 (0.39%), 182 (0.36%), and 158 (0.32%) publications respectively. The year 1990 has the minimum number of publications 158 (0.32%) as a comparative study reveals (Table 1). Last ten years shows a remarkable growth on coronary artery disease research i.e. 2010 to 2019, which is appreciable towards scientific community.

The exponential growth rate of publications over the years has been shown in Table1. The exponential growth rate has been calculated based on previous years. The exponential growth rate over three decades ranges between -0.008342 (1993) and 0.017515 (1991). The exponential growth rate is accounted for a maximum during the year 1991. It indicates that awareness and more focus on coronary artery disease is felt severe year after year in this study period.

### 4.1.1 Analysis of Year Wise Publication Distribution of BRICS on CAD

The BRICS countries research output in coronary artery disease research cumulated to 50036 publications in 30 years during 1990-2019, and they increased from 158 in the year 1990 to 5701 publications in the year 2019, registering 112.70% growth per annum. The share of Brazilian publications in BRICS output is 6218 (12.43%) during 1990-19, which increased from 31 to 502 from 1990 to 2019, registering 9.73% of growth per annum. The share of Russian scientists in the study is 5058 (10.11%) during three decades, which increased from 14 in the year 1990 to 377 in the year 2019 registered 11.60% of growth per year. Indian publications share 4706 (9.41%) which increased from 27 to 394 from 1990 to 2019 and registered 9.34% of growth. The share of the Peoples Republic of China output is 32770 (65.49%) during 1990-2019, which increased from 75 publications to 4356 publication during thirty years registered 14.50% of growth per annum. The share of South African publications in BRICS output is 1284 (2.57%) which increased from 11 in the year 1990 to 72 in the year 2019, registering 6.46% of growth per annum.

**Table 1A: Year Wise Publication Distribution of BRICS on CAD**

| S.No. | Year | Brazil | Russia | India | China | South Africa | Total |
|---|---|---|---|---|---|---|---|
| 1 | 1990 | 31 | 14 | 27 | 75 | 11 | 158 |
| 2 | 1991 | 55 | 45 | 39 | 98 | 29 | 266 |
| 3 | 1992 | 33 | 54 | 30 | 77 | 40 | 234 |
| 4 | 1993 | 14 | 70 | 25 | 39 | 34 | 182 |
| 5 | 1994 | 22 | 102 | 26 | 30 | 16 | 196 |
| 6 | 1995 | 32 | 78 | 24 | 38 | 25 | 197 |
| 7 | 1996 | 40 | 115 | 45 | 55 | 36 | 291 |
| 8 | 1997 | 49 | 129 | 49 | 37 | 33 | 297 |
| 9 | 1998 | 38 | 102 | 53 | 83 | 35 | 311 |
| 10 | 1999 | 57 | 98 | 45 | 112 | 31 | 343 |
| 11 | 2000 | 73 | 124 | 55 | 115 | 32 | 399 |
| 12 | 2001 | 66 | 86 | 55 | 127 | 26 | 360 |
| 13 | 2002 | 91 | 148 | 58 | 173 | 20 | 490 |
| 14 | 2003 | 106 | 132 | 71 | 228 | 22 | 559 |
| 15 | 2004 | 143 | 178 | 96 | 298 | 22 | 737 |
| 16 | 2005 | 173 | 167 | 110 | 321 | 23 | 794 |
| 17 | 2006 | 208 | 154 | 104 | 496 | 32 | 994 |
| 18 | 2007 | 283 | 172 | 143 | 579 | 47 | 1224 |
| 19 | 2008 | 254 | 169 | 208 | 819 | 36 | 1486 |
| 20 | 2009 | 311 | 195 | 223 | 1014 | 56 | 1799 |
| 21 | 2010 | 363 | 220 | 250 | 1275 | 41 | 2149 |
| 22 | 2011 | 390 | 238 | 258 | 1471 | 43 | 2400 |
| 23 | 2012 | 373 | 222 | 304 | 1759 | 69 | 2727 |
| 24 | 2013 | 404 | 218 | 316 | 2332 | 66 | 3336 |
| 25 | 2014 | 357 | 230 | 331 | 2700 | 67 | 3685 |
| 26 | 2015 | 406 | 276 | 270 | 3108 | 70 | 4130 |
| 27 | 2016 | 409 | 289 | 412 | 3385 | 82 | 4577 |
| 28 | 2017 | 453 | 300 | 313 | 3668 | 88 | 4822 |
| 29 | 2018 | 482 | 356 | 372 | 3902 | 80 | 5192 |
| 30 | 2019 | 502 | 377 | 394 | 4356 | 72 | 5701 |
| Total | | 6218 | 5058 | 4706 | 32770 | 1284 | 50036 |
| Percentage | | 12.43 | 10.11 | 9.41 | 65.49 | 2.57 | 100.00 |

It is noteworthy that the Peoples Republic of China has contributed the most number of publications 32770 (65.49%) during the study period and also registered the highest growth rate of 14.50 during 1990-2019.

### 4.2 Analysis of Exponential Growth Rate in Block Years

**Table 2: Exponential Growth Rate in Block Years**

| S.No. | Year | Publication | Cum. Publication | Percentage of Publication | Percentage of Cum. | Exponential Growth Rate |
|---|---|---|---|---|---|---|
| 1 | 1990-1994 | 1036 | 1036 | 2.07 | 2.07 | -- |
| 2 | 1995-1999 | 1439 | 2475 | 2.88 | 4.95 | 0.120793 |
| 3 | 2000-2004 | 2545 | 5020 | 5.09 | 10.03 | 0.198641 |
| 4 | 2005-2009 | 6297 | 11317 | 12.58 | 22.62 | 0.178209 |
| 5 | 2010-2014 | 14297 | 25614 | 28.57 | 51.19 | 0.113031 |
| 6 | 2015-2019 | 24422 | 50036 | 48.81 | 100.00 | 0.154251 |
| Total | | 50036 | | 100.00 | | |

In order to examine the growth trend, the whole study period of 30 years has been divided into six block periods. Each block year comprises of five years. The growth of research output on coronary artery disease research is presented in table 2 in six block periods, such as 1990-1994, 1995-1999, 2000-2004, 2005-2009, 2010-2014, and 2015-2019. The five-year cumulative output increased from 1036, 1439, 2545, 6297, 14297, and 24422 publications from 1990-1994, 1995-1999, 2000-2004, 2005-2009, 2010-2014, and 2015-2019 registering 2.07%, 2.88%, 5.09%, 12.58%, 28.57%, and 48.81% growth respectively.

The exponential growth rate is very high in the third block period 2000-2004 (0.198641), so this table indicates that the exponential growth rate is gradually increasing and then a slight fluctuation in growth is observed in the fourth block (2005-2009) and again decreased in growth is noted in the fifth block (2010-2014) but the next block period, i.e. (2015-2019) have not shown remarkable growth.

It is observed from Table 2 that the block year 2015-2019 has more publications compared to the other five-block years. It shows that in recent years, the growth of literature in coronary artery disease increases and alarms the society the effects of the disease in BRICS nations which is generally increasing. It indicates that coronary artery disease has a very high exponential growth rate in the second block.

## 4.3 Analysis of Relative Growth Rate and Doubling Time of CAD Publications

**Table 3: Relative Growth Rate and Doubling Time of Publications**

| S.No. | Year | Publication | Cum. Publication | W1 | W2 | RT (p) | Mean RP (p) | Dt (p) | Mean Dt (p) |
|---|---|---|---|---|---|---|---|---|---|
| 1 | 1990 | 158 | 158 | 5.0626 | 5.0626 | 0.0000 | 0.9389 | -- | 0.6052 |
| 2 | 1991 | 266 | 424 | 5.5835 | 6.0497 | 0.4662 | | 1.4864 | |
| 3 | 1992 | 234 | 658 | 5.4553 | 6.4892 | 1.0339 | | 0.6703 | |
| 4 | 1993 | 182 | 840 | 5.2040 | 6.7334 | 1.5294 | | 0.4531 | |
| 5 | 1994 | 196 | 1036 | 5.2781 | 6.9431 | 1.6650 | | 0.4162 | |
| 6 | 1995 | 197 | 1233 | 5.2832 | 7.1172 | 1.8340 | 1.8409 | 0.3779 | 0.3778 |
| 7 | 1996 | 291 | 1524 | 5.6733 | 7.3291 | 1.6558 | | 0.4185 | |
| 8 | 1997 | 297 | 1821 | 5.6937 | 7.5071 | 1.8134 | | 0.3822 | |
| 9 | 1998 | 311 | 2132 | 5.7398 | 7.6648 | 1.9250 | | 0.3600 | |
| 10 | 1999 | 343 | 2475 | 5.8377 | 7.8140 | 1.9763 | | 0.3507 | |
| 11 | 2000 | 399 | 2874 | 5.9890 | 7.9635 | 1.9745 | 2.0306 | 0.3510 | 0.3420 |
| 12 | 2001 | 360 | 3234 | 5.8861 | 8.0815 | 2.1954 | | 0.3157 | |
| 13 | 2002 | 490 | 3724 | 6.1944 | 8.2226 | 2.0281 | | 0.3417 | |
| 14 | 2003 | 559 | 4283 | 6.3261 | 8.3624 | 2.0363 | | 0.3403 | |
| 15 | 2004 | 737 | 5020 | 6.6026 | 8.5212 | 1.9186 | | 0.3612 | |
| 16 | 2005 | 794 | 5814 | 6.6771 | 8.6680 | 1.9909 | 1.8985 | 0.3481 | 0.3653 |
| 17 | 2006 | 994 | 6808 | 6.9017 | 8.8259 | 1.9241 | | 0.3602 | |
| 18 | 2007 | 1224 | 8032 | 7.1099 | 8.9912 | 1.8813 | | 0.3684 | |
| 19 | 2008 | 1486 | 9518 | 7.3038 | 9.1609 | 1.8571 | | 0.3732 | |
| 20 | 2009 | 1799 | 11317 | 7.4950 | 9.3341 | 1.8391 | | 0.3768 | |
| 21 | 2010 | 2149 | 13466 | 7.6728 | 9.5079 | 1.8352 | 1.8931 | 0.3776 | 0.3662 |
| 22 | 2011 | 2400 | 15866 | 7.7832 | 9.6719 | 1.8887 | | 0.3669 | |
| 23 | 2012 | 2727 | 18593 | 7.9110 | 9.8305 | 1.9196 | | 0.3610 | |
| 24 | 2013 | 3336 | 21929 | 8.1125 | 9.9956 | 1.8830 | | 0.3680 | |
| 25 | 2014 | 3685 | 25614 | 8.2120 | 10.1509 | 1.9389 | | 0.3574 | |
| 26 | 2015 | 4130 | 29744 | 8.3260 | 10.3004 | 1.9744 | 2.0800 | 0.3510 | 0.3336 |
| 27 | 2016 | 4577 | 34321 | 8.4288 | 10.4435 | 2.0147 | | 0.3440 | |
| 28 | 2017 | 4822 | 39143 | 8.4809 | 10.5750 | 2.0940 | | 0.3309 | |
| 29 | 2018 | 5192 | 44335 | 8.5549 | 10.6995 | 2.1447 | | 0.3231 | |
| 30 | 2019 | 5701 | 50036 | 8.6484 | 10.8205 | 2.1721 | | 0.3190 | |
| Total | | 50036 | | | | | 1.7803 | 11.9507 | 0.3984 |

The analysis of growth rate in coronary artery disease research output is one of the essential aspects of the discussion. The present analysis aims to identify the trends and growth of prospects in the research. However, an increase in coronary artery disease research has made it extremely difficult for scientists to keep in touch with the recent advances in their fields. The growth rate of research on coronary artery disease is determined by calculating the relative growth rates and doubling the publications' time. In the research design, the details of this model have been elaborated.

Table 3 depicts relative growth rate data and doubling time for total research output on coronary artery disease. It is observed that relative growth rates have progressively increased from 0.4662 in 1990 to 2.1721 in the year 2019. The whole study period 'mean relative growth rate' is 1.7803.

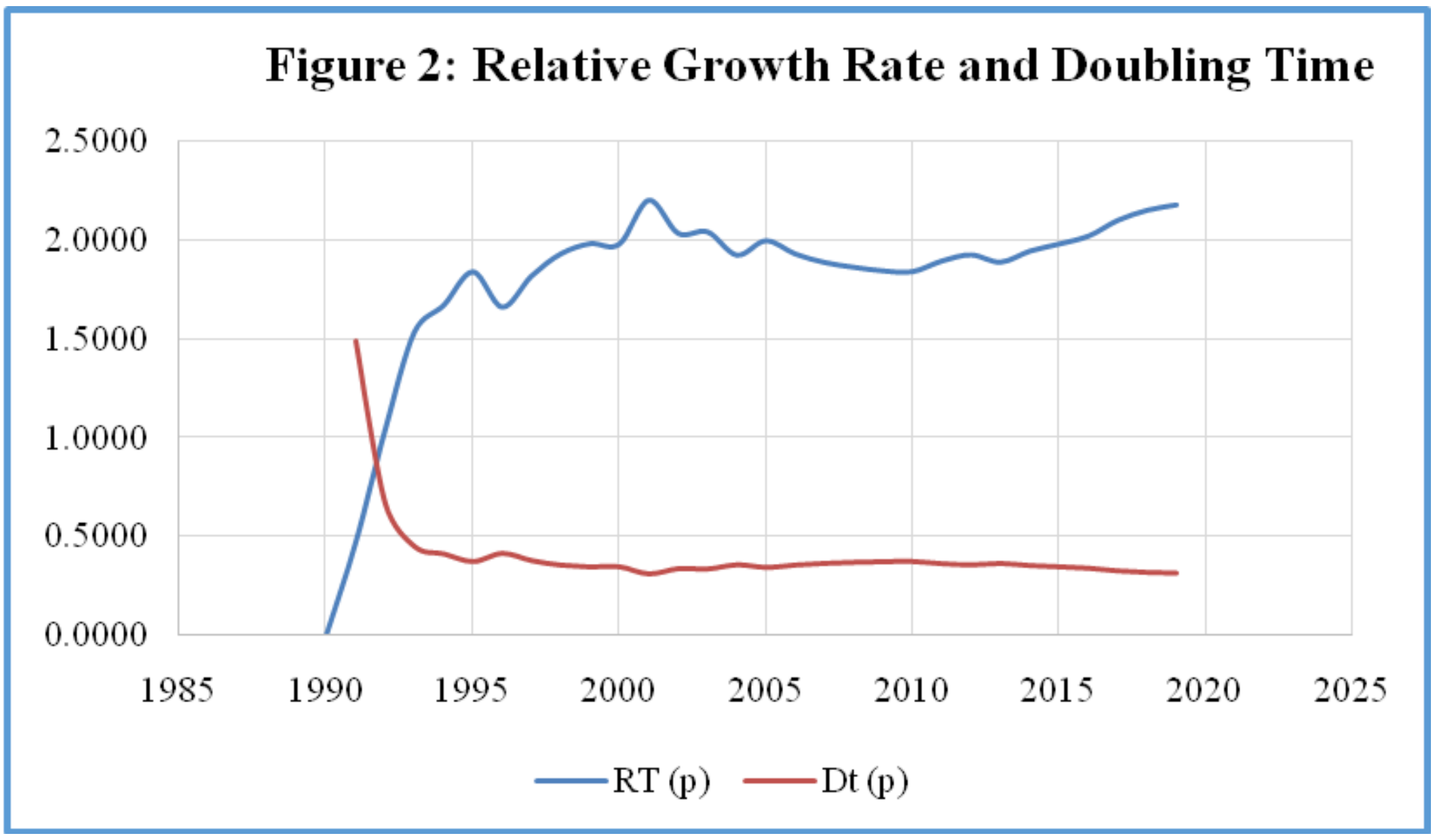

The doubling time for publications of all sources in coronary artery disease research output has decreased from 1.4864 in 1990 to 0.3190 in 2019. During the study period, the doubling time value is 11.9507. The whole study period 'mean doubling time' has been calculated as 0.3984.

The relative growth rate has shown an increasing trend, which means the rate of increase is high in terms of segment, and this has been highlighted by doubling time for publications, which is less than the relative growth rate. The study is substantiated by (Janaarthanan, Nithyanandham, & Natarajan, 2019) as there is increasing in relative growth rate and decreasing in doubling time applied on Osteoporosis disease in Children. ***Hence, the hypothesis that the relative growth rate of total scientific publications shows an increasing trend and the doubling time for*** ***publications reflects a decreasing trend as noted has been substantiated. Hence, the first formulated hypothesis is proved.***

### 4.4 Analysis of Relative Growth Rate & Doubling time Vs. Bock Years

**Table 4: Block Wise Relative Growth Rate and Doubling Time**

| S. No. | Year | Publication | Percentage of Publication | Cum. Publication | Percentage of Cum. | W1 | W2 | RGR | Dt |
|---|---|---|---|---|---|---|---|---|---|
| 1 | 1990-1994 | 1036 | 2.07 | 1036 | 2.07 | 6.9431 | 6.9431 | 0.0000 | |
| 2 | 1995-1999 | 1439 | 2.88 | 2475 | 4.95 | 7.2717 | 7.8140 | 0.5423 | 1.2779 |
| 3 | 2000-2004 | 2545 | 5.09 | 5020 | 10.03 | 7.8419 | 8.5212 | 0.6793 | 1.0202 |
| 4 | 2005-2009 | 6297 | 12.58 | 11317 | 22.62 | 8.7478 | 9.3341 | 0.5862 | 1.1821 |
| 5 | 2010-2014 | 14297 | 28.57 | 25614 | 51.19 | 9.5678 | 10.1509 | 0.5831 | 1.1885 |
| 6 | 2015-2019 | 24422 | 48.81 | 50036 | 100.00 | 10.1032 | 10.8205 | 0.7173 | 0.9662 |
| Total | | 50036 | 100.00 | | | | | 0.6216 | 1.1148 |

For the purpose of the analysis concerning the study of research publications with 'relative growth rate,' the study period has been grouped into six block periods comprising five years per group. Accordingly, the global research output growth rate is presented in the table. It is examined from Table 4 that there is an increasing trend in the quantum of relative growth rate from one block period to the next period. The maximum increase is observed in the block year from 2015 to 2019 (0.7173).

The mean relative growth rate for the whole study period output is 0.6216. The doubling time for publications has observed a decreasing trend, i.e., (1.2779 to 0.9662). The doubling time for publications for the entire period has been computed as 1.1148.

It is observed from Table 4 that the block year 2015-2019 has more growth rate compared to other block years. It is shown that in recent years, the growth of research in coronary artery disease is getting increasing. Nevertheless, it is doubling time for publications, which is decreasing in the block years.

To achieve the second objective of examining Relative growth rate and Doubling time on Coronary Artery Disease research, the analysis of Relative growth rate and doubling time is carried out and it is explained in tables 3 and 4.

## 4.5 Time Series Analysis of Research Productivity on CAD Literature

| Table 5: Time Series Analysis of Research Productivity on Coronary Artery Disease | | | | | |
|---|---|---|---|---|---|
| S.No. | Year | Publication (Y) | X | $X^2$ | XY |
| 1 | 1990 | 158 | -14 | 196 | -2212 |
| 2 | 1991 | 266 | -13 | 169 | -3458 |
| 3 | 1992 | 234 | -12 | 144 | -2808 |
| 4 | 1993 | 182 | -11 | 121 | -2002 |
| 5 | 1994 | 196 | -10 | 100 | -1960 |
| 6 | 1995 | 197 | -9 | 81 | -1773 |
| 7 | 1996 | 291 | -8 | 64 | -2328 |
| 8 | 1997 | 297 | -7 | 49 | -2079 |
| 9 | 1998 | 311 | -6 | 36 | -1866 |
| 10 | 1999 | 343 | -5 | 25 | -1715 |
| 11 | 2000 | 399 | -4 | 16 | -1596 |
| 12 | 2001 | 360 | -3 | 9 | -1080 |
| 13 | 2002 | 490 | -2 | 4 | -980 |
| 14 | 2003 | 559 | -1 | 1 | -559 |
| 15 | 2004 | 737 | 0 | 0 | 0 |
| 16 | 2005 | 794 | 1 | 1 | 794 |
| 17 | 2006 | 994 | 2 | 4 | 1988 |
| 18 | 2007 | 1224 | 3 | 9 | 3672 |
| 19 | 2008 | 1486 | 4 | 16 | 5944 |
| 20 | 2009 | 1799 | 5 | 25 | 8995 |
| 21 | 2010 | 2149 | 6 | 36 | 12894 |
| 22 | 2011 | 2400 | 7 | 49 | 16800 |
| 23 | 2012 | 2727 | 8 | 64 | 21816 |
| 24 | 2013 | 3336 | 9 | 81 | 30024 |
| 25 | 2014 | 3685 | 10 | 100 | 36850 |
| 26 | 2015 | 4130 | 11 | 121 | 45430 |
| 27 | 2016 | 4577 | 12 | 144 | 54924 |
| 28 | 2017 | 4822 | 13 | 169 | 62686 |
| 29 | 2018 | 5192 | 14 | 196 | 72688 |
| 30 | 2019 | 5701 | 15 | 225 | 85515 |
| Total | | 50036 | | | |

In the present analysis, the straight-line equation was applied to arrive at a future growth projection under Time Series Analysis.

Straight line equation Yc = a +bX

$$a = \frac{\sum Y}{N}$$

$$a = \frac{50036}{30}$$

$$a = 1667.86667$$

$$b = \frac{\sum XY}{\sum X^2}$$

$$b = \frac{434604}{2255}$$

$$b = 192.729047$$

Estimated literature in 2022 is when X = 2022-2004

X = 18

Yc= a+bX

Yc = 1667.86667+ (192.729047× 18)

Yc = 1667.86667+3469.122846

Yc = 5136.989516

Estimated literature for 2037 is when X = 2037- 2004

X = 33

Yc = a +b X

Yc = 1667.86667+ (192.729047× 33)

Yc = 1667.86667+6360.058551

Yc = 8027.925221

On the application of the formula of time series analysis and subsequently, from the research obtained separately for the years 2022 and 2037, it is found that the future growth trend in coronary artery disease output is assumed to be slow during the years to follow. It will be 5137 in the year 2022, and it is estimated to be 8028 in the year 2037. The inference is that there is not a satisfied amount of growth assumed at the BRICS level literature research output on coronary artery disease.

## 4.6 Analysis of Page Wise Distribution of CAD Literature

**Table 6: Page Wise Distribution of Coronary Artery Disease**

| Year/ Pages | 1 | 2 | 3 | 4 | 5 | 6 | 7 | 8 | 9 | 10 and > 10 | Total |
|---|---|---|---|---|---|---|---|---|---|---|---|
| 1990 | 8 | 20 | 132 | 180 | 85 | 54 | 56 | 24 | 18 | 112 | 689 |
| 1991 | 4 | 26 | 144 | 308 | 235 | 126 | 126 | 104 | 63 | 252 | 1388 |
| 1992 | 6 | 16 | 114 | 220 | 220 | 126 | 98 | 112 | 54 | 395 | 1361 |
| 1993 | 8 | 10 | 66 | 148 | 130 | 150 | 140 | 104 | 54 | 394 | 1204 |
| 1994 | 5 | 10 | 93 | 140 | 210 | 168 | 91 | 96 | 99 | 168 | 1080 |
| 1995 | 5 | 14 | 69 | 144 | 160 | 156 | 154 | 88 | 108 | 296 | 1194 |
| 1996 | 22 | 26 | 93 | 212 | 245 | 210 | 154 | 168 | 90 | 426 | 1646 |
| 1997 | 33 | 22 | 78 | 172 | 210 | 258 | 189 | 200 | 153 | 378 | 1693 |
| 1998 | 25 | 30 | 81 | 228 | 225 | 210 | 182 | 168 | 153 | 527 | 1829 |
| 1999 | 20 | 30 | 72 | 212 | 265 | 306 | 315 | 232 | 198 | 357 | 2007 |
| 2000 | 26 | 18 | 72 | 264 | 315 | 372 | 315 | 288 | 135 | 693 | 2498 |
| 2001 | 31 | 18 | 57 | 160 | 330 | 300 | 308 | 304 | 153 | 581 | 2242 |
| 2002 | 37 | 30 | 99 | 264 | 425 | 474 | 357 | 312 | 198 | 783 | 2979 |
| 2003 | 55 | 38 | 120 | 296 | 380 | 540 | 434 | 408 | 180 | 910 | 3361 |
| 2004 | 87 | 42 | 111 | 332 | 470 | 756 | 658 | 504 | 432 | 1073 | 4465 |
| 2005 | 98 | 80 | 114 | 264 | 490 | 630 | 826 | 632 | 459 | 1287 | 4880 |
| 2006 | 116 | 62 | 102 | 268 | 655 | 1014 | 945 | 832 | 738 | 1579 | 6311 |
| 2007 | 88 | 98 | 129 | 384 | 660 | 1206 | 1365 | 1080 | 855 | 2316 | 8181 |
| 2008 | 130 | 112 | 144 | 536 | 980 | 1302 | 1323 | 1360 | 1143 | 2790 | 9820 |
| 2009 | 159 | 138 | 183 | 472 | 1180 | 1614 | 1799 | 1576 | 1224 | 3994 | 12339 |
| 2010 | 219 | 206 | 249 | 496 | 1270 | 1980 | 1974 | 1808 | 1683 | 4521 | 14406 |
| 2011 | 161 | 170 | 237 | 572 | 1530 | 2142 | 2576 | 2072 | 1962 | 5431 | 16853 |
| 2012 | 182 | 174 | 198 | 592 | 1400 | 2244 | 2807 | 2608 | 2367 | 7810 | 20382 |
| 2013 | 295 | 208 | 273 | 592 | 1560 | 2568 | 3143 | 3256 | 2970 | 9850 | 24715 |
| 2014 | 274 | 214 | 195 | 460 | 1480 | 2922 | 3458 | 3760 | 3771 | 12438 | 28972 |
| 2015 | 257 | 190 | 234 | 568 | 1395 | 2976 | 3962 | 4136 | 3825 | 16554 | 34097 |
| 2016 | 268 | 282 | 252 | 544 | 1395 | 3000 | 4123 | 4600 | 4761 | 19346 | 38571 |
| 2017 | 230 | 216 | 153 | 484 | 1280 | 2892 | 4382 | 5144 | 4716 | 24288 | 43785 |
| 2018 | 325 | 260 | 174 | 384 | 1355 | 2844 | 4865 | 4968 | 5355 | 26768 | 47298 |
| 2019 | 226 | 348 | 201 | 436 | 1200 | 2748 | 4711 | 5600 | 5778 | 30222 | 51470 |
| Total | 3400 | 3108 | 4239 | 10332 | 21735 | 36288 | 45836 | 46544 | 43695 | 176539 | 391716 |

During the study period, the number of articles published was 50036. They were represented in 391716 pages in many journals. The highest number of pages contributed by authors through their research articles was found in 2019, having 51470 pages, followed by the year 2018 contributed in 47298 pages. The year 1990 recorded least number of pages contributed by the authors. The year 2019 seems to be the most productive in terms of length of pages as it has also shown more number of pages, i.e., more than ten pages.

It is observed from the table that in terms of the length of pages contributed the growth is observed uneven from 1990 to 1993. Yet it was gradually increasing from 1080 in 1994 to 2498 in 2000, and it falls to 2242 in the year 2001. The gradual increase is noted from 2002 onwards. During the three decades of study, the average number of pages used by the researchers is 7.83. It is highlighted that the average number of pages gets increased from 4.36 in 1990 to 9.03 in 2019 indicating that the scientists are using a more significant number of pages to communicate their research in journals with the passage of time.

### 4.7 Analysis of Average Pages per Paper in Coronary Artery Disease Research

It is observed from Table 7 that there is a fluctuation trend in the study period. On the other hand, after the year 2012, the number of pages crossed twenty thousand. The highest number of pages was noted in the year 2019, having 51470 pages. The average number of pages per contribution exhibits a fluctuation trend. In other words, it was 6.62 in the year 1993, but in the year 1999, it was 5.58 per contribution. The total average number of pages per contribution is 7.83.

During the entire study period, coronary artery disease research articles ranged from 4 to 9 pages. Average pages during the thirty years of research come to 7.83 pages. During the three decades of study, the average number of pages used by the researchers is 7.83. It is highlighted that the average number of pages gets increased from 4.36 in 1990 to 9.03 in 2019 indicating that the scientists are finding more facts and that the same are communicated elaborately in the form of articles and get them published in journals with the passage of time.

| Table 7: Average Pages Per Paper in Coronary Artery Disease | | | | | |
|---|---|---|---|---|---|
| S.No. | Year | No. of Articles | No. of Pages | % of Pages | Average Pages Per Paper |
| 1 | 1990 | 158 | 689 | 0.18 | 4.36 |
| 2 | 1991 | 266 | 1388 | 0.35 | 5.22 |
| 3 | 1992 | 234 | 1361 | 0.35 | 5.82 |
| 4 | 1993 | 182 | 1204 | 0.31 | 6.62 |
| 5 | 1994 | 196 | 1080 | 0.28 | 5.51 |
| 6 | 1995 | 197 | 1194 | 0.30 | 6.06 |
| 7 | 1996 | 291 | 1646 | 0.42 | 5.66 |
| 8 | 1997 | 297 | 1693 | 0.43 | 5.70 |
| 9 | 1998 | 311 | 1829 | 0.47 | 5.88 |
| 10 | 1999 | 343 | 2007 | 0.51 | 5.85 |
| 11 | 2000 | 399 | 2498 | 0.64 | 6.26 |
| 12 | 2001 | 360 | 2242 | 0.57 | 6.23 |
| 13 | 2002 | 490 | 2979 | 0.76 | 6.08 |
| 14 | 2003 | 559 | 3361 | 0.86 | 6.01 |
| 15 | 2004 | 737 | 4465 | 1.14 | 6.06 |
| 16 | 2005 | 794 | 4880 | 1.25 | 6.15 |
| 17 | 2006 | 994 | 6311 | 1.61 | 6.35 |
| 18 | 2007 | 1224 | 8181 | 2.09 | 6.68 |
| 19 | 2008 | 1486 | 9820 | 2.51 | 6.61 |
| 20 | 2009 | 1799 | 12339 | 3.15 | 6.86 |
| 21 | 2010 | 2149 | 14406 | 3.68 | 6.70 |
| 22 | 2011 | 2400 | 16853 | 4.30 | 7.02 |
| 23 | 2012 | 2727 | 20382 | 5.20 | 7.47 |
| 24 | 2013 | 3336 | 24715 | 6.31 | 7.41 |
| 25 | 2014 | 3685 | 28972 | 7.40 | 7.86 |
| 26 | 2015 | 4130 | 34097 | 8.70 | 8.26 |
| 27 | 2016 | 4577 | 38571 | 9.85 | 8.43 |
| 28 | 2017 | 4822 | 43785 | 11.18 | 9.08 |
| 29 | 2018 | 5192 | 47298 | 12.07 | 9.11 |
| 30 | 2019 | 5701 | 51470 | 13.14 | 9.03 |
| Total | | 50036 | 391716 | 100.00 | 7.83 |

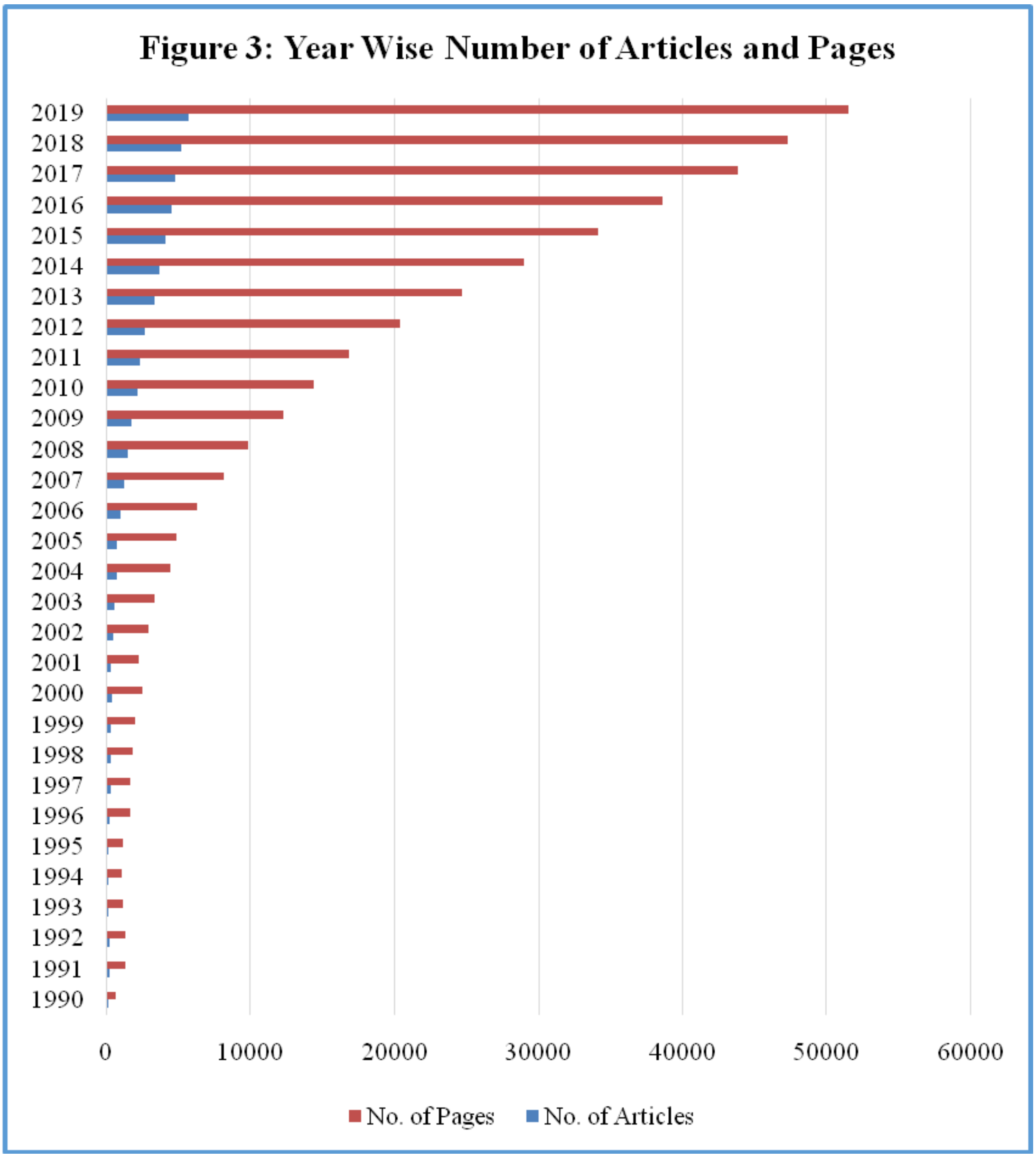

## 4.8 Ratio of Growth with BRICS Countries

**Table 8: Ratio of Growth of BRICS Countries**

| S.No. | Country | TP | % | Brazil | Russia | India | China | South Africa |
|---|---|---|---|---|---|---|---|---|
| 1 | Peoples R China | 32770 | 65.49 | 0.19 | 0.15 | 0.14 | 1.00 | 0.04 |
| 2 | Brazil | 6218 | 12.43 | 1.00 | 0.81 | 0.76 | 5.27 | 0.21 |
| 3 | Russia | 5058 | 10.11 | 1.23 | 1.00 | 0.93 | 6.48 | 0.25 |
| 4 | India | 4706 | 9.41 | 1.32 | 1.07 | 1.00 | 6.96 | 0.27 |
| 5 | South Africa | 1284 | 2.57 | 4.84 | 3.94 | 3.67 | 25.52 | 1.00 |
| Total | | 50036 | 100.00 | | | | | |

The table reveals the Growth of Publications of the BRICS countries. All the BRICS countries have been compared with each other in terms of total publication output. The ratio has been calculated by dividing the contribution of one country with that of other BRICS countries, that is, Brazil, Russia, India, China, and South Africa. The Peoples Republic of China has the highest ratio of growth in this table, as it contributed to the largest number of publications among BRICS Countries, followed by Brazil, Russia, India, and South Africa.

### 4.9 Analysis of Collaboration of Publications on CAD

Another standard analysis involves drawing up a list of the most productive countries, which have collaborated with BRICS countries in a CAD field. In this regard, it is usual to include a ranking of those who have collaborated in the most significant number of publications. The procedure used to identify the most productive countries consists of obtaining the authors' affiliation who have taken part in writing the documents. To get a more accurate analysis, all the authors' affiliations rather than only that of the first author are to be pursued.

#### 4.9.1 Analysis of Top Ten Collaborating Countries with Brazil

**Table 9A: Collaboration of Brazil**

| S.No. | Collaborating Countries | Publications | % | Cumulative | Cum. % | Av. Collaboration Per Year |
|---|---|---|---|---|---|---|
| 1 | USA | 2837 | 45.68 | 2837 | 45.63 | 94.57 |
| 2 | UK | 991 | 15.96 | 3828 | 61.56 | 33.03 |
| 3 | Netherlands | 328 | 5.28 | 4156 | 66.84 | 10.93 |
| 4 | Ireland | 302 | 4.86 | 4458 | 71.70 | 10.07 |
| 5 | Switzerland | 158 | 2.54 | 4616 | 74.24 | 5.27 |
| 6 | Germany | 93 | 1.50 | 4709 | 75.73 | 3.10 |
| 7 | France | 47 | 0.76 | 4756 | 76.49 | 1.57 |
| 8 | U Arab Emirates | 36 | 0.58 | 4792 | 77.07 | 1.20 |
| 9 | Italy | 35 | 0.56 | 4827 | 77.63 | 1.17 |
| 10 | Spain | 24 | 0.39 | 4851 | 78.02 | 0.80 |
| 11 | Other Countries | 1367 | 22.01 | 6218 | 100.00 | 45.57 |

The table reveals the country-wise collaboration of Brazil on coronary artery disease research output during the study period. Overall, 6218 records were published in coronary artery disease research from Brazil during three decades of study from 1990 to 2019. The top 10 collaborating countries are listed in the table for analysis. Among them United States of America collaborated 2837 (45.68%) publications, followed by UK 991 (15.96%), Netherlands 328 (5.28%), Ireland 302 (4.86%), Switzerland 158 (2.54%), Germany 93 (1.50%), France 47 (0.76%), United Arab Emirates 36 (0.58%), Italy 35 (0.56%), Spain 24 (0.39%), and other countries collaborated 1367 (22.01%) publications with Brazil. It is to be noted that Brazilian scientists contributed 1187 solo publications and without collaborating other countries.

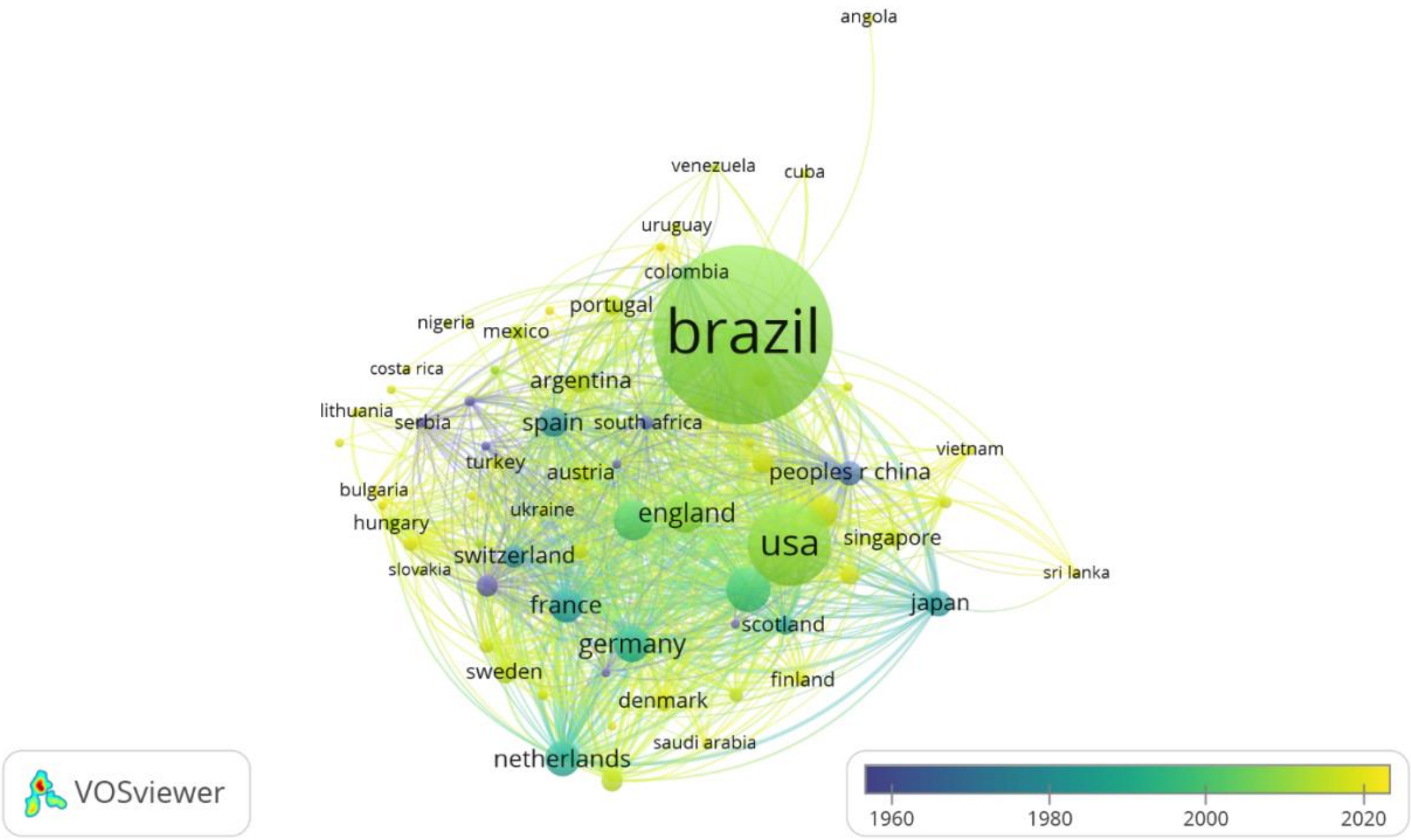

**Figure 4: Collaborating Countries with Brazil**

It can be seen from the table that average publication of collaborated country-wise reveals that United States of America has shown the maximum average number of publications per year, i.e., 94.57. The UK has an average number of publications with 33.03, followed by Netherlands 10.93, Ireland 10.07, Switzerland 5.27, and the rest of the countries have less than five as the average number of publications per year. The United States of America is the highest collaborating nation in the case of an average number of publications per year.

The United States of America emerged as the top collaborative contributing country in coronary artery disease research with Brazil. The UK ranked the second,

Netherlands, which ranked the third, and Ireland is ranked the fourth. Switzerland is ranked in fifth place. This analysis helps in identifying the countries which have taken up the research work in the field. The developed countries are concentrating more on coronary artery disease research than developing and underdeveloped countries. The average number of publications from each country has been calculated by dividing the total publications by the number of years (30), and the same is shown in the table. The above table revealed that the United Arab Emirates, a developing country, is also contributing to the research output on coronary artery disease as par with developed countries.

### 4.9.2 Analysis of Top Ten Collaborating Countries with Russia

**Table 9B: Collaboration of Russia**

| S.No. | Collaborating Country | Publications | % | Cumulative | Cum.% | Av. Collaboration Per Year |
|---|---|---|---|---|---|---|
| 1 | USA | 1535 | 30.35 | 1535 | 30.35 | 51.17 |
| 2 | UK | 471 | 9.31 | 2006 | 39.66 | 15.70 |
| 3 | Ireland | 202 | 3.99 | 2208 | 43.65 | 6.73 |
| 4 | Netherlands | 128 | 2.53 | 2336 | 46.18 | 4.27 |
| 5 | Switzerland | 125 | 2.47 | 2461 | 48.66 | 4.17 |
| 6 | Germany | 55 | 1.09 | 2516 | 49.74 | 1.83 |
| 7 | U Arab Emirates | 29 | 0.57 | 2545 | 50.32 | 0.97 |
| 8 | Italy | 19 | 0.38 | 2564 | 50.69 | 0.63 |
| 9 | New Zealand | 10 | 0.20 | 2574 | 50.89 | 0.33 |
| 10 | France | 10 | 0.20 | 2584 | 51.09 | 0.33 |
| 11 | Other Countries | 2474 | 48.91 | 5058 | 100.00 | 82.47 |

The table depicts the country-wise collaboration of Russia on coronary artery disease research output during the study period. Overall, 5058 records were published in coronary artery disease research from Russia during three decades of study from 1990 to 2019. The top 10 collaborating countries are listed in the table for analysis. Among them United States of America collaborated 1535 (30.35%) publications, followed by UK 471 (9.31%), Ireland 202 (3.99%), Netherlands128 (2.53%), Switzerland 125 (2.47%), Germany 55 (1.09%), United Arab Emirates 29 (0.57%), Italy 19 (0.38%), New Zealand10 (0.20%), France 10 (0.20%), and other countries collaborated 2474 (48.91%) publications with Russia. It is to be noted that Russian scientists contributed 2381 publications without collaborating other countries.

It can be seen from the table that collaborated country-wise average publications from the United States of America has shown the maximum average number of publications per year, i.e., 51.17. The UK has an average number of publications with 15.70, followed by Ireland 6.73, Netherlands 4.27, Switzerland 4.17, and the rest of the countries have less than four as the average number of publications per year. The United States of America is the highest collaborating nation in the case of an average number of publications per year.

The United States of America emerged as the top collaborative contributing country in coronary artery disease research with Russia. The UK ranked the second, Ireland, which ranked the third, and the Netherlands is ranked fourth. Switzerland is ranked in fifth place. This analysis helps in identifying the countries which have taken up the research work in the field. The developed countries are concentrating more on coronary artery disease research than developing and underdeveloped countries. The average number of publications from each country has been calculated by dividing the total publications by the number of years 30, and the same is shown in the table. The above table revealed that the United Arab Emirates, a developing country, is also contributing to the research output on coronary artery disease as par with developed countries.

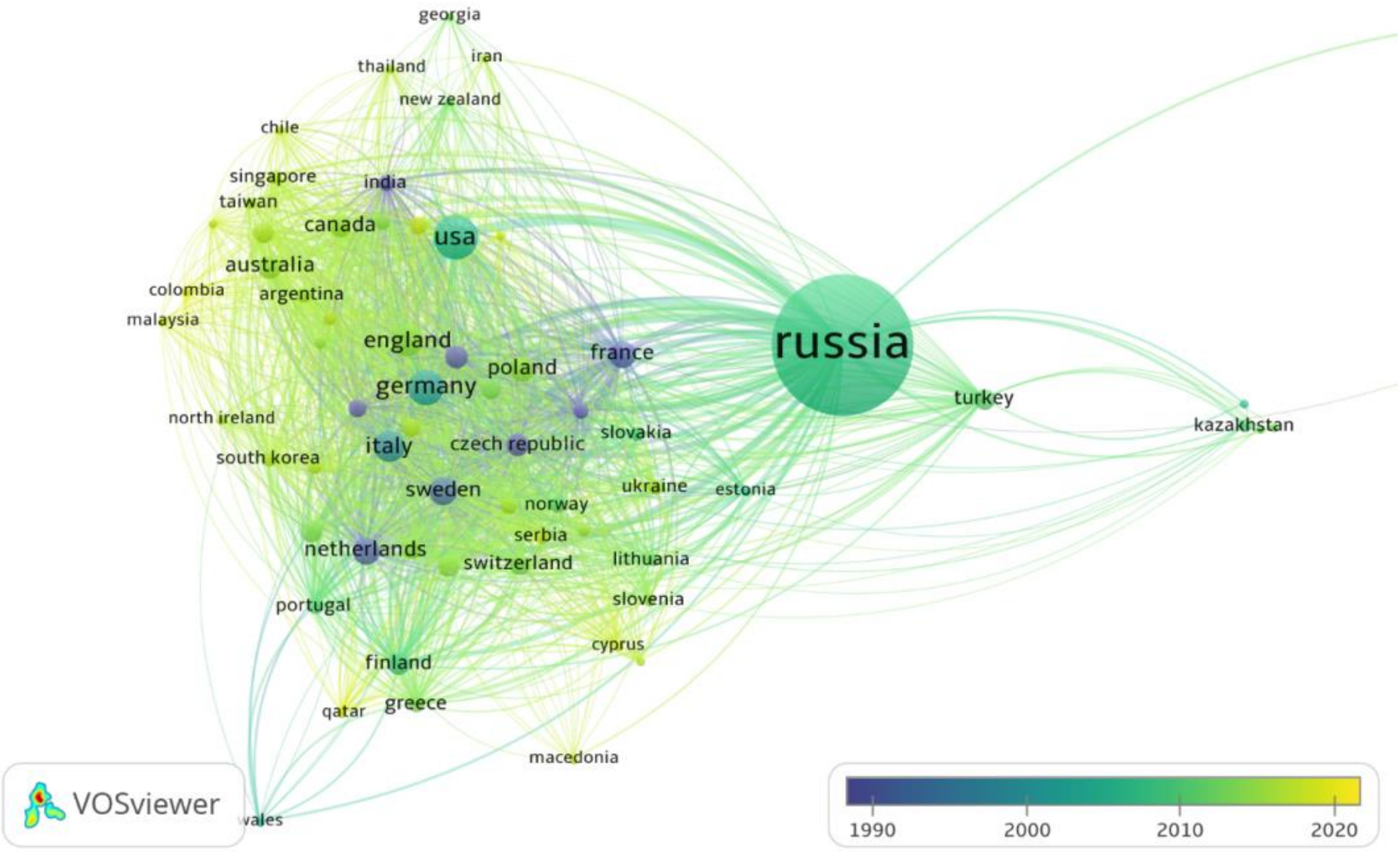

**Figure 5: Collaborating Countries with Russia**

### 4.9.3 Analysis of Top Ten Collaborating Countries with India

**Table 9C: Collaboration of India**

| S.No. | Collaborating Country | Publications | % | Cumulative | Cum. % | Av. Collaboration Per Year |
|---|---|---|---|---|---|---|
| 1 | USA | 1739 | 37.02 | 1739 | 36.95 | 57.97 |
| 2 | UK | 869 | 18.50 | 2608 | 55.42 | 28.97 |
| 3 | Netherlands | 470 | 10.00 | 3078 | 65.41 | 15.67 |
| 4 | Ireland | 226 | 4.81 | 3304 | 70.21 | 7.53 |
| 5 | Switzerland | 123 | 2.62 | 3427 | 72.82 | 4.10 |
| 6 | Germany | 113 | 2.41 | 3540 | 75.22 | 3.77 |
| 7 | U Arab Emirates | 65 | 1.38 | 3605 | 76.60 | 2.17 |
| 8 | France | 53 | 1.13 | 3658 | 77.73 | 1.77 |
| 9 | Italy | 43 | 0.92 | 3701 | 78.64 | 1.43 |
| 10 | Japan | 36 | 0.77 | 3737 | 79.41 | 1.20 |
| 11 | Other Countries | 969 | 20.63 | 4706 | 100.00 | 32.30 |

The table explores the country-wise collaboration of India on coronary artery disease research output during the study period. Overall, 4706 records were published in coronary artery disease research from India during three decades of study from 1990 to 2019. The top 10 collaborating countries are listed in the table for analysis. Among them United States of America collaborated 1739 (37.02%) publications, followed by UK 869 (18.50%), Netherlands 470 (10.00%), Ireland 226 (4.81%), Switzerland 123 (2.62%), Germany 113 (2.41%), United Arab Emirates 65 (1.38%), France 53 (1.13%), Italy 43 (0.92%), Japan 36 (0.77%), and other countries collaborated 969 (20.63%) publications with India. It is to be noted that Indian scientists contributed 655 publications without collaborating other countries.

It can be seen from this table that collaborated country-wise average publications from them United States of America has shown the maximum average number of publications per year, i.e., 57.97. The UK has an average number of publications with 28.97, followed by Netherlands 15.67, Ireland 7.53, Switzerland 4.10, and the rest of the countries have less than four as the average number of publications per year. The United States of America is the highest collaborating nation in the case of an average number of publications per year.

The United States of America emerged as the top collaborative contributing country in coronary artery disease research with India. The UK ranked the second, Netherlands, which ranked the third, and Ireland is ranked fourth. Switzerland is ranked in fifth place. This analysis helps in identifying the countries which have taken up the research work in the field. The developed countries are concentrating more on coronary artery disease research than developing and underdeveloped countries. The average number of publications from each country has been calculated by dividing the total publications by the number of years (30), and the same is shown in the table. The above table revealed that the United Arab Emirates, a developing country, is also contributing to the research output on coronary artery disease as par with developed countries.

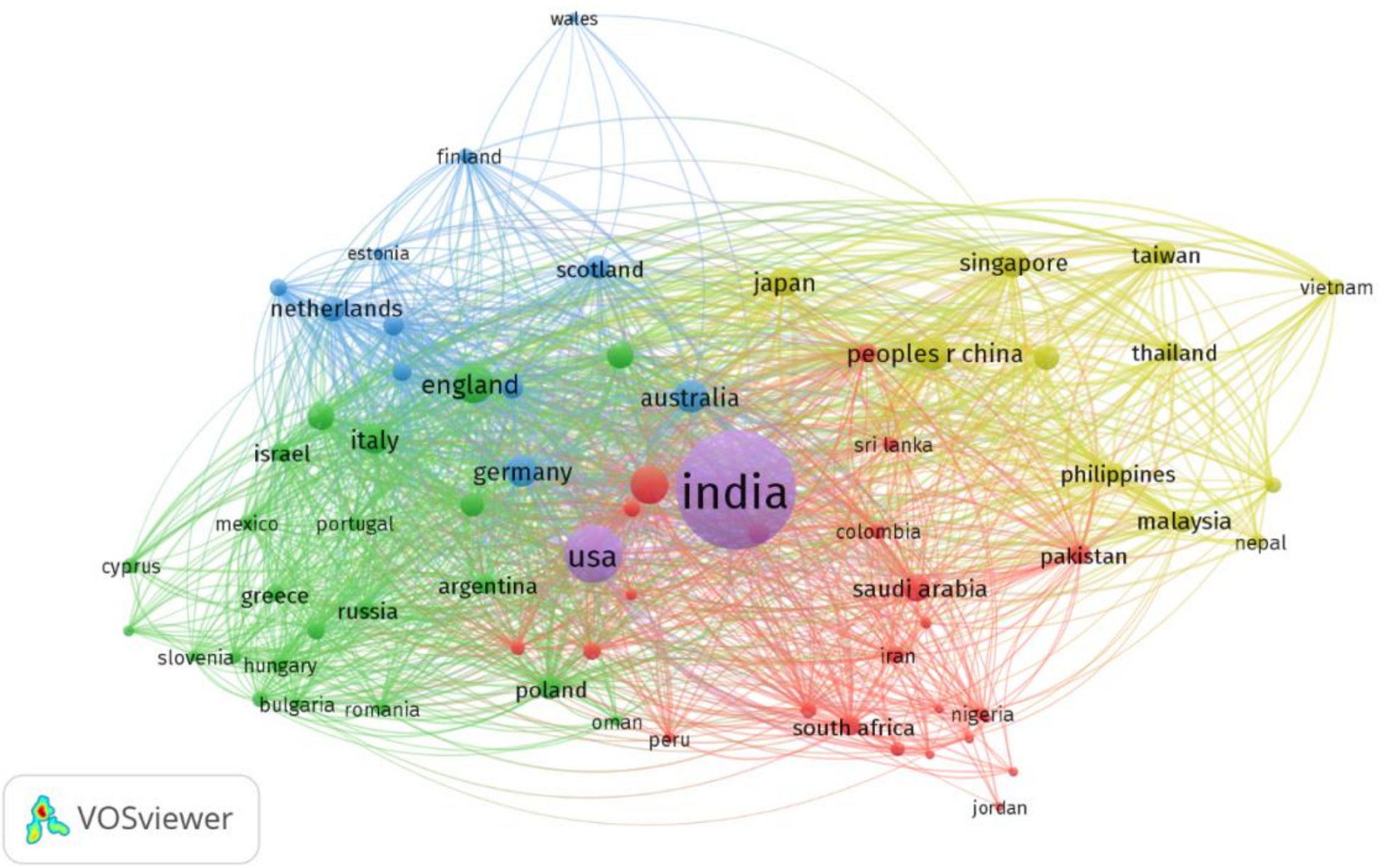

**Figure 6: Collaborating Countries with India**

### 4.9.4 Analysis of Top Ten Collaborating Countries with Peoples Republic of China

The table demonstrates the country-wise collaboration of the Peoples Republic of China on coronary artery disease research output during the study period. Overall, 32770 records were published in coronary artery disease research from the Peoples Republic of China during three decades of study from 1990 to 2019. The top 10 collaborating countries are listed in the table for analysis. Among them, United States of America collaborated 13982 (42.67%) publications, followed by UK 6975 (21.28%), Netherlands 2489 (7.60%), Switzerland 1666 (5.08%), Ireland 1281 (3.91%), Greece 987 (3.01%), India 761 (2.32%), Germany 667 (2.04%), Italy 443 (1.35%), Japan 423 (1.29%), and other countries collaborated 3096 (9.45%) publications with Peoples Republic of China. It is to be noted that Chinese scientists contributed 1078 publications without collaborating other countries.

**Table 9D: Collaboration of the Peoples Republic of China**

| S.No. | Collaborating Country | Publications | % | Cumulative | Cum.% | Av. Collaboration Per Year |
|---|---|---|---|---|---|---|
| 1 | USA | 13982 | 42.67 | 13982 | 42.67 | 466.07 |
| 2 | UK | 6975 | 21.28 | 20957 | 63.95 | 232.50 |
| 3 | Netherlands | 2489 | 7.60 | 23446 | 71.55 | 82.97 |
| 4 | Switzerland | 1666 | 5.08 | 25112 | 76.63 | 55.53 |
| 5 | Ireland | 1281 | 3.91 | 26393 | 80.54 | 42.70 |
| 6 | Greece | 987 | 3.01 | 27380 | 83.55 | 32.90 |
| 7 | India | 761 | 2.32 | 28141 | 85.87 | 25.37 |
| 8 | Germany | 667 | 2.04 | 28808 | 87.91 | 22.23 |
| 9 | Italy | 443 | 1.35 | 29251 | 89.26 | 14.77 |
| 10 | Japan | 423 | 1.29 | 29674 | 90.55 | 14.10 |
| 11 | Other Countries | 3096 | 9.45 | 32770 | 100.00 | 103.20 |

It can be seen from the table that collaborated country-wise average publications from the United States of America has shown the maximum average number of publications per year, i.e., 466.07. The UK has an average number of publications with 232.50, followed by Netherlands 82.97, Switzerland 55.53, Ireland 42.70 and the rest of the countries have less than 40 as the average number of publications per year. The United States of America is the highest collaborating nation in the case of an average number of publications per year.

The United States of America emerged as the top collaborative contributing country in coronary artery disease research with the Peoples Republic of China. The UK ranked the second, Netherlands, which ranked the third, and Switzerland is ranked fourth. Ireland is ranked in fifth place. This analysis helps in identifying the countries which have taken up the research work in the field. The developed countries are concentrating more on coronary artery disease research than developing and underdeveloped countries. The average number of publications from each country has been calculated by dividing the total publications by the number of years 30, and the same is shown in the table. The above table revealed that India, a developing country, is also contributing to the research output on coronary artery disease as par with developed countries.

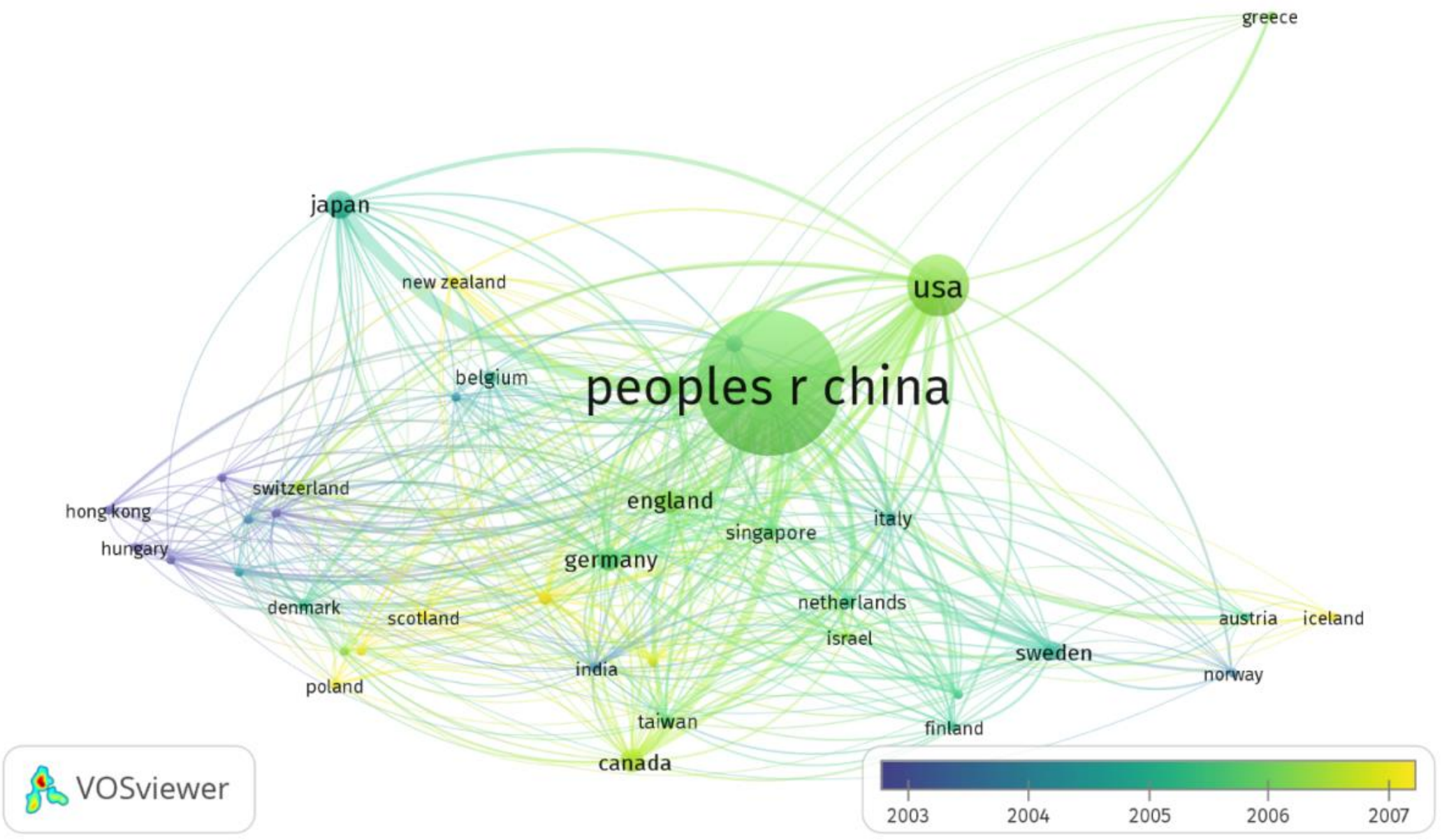

**Figure 7: Collaborating Countries with Peoples Republic of China**

### 4.9.5 Analysis of Top Ten Collaborating Countries with South Africa

**Table 9E: Collaboration of South Africa**

| S.No. | Collaborating Country | Publications | % | Cumulative | Cum.% | Av. Collaboration Per Year |
|---|---|---|---|---|---|---|
| 1 | USA | 523 | 40.73 | 523 | 40.73 | 17.43 |
| 2 | UK | 266 | 20.72 | 789 | 61.45 | 8.87 |
| 3 | Netherlands | 79 | 6.15 | 868 | 67.60 | 2.63 |
| 4 | Ireland | 38 | 2.96 | 906 | 70.56 | 1.27 |
| 5 | Germany | 34 | 2.65 | 940 | 73.21 | 1.13 |
| 6 | Switzerland | 30 | 2.34 | 970 | 75.55 | 1.00 |
| 7 | Italy | 17 | 1.32 | 987 | 76.87 | 0.57 |
| 8 | Canada | 15 | 1.17 | 1002 | 78.04 | 0.50 |
| 9 | U Arab Emirates | 7 | 0.55 | 1009 | 78.58 | 0.23 |
| 10 | France | 6 | 0.47 | 1015 | 79.05 | 0.20 |
| 11 | Other Countries | 269 | 20.95 | 1284 | 100.00 | 8.97 |

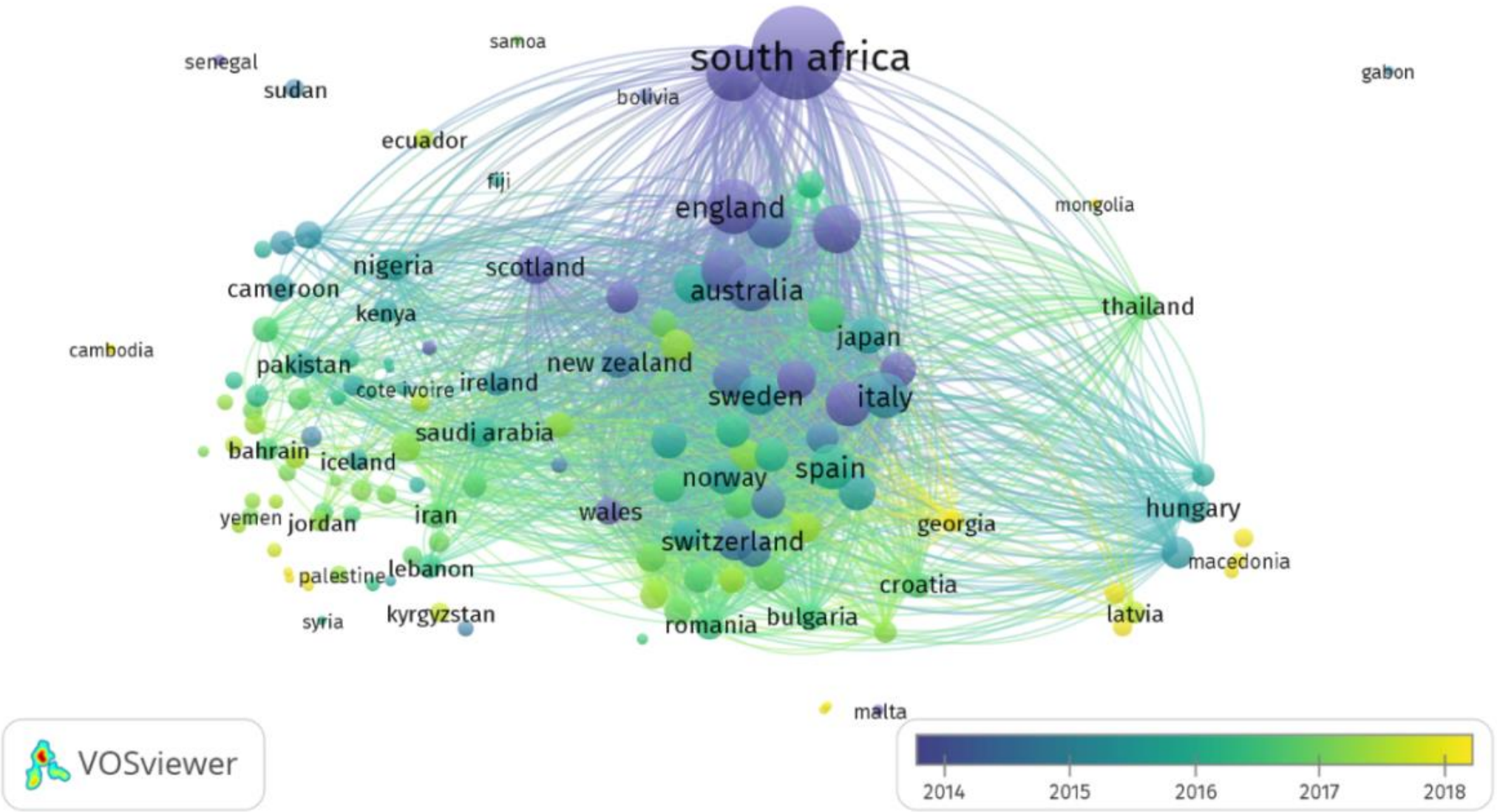

**Figure 8: Collaborating Countries with South Africa**

The table highlights the country-wise collaboration of South Africa on coronary artery disease research output during the study period. Overall, 1284 records were published in coronary artery disease research from South Africa during three decades of study from 1990 to 2019. The top 10 collaborating countries are listed in the table for analysis. Among the United States of America collaborated 523 (40.73%) publications, followed by UK 266 (20.72%), Netherlands 79 (6.15%), Ireland 38 (2.96%), Germany 34 (2.65%), Switzerland 30 (2.34%), Italy 17 (1.32%), Canada 15 (1.17%), United Arab Emirates 7 (0.55%), France 6 (0.47%), and other countries collaborated 269 (20.95%) publications with South Africa. It is to be noted that South African scientists contributed 198 publications without collaborating other countries.

It can be seen from this table that collaborated country-wise average publications from the United States of America has shown the maximum average number of publications per year, i.e., 17.43. The UK has an average number of publications with 8.87, followed by Netherlands 2.63, Ireland 1.27, Germany 1.13, Switzerland 1.00, and the rest of the countries have less than one as the average number of publications per year. The United States of America is the highest collaborating nation in the case of an average number of publications per year.

The United States of America emerged as the top collaborative contributing country in coronary artery disease research with South Africa. The UK ranked the second, Netherlands, which ranked the third, and Ireland is ranked fourth. Germany is ranked in fifth place. This analysis helps in identifying the countries which have taken up the research work in the field. The developed countries are concentrating more on coronary artery disease research than developing and underdeveloped countries. The average number of publications from each country has been calculated by dividing the total publications by the number of years 30, and the same is shown in the table. The above table revealed that the United Arab Emirates, a developing country, is also contributing to the research output on coronary artery disease as par with developed countries.

### 4.9.6 Analysis of Overall Collaboration of BRICS with Other Countries

**Table 9F: Overall Collaboration of BRICS**

| S.No. | Collaborating Country | Brazil | Russia | India | China | South Africa | Total | Percentage |
|---|---|---|---|---|---|---|---|---|
| 1 | USA | 2837 | 1535 | 1739 | 13982 | 523 | 20616 | 41.20 |
| 2 | UK | 991 | 471 | 869 | 6975 | 266 | 9572 | 19.13 |
| 3 | Netherlands | 328 | 128 | 470 | 2489 | 79 | 3494 | 6.98 |
| 4 | Switzerland | 158 | 125 | 123 | 1666 | 30 | 2102 | 4.20 |
| 5 | Ireland | 302 | 202 | 226 | 1281 | 38 | 2049 | 4.10 |
| 6 | Germany | 93 | 55 | 113 | 667 | 34 | 962 | 1.92 |
| 7 | Italy | 35 | 19 | 43 | 443 | 17 | 557 | 1.11 |
| 8 | France | 47 | 10 | 53 | 222 | 6 | 338 | 0.68 |
| 9 | U Arab Emirates | 36 | 29 | 65 | 232 | 7 | 369 | 0.74 |
| 10 | Spain | 24 | 1 | 25 | 22 | 3 | 75 | 0.15 |
| 11 | Other Countries | 1367 | 2483 | 980 | 4791 | 281 | 9902 | 19.79 |
| Grand Total | | 6218 | 5058 | 4706 | 32770 | 1284 | 50036 | 100.00 |

About 60 other countries collaborated with BRICS nations in 50036 coronary artery disease research papers during 1990-2019. These 50036 papers together registered 758573 citations, with 15.16 citations per paper. The USA, among foreign countries, contributed the largest share 20616 (41.20%) to BRICS international collaborative papers in coronary artery disease research, followed by UK 9572 (19.13%), Netherlands 3494 (6.98%), Switzerland 2102 (4.20%), Ireland 2049 (4.10%), Germany 962 (1.92%), Italy 557 (1.11%), United Arab Emirates, France, and Spain contributed less than one per cent each during 1990-2019.

The investigation reveals that the share of BRICS international collaborative publications (ICP) in overall output in coronary artery disease research is 20616 (41.20%) with the USA during 1990-2019 and other countries share is 9902 (19.79%) during the study period. It is to be noted that there is a high level of collaboration on coronary artery disease literature from BRICS nations.

### 4.9.7 Analysis of Overall Collaboration Among BRICS (Inter Collaboration)

**Table 9G: Overall Collaboration Among BRICS**

| S.No. | Collaborating Country | Brazil | Russia | India | China | South Africa | Total | Percentage |
|---|---|---|---|---|---|---|---|---|
| 1 | Brazil | 1187* | 2 | 12 | 238 | 1 | 1440 | 2.88 |
| 2 | Russia | 0 | 2381* | 2 | 0 | 0 | 2383 | 4.76 |
| 3 | India | 4 | 1 | 655* | 761 | 3 | 1424 | 2.85 |
| 4 | China | 8 | 8 | 11 | 1078* | 0 | 1105 | 2.21 |
| 5 | South Africa | 1 | 0 | 5 | 6 | 198* | 210 | 0.42 |
| Total | | 1200 | 2392 | 685 | 2083 | 202 | 6562 | 13.11 |
| BRICS Collaboration | | 13 | 11 | 30 | 1005 | 4 | 1063 | 2.12 |

*Sole output of country not calculated in total collaboration

The collaboration within BRICS nations is 1063 (2.12%) of the total output publications during thirty years of the period from 1990 through 2019. The Peoples Republic of China collaborated 1005 research papers with other four nations contributing 761 research papers with India, 238 with Brazil, six research papers with South Africa and no publications with Russia. India collaborated 12 research papers with Brazil, 11 with Peoples Republic of China, 5 with South Africa, and 2 with Russia. Brazil collaborated eight papers with the Peoples Republic of China, four papers with India, one paper with South Africa and zero collaboration with Russia. In the same way, Russia has 11 collaborative papers, having 8 with the Peoples Republic of China, two papers with Brazil, and one paper with India. Similarly, South Africa collaborated only four research papers with other BRICS nations collaborated three papers with India and one paper with Brazil and did not have any paper with other nations among BRICS.

## 4.10 BRICS Countries with H-Index and CPP

**Table 10: BRICS Countries with H-Index & CPP**

| S.No. | h-index | Countries | Citation Sum within h-core | All Citations | All Articles | CPP |
|---|---|---|---|---|---|---|
| 1 | 174 | Peoples R China | 82354 | 480352 | 32770 | 14.66 |
| 2 | 148 | Brazil | 77838 | 156941 | 6218 | 25.24 |
| 3 | 112 | India | 60968 | 116804 | 4706 | 24.82 |
| 4 | 103 | South Africa | 52183 | 71044 | 1284 | 55.33 |
| 5 | 90 | Russia | 45881 | 65585 | 5058 | 12.97 |
| Total | 627 | | 319224 | 890726 | 50036 | 17.80 |

It is noted the Peoples Republic of China is the most productive country in terms of h-index, and the country has collaborated the highest number of articles within BRICS countries. Brazil collaborated 6218 publications with 156941 citations in its credit and producing 25.24 citations per paper. The Peoples Republic of China, which has contributed the most significant number of publications among BRICS countries (32770), is having 14.66 citations per paper.

It is analyzed from the table, citation per paper BRICS countries are in the ranges from 12.97 to 55.33. Peoples Republic of China, Brazil, India, South Africa and Russia are ranked in terms of h-index. It is also observed that South Africa and Brazil have contributed a smaller number of publications in the study, but they are on top with respect to citation per papers. Russian research has the least number of citation and having a smaller number of citations per paper as well as 12.97. The overall citation per paper is 17.80 of BRICS countries.

## 4.11 Analysis of Activity Index of BRICS Countries

To achieve the third objective of to compare and measure the analysis of country-wise Coronary Artery Disease research output performance, the country wise analysis is carried out and it is explained in tables 8, 9A, 9B, 9C, 9D, 9E, 9F, 9G, and 10.

Activity index characterizes the relative research effort of a country in a given field, and it is explained as:

Activity Index suggested by (Price, 1981) and elaborated by (Karki & Garg, 1997) has been used. To measure the relative research effort of a country in a given field. Mathematically:

$$CAI = \left[ \frac{\left( \frac{C_i}{C_o} \right)}{\left( \frac{W_i}{W_o} \right)} \right] X100$$

Where as

$C_i$ = individual Country output in the year i

$C_o$ = Total of Individual Country output

$W_i$ = World output in the year i

$W_o$ = Total output

The Coronary Artery Disease research published country-wise year block periods along with the Activity Index is presented in Table 11.

In Table 11, the Activity Index for BRICS countries has been calculated. It is to analyze how the BRICS country's research performance change over different years. A comparison of the BRICS country's research performance with the world's research performance has been made using Activity Index calculation. The table indicates the average production of coronary artery disease research of each individual country per year and the total average production of coronary artery disease research output in BRICS. The activity index of BRICS countries from the year 2010 onwards has shown a growth except for South Africa. It remains low from the year 1990 to 2009 of the cumulated output of all the countries together. Brazil shows an increasing trend from the year 2007 as from 1991 to 2006 its activity index is under activity.

Russia's data show fluctuation up to the year 2014 and from 2015 it has shown an increasing trend. India activity index up to the year 2007 is low and from 2008 up to 2019 has shown positive activity index. China's growth has been growing from the year 2012 onwards, and it is low from the year 1990-2011. South Africa has a positive activity index from 1990 to 1993, and it decreased and then it fluctuates from the year 1994 onwards.

To achieve the fourth objective of measuring research productivity through the Activity index concerning countries during the study period, the analysis of activity index is carried out and it is interpreted in table 11.

**Table 11: Activity Index of BRICS Countries**

| S.No. | Year | Global Output | BRICS Output | Activity Index | Brazil | Activity Index | Russia | Activity Index | India | Activity Index | China | Activity Index | South Africa | Activity Index |
|---|---|---|---|---|---|---|---|---|---|---|---|---|---|---|
| 1 | 1990 | 1808 | 158 | 71.41 | 31 | 112.75 | 14 | 62.60 | 27 | 129.76 | 75 | 51.76 | 11 | 193.75 |
| 2 | 1991 | 5429 | 266 | 40.04 | 55 | 66.62 | 45 | 67.01 | 39 | 62.42 | 98 | 22.52 | 29 | 170.11 |
| 3 | 1992 | 5927 | 234 | 32.26 | 33 | 36.61 | 54 | 73.65 | 30 | 43.98 | 77 | 16.21 | 40 | 214.92 |
| 4 | 1993 | 6234 | 182 | 23.86 | 14 | 14.77 | 70 | 90.77 | 25 | 34.84 | 39 | 7.81 | 34 | 173.68 |
| 5 | 1994 | 6754 | 196 | 23.72 | 22 | 21.42 | 102 | 122.09 | 26 | 33.45 | 30 | 5.54 | 16 | 75.44 |
| 6 | 1995 | 7254 | 197 | 22.19 | 32 | 29.01 | 78 | 86.93 | 24 | 28.75 | 38 | 6.54 | 25 | 109.75 |
| 7 | 1996 | 7803 | 291 | 30.48 | 40 | 33.71 | 115 | 119.14 | 45 | 50.11 | 55 | 8.80 | 36 | 146.92 |
| 8 | 1997 | 9154 | 297 | 26.51 | 49 | 35.20 | 129 | 113.92 | 49 | 46.51 | 37 | 5.04 | 33 | 114.80 |
| 9 | 1998 | 9379 | 311 | 27.10 | 38 | 26.64 | 102 | 87.92 | 53 | 49.10 | 83 | 11.04 | 35 | 118.84 |
| 10 | 1999 | 9908 | 343 | 28.29 | 57 | 37.83 | 98 | 79.96 | 45 | 39.46 | 112 | 14.10 | 31 | 99.64 |
| 11 | 2000 | 10630 | 399 | 30.67 | 73 | 45.16 | 124 | 94.30 | 55 | 44.96 | 115 | 13.50 | 32 | 95.87 |
| 12 | 2001 | 10350 | 360 | 28.42 | 66 | 41.93 | 86 | 67.17 | 55 | 46.17 | 127 | 15.31 | 26 | 80.00 |
| 13 | 2002 | 10500 | 490 | 38.14 | 91 | 56.99 | 148 | 113.95 | 58 | 48.00 | 173 | 20.56 | 20 | 60.66 |
| 14 | 2003 | 11525 | 559 | 39.64 | 106 | 60.48 | 132 | 92.59 | 71 | 53.53 | 228 | 24.68 | 22 | 60.79 |
| 15 | 2004 | 12602 | 737 | 47.79 | 143 | 74.62 | 178 | 114.19 | 96 | 66.19 | 298 | 29.51 | 22 | 55.59 |
| 16 | 2005 | 13274 | 794 | 48.88 | 173 | 85.71 | 167 | 101.71 | 110 | 72.00 | 321 | 30.17 | 23 | 55.18 |
| 17 | 2006 | 14061 | 994 | 57.77 | 208 | 97.28 | 154 | 88.54 | 104 | 64.27 | 496 | 44.02 | 32 | 72.47 |
| 18 | 2007 | 14842 | 1224 | 67.39 | 283 | 125.39 | 172 | 93.68 | 143 | 83.72 | 579 | 48.68 | 47 | 100.84 |
| 19 | 2008 | 16001 | 1486 | 75.89 | 254 | 104.39 | 169 | 85.38 | 208 | 112.95 | 819 | 63.87 | 36 | 71.65 |
| 20 | 2009 | 17194 | 1799 | 85.50 | 311 | 118.94 | 195 | 91.68 | 223 | 112.69 | 1014 | 73.59 | 56 | 103.72 |
| 21 | 2010 | 17502 | 2149 | 100.34 | 363 | 136.39 | 220 | 101.62 | 250 | 124.11 | 1275 | 90.90 | 41 | 74.60 |
| 22 | 2011 | 18510 | 2400 | 105.96 | 390 | 138.55 | 238 | 103.95 | 258 | 121.11 | 1471 | 99.16 | 43 | 73.98 |
| 23 | 2012 | 18924 | 2727 | 117.76 | 373 | 129.62 | 222 | 94.84 | 304 | 139.58 | 1759 | 115.98 | 69 | 116.11 |
| 24 | 2013 | 21037 | 3336 | 129.59 | 404 | 126.29 | 218 | 83.77 | 316 | 130.52 | 2332 | 138.32 | 66 | 99.91 |
| 25 | 2014 | 20598 | 3685 | 146.20 | 357 | 113.97 | 230 | 90.27 | 331 | 139.63 | 2700 | 163.56 | 67 | 103.59 |
| 26 | 2015 | 21551 | 4130 | 156.61 | 406 | 123.89 | 276 | 103.53 | 270 | 108.86 | 3108 | 179.95 | 70 | 103.44 |
| 27 | 2016 | 22254 | 4577 | 168.07 | 409 | 120.86 | 289 | 104.98 | 412 | 160.86 | 3385 | 189.80 | 82 | 117.34 |
| 28 | 2017 | 22454 | 4822 | 175.49 | 453 | 132.67 | 300 | 108.01 | 313 | 121.12 | 3668 | 203.83 | 88 | 124.81 |
| 29 | 2018 | 22478 | 5192 | 188.76 | 482 | 141.01 | 356 | 128.03 | 372 | 143.80 | 3902 | 216.60 | 80 | 113.34 |
| 30 | 2019 | 22959 | 5701 | 202.92 | 502 | 143.78 | 377 | 132.75 | 394 | 149.11 | 4356 | 236.74 | 72 | 99.87 |
| Total | | 408896 | 50036 | 100.00 | 6218 | 100.00 | 5058 | 100.00 | 4706 | 100.00 | 32770 | 100.00 | 1284 | 100.00 |

## 4.12 Analysis of Document Type

| Table 12: Document Wise Publications | | | | | |
|---|---|---|---|---|---|
| S.No. | Document Type | Records | Percentage | Cum. Publication | Cum. Percentage |
| 1 | Article | 39873 | 79.689 | 39873 | 79.689 |
| 2 | Review | 4102 | 8.198 | 43975 | 87.887 |
| 3 | Meeting Abstract | 3946 | 7.886 | 47921 | 95.773 |
| 4 | Letter | 708 | 1.415 | 48629 | 97.188 |
| 5 | Editorial Material | 554 | 1.107 | 49183 | 98.295 |
| 6 | Article; Proceedings Paper | 496 | 0.991 | 49679 | 99.287 |
| 7 | Article; Early Access | 127 | 0.254 | 49806 | 99.540 |
| 8 | Correction | 59 | 0.118 | 49865 | 99.658 |
| 9 | Article; Retracted Publication | 44 | 0.088 | 49909 | 99.746 |
| 10 | Note | 40 | 0.080 | 49949 | 99.826 |
| 11 | Article; Book Chapter | 24 | 0.048 | 49973 | 99.874 |
| 12 | Review; Early Access | 24 | 0.048 | 49997 | 99.922 |
| 13 | Review; Book Chapter | 22 | 0.044 | 50019 | 99.966 |
| 14 | News Item | 5 | 0.010 | 50024 | 99.976 |
| 15 | Reprint | 2 | 0.004 | 50026 | 99.980 |
| 16 | Review; Retracted Publication | 2 | 0.004 | 50028 | 99.984 |
| 17 | Editorial Material; Early Access | 2 | 0.004 | 50030 | 99.988 |
| 18 | Discussion | 2 | 0.004 | 50032 | 99.992 |
| 19 | Biographical-Item | 1 | 0.002 | 50033 | 99.994 |
| 20 | Article; Data Paper | 1 | 0.002 | 50034 | 99.996 |
| 21 | Retraction | 1 | 0.002 | 50035 | 99.998 |
| 22 | Editorial Material; Retracted Publication | 1 | 0.002 | 50036 | 100.000 |
| Total | | 50036 | 100.000 | | |

Table 12 shows the document type of distributions. It could be seen clearly from the table that article type of document have shown a predominant contribution (79.689%), and it occupies the first position concerning the total number of publications reported during the study period. The Review as a source on coronary artery disease is a productive output, follows next in order 4102 (8.189%) in terms of the full document of publication output found in this analysis. The Meeting Abstract is another type of document of productive research output that takes third to share 3946 (7.886%) output concerning the total number of publications examined in the study. The form Letter as a document of publication output slips down to fourth to 708 (1.415%) of output performance.

The Editorial Material as a document publication output takes the fifth position 554 (1.107%) of the analysis. The remaining document types are Article; Proceedings Paper 496 (0.991%), Article; Early Access 127 (0.254%), Correction 59 (0.118%), Article; Retracted Publication 44 (0.088%), Note 40 (0.080%), Article; Book Chapter 24 (0.048%), Review; Early Access 24 (0.048%), Review; Book Chapter 22 (0.044%), News Item 5 (0.010%), Reprint, Review; Retracted Publication, Editorial Material; Early Access, Discussion 2 (0.004%), Biographical-Item, Retraction, and Editorial Material; Retracted Publication 1 (0.002%).

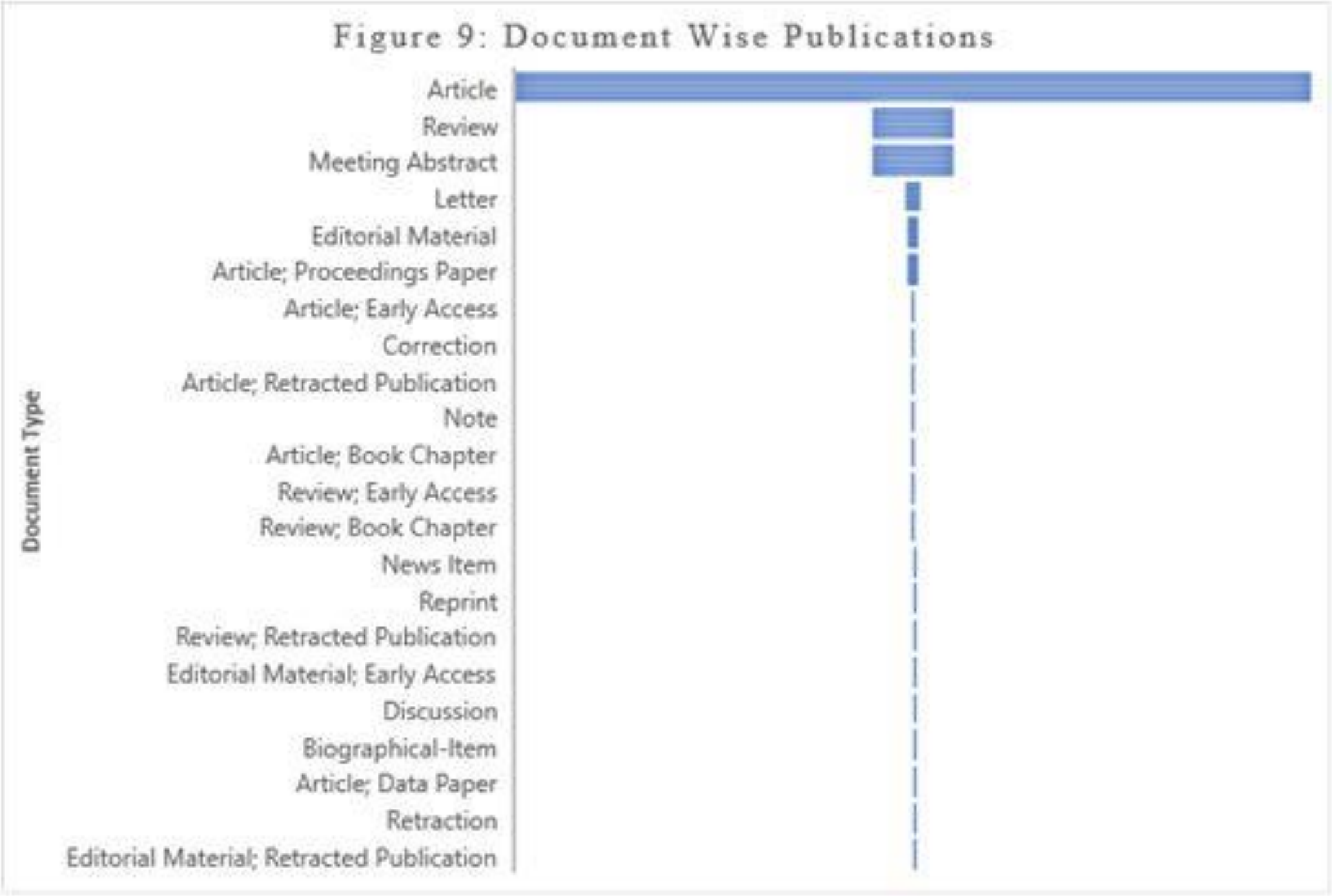

It could be deducted from the above discussion that journal articles are predominating over other sources of publications. The majority of the work on coronary artery disease by scientists preferred to publish their research papers in the form of articles. The outcome is supported by another research conducted on Aromatic Plants in which researchers also prefer to publish their research in the form of articles (Suresh, 2019). The other preferred forms of publications among the researchers are reviews, meeting abstracts, and letters. ***The Journal source of*** ***distribution on coronary artery disease output involves a predominant spot compared to other sources of productions. The second hypothesis has been proved.***

## 4.13 Analysis of Document Type Wise of Publications of BRICS Countries

**Table 13: Document Type Wise of Publications of BRICS**

| Country | Brazil | Russia | India | China | South Africa | Total |
|---|---|---|---|---|---|---|
| Article | 4686 | 3418 | 3315 | 27475 | 979 | 39873 |
| Review | 521 | 530 | 535 | 2351 | 165 | 4102 |
| Meeting Abstract | 573 | 980 | 539 | 1791 | 63 | 3946 |
| Letter | 113 | 7 | 129 | 446 | 13 | 708 |
| Editorial Material | 156 | 58 | 93 | 204 | 43 | 554 |
| Article; Proceedings Paper | 129 | 52 | 57 | 241 | 17 | 496 |
| Article; Early Access | 14 | 5 | 11 | 95 | 2 | 127 |
| Correction | 4 | 1 | 4 | 50 | 0 | 59 |
| Article; Retracted Publication | 1 | 0 | 4 | 39 | 0 | 44 |
| Note | 3 | 2 | 5 | 29 | 1 | 40 |
| Review; Early Access | 1 | 0 | 6 | 17 | 0 | 24 |
| Article; Book Chapter | 6 | 2 | 2 | 13 | 1 | 24 |
| Review; Book Chapter | 6 | 1 | 3 | 12 | 0 | 22 |
| News Item | 3 | 2 | 0 | 0 | 0 | 5 |
| Reprint | 0 | 0 | 2 | 0 | 0 | 2 |
| Review; Retracted Publication | 1 | 0 | 0 | 1 | 0 | 2 |
| Editorial Material; Early Access | 1 | 0 | 0 | 1 | 0 | 2 |
| Discussion | 0 | 0 | 1 | 1 | 0 | 2 |
| Article; Data Paper | 0 | 0 | 0 | 1 | 0 | 1 |
| Retraction | 0 | 0 | 0 | 1 | 0 | 1 |
| Editorial Material; Retracted Publication | 0 | 0 | 0 | 1 | 0 | 1 |
| Biographical Item | 0 | 0 | 0 | 1 | 0 | 1 |
| Total | 6218 | 5058 | 4706 | 32770 | 1284 | 50036 |

The document type analysis on coronary artery disease research further analyzed of BRICS countries. The country-wise distribution of the document form is demonstrated in table 13.

In terms of Article type, the People Republic China shows higher publications (27475), followed by Brazil (4686), Russia (3418), India (3315), and South Africa (979 and it is evident that the total article accumulated to 39873 over three decades of study. Based on the Review, the People Republic of China again shows a higher record (2351), followed by India (535), Russia has 530, Brazil has 521 and South Africa is having 165 and the total is 4102.

Based on Meeting Abstract, the People Republic of China shows higher publications (1791), followed by Russia (980), Brazil has 573, India has 539, and South Africa contributed 63 in Meeting Abstract publications. Again, in terms of Letter document type, People Republic China shows higher records (446), followed by India (129), and Brazil (113), and other two countries Letter records are below 100. The minimum record of letter document type is of Russia (7). The remaining document types Editorial Material shows the highest records published by the People Republic China (204) and lowest records by South Africa as 43. Article; Proceedings shows highest records published by People Republic China (241) and lowest records published by South Africa (17).

Other document types have shown less contribution to coronary artery disease research irrespective of the country. This table concluded that the People Republic of China (32770) country leads in document publications. The minimum document publication contributed by South Africa (1284) in this table analysis.

### 4.14 Analysis of Year Wise Document Type of Publications

The below table 14 presents the world coronary artery disease research publications which have been grouped into 30 years broad categories with year-wise distributions from which they have been called such as Article, Review, Meeting Abstract, Letter, Editorial Material, Article; Proceedings Paper, Article; Early Access, Correction, etc. for thirty years.

It could be seen from the analysis of data presented in table 14 that the Article shows the highest value records 39873, within this journal article 2019 shows maximum records published (4446) and minimum records 130 in the year of 1990.
The document type 'Review' shows that the highest records were published in the year 2019 (612), and the lowest records were published in 1993 & 1995 (7). While 'Meeting Abstract,' which shows the highest records published in the year 2018 (362) and lowest Meeting Abstract publications in the year 1991 (2). The 'Letter' document type shows that the highest records published in 2015 (79) and the lowest 'Meeting Abstract' publication in the year 1994 (0).

## Table 14: Year Wise Document Type of Publications

| Document Type | 1990 | 1991 | 1992 | 1993 | 1994 | 1995 | 1996 | 1997 | 1998 | 1999 | 2000 | 2001 | 2002 | 2003 | 2004 | 2005 | 2006 | 2007 | 2008 | 2009 | 2010 | 2011 | 2012 | 2013 | 2014 | 2015 | 2016 | 2017 | 2018 | 2019 | Total |
|---|---|---|---|---|---|---|---|---|---|---|---|---|---|---|---|---|---|---|---|---|---|---|---|---|---|---|---|---|---|---|---|
| Article | 130 | 241 | 203 | 150 | 162 | 167 | 235 | 215 | 237 | 281 | 320 | 294 | 379 | 428 | 557 | 592 | 762 | 983 | 1165 | 1409 | 1639 | 1955 | 2198 | 2655 | 2994 | 3386 | 3726 | 3902 | 4062 | 4446 | **39873** |
| Review | 8 | 12 | 14 | 7 | 15 | 7 | 12 | 26 | 17 | 19 | 26 | 19 | 31 | 36 | 45 | 47 | 47 | 78 | 90 | 118 | 132 | 163 | 217 | 225 | 272 | 308 | 368 | 526 | 605 | 612 | **4102** |
| Meeting Abstract | 6 | 2 | 6 | 6 | 4 | 4 | 20 | 31 | 25 | 20 | 23 | 31 | 42 | 60 | 90 | 115 | 128 | 104 | 142 | 188 | 279 | 214 | 209 | 338 | 304 | 276 | 344 | 273 | 362 | 300 | **3946** |
| Letter | 1 | 3 | 2 | 1 | 0 | 3 | 6 | 2 | 8 | 7 | 4 | 5 | 4 | 4 | 10 | 8 | 11 | 20 | 31 | 25 | 40 | 28 | 46 | 59 | 70 | 79 | 51 | 45 | 64 | 71 | **708** |
| Editorial Material | 0 | 1 | 2 | 2 | 1 | 3 | 3 | 6 | 4 | 3 | 2 | 2 | 3 | 7 | 6 | 11 | 9 | 14 | 22 | 28 | 34 | 17 | 35 | 34 | 27 | 42 | 55 | 40 | 67 | 74 | **554** |
| Article; Proceedings Paper | 4 | 5 | 5 | 3 | 7 | 4 | 15 | 16 | 20 | 11 | 22 | 9 | 30 | 22 | 26 | 19 | 34 | 23 | 33 | 27 | 19 | 17 | 11 | 12 | 7 | 26 | 16 | 20 | 14 | 19 | **496** |
| Article; Early Access | 0 | 0 | 0 | 0 | 0 | 0 | 0 | 0 | 0 | 0 | 0 | 0 | 0 | 0 | 0 | 0 | 0 | 0 | 0 | 0 | 0 | 0 | 0 | 0 | 0 | 0 | 0 | 1 | 0 | 126 | **127** |
| Correction | 0 | 0 | 0 | 0 | 0 | 0 | 0 | 1 | 0 | 1 | 2 | 0 | 0 | 2 | 0 | 1 | 0 | 1 | 0 | 1 | 3 | 1 | 2 | 4 | 1 | 5 | 8 | 8 | 7 | 11 | **59** |
| Article; Retracted Publication | 0 | 0 | 0 | 0 | 0 | 0 | 0 | 0 | 0 | 1 | 0 | 0 | 0 | 0 | 3 | 1 | 2 | 1 | 3 | 3 | 0 | 1 | 6 | 5 | 7 | 4 | 4 | 3 | 0 | 0 | **44** |
| Note | 9 | 2 | 2 | 12 | 6 | 9 | 0 | 0 | 0 | 0 | 0 | 0 | 0 | 0 | 0 | 0 | 0 | 0 | 0 | 0 | 0 | 0 | 0 | 0 | 0 | 0 | 0 | 0 | 0 | 0 | **40** |
| Review; Early Access | 0 | 0 | 0 | 0 | 0 | 0 | 0 | 0 | 0 | 0 | 0 | 0 | 0 | 0 | 0 | 0 | 0 | 0 | 0 | 0 | 0 | 0 | 0 | 0 | 0 | 0 | 0 | 0 | 0 | 24 | **24** |
| Article; Book Chapter | 0 | 0 | 0 | 0 | 0 | 0 | 0 | 0 | 0 | 0 | 0 | 0 | 0 | 0 | 0 | 0 | 0 | 0 | 0 | 0 | 1 | 0 | 0 | 0 | 0 | 1 | 1 | 2 | 9 | 10 | **24** |
| Review; Book Chapter | 0 | 0 | 0 | 0 | 0 | 0 | 0 | 0 | 0 | 0 | 0 | 0 | 0 | 0 | 0 | 0 | 0 | 0 | 0 | 0 | 2 | 2 | 2 | 2 | 3 | 2 | 1 | 2 | 1 | 5 | **22** |
| News Item | 0 | 0 | 0 | 0 | 0 | 0 | 0 | 0 | 0 | 0 | 0 | 0 | 0 | 0 | 0 | 0 | 1 | 0 | 0 | 0 | 0 | 0 | 0 | 0 | 0 | 1 | 2 | 0 | 0 | 1 | **5** |
| Discussion | 0 | 0 | 0 | 1 | 1 | 0 | 0 | 0 | 0 | 0 | 0 | 0 | 0 | 0 | 0 | 0 | 0 | 0 | 0 | 0 | 0 | 0 | 0 | 0 | 0 | 0 | 0 | 0 | 0 | 0 | **2** |
| Reprint | 0 | 0 | 0 | 0 | 0 | 0 | 0 | 0 | 0 | 0 | 0 | 0 | 0 | 0 | 0 | 0 | 0 | 0 | 0 | 0 | 0 | 1 | 1 | 0 | 0 | 0 | 0 | 0 | 0 | 0 | **2** |
| Review; Retracted Publication | 0 | 0 | 0 | 0 | 0 | 0 | 0 | 0 | 0 | 0 | 0 | 0 | 0 | 0 | 0 | 0 | 0 | 0 | 0 | 0 | 0 | 1 | 0 | 1 | 0 | 0 | 0 | 0 | 0 | 0 | **2** |
| Editorial Material; Early Access | 0 | 0 | 0 | 0 | 0 | 0 | 0 | 0 | 0 | 0 | 0 | 0 | 0 | 0 | 0 | 0 | 0 | 0 | 0 | 0 | 0 | 0 | 0 | 0 | 0 | 0 | 0 | 0 | 0 | 2 | **2** |
| Retraction | 0 | 0 | 0 | 0 | 0 | 0 | 0 | 0 | 0 | 0 | 0 | 0 | 0 | 0 | 0 | 0 | 0 | 0 | 0 | 0 | 0 | 0 | 0 | 0 | 0 | 0 | 1 | 0 | 0 | 0 | **1** |
| Biographical-Item | 0 | 0 | 0 | 0 | 0 | 0 | 0 | 0 | 0 | 0 | 0 | 0 | 1 | 0 | 0 | 0 | 0 | 0 | 0 | 0 | 0 | 0 | 0 | 0 | 0 | 0 | 0 | 0 | 0 | 0 | **1** |
| Editorial Material; Retracted Publication | 0 | 0 | 0 | 0 | 0 | 0 | 0 | 0 | 0 | 0 | 0 | 0 | 0 | 0 | 0 | 0 | 0 | 0 | 0 | 0 | 0 | 0 | 0 | 1 | 0 | 0 | 0 | 0 | 0 | 0 | **1** |
| Article; Data Paper | 0 | 0 | 0 | 0 | 0 | 0 | 0 | 0 | 0 | 0 | 0 | 0 | 0 | 0 | 0 | 0 | 0 | 0 | 0 | 0 | 0 | 0 | 0 | 0 | 0 | 0 | 0 | 0 | 1 | 0 | **1** |
| **Total** | **158** | **266** | **234** | **182** | **196** | **197** | **291** | **297** | **311** | **343** | **399** | **360** | **490** | **559** | **737** | **794** | **994** | **1224** | **1486** | **1799** | **2149** | **2400** | **2727** | **3336** | **3685** | **4130** | **4577** | **4822** | **5192** | **5701** | **50036** |

The 'Editorial Material' shows the highest records published in the year 2019 (74) and lowest 'Editorial Material' published in the year 1990 (0).

It is observed in the study 'Article; Proceedings Paper' shows that the highest records were published in the year 2006 (34) and the lowest records were published in the year 1993 (3). Based on Article; Early Access shows that the highest records were published in the year 2019. 'Correction' shows the highest records published in the year 2019 (11). The document type 'Article; Retracted Publication' shows that the highest records were published in 2014 (7) and no publications in the year 1990 to 1998, 2000 to 2003, 2010, 2018, and 2019. Regarding the 'Note' publications, the highest is in the year 1993. On the other hand, 'Review; Early Access' shows the highest publications in 2019 (24) and the lowest publications from 1990 to 2018 (0).

Based on Article; Book Chapter shows that highest publications in the year of 2018 (10) and no publications from 1990 to 2009, 2011 to 2014. The remaining document types have less than 24 publications.

### 4.15 Analysis of Language Wise Publication

**Table 15: Language Wise Publications**

| S.No. | Language | Publications | Percentage | Cum. Records | Cum. % |
|---|---|---|---|---|---|
| 1 | English | 46660 | 93.253 | 46660 | 93.253 |
| 2 | Russian | 2874 | 5.744 | 49534 | 98.997 |
| 3 | Portuguese | 376 | 0.751 | 49910 | 99.748 |
| 4 | Chinese | 70 | 0.140 | 49980 | 99.888 |
| 5 | Spanish | 44 | 0.088 | 50024 | 99.976 |
| 6 | French | 5 | 0.010 | 50029 | 99.986 |
| 7 | German | 4 | 0.008 | 50033 | 99.994 |
| 8 | Czech | 1 | 0.002 | 50034 | 99.996 |
| 9 | Serbian | 1 | 0.002 | 50035 | 99.998 |
| 10 | Japanese | 1 | 0.002 | 50036 | 100.000 |
| | | 50036 | 100.000 | | |

This analysis of the language-wise distribution of research output in any field is one of the key factors of the communication of research information. The researchers worldwide do not know all languages. Generally, English is a medium of research communication as it is widely recognized all over the world. However, a few research papers have been published in other languages. In this study, the quantitative study of language-wise production, an attempt has been made to show the results. This type of analysis enables one to identify the most preferred language in publishing coronary artery disease research output. Table 15 presents data of ten languages through which brought out the coronary artery disease research output.

It can be seen from table 15 and figure 10 that English has been used as a significant communication language for coronary artery disease publications. Nearly 93.532% of publications appear in the English language and dominates in the first place out of ten languages, followed by Russian (5.744%), Portuguese (0.751%), Chinese (0.140%), and Spanish (0.088%). The remaining language's contributions are less than six articles in coronary artery disease. The results are substantiated by (Maghsoudi et al., 2020), which also investigated that the English language is preferred almost 95% of scientists to communicate their scholarly output.

Figure 10: Language Wise Publications

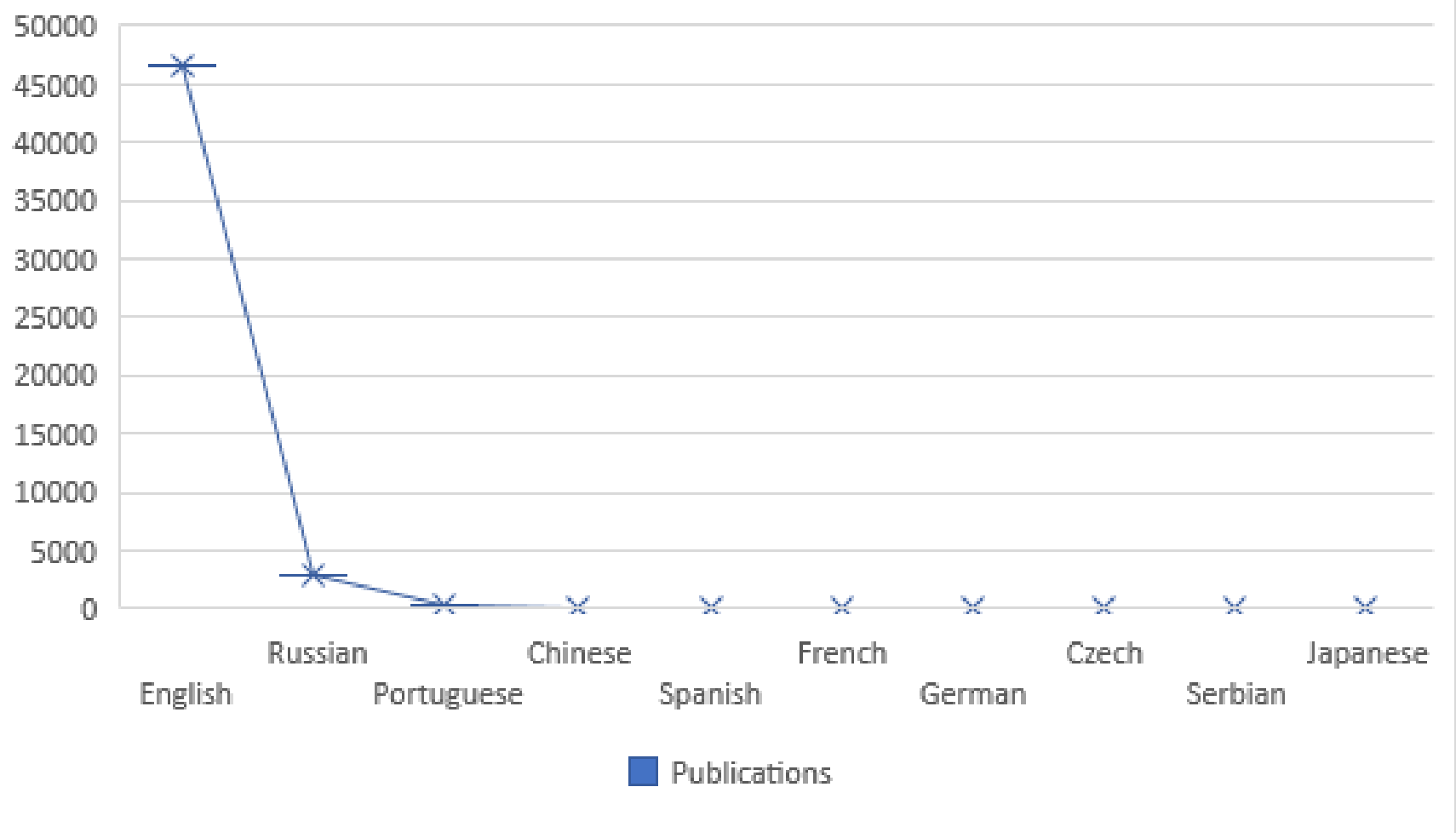

The table demonstrates that the English language (93.253%) is the prime language channel for coronary artery disease research productivity. It dominates in the first place out of ten languages; the remaining languages were in regional languages. Russian, Portuguese, and Chinese are other languages through which literature on coronary artery disease found have been brought out.

To achieve the fifth objective of determining the document wise research concentrations in Coronary Artery Disease research in the study period, the analysis of document wise research is carried out and it is examined in tables 13 and 14.

## 4.16 Year Wise Distribution of Language

**Table 16: Language Wise Distribution of Publications Per Year**

| **Languages** | 1990 | 1991 | 1992 | 1993 | 1994 | 1995 | 1996 | 1997 | 1998 | 1999 | 2000 | 2001 | 2002 | 2003 | 2004 | Total |
|---|---|---|---|---|---|---|---|---|---|---|---|---|---|---|---|---|
| English | 56 | 113 | 124 | 128 | 116 | 138 | 210 | 197 | 231 | 262 | 298 | 295 | 378 | 451 | 606 | 3603 |
| Russian | 95 | 148 | 104 | 52 | 76 | 55 | 78 | 92 | 80 | 74 | 96 | 58 | 105 | 100 | 124 | 1337 |
| Portuguese | 0 | 0 | 1 | 0 | 0 | 2 | 2 | 7 | 0 | 5 | 3 | 6 | 2 | 1 | 5 | 34 |
| Chinese | 4 | 5 | 4 | 1 | 3 | 2 | 0 | 0 | 0 | 1 | 0 | 1 | 3 | 3 | 1 | 28 |
| Spanish | 1 | 0 | 1 | 0 | 0 | 0 | 1 | 0 | 0 | 1 | 2 | 0 | 1 | 4 | 1 | 12 |
| French | 0 | 0 | 0 | 0 | 0 | 0 | 0 | 0 | 0 | 0 | 0 | 0 | 1 | 0 | 0 | 1 |
| German | 1 | 0 | 0 | 1 | 1 | 0 | 0 | 1 | 0 | 0 | 0 | 0 | 0 | 0 | 0 | 4 |
| Japanese | 1 | 0 | 0 | 0 | 0 | 0 | 0 | 0 | 0 | 0 | 0 | 0 | 0 | 0 | 0 | 1 |
| Czech | 0 | 0 | 0 | 0 | 0 | 0 | 0 | 0 | 0 | 0 | 0 | 0 | 0 | 0 | 0 | 0 |
| Serbian | 0 | 0 | 0 | 0 | 0 | 0 | 0 | 0 | 0 | 0 | 0 | 0 | 0 | 0 | 0 | 0 |
| | 158 | 266 | 234 | 182 | 196 | 197 | 291 | 297 | 311 | 343 | 399 | 360 | 490 | 559 | 737 | 5020 |

| **Languages** | 2005 | 2006 | 2007 | 2008 | 2009 | 2010 | 2011 | 2012 | 2013 | 2014 | 2015 | 2016 | 2017 | 2018 | 2019 | Total | Grand Total |
|---|---|---|---|---|---|---|---|---|---|---|---|---|---|---|---|---|---|
| English | 669 | 885 | 1038 | 1316 | 1621 | 1960 | 2232 | 2562 | 3206 | 3607 | 4035 | 4487 | 4738 | 5105 | 5596 | 43057 | 46660 |
| Russian | 109 | 97 | 113 | 119 | 116 | 120 | 139 | 137 | 107 | 68 | 82 | 80 | 71 | 79 | 100 | 1537 | 2874 |
| Portuguese | 10 | 5 | 66 | 45 | 57 | 61 | 24 | 25 | 18 | 8 | 7 | 4 | 7 | 3 | 2 | 342 | 376 |
| Chinese | 3 | 1 | 0 | 3 | 3 | 4 | 2 | 3 | 3 | 2 | 6 | 3 | 5 | 2 | 2 | 42 | 70 |
| Spanish | 2 | 6 | 5 | 3 | 2 | 4 | 3 | 0 | 2 | 0 | 0 | 1 | 1 | 3 | 0 | 32 | 44 |
| French | 1 | 0 | 2 | 0 | 0 | 0 | 0 | 0 | 0 | 0 | 0 | 0 | 0 | 0 | 1 | 4 | 5 |
| German | 0 | 0 | 0 | 0 | 0 | 0 | 0 | 0 | 0 | 0 | 0 | 0 | 0 | 0 | 0 | 0 | 4 |
| Japanese | 0 | 0 | 0 | 0 | 0 | 0 | 0 | 0 | 0 | 0 | 0 | 0 | 0 | 0 | 0 | 0 | 1 |
| Czech | 0 | 0 | 0 | 0 | 0 | 0 | 0 | 0 | 0 | 0 | 0 | 1 | 0 | 0 | 0 | 1 | 1 |
| Serbian | 0 | 0 | 0 | 0 | 0 | 0 | 0 | 0 | 0 | 0 | 0 | 1 | 0 | 0 | 0 | 1 | 1 |
| Total | 794 | 994 | 1224 | 1486 | 1799 | 2149 | 2400 | 2727 | 3336 | 3685 | 4130 | 4577 | 4822 | 5192 | 5701 | 45016 | 50036 |
| | 952 | 1260 | 1458 | 1668 | 1995 | 2346 | 2691 | 3024 | 3647 | 4028 | 4529 | 4937 | 5312 | 5751 | 6438 | 50036 | |

Table 16 shows that year-wise language distributions. The analysis reveals that English has the highest output in 2019 and the minimum output in 1990. Russian has a maximum output in the year 2011 and the lowest output in the year 1993. Portuguese has the most incredible output in the year 2007 and lowest output in the years 1990, 1991, 1993, 1994, and 1998; Chinese has maximum output records in the year 2015 and lowest output records in the year 1996, 1997, 1998 and 2000; Spanish has the higher output respectively in the year 2006 and many years has zero output.

The English language publications are increasing over the years, indicating that those who have contributed to their native language also started publishing their research output in English. The year 2019 has recorded the highest publications in the English language (5596).

### 4.17 Language Wise Distributions of Publications of BRICS Country

The study of Language-wise distribution of BRICS countries on coronary artery disease research analyzed Language-wise distributions. The Language-wise distribution of the countries is revealed in Table 17.

This analysis reveals that the Peoples Republic of China has the highest output in the English language (32694), followed by Chinese (70), French (2), Portuguese, Serbian and Spanish (1) each. The study indicates that Brazil has the highest publications in English (5799), followed by Portuguese (375), Spanish (41), and Czech language having one publication. It is also observed that Russia has the highest publications in English (2180), followed by Russian (2874) and the Spanish language has one publication.

The investigation shows that India has the highest publications in English languages (4704), followed by French and Spanish (1) each, and have no publications in other languages. It is also revealed that South Africa has the highest publications in the English language (1283), followed by Germany, having (1) publication and no contribution in other languages. It is noted that English remains the medium of scientific communication of authors from BRICS countries as a whole.

To achieve the sixth objective of evaluating the language-wise distribution in Coronary Artery Disease research, the analysis of language-wise distribution is carried out and it is revealed in tables 15, 16, and 17.

| Table 17: Language Wise Distribution of Language of Publications | | | | | | | | | | | |
|---|---|---|---|---|---|---|---|---|---|---|---|
| Countries | Chinese | Czech | English | French | German | Portuguese | Russian | Serbian | Spanish | Japanese | Total |
| Peoples R China | 70 | 0 | 32694 | 2 | 0 | 1 | 0 | 1 | 1 | 1 | 32770 |
| Brazil | 0 | 1 | 5799 | 2 | 0 | 375 | 0 | 0 | 41 | 0 | 6218 |
| Russia | 0 | 0 | 2180 | 0 | 3 | 0 | 2874 | 0 | 1 | 0 | 5058 |
| India | 0 | 0 | 4704 | 1 | 0 | 0 | 0 | 0 | 1 | 0 | 4706 |
| South Africa | 0 | 0 | 1283 | 0 | 1 | 0 | 0 | 0 | 0 | 0 | 1284 |
| Total | 70 | 1 | 46660 | 5 | 4 | 376 | 2874 | 1 | 44 | 1 | 50036 |

## 4.18 Authorship Pattern on Coronary Artery Disease

Authorship pattern for the literature on coronary artery disease has also been examined. The study of authorship patterns of productivity is one of the crucial aspects of the scientometric analysis. It is necessary to concentrate on authorship patterns to assess the research contributions in any field and coronary artery disease is not an exception.

There are various studies regarding authorship productivity and attempt has been to conduct some quantitative aspects which can be highlighted below:

- Authorship pattern
- Collaboration Index (CI)
- Degree of Collaboration (DC)
- Collaboration Coefficient (CC)
- Modified Collaborative Coefficient (MCC)
- Co-authorship Index (CAI)

The table demonstrates the authorship pattern in coronary artery disease research output. It could be noted that six authorship pattern published 6642 papers on coronary artery disease research, constituting 13.27% of the total publications. It is observed that five authorship pattern of authors published 6146 papers on coronary artery disease research, constituting 12.28% of the total publications. It is observed that seven authors published 5588 papers on coronary artery disease research, consisting of 11.17% of the total publications. It is noted that four authors published 5387 papers in coronary artery disease research, constituting 10.77% of the total publications, followed by eight authors pattern published 4719 papers on coronary artery disease research, constituting 9.43% of the total publications. The remaining authors published less than 9 per cent of publications on CAD.

It could be seen that the six authors' contributions ranked first in order (13.27%) with respect to the total number of output during the study period of analysis. It is clear from the table and Figure 11 that nearly 97.91% of research output published collaboratively either by double authors and more than two authors in the case of BRICS countries publications on CAD. Single authors' contribution has been noted in about 2.09% of publications.

**Table 18: Authorship Pattern**

| Authorship Pattern | Publications | Percentage |
|---|---|---|
| Single Author | 1044 | 2.09 |
| Double Authors | 2900 | 5.80 |
| Three Authors | 4249 | 8.49 |
| Four Authors | 5387 | 10.77 |
| Five Authors | 6146 | 12.28 |
| Six Authors | 6642 | 13.27 |
| Seven Authors | 5588 | 11.17 |
| Eight Authors | 4719 | 9.43 |
| Nine Authors | 3593 | 7.18 |
| Ten and > 10 Authors | 9768 | 19.52 |
| Total | 50036 | 100.00 |

**Figure 11: Authorship Pattern**

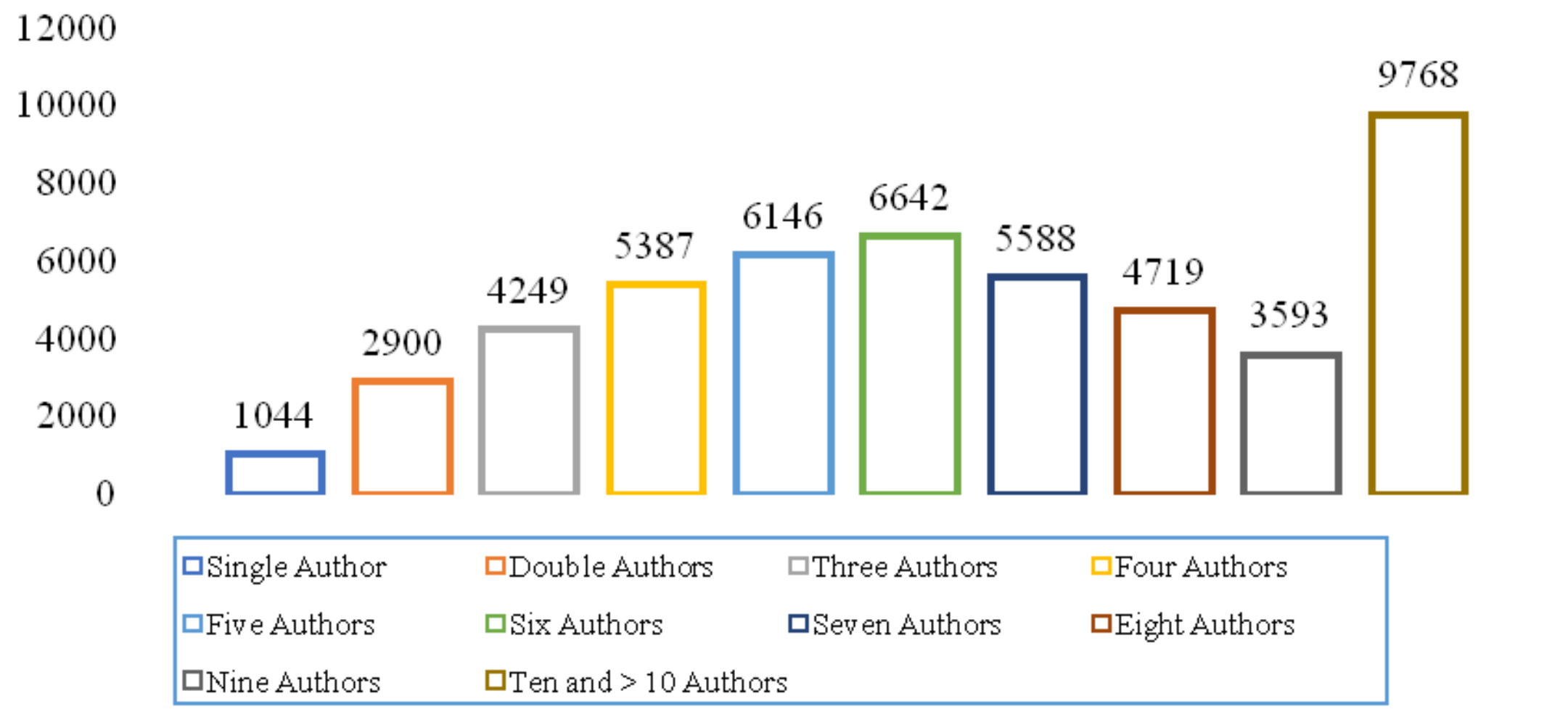

## 4.19 Year Wise Authorship Pattern on CAD of BRICS Countries

### 4.19.1 Year Wise Authorship Pattern of Brazil

**Table 19A: Year Wise Authorship Pattern of Brazil on CAD**

| Year | 1 | 2 | 3 | 4 | 5 | 6 | 7 | 8 | 9 | 10 and > 10 | Total | % |
|---|---|---|---|---|---|---|---|---|---|---|---|---|
| 1990 | 1 | 0 | 1 | 1 | 2 | 1 | 0 | 0 | 0 | 0 | 6 | 0.10 |
| 1991 | 1 | 2 | 3 | 2 | 1 | 1 | 0 | 0 | 0 | 1 | 11 | 0.18 |
| 1992 | 1 | 2 | 0 | 1 | 1 | 1 | 1 | 2 | 2 | 2 | 13 | 0.21 |
| 1993 | 0 | 1 | 2 | 2 | 2 | 0 | 4 | 2 | 0 | 1 | 14 | 0.23 |
| 1994 | 0 | 0 | 1 | 5 | 4 | 5 | 2 | 1 | 2 | 2 | 22 | 0.35 |
| 1995 | 3 | 3 | 1 | 5 | 4 | 2 | 10 | 2 | 0 | 2 | 32 | 0.51 |
| 1996 | 2 | 2 | 7 | 10 | 3 | 5 | 3 | 3 | 1 | 4 | 40 | 0.64 |
| 1997 | 2 | 7 | 6 | 5 | 5 | 5 | 6 | 5 | 4 | 4 | 49 | 0.79 |
| 1998 | 2 | 2 | 3 | 2 | 8 | 5 | 5 | 5 | 1 | 5 | 38 | 0.61 |
| 1999 | 1 | 1 | 5 | 14 | 7 | 9 | 12 | 3 | 5 | 7 | 64 | 1.03 |
| 2000 | 3 | 5 | 6 | 9 | 8 | 17 | 11 | 5 | 3 | 6 | 73 | 1.17 |
| 2001 | 0 | 6 | 4 | 5 | 15 | 12 | 7 | 3 | 5 | 9 | 66 | 1.06 |
| 2002 | 0 | 8 | 8 | 13 | 16 | 10 | 10 | 10 | 3 | 13 | 91 | 1.46 |
| 2003 | 3 | 7 | 10 | 13 | 9 | 19 | 17 | 10 | 8 | 11 | 107 | 1.72 |
| 2004 | 3 | 8 | 6 | 15 | 24 | 23 | 14 | 14 | 8 | 23 | 138 | 2.22 |
| 2005 | 1 | 5 | 12 | 21 | 32 | 10 | 22 | 23 | 10 | 38 | 174 | 2.80 |
| 2006 | 3 | 7 | 20 | 20 | 22 | 33 | 27 | 38 | 15 | 22 | 207 | 3.33 |
| 2007 | 9 | 11 | 26 | 23 | 37 | 52 | 34 | 26 | 23 | 48 | 289 | 4.65 |
| 2008 | 5 | 15 | 35 | 29 | 32 | 38 | 34 | 28 | 19 | 41 | 276 | 4.44 |
| 2009 | 13 | 16 | 24 | 29 | 50 | 53 | 31 | 23 | 18 | 56 | 313 | 5.03 |
| 2010 | 14 | 19 | 26 | 38 | 49 | 47 | 52 | 29 | 23 | 71 | 368 | 5.92 |
| 2011 | 5 | 19 | 37 | 48 | 55 | 47 | 38 | 41 | 23 | 78 | 391 | 6.29 |
| 2012 | 5 | 16 | 27 | 43 | 41 | 46 | 37 | 46 | 32 | 83 | 376 | 6.05 |
| 2013 | 3 | 28 | 26 | 35 | 46 | 38 | 49 | 51 | 24 | 123 | 423 | 6.80 |
| 2014 | 6 | 17 | 26 | 28 | 36 | 52 | 43 | 40 | 36 | 104 | 388 | 6.24 |
| 2015 | 6 | 22 | 17 | 32 | 40 | 60 | 31 | 39 | 38 | 131 | 416 | 6.69 |
| 2016 | 8 | 22 | 36 | 42 | 40 | 55 | 38 | 35 | 26 | 111 | 413 | 6.64 |
| 2017 | 7 | 12 | 25 | 32 | 36 | 49 | 35 | 44 | 37 | 176 | 453 | 7.29 |
| 2018 | 5 | 30 | 28 | 36 | 36 | 49 | 36 | 45 | 41 | 163 | 469 | 7.54 |
| 2019 | 14 | 30 | 26 | 29 | 41 | 50 | 47 | 42 | 31 | 188 | 498 | 8.01 |
| Total | 126 | 323 | 454 | 587 | 702 | 794 | 656 | 615 | 438 | 1523 | 6218 | 100.00 |
| % | 2.03 | 5.19 | 7.30 | 9.44 | 11.29 | 12.77 | 10.55 | 9.89 | 7.04 | 24.49 | 100.00 | |

The authorship pattern is indicated in the table from 1990 to 2019. The researcher has categorized the authorship pattern as a single author, double authors, three authors, four authors, five authors, six authors, seven authors, eight authors, nine authors, and ten and above authors produced by year wise.

There is an indication of the degree of collaboration between authors who have contributed to the study. It is evident that there is a tremendous amount of collaboration indicated in the table as 97.97 per cent have two or more authors, and a single authorship pattern has produced about 2.03 per cent. The analysis reveals that in the year 2015, the highest articles were produced by six authors (60) and the lowest output by one author (6). In 2017, six authors produced the highest articles (49) and the lowest publication by one author (7). In 2018 the highest articles were produced by six authors (49), and the lowest publications by one author (5). From the year 2007 up to 2019, it is observed that ten and more than ten authors have contributed much of the publications, as is evident from the table above.

It could be concluded that the six authors' team (12.77%) has shown the highest productivity, followed by five authors (11.29%) team apart from 'ten and greater than ten' which has 24.49% of the share and minimum productivity produced by single authors' team (2.03%).

### 4.19.2 Year Wise Authorship Pattern of Russia

**Table 19B: Year Wise Authorship Pattern of Russia on CAD**

| Year | 1 | 2 | 3 | 4 | 5 | 6 | 7 | 8 | 9 | 10 and > 10 | Total | % |
|---|---|---|---|---|---|---|---|---|---|---|---|---|
| 1991 | 0 | 0 | 0 | 0 | 0 | 1 | 0 | 0 | 0 | 0 | 1 | 0.02 |
| 1992 | 3 | 4 | 8 | 10 | 7 | 4 | 5 | 0 | 0 | 0 | 41 | 0.81 |
| 1993 | 7 | 14 | 16 | 7 | 11 | 9 | 4 | 1 | 0 | 2 | 71 | 1.40 |
| 1994 | 12 | 16 | 12 | 18 | 15 | 18 | 3 | 5 | 0 | 4 | 103 | 2.04 |
| 1995 | 5 | 9 | 7 | 9 | 14 | 14 | 10 | 5 | 3 | 7 | 83 | 1.64 |
| 1996 | 14 | 15 | 24 | 19 | 12 | 10 | 10 | 5 | 4 | 5 | 118 | 2.33 |
| 1997 | 10 | 15 | 22 | 29 | 17 | 17 | 8 | 5 | 0 | 6 | 129 | 2.55 |
| 1998 | 12 | 13 | 20 | 25 | 14 | 12 | 8 | 7 | 2 | 3 | 116 | 2.29 |
| 1999 | 10 | 18 | 18 | 15 | 13 | 9 | 6 | 8 | 2 | 5 | 104 | 2.06 |
| 2000 | 13 | 15 | 16 | 15 | 30 | 16 | 10 | 4 | 7 | 5 | 131 | 2.59 |
| 2001 | 9 | 12 | 17 | 14 | 11 | 9 | 8 | 0 | 2 | 6 | 88 | 1.74 |
| 2002 | 13 | 16 | 29 | 27 | 13 | 20 | 13 | 8 | 4 | 11 | 154 | 3.04 |
| 2003 | 11 | 21 | 26 | 19 | 20 | 22 | 7 | 4 | 2 | 9 | 141 | 2.79 |
| 2004 | 16 | 23 | 22 | 27 | 24 | 19 | 22 | 9 | 12 | 7 | 181 | 3.58 |
| 2005 | 18 | 17 | 20 | 30 | 18 | 20 | 15 | 12 | 8 | 9 | 167 | 3.30 |
| 2006 | 8 | 18 | 31 | 20 | 20 | 14 | 11 | 11 | 7 | 16 | 156 | 3.08 |
| 2007 | 14 | 18 | 27 | 24 | 25 | 24 | 19 | 6 | 4 | 13 | 174 | 3.44 |
| 2008 | 12 | 19 | 19 | 23 | 33 | 16 | 24 | 11 | 8 | 10 | 175 | 3.46 |
| 2009 | 17 | 26 | 25 | 25 | 29 | 25 | 17 | 15 | 4 | 17 | 200 | 3.95 |
| 2010 | 22 | 29 | 28 | 30 | 31 | 25 | 21 | 12 | 9 | 19 | 226 | 4.47 |
| 2011 | 18 | 20 | 33 | 38 | 29 | 33 | 17 | 19 | 10 | 22 | 239 | 4.73 |
| 2012 | 22 | 24 | 31 | 24 | 27 | 20 | 25 | 17 | 12 | 24 | 226 | 4.47 |
| 2013 | 14 | 30 | 30 | 31 | 29 | 27 | 25 | 14 | 4 | 15 | 219 | 4.33 |
| 2014 | 10 | 19 | 31 | 36 | 26 | 29 | 22 | 13 | 15 | 31 | 232 | 4.59 |
| 2015 | 11 | 32 | 24 | 40 | 47 | 34 | 22 | 13 | 13 | 41 | 277 | 5.48 |
| 2016 | 10 | 23 | 41 | 37 | 37 | 29 | 24 | 16 | 14 | 58 | 289 | 5.71 |
| 2017 | 6 | 19 | 33 | 44 | 43 | 37 | 27 | 23 | 18 | 51 | 301 | 5.95 |
| 2018 | 13 | 29 | 28 | 54 | 53 | 38 | 25 | 26 | 19 | 66 | 351 | 6.94 |
| 2019 | 9 | 24 | 24 | 53 | 54 | 39 | 31 | 27 | 17 | 87 | 365 | 7.22 |
| Total | 339 | 538 | 662 | 743 | 702 | 590 | 439 | 296 | 200 | 549 | 5058 | 100.00 |
| % | 6.70 | 10.64 | 13.09 | 14.69 | 13.88 | 11.66 | 8.68 | 5.85 | 3.95 | 10.85 | 100.00 | |

Table investigates the year-wise distribution of authorship pattern of coronary artery disease literature contributed by Russian scientists. Out of 5058 papers, the authorship pattern up to 9 authors results in a total of 4509 research output remaining 549 papers have been published by more than ten authors. Single author contributions are accounted for 339 (6.70%) during the study period. The highest percentage of 14.69% is recorded by four authors followed by five and three authors showing 13.88 and 13.09 percentages respectively. However, seven, eight and nine authors have contributed less than ten percentages in this study. This analysis of results shows that individual contribution is not at the rate of appreciation compared to collaborative research in the field of CAD literature research. The number of authors engaging collaborative research is found increasing year by year from 1991 to 2019, ranging from 39 to 356. It can be noticed that 6.70 % of authors/scientists collectively contribute papers in the field of CAD literature individually.

It could be concluded that the four authors' team (14.69%) has shown the highest productivity, followed by five authors (13.88%) team apart from 'ten and greater than ten' which has 10.85% of the share and minimum productivity produced by single authors' team (6.70%).

### 4.19.3 Year Wise Authorship Pattern of India

The authorship pattern is indicated in the table from 1990 to 2019. The researcher has categorized the authorship pattern as a single author, double authors, three authors, four authors, five authors, six authors, seven authors, eight authors, nine authors, and ten and above authors produced by year wise.

There is an indication of the degree of collaboration between authors who have contributed to the study. It is evident that there is a tremendous amount of collaboration indicated in the table as 95.94 per cent have two or more authors, and a single authorship pattern has produced about 4.06 per cent. It is observed that the year 2016 has contributed the highest number of papers 422 (8.97%) followed by the year 2019, 389 (8.27%). The year 1990 to 1996 have contributed less than one per cent of publications in the study. From the year 2017 up to 2019, it is observed that ten and more than ten authors have contributed much of the publication, as is evident from the table below.

It could be concluded that the four authors' team (16.23%) has shown the highest productivity, followed by three authors (15.38%) team apart from 'ten and greater than ten' which has 12.03% of the share and minimum productivity produced by single authors' team (4.06%).

| Table 19C: Year Wise Authorship Pattern of India on CAD | | | | | | | | | | | |
|---|---|---|---|---|---|---|---|---|---|---|---|
| **Year** | **1** | **2** | **3** | **4** | **5** | **6** | **7** | **8** | **9** | **10 and > 10** | **Total** | **%** |
| 1990 | 2 | 3 | 3 | 6 | 1 | 0 | 0 | 0 | 0 | 1 | 16 | 0.34 |
| 1991 | 1 | 7 | 6 | 3 | 7 | 0 | 0 | 0 | 0 | 0 | 24 | 0.51 |
| 1992 | 2 | 1 | 6 | 5 | 4 | 3 | 0 | 0 | 0 | 0 | 21 | 0.45 |
| 1993 | 2 | 2 | 11 | 3 | 3 | 1 | 2 | 0 | 0 | 1 | 25 | 0.53 |
| 1994 | 1 | 4 | 3 | 10 | 4 | 2 | 0 | 0 | 0 | 2 | 26 | 0.55 |
| 1995 | 1 | 5 | 2 | 3 | 3 | 3 | 1 | 2 | 0 | 4 | 24 | 0.51 |
| 1996 | 1 | 5 | 15 | 10 | 6 | 0 | 3 | 3 | 1 | 1 | 45 | 0.96 |
| 1997 | 5 | 8 | 6 | 6 | 10 | 9 | 3 | 2 | 0 | 0 | 49 | 1.04 |
| 1998 | 5 | 7 | 6 | 8 | 11 | 7 | 4 | 2 | 1 | 2 | 53 | 1.13 |
| 1999 | 2 | 9 | 9 | 7 | 10 | 7 | 2 | 1 | 0 | 2 | 49 | 1.04 |
| 2000 | 1 | 10 | 5 | 5 | 14 | 12 | 4 | 1 | 1 | 2 | 55 | 1.17 |
| 2001 | 6 | 9 | 9 | 6 | 9 | 9 | 3 | 3 | 0 | 3 | 57 | 1.21 |
| 2002 | 2 | 11 | 9 | 10 | 10 | 9 | 2 | 3 | 3 | 1 | 60 | 1.27 |
| 2003 | 7 | 10 | 12 | 16 | 9 | 5 | 4 | 5 | 1 | 3 | 72 | 1.53 |
| 2004 | 3 | 17 | 13 | 22 | 12 | 16 | 9 | 5 | 1 | 3 | 101 | 2.15 |
| 2005 | 9 | 10 | 14 | 18 | 25 | 13 | 11 | 6 | 1 | 4 | 111 | 2.36 |
| 2006 | 3 | 16 | 12 | 16 | 17 | 14 | 10 | 3 | 1 | 11 | 103 | 2.19 |
| 2007 | 11 | 13 | 28 | 30 | 18 | 14 | 8 | 2 | 3 | 16 | 143 | 3.04 |
| 2008 | 11 | 28 | 36 | 36 | 31 | 23 | 10 | 15 | 6 | 16 | 212 | 4.50 |
| 2009 | 13 | 36 | 33 | 34 | 36 | 24 | 15 | 17 | 7 | 9 | 224 | 4.76 |
| 2010 | 14 | 40 | 49 | 57 | 36 | 15 | 10 | 5 | 10 | 19 | 255 | 5.42 |
| 2011 | 6 | 33 | 40 | 46 | 31 | 30 | 29 | 17 | 5 | 28 | 265 | 5.63 |
| 2012 | 10 | 36 | 38 | 47 | 47 | 36 | 24 | 24 | 11 | 34 | 307 | 6.52 |
| 2013 | 7 | 45 | 46 | 46 | 53 | 45 | 19 | 15 | 11 | 30 | 317 | 6.74 |
| 2014 | 3 | 34 | 48 | 51 | 50 | 37 | 29 | 18 | 16 | 44 | 330 | 7.01 |
| 2015 | 12 | 22 | 43 | 46 | 37 | 38 | 20 | 8 | 6 | 40 | 272 | 5.78 |
| 2016 | 14 | 64 | 65 | 62 | 45 | 39 | 22 | 29 | 20 | 62 | 422 | 8.97 |
| 2017 | 12 | 29 | 44 | 44 | 38 | 26 | 17 | 24 | 16 | 62 | 312 | 6.63 |
| 2018 | 9 | 48 | 62 | 49 | 27 | 35 | 32 | 15 | 19 | 71 | 367 | 7.80 |
| 2019 | 16 | 38 | 51 | 62 | 28 | 40 | 31 | 22 | 6 | 95 | 389 | 8.27 |
| Total | 191 | 600 | 724 | 764 | 632 | 512 | 324 | 247 | 146 | 566 | 4706 | 100.00 |
| % | 4.06 | 12.75 | 15.38 | 16.23 | 13.43 | 10.88 | 6.88 | 5.25 | 3.10 | 12.03 | 100.00 | |

### 4.19.4 Year Wise Authorship Pattern of Peoples Republic of China

The table demonstrates the year-wise distribution of authorship pattern of coronary artery disease literature contributed by Peoples Republic of China scientists. Out of 32770 papers, the authorship pattern up to 9 authors results in a total of 25917 research output remaining 6853 papers have been published by more than ten authors. Single author contributions are accounted to 294 (0.90%) during the study period.

| Table 19D: Year Wise Authorship Pattern of the Peoples Republic of China on CAD | | | | | | | | | | | |
|---|---|---|---|---|---|---|---|---|---|---|---|
| **Year** | **1** | **2** | **3** | **4** | **5** | **6** | **7** | **8** | **9** | **10 and > 10** | **Total** | **%** |
| 1990 | 1 | 1 | 5 | 0 | 3 | 0 | 1 | 1 | 0 | 0 | 12 | 0.04 |
| 1991 | 2 | 2 | 7 | 4 | 3 | 4 | 2 | 0 | 1 | 1 | 26 | 0.08 |
| 1992 | 1 | 7 | 4 | 3 | 3 | 8 | 1 | 0 | 1 | 0 | 28 | 0.09 |
| 1993 | 2 | 4 | 10 | 2 | 6 | 7 | 4 | 4 | 0 | 0 | 39 | 0.12 |
| 1994 | 0 | 5 | 5 | 2 | 7 | 6 | 3 | 1 | 0 | 1 | 30 | 0.09 |
| 1995 | 0 | 8 | 8 | 8 | 5 | 0 | 1 | 1 | 1 | 3 | 35 | 0.11 |
| 1996 | 0 | 11 | 10 | 11 | 9 | 5 | 1 | 3 | 2 | 3 | 55 | 0.17 |
| 1997 | 1 | 2 | 8 | 3 | 4 | 4 | 5 | 3 | 3 | 4 | 37 | 0.11 |
| 1998 | 7 | 5 | 14 | 12 | 16 | 12 | 5 | 2 | 2 | 1 | 76 | 0.23 |
| 1999 | 5 | 13 | 18 | 20 | 15 | 24 | 8 | 3 | 1 | 5 | 112 | 0.34 |
| 2000 | 3 | 11 | 14 | 27 | 20 | 14 | 5 | 6 | 4 | 8 | 112 | 0.34 |
| 2001 | 1 | 12 | 28 | 19 | 19 | 21 | 8 | 6 | 6 | 3 | 123 | 0.38 |
| 2002 | 2 | 12 | 25 | 27 | 26 | 23 | 21 | 9 | 5 | 11 | 161 | 0.49 |
| 2003 | 5 | 18 | 27 | 27 | 41 | 28 | 34 | 15 | 10 | 13 | 218 | 0.67 |
| 2004 | 5 | 19 | 23 | 34 | 47 | 55 | 46 | 22 | 10 | 23 | 284 | 0.87 |
| 2005 | 7 | 20 | 35 | 45 | 50 | 51 | 30 | 29 | 18 | 24 | 309 | 0.94 |
| 2006 | 6 | 31 | 53 | 61 | 63 | 56 | 53 | 61 | 30 | 73 | 487 | 1.49 |
| 2007 | 7 | 24 | 34 | 71 | 81 | 75 | 75 | 65 | 43 | 82 | 557 | 1.70 |
| 2008 | 9 | 31 | 75 | 87 | 107 | 110 | 120 | 73 | 79 | 95 | 786 | 2.40 |
| 2009 | 12 | 38 | 76 | 98 | 133 | 146 | 145 | 107 | 74 | 164 | 993 | 3.03 |
| 2010 | 19 | 56 | 86 | 134 | 160 | 201 | 173 | 124 | 121 | 185 | 1259 | 3.84 |
| 2011 | 12 | 50 | 101 | 160 | 188 | 208 | 214 | 147 | 121 | 240 | 1441 | 4.40 |
| 2012 | 24 | 64 | 128 | 170 | 197 | 260 | 242 | 198 | 140 | 324 | 1747 | 5.33 |
| 2013 | 13 | 103 | 149 | 211 | 258 | 346 | 314 | 267 | 187 | 474 | 2322 | 7.09 |
| 2014 | 28 | 115 | 162 | 245 | 314 | 389 | 341 | 283 | 251 | 562 | 2690 | 8.21 |
| 2015 | 27 | 98 | 173 | 261 | 384 | 459 | 394 | 348 | 290 | 680 | 3114 | 9.50 |
| 2016 | 17 | 108 | 214 | 295 | 394 | 483 | 433 | 375 | 289 | 807 | 3415 | 10.42 |
| 2017 | 23 | 116 | 222 | 327 | 442 | 523 | 399 | 420 | 352 | 908 | 3732 | 11.39 |
| 2018 | 28 | 135 | 244 | 371 | 496 | 494 | 456 | 443 | 372 | 962 | 4001 | 12.21 |
| 2019 | 27 | 161 | 290 | 382 | 483 | 612 | 563 | 490 | 364 | 1197 | 4569 | 13.94 |
| Total | 294 | 1280 | 2248 | 3117 | 3974 | 4624 | 4097 | 3506 | 2777 | 6853 | 32770 | 100.00 |
| % | 0.90 | 3.91 | 6.86 | 9.51 | 12.13 | 14.11 | 12.50 | 10.70 | 8.47 | 20.91 | 100.00 | |

The highest percentage of 14.11% is recorded by six authors followed by seven and five authors showing 12.50 and 12.13 percentages respectively. However, two, three, four and nine authors have contributed less than ten percentages in this study. This analysis of results shows that individual contribution is not at the rate of appreciation compared to collaborative research in the field of CAD literature research. The number of authors engaging collaborative research is found increasing year by year from 1990 to 2019, ranging from 11 to 4542. It can be noticed that 0.90 % of authors/scientists collectively contribute papers in the field of CAD literature individually.

It could be concluded that the six authors' team (14.11%) has shown the highest productivity, followed by seven authors (12.50%) team apart from 'ten and greater than ten' which has 20.91% of the share and minimum productivity produced by single authors' team (0.90%).

### 4.19.5 Year Wise Authorship Pattern of South Africa

**Table 19E: Year Wise Authorship Pattern of South Africa on CAD**

| Year | 1 | 2 | 3 | 4 | 5 | 6 | 7 | 8 | 9 | 10and > 10 | Total | % |
|---|---|---|---|---|---|---|---|---|---|---|---|---|
| 1990 | 1 | 2 | 1 | 2 | 3 | 0 | 0 | 0 | 0 | 0 | 9 | 0.70 |
| 1991 | 5 | 2 | 4 | 6 | 3 | 3 | 2 | 1 | 1 | 2 | 29 | 2.26 |
| 1992 | 4 | 7 | 10 | 9 | 3 | 4 | 2 | 0 | 1 | 0 | 40 | 3.12 |
| 1993 | 4 | 6 | 7 | 6 | 3 | 3 | 0 | 1 | 1 | 3 | 34 | 2.65 |
| 1994 | 3 | 4 | 2 | 4 | 2 | 0 | 1 | 0 | 0 | 0 | 16 | 1.25 |
| 1995 | 2 | 9 | 2 | 3 | 1 | 2 | 3 | 0 | 0 | 3 | 25 | 1.95 |
| 1996 | 6 | 6 | 3 | 7 | 4 | 4 | 2 | 3 | 0 | 1 | 36 | 2.80 |
| 1997 | 4 | 7 | 3 | 3 | 5 | 7 | 3 | 0 | 1 | 0 | 33 | 2.57 |
| 1998 | 3 | 6 | 6 | 5 | 6 | 4 | 2 | 1 | 0 | 2 | 35 | 2.73 |
| 1999 | 5 | 5 | 4 | 5 | 4 | 1 | 3 | 1 | 0 | 3 | 31 | 2.41 |
| 2000 | 5 | 3 | 5 | 5 | 4 | 5 | 0 | 1 | 0 | 4 | 32 | 2.49 |
| 2001 | 2 | 2 | 5 | 2 | 3 | 4 | 3 | 0 | 2 | 3 | 26 | 2.02 |
| 2002 | 2 | 5 | 2 | 8 | 1 | 1 | 0 | 1 | 0 | 0 | 20 | 1.56 |
| 2003 | 6 | 2 | 4 | 3 | 0 | 2 | 0 | 2 | 1 | 2 | 22 | 1.71 |
| 2004 | 2 | 1 | 6 | 4 | 2 | 3 | 1 | 1 | 1 | 1 | 22 | 1.71 |
| 2005 | 0 | 5 | 5 | 1 | 2 | 4 | 3 | 1 | 1 | 1 | 23 | 1.79 |
| 2006 | 1 | 6 | 4 | 3 | 5 | 2 | 6 | 3 | 2 | 0 | 32 | 2.49 |
| 2007 | 1 | 8 | 11 | 6 | 4 | 5 | 4 | 1 | 1 | 7 | 48 | 3.74 |
| 2008 | 5 | 5 | 6 | 5 | 4 | 0 | 4 | 2 | 1 | 4 | 36 | 2.80 |
| 2009 | 5 | 10 | 8 | 8 | 4 | 5 | 2 | 4 | 2 | 8 | 56 | 4.36 |
| 2010 | 3 | 5 | 4 | 6 | 5 | 2 | 2 | 4 | 0 | 10 | 41 | 3.19 |
| 2011 | 2 | 3 | 8 | 10 | 7 | 3 | 0 | 4 | 0 | 6 | 43 | 3.35 |
| 2012 | 5 | 9 | 11 | 11 | 9 | 9 | 1 | 2 | 0 | 13 | 70 | 5.45 |
| 2013 | 4 | 7 | 6 | 8 | 6 | 8 | 1 | 2 | 2 | 21 | 65 | 5.06 |
| 2014 | 4 | 6 | 5 | 8 | 3 | 6 | 10 | 2 | 2 | 22 | 68 | 5.30 |
| 2015 | 1 | 7 | 6 | 6 | 5 | 10 | 2 | 5 | 1 | 27 | 70 | 5.45 |
| 2016 | 2 | 5 | 4 | 8 | 13 | 4 | 4 | 7 | 0 | 35 | 82 | 6.39 |
| 2017 | 0 | 5 | 10 | 15 | 14 | 3 | 5 | 2 | 2 | 32 | 88 | 6.85 |
| 2018 | 4 | 6 | 6 | 4 | 4 | 9 | 3 | 2 | 4 | 38 | 80 | 6.23 |
| 2019 | 3 | 5 | 3 | 5 | 7 | 9 | 3 | 2 | 6 | 29 | 72 | 5.61 |
| Total | 94 | 159 | 161 | 176 | 136 | 122 | 72 | 55 | 32 | 277 | 1284 | 100.00 |
| % | 7.32 | 12.38 | 12.54 | 13.71 | 10.59 | 9.50 | 5.61 | 4.28 | 2.49 | 21.57 | 100.00 | 0.00 |

To achieve the seventh objective of assessing the nature of the authorship pattern and find out the degree of collaboration, the analysis of authorship pattern and degree of collaboration is carried out and it is elucidated in tables 18, 19A, 19B, 19C, 19D, 19E, 20, 21, 22, 23 and 24.

The authorship pattern is indicated in the table from 1990 to 2019. The researcher has categorized the authorship pattern as a single author, double authors, three authors, four authors, five authors, six authors, seven authors, eight authors, nine authors, and ten and above authors produced by year wise.

The table explored the overall and thirty years wise distribution of authorship trend. It is evident from the table that only 7.32 per cent publications were single-authored publications while rest of 92.68 had two or more authors. The maximum number of publications were more than ten authored publications (21.57 %) followed by four authored publications (13.71%), three authored (12.54%), two authored (12.38%), and five authored publications (10.59 %). Six to ten authored publications accounted for 24.56 per cent, while more than nine authored publications accounted for 21.88 per cent.

It could be concluded that the four authors' team (13.71%) has shown the highest productivity, followed by three authors (12.54%) team apart from 'ten and greater than ten' which has 21.57% of the share and minimum productivity produced by nine authors' team (2.49%).

### 4.20 Single Vs. Multiple Authors & Degree of Collaboration on CAD

The study provides an idea about the degree of collaboration in the CAD field for thirty years. However, rather than merely knowing the degree of collaboration in a given year, it is more interesting to compare this collaboration index across the years. In the present example, these indices took values from 0.97 in 1990 to 0.99 in 2019. This means that publications about coronary artery disease research for the time period selected (1990-2019) result from collaborative work, with an average of about five to six authors working together on the same article.

The data from the below table indicates the degree of collaboration in the research output of coronary artery disease research that the overall degree of collaboration is 0.98 during the period. Out of the total 50036 publications, 97.91% are published under a collaborative venture of publication in coronary artery disease research. It could be seen clearly from the below analysis that the degree of collaboration in publishing research output on coronary artery disease research has shown an increasing trend during the research period. Based on this study, the result of the degree of collaboration DC = 0.98, which clearly supports that 98% of publications are brought out with an effort of collaborative authors rather than the individual author and is substantiated by (Suresh & Thanuskodi, 2019) in which 92% of publications are of collaborative in nature. ***There has been an increasing pattern in collaboration on coronary artery disease research, examined in recent years. Hence, the third hypothesis has been proved.***

**Table 20: Single vs Multiple Authors & Degree of Collaboration of CAD**

| Year | Single Authors | | Multiple Authors | | Total | Degree of collaboration |
|---|---|---|---|---|---|---|
| | Output | % | Output | % | | |
| 1990 | 5 | 0.48 | 153 | 0.31 | 158 | 0.97 |
| 1991 | 9 | 0.86 | 257 | 0.52 | 266 | 0.97 |
| 1992 | 11 | 1.05 | 223 | 0.46 | 234 | 0.95 |
| 1993 | 15 | 1.44 | 167 | 0.34 | 182 | 0.92 |
| 1994 | 16 | 1.53 | 180 | 0.37 | 196 | 0.92 |
| 1995 | 11 | 1.05 | 186 | 0.38 | 197 | 0.94 |
| 1996 | 23 | 2.20 | 268 | 0.55 | 291 | 0.92 |
| 1997 | 22 | 2.11 | 275 | 0.56 | 297 | 0.93 |
| 1998 | 29 | 2.78 | 282 | 0.58 | 311 | 0.91 |
| 1999 | 23 | 2.20 | 320 | 0.65 | 343 | 0.93 |
| 2000 | 25 | 2.39 | 374 | 0.76 | 399 | 0.94 |
| 2001 | 18 | 1.72 | 342 | 0.70 | 360 | 0.95 |
| 2002 | 19 | 1.82 | 471 | 0.96 | 490 | 0.96 |
| 2003 | 32 | 3.07 | 527 | 1.08 | 559 | 0.94 |
| 2004 | 29 | 2.78 | 708 | 1.45 | 737 | 0.96 |
| 2005 | 35 | 3.35 | 759 | 1.55 | 794 | 0.96 |
| 2006 | 21 | 2.01 | 973 | 1.99 | 994 | 0.98 |
| 2007 | 42 | 4.02 | 1182 | 2.41 | 1224 | 0.97 |
| 2008 | 42 | 4.02 | 1444 | 2.95 | 1486 | 0.97 |
| 2009 | 60 | 5.75 | 1739 | 3.55 | 1799 | 0.97 |
| 2010 | 72 | 6.90 | 2077 | 4.24 | 2149 | 0.97 |
| 2011 | 43 | 4.12 | 2357 | 4.81 | 2400 | 0.98 |
| 2012 | 66 | 6.32 | 2661 | 5.43 | 2727 | 0.98 |
| 2013 | 41 | 3.93 | 3295 | 6.73 | 3336 | 0.99 |
| 2014 | 51 | 4.89 | 3634 | 7.42 | 3685 | 0.99 |
| 2015 | 57 | 5.46 | 4073 | 8.31 | 4130 | 0.99 |
| 2016 | 51 | 4.89 | 4526 | 9.24 | 4577 | 0.99 |
| 2017 | 48 | 4.60 | 4774 | 9.74 | 4822 | 0.99 |
| 2018 | 59 | 5.65 | 5133 | 10.48 | 5192 | 0.99 |
| 2019 | 69 | 6.61 | 5632 | 11.50 | 5701 | 0.99 |
| Total | 1044 | 100.00 | 48992 | 100.00 | 50036 | 0.98 |

**Figure 12: Single Vs. Multiple Authors**

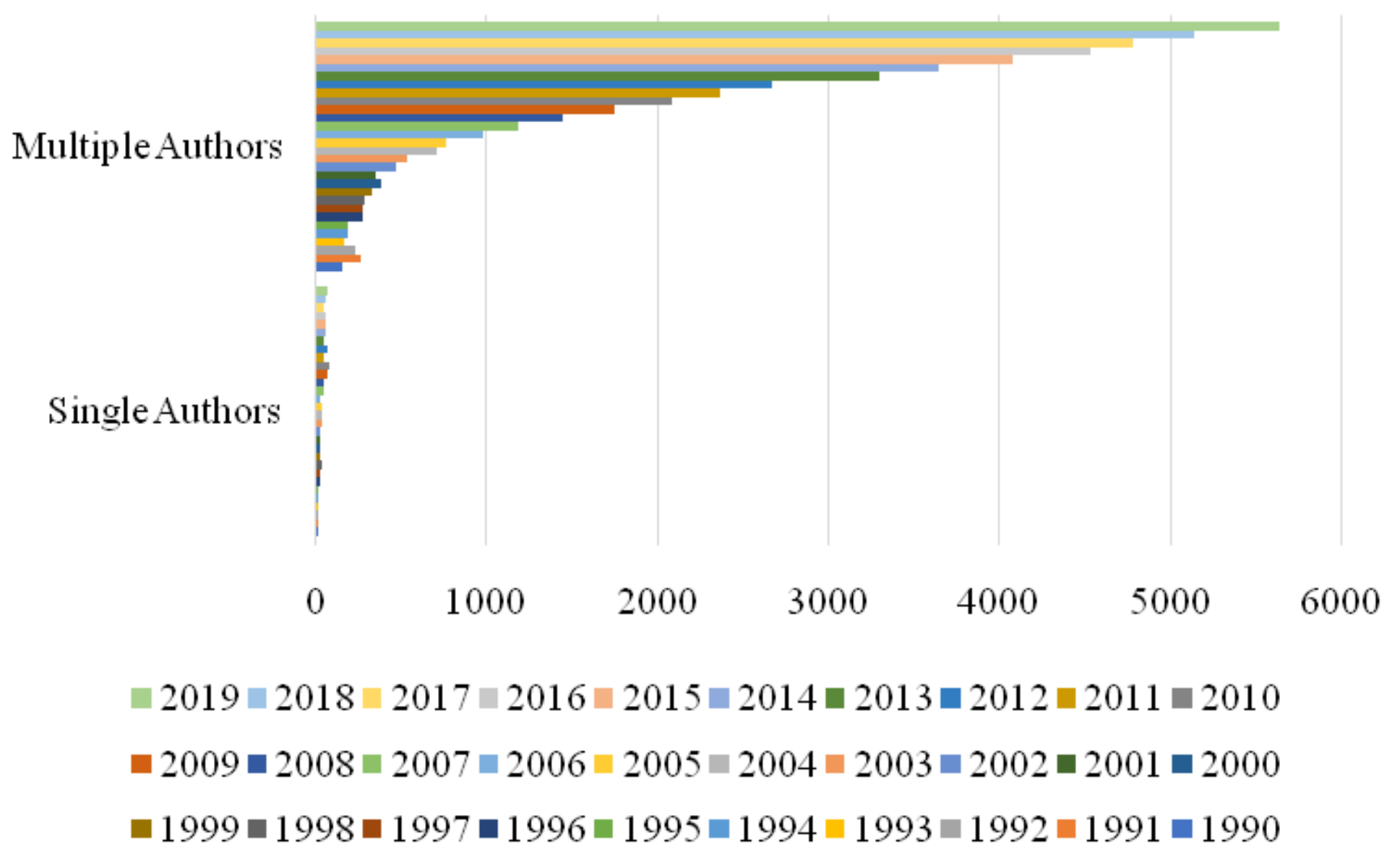

## 4.21 Authorship Pattern in Block Years

**Table 21: Authorship Pattern in Block Years**

| S.No. | Authorship Pattern | Block Years | | | | | | Total |
|---|---|---|---|---|---|---|---|---|
| | | 1990-1994 | 1995-1999 | 2000-2004 | 2005-2009 | 2010-2014 | 2015-2019 | |
| 1 | Single Author | 56 | 108 | 123 | 200 | 273 | 284 | 1044 |
| 2 | Double Authors | 170 | 190 | 265 | 439 | 811 | 1025 | 2900 |
| 3 | Three Authors | 218 | 222 | 334 | 656 | 1141 | 1678 | 4249 |
| 4 | Four Authors | 190 | 238 | 372 | 774 | 1520 | 2293 | 5387 |
| 5 | Five Authors | 156 | 202 | 384 | 848 | 1743 | 2813 | 6146 |
| 6 | Six Authors | 111 | 173 | 376 | 831 | 1972 | 3179 | 6642 |
| 7 | Seven Authors | 59 | 122 | 261 | 730 | 1743 | 2673 | 5588 |
| 8 | Eight Authors | 30 | 74 | 148 | 586 | 1402 | 2479 | 4719 |
| 9 | Nine Authors | 16 | 34 | 95 | 390 | 1067 | 1991 | 3593 |
| 10 | Ten and > 10 Authors | 30 | 76 | 187 | 843 | 2625 | 6007 | 9768 |
| **Total** | | 1036 | 1439 | 2545 | 6297 | 14297 | 24422 | 50036 |
| Total Authors | | 4603 | 7087 | 13445 | 38616 | 103020 | 217987 | 384758 |
| DC | | 0.95 | 0.92 | 0.95 | 0.97 | 0.98 | 0.99 | 0.98 |
| CI | | 4.44 | 4.92 | 5.28 | 6.13 | 7.21 | 8.93 | 7.69 |
| CC | | 0.68 | 0.69 | 0.73 | 0.77 | 0.80 | 0.81 | 0.79 |
| MCC | | 0.68 | 0.69 | 0.73 | 0.77 | 0.80 | 0.81 | 0.79 |

DC-Degree of Collaboration, CI – Collaborative Index, CC- Collaborative Coefficient, MCC – Modified Collaborative Coefficient

A count of the number of authors who have contributed to the study offers some indication of the degree of collaboration between authors. The study report that just a minimal number of the articles included in the study (2.09 per cent) have a single author, while the other half (97.91 per cent) have two or more contributing authors. This is further supported by the study conducted on biochemistry research in which 97.46% of the literature is contributed by multi-authors (Sudhier & Dileepkumar, 2020).

Table 21 reveals that the authorship pattern of block-wise distributions. The degree of collaboration ranges from 0.95 and 0.99. This signifies that there is an existence of collaborative research in Coronary Artery Disease among the BRICS countries. The collaborative index ranges between 4.44 and 8.93. It is observed that the collaborative coefficient ranges from 0.68 and 0.81. The modified collaborative co-efficient is between the range of 0.68 and 0.81. From the table, it is observed that the authorship pattern indicators are in increasing trends in the block years. This illustrates, once again, the increasing interest in teamwork.

### 4.22 Collaborative Indices on CAD

The collaborative Index (CI) ranges between 3.80 and 10.99 during the research period of 1990 to 2019. CI is determined minimum during the year 1990. It is highest in the year 2019. Therefore, it can be found that the collaborative Index is improving from 1990 onwards.

The degree of collaboration ranges between 0.88 and 0.99 from the year 1990 to 2019. The degree of collaboration is found minimum during the year 1990 and steadily increasing up to 2019 to 0.99. It has got an increase from the year 2007 to 2018. Therefore, it can be concluded that the degree of collaboration is in an upward direction from 1990.

The Co-authorship Index (CAI) ranges between 89.92 and 101.19 during the study period of 1990 to 2019. CAI is the lowest value in the year 1990. Moreover, it is the highest value in the year 2017. Therefore, it can be revealed that the Co-authorship index is showing an increasing trend from 1990 to 2019.

The collaboration coefficient (CC) ranges between 0.62786 and 0.81674 during the study period of 1990 to 2019. CC is found minimum in the year 1990, and it is increasing in the year to follow up to 2019 and is highest in the year 2017. Therefore, it can be understood that the collaborative coefficient is also showing an increasing trend from 1990 onwards.

Modified Collaborative Coefficient (MCC) is calculated to overcome the collaborative coefficient limitation, which ranges from 0.63186 to 0.81691. The value of MCC is lowest during the year 1990, and it is highest in the year 2017.

**Table 22: Collaborative Indices**

| S.No. | Year | Publication | Authors | CI | DC | CAI | CC | MCC | MCC-CC |
|---|---|---|---|---|---|---|---|---|---|
| 1 | 1990 | 158 | 600 | 3.80 | 0.88 | 89.92 | 0.62786 | 0.63186 | 0.00400 |
| 2 | 1991 | 266 | 1099 | 4.13 | 0.91 | 92.98 | 0.65700 | 0.65948 | 0.00248 |
| 3 | 1992 | 234 | 974 | 4.16 | 0.91 | 93.03 | 0.65320 | 0.65600 | 0.00280 |
| 4 | 1993 | 182 | 1051 | 5.77 | 0.92 | 93.78 | 0.66551 | 0.66918 | 0.00368 |
| 5 | 1994 | 196 | 879 | 4.48 | 0.92 | 93.86 | 0.68027 | 0.68376 | 0.00349 |
| 6 | 1995 | 197 | 1012 | 5.14 | 0.94 | 96.50 | 0.70796 | 0.71157 | 0.00361 |
| 7 | 1996 | 291 | 1342 | 4.61 | 0.92 | 94.13 | 0.67717 | 0.67950 | 0.00234 |
| 8 | 1997 | 297 | 1399 | 4.71 | 0.93 | 94.64 | 0.69382 | 0.69616 | 0.00234 |
| 9 | 1998 | 311 | 1683 | 5.41 | 0.91 | 93.00 | 0.68552 | 0.68773 | 0.00221 |
| 10 | 1999 | 343 | 1651 | 4.81 | 0.94 | 96.54 | 0.70776 | 0.70982 | 0.00207 |
| 11 | 2000 | 399 | 2021 | 5.07 | 0.94 | 95.80 | 0.71570 | 0.71750 | 0.00180 |
| 12 | 2001 | 360 | 1844 | 5.12 | 0.95 | 97.10 | 0.71845 | 0.72045 | 0.00200 |
| 13 | 2002 | 490 | 2530 | 5.16 | 0.96 | 98.24 | 0.73348 | 0.73498 | 0.00150 |
| 14 | 2003 | 559 | 2936 | 5.25 | 0.94 | 96.36 | 0.72248 | 0.72377 | 0.00129 |
| 15 | 2004 | 737 | 4114 | 5.58 | 0.96 | 98.18 | 0.74972 | 0.75073 | 0.00102 |
| 16 | 2005 | 794 | 4561 | 5.74 | 0.95 | 97.57 | 0.75068 | 0.75163 | 0.00095 |
| 17 | 2006 | 994 | 6048 | 6.08 | 0.98 | 99.94 | 0.77265 | 0.77343 | 0.00078 |
| 18 | 2007 | 1224 | 7578 | 6.19 | 0.96 | 98.62 | 0.76909 | 0.76972 | 0.00063 |
| 19 | 2008 | 1486 | 9047 | 6.09 | 0.97 | 99.32 | 0.77172 | 0.77224 | 0.00052 |
| 20 | 2009 | 1799 | 11382 | 6.33 | 0.97 | 98.85 | 0.77213 | 0.77256 | 0.00043 |
| 21 | 2010 | 2149 | 13600 | 6.33 | 0.97 | 98.78 | 0.77195 | 0.77231 | 0.00036 |
| 22 | 2011 | 2400 | 16233 | 6.76 | 0.98 | 100.37 | 0.79165 | 0.79197 | 0.00033 |
| 23 | 2012 | 2727 | 19110 | 7.01 | 0.98 | 99.73 | 0.79069 | 0.79098 | 0.00029 |
| 24 | 2013 | 3336 | 24577 | 7.37 | 0.99 | 100.95 | 0.79996 | 0.80020 | 0.00024 |
| 25 | 2014 | 3685 | 29500 | 8.01 | 0.99 | 100.79 | 0.80518 | 0.80539 | 0.00022 |
| 26 | 2015 | 4130 | 31775 | 7.69 | 0.99 | 100.80 | 0.81062 | 0.81082 | 0.00020 |
| 27 | 2016 | 4577 | 38678 | 8.45 | 0.99 | 101.07 | 0.80905 | 0.80923 | 0.00018 |
| 28 | 2017 | 4822 | 40623 | 8.42 | 0.99 | 101.19 | 0.81674 | 0.81691 | 0.00017 |
| 29 | 2018 | 5192 | 44281 | 8.53 | 0.99 | 101.06 | 0.81125 | 0.81141 | 0.00016 |
| 30 | 2019 | 5701 | 62630 | 10.99 | 0.99 | 100.95 | 0.81390 | 0.81404 | 0.00014 |
| Total | | 50036 | 384758 | 7.69 | 0.98 | 100.00 | | | |

## 4.23 Co-Authorship Index on CAD

Another possible way of analyzing author collaboration patterns is the co-authorship index (CAI). This index is obtained by calculating the number of single-, two-, multi-, and mega-authored papers for different nations or different sub-disciplines. The following formula gives the CAI:

$$CAI = \left[ \frac{\left( \frac{N_{ij}}{N_{io}} \right)}{\left( \frac{N_{oj}}{N_{oo}} \right)} \right] \times 100$$

$N_{ij}$: Number of papers having j authors in a block I

$N_{io}$: Total Output of Block I

$N_{oj}$: Number of papers having j authors for all blocks;

$N_{oo}$: Total number of papers for all authors and all blocks

J=1, 2, 3, > 4

CAI = 100 implies that co-authorship in a particular block for a particular type of authorship corresponds to the world average, CAI >100 reflects higher than average co-authorship effort and CAI < 100 lower than average co-authorship effort in a particular block for a particular type of authorship.

The pattern of co-authorship index is also calculated year-wise, and the same is shown in table 23. It is found from the table that the co-authorship index for single-author papers was 98.90 in the year 1990, which increased to 100.89 in the year 2019. Subsequently, it shows the inclining trend wherein it was 99.66 in the year 2012.

It is revealed from the table 23 that publications follow the same pattern of increasing CAI from two to four authorship patterns except for five and above five authors, which is decreasing from the year 1990 to 2019. CAI of single authorship pattern gets 100 from the year 2011, and it goes down next year, i.e., 2012, and then again it attained 100 and followed the same pattern up to 2019. CAI of two authorship pattern increased to 100 from the year 2011 also, and it remains a hundred plus up to 2019 except the year 2013 as it shows decline (99.57). CAI of third and fourth authorship pattern has been increased from the year 2012, and it remains continuously up to the year 2019. The CAI of the fifth authorship pattern has a different magnitude, up to the year 2011, it got 100 plus CAI and then decreases to below 100 up to 2019.

The inference from the below table depicts that the CAI pattern increases regarding single, two, three, and four authorships, but five and above five authorship pattern has a decreasing trend in terms of CAI.

**Table 23: Year Wise Co-Authorship Index of CAD**

| Year | Single | CAI | Two | CAI | Three | CAI | Four | CAI | Five & Above Five | CAI | Total |
|---|---|---|---|---|---|---|---|---|---|---|---|
| 1990 | 5 | 98.90 | 25 | 89.36 | 40 | 81.61 | 39 | 84.40 | 49 | 254.19 | 158 |
| 1991 | 9 | 98.68 | 45 | 88.19 | 64 | 82.99 | 44 | 93.53 | 104 | 224.40 | 266 |
| 1992 | 11 | 97.33 | 44 | 86.19 | 45 | 88.26 | 48 | 89.08 | 86 | 233.04 | 234 |
| 1993 | 15 | 93.71 | 27 | 90.40 | 46 | 81.66 | 20 | 99.75 | 74 | 218.64 | 182 |
| 1994 | 16 | 93.79 | 29 | 90.45 | 23 | 96.46 | 39 | 89.77 | 89 | 201.15 | 196 |
| 1995 | 11 | 96.43 | 34 | 87.83 | 20 | 98.19 | 28 | 96.14 | 104 | 173.94 | 197 |
| 1996 | 23 | 94.06 | 39 | 91.93 | 59 | 87.12 | 57 | 90.11 | 113 | 225.38 | 291 |
| 1997 | 22 | 94.57 | 39 | 92.21 | 45 | 92.72 | 46 | 94.71 | 145 | 188.57 | 297 |
| 1998 | 29 | 92.61 | 32 | 95.23 | 48 | 92.41 | 51 | 93.69 | 151 | 189.56 | 311 |
| 1999 | 23 | 95.28 | 46 | 91.92 | 50 | 93.35 | 56 | 93.77 | 168 | 187.99 | 343 |
| 2000 | 25 | 95.73 | 44 | 94.45 | 46 | 96.68 | 60 | 95.21 | 224 | 161.60 | 399 |
| 2001 | 18 | 97.02 | 41 | 94.06 | 64 | 89.85 | 45 | 98.06 | 192 | 171.95 | 360 |
| 2002 | 19 | 98.17 | 52 | 94.89 | 71 | 93.45 | 87 | 92.17 | 261 | 172.20 | 490 |
| 2003 | 32 | 96.28 | 59 | 94.95 | 80 | 93.64 | 76 | 96.83 | 312 | 162.81 | 559 |
| 2004 | 29 | 98.11 | 69 | 96.21 | 73 | 98.46 | 104 | 96.25 | 462 | 137.48 | 737 |
| 2005 | 35 | 97.63 | 58 | 98.40 | 87 | 97.31 | 117 | 95.55 | 497 | 137.82 | 794 |
| 2006 | 21 | 99.97 | 77 | 97.93 | 133 | 94.66 | 121 | 98.42 | 642 | 130.48 | 994 |
| 2007 | 42 | 98.63 | 77 | 99.47 | 117 | 98.83 | 157 | 97.69 | 831 | 118.30 | 1224 |
| 2008 | 42 | 99.24 | 98 | 99.15 | 162 | 97.37 | 183 | 98.26 | 1001 | 120.26 | 1486 |
| 2009 | 60 | 98.72 | 129 | 98.54 | 157 | 99.74 | 196 | 99.86 | 1257 | 111.01 | 1799 |
| 2010 | 72 | 98.71 | 151 | 98.69 | 183 | 99.97 | 255 | 98.77 | 1488 | 113.33 | 2149 |
| 2011 | 43 | 100.30 | 127 | 100.54 | 211 | 99.67 | 295 | 98.29 | 1724 | 103.78 | 2400 |
| 2012 | 66 | 99.66 | 147 | 100.43 | 225 | 100.26 | 286 | 100.31 | 2003 | 97.82 | 2727 |
| 2013 | 41 | 100.88 | 207 | 99.57 | 261 | 100.73 | 323 | 101.21 | 2504 | 91.89 | 3336 |
| 2014 | 51 | 100.72 | 179 | 101.00 | 261 | 101.54 | 361 | 101.09 | 2833 | 85.19 | 3685 |
| 2015 | 57 | 100.72 | 181 | 101.50 | 253 | 102.59 | 374 | 101.92 | 3265 | 77.17 | 4130 |
| 2016 | 51 | 100.99 | 220 | 101.05 | 350 | 100.92 | 433 | 101.46 | 3523 | 84.85 | 4577 |
| 2017 | 48 | 101.11 | 171 | 102.39 | 326 | 101.89 | 450 | 101.61 | 3827 | 76.03 | 4822 |
| 2018 | 59 | 100.97 | 221 | 101.63 | 361 | 101.68 | 510 | 101.06 | 4041 | 81.68 | 5192 |
| 2019 | 69 | 100.89 | 232 | 101.83 | 388 | 101.84 | 526 | 101.73 | 4486 | 78.53 | 5701 |
| Total | 1044 | | 2900 | | 4249 | | 5387 | | 36456 | | 50036 |

## 4.24 Year Wise Benchmark for Co-Authorship Index on CAD

**Table 24: Year Wise Benchmark for Co-Authorship Index**

| Year | CAI Single Author | CAI Two Author | CAI Three Author | CAI Four Author | CAI Five & Above Five |
|---|---|---|---|---|---|
| 1990 | -- | -- | -- | -- | ++ |
| 1991 | -- | -- | -- | -- | ++ |
| 1992 | -- | -- | -- | -- | ++ |
| 1993 | -- | -- | -- | -- | ++ |
| 1994 | -- | -- | -- | -- | ++ |
| 1995 | -- | -- | -- | -- | ++ |
| 1996 | -- | -- | -- | -- | ++ |
| 1997 | -- | -- | -- | -- | ++ |
| 1998 | -- | -- | -- | -- | ++ |
| 1999 | -- | -- | -- | -- | ++ |
| 2000 | -- | -- | -- | -- | ++ |
| 2001 | -- | -- | -- | -- | ++ |
| 2002 | -- | -- | -- | -- | ++ |
| 2003 | -- | -- | -- | -- | ++ |
| 2004 | -- | -- | -- | -- | ++ |
| 2005 | -- | -- | -- | -- | ++ |
| 2006 | -- | -- | -- | -- | ++ |
| 2007 | -- | -- | -- | -- | ++ |
| 2008 | -- | -- | -- | -- | ++ |
| 2009 | -- | -- | -- | -- | ++ |
| 2010 | -- | -- | -- | -- | ++ |
| 2011 | ++ | ++ | -- | -- | ++ |
| 2012 | -- | ++ | ++ | ++ | -- |
| 2013 | ++ | -- | ++ | ++ | -- |
| 2014 | ++ | ++ | ++ | ++ | -- |
| 2015 | ++ | ++ | ++ | ++ | -- |
| 2016 | ++ | ++ | ++ | ++ | -- |
| 2017 | ++ | ++ | ++ | ++ | -- |
| 2018 | ++ | ++ | ++ | ++ | -- |
| 2019 | ++ | ++ | ++ | ++ | -- |

In order to identify the priority status of the research productivity index, the values are replaced with the Benchmark (symbol). CAI has been further simplified as symbolic representation as CAI = 100 for the usual average of co-authorship index then the value of more than 99 value is called as above average as ++, Less than 100 value is called as below average of CAI as - - and the same is shown in table 24.

### 4.25 Analysis of Prolific Authors on CAD from BRICS Countries

In this research period from 1990 to 2019, 384758 scientists have contributed 50036 articles over 3066 journals on coronary artery disease research. According to this, the study has taken most productive authors and their published numbers of records and were ranked according to their publications. The authors contributed the highest publications in coronary artery disease scientific literature among the BRICS countries have been included in the tables below. The first ten authors included were identified as the most productive contributors to coronary artery disease output in tables from 25A to 25E.

#### 4.25.1 Prolific Authors from Brazil

**Table 25A: Prolific Authors on CAD from Brazil**

| S.No. | Authors | Affiliated Institutions | Publications |
|---|---|---|---|
| 1 | Ramires JAF | Universidade de Sao Paulo | 231 |
| 2 | Santos RD | Hospital Israelita Albert Einstein | 146 |
| 3 | Hueb W | Universidade de Sao Paulo | 137 |
| 4 | Kalil R | Institute Cardiol Do Rio Grande Do Sul | 117 |
| 5 | Pereira AC | University Sao Paulo | 110 |
| 6 | Rochitte CE | Universidade de Sao Paulo | 99 |
| 7 | Cesar LAM | University Sao Paulo | 96 |
| 8 | Nicolau JC | Universidade de Sao Paulo | 94 |
| 9 | Maranhao RC | University Sao Paulo | 93 |
| 10 | Abizaid A | Instituto Dante Pazzanese de Cardiologia | 93 |

The list of ten top authors who produced the highest contribution to research output on CAD in Brazil is given in the table. In terms of a number of publications, Ramires JAF from Universidade de Sao Paulo is the most productive author with 231 publications followed by Santos RD 146 from Hospital Israelita Albert Einstein, Hueb W 137 from Universidade de Sao Paulo, Kalil R 117 publications from Institute Cardiol Do Rio Grande Do Sul and Pereira AC contributed 110 publications, the author is from University Sao Paulo. It is also noted that 5 out of 10 prolific authors contributed more than a hundred research publications each while the rest five authors contributed more than 90 publications each.

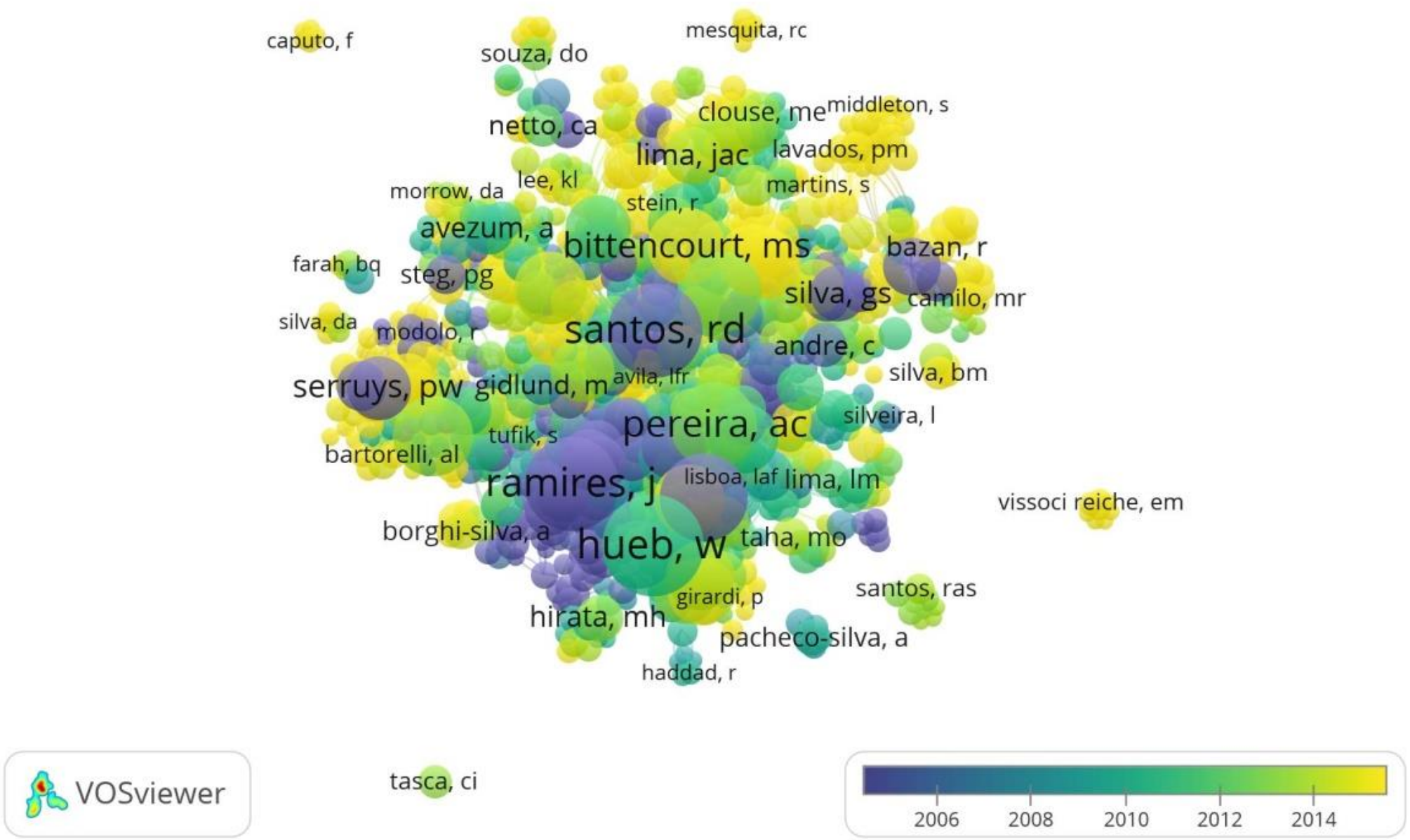

**Figure 13: Prolific Authors from Brazil**

It could be concluded from the above analysis, the authors “Ramires JAF” from Universidade de Sao Paulo, “Santos RD” from Hospital Israelita Albert Einstein, and “Hueb W” from Universidade de Sao Paulo have contributed the greatest number of publications and were identified the most productive authors on coronary artery disease research output from Brazil.

### 4.25.2 Prolific Authors from Russia

| Table 25B: Prolific Authors on CAD from Russia | | | |
|---|---|---|---|
| S.No. | Authors | Affiliated Institutions | Publications |
| 1 | Barbarash OL | Science& Res Inst Complex Problems Cardiology | 100 |
| 2 | Skvortsova VI | Russian State Med University | 93 |
| 3 | Orekhov AN | Research Institute of Human Morphology | 84 |
| 4 | Belenkov YN | Sechenov First Moscow State Med University | 83 |
| 5 | Sidorenko BA | Centre State Medical Academy | 78 |
| 6 | Kukharchuk VV | National Medical Research Center of Cardiology | 63 |
| 7 | Gratsiansky NA | Research Institute Physical Chem Medicine | 55 |
| 8 | Deev AD | National Research Centre Preventive Medicine | 52 |
| 9 | Masenko VP | National Medical Research Center of Cardiology | 49 |
| 10 | Pokushalov E | National Medical Research Centre | 47 |

The list of top ten authors who produced the highest contribution to research output on CAD in Russia is given in the table. In terms of the number of publications, Barbarash OL from Science & Research Institute Complex Problems Cardiology is the most productive author with 100 publications, followed by Skvortsova VI 93 from Russian State Medical University, Orekhov AN 84 from Research Institute of Human Morphology, and Belenkov YN 83 publications from Sechenov First Moscow State Med University. It is also noted that 1 out of 10 prolific authors contributed more than a hundred research publications, while nine authors contributed less than 100 journals articles each.

It could be concluded from the above analysis, the authors "Barbarash OL" from Science & Research Institute Complex Problems Cardiology, "Skvortsova VI" from Russian State Medical University, and "Orekhov AN" from Research Institute of Human Morphology have contributed the more number of publications and were identified the most productive authors on coronary artery disease research output from Russia.

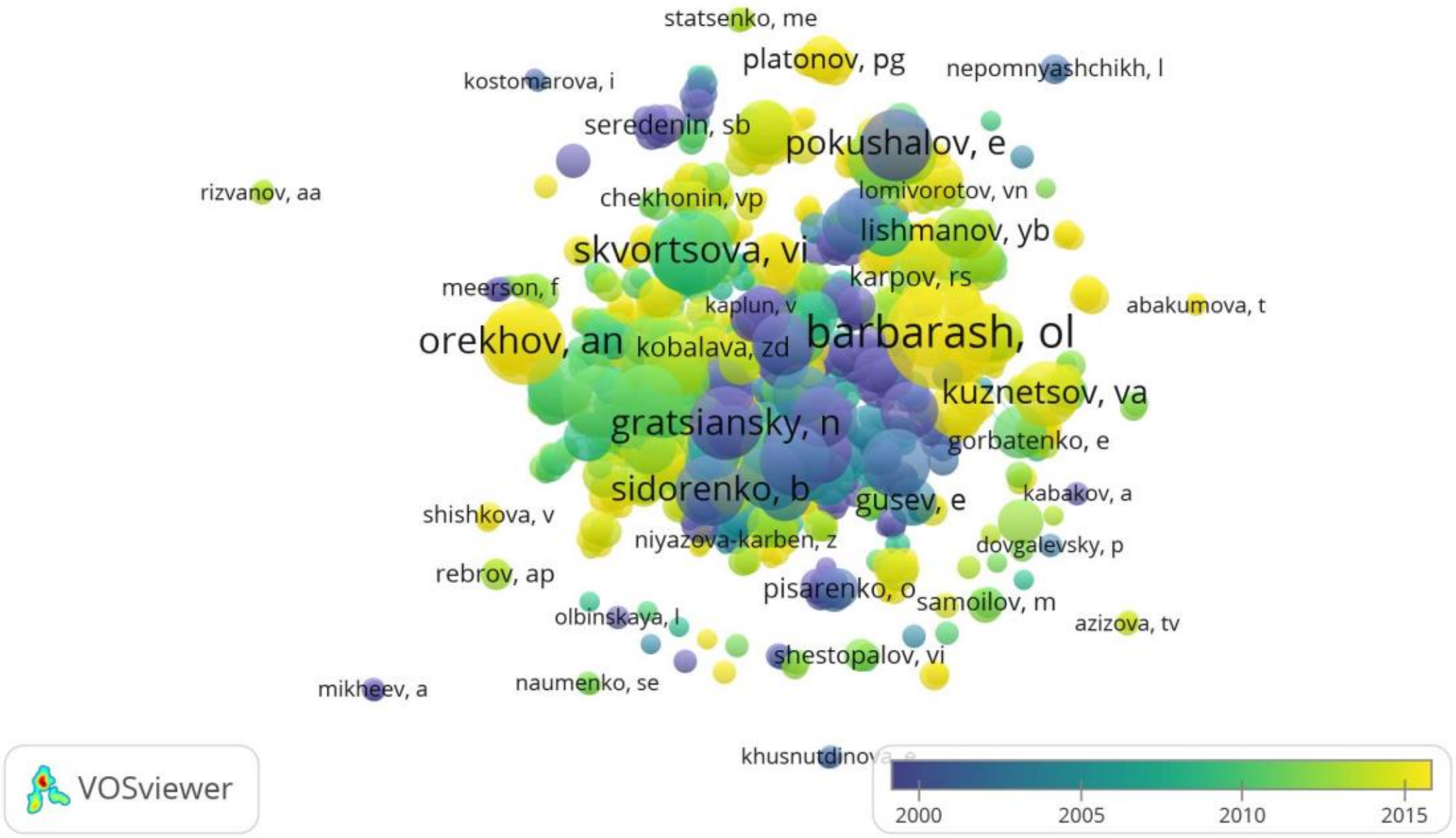

**Figure 14: Prolific Authors from Russia**

### 4.25.3 Prolific Authors from India

| Table 25C: Prolific Authors on CAD from India | | | |
|---|---|---|---|
| S.No. | Authors | Affiliated Institutions | Publications |
| 1 | Kumar A | Sanjay Gandhi Postgraduate Institute of Medical Sciences | 147 |
| 2 | Kumar S | Indian Institute of Technology (IIT) - Guwahati | 100 |
| 3 | Prasad K | All India Institute of Medical Sciences (AIIMS) New Delhi | 88 |
| 4 | Sharma A | Zydus Hospital, Ahmadabad, Gujarat | 78 |
| 5 | Kaul S | Nizam's Institute of Medical Sciences | 75 |
| 6 | Mohan V | Madras Diabetes Research Foundation | 74 |
| 7 | Singh N | All India Institute of Medical Sciences (AIIMS) New Delhi | 68 |
| 8 | Singh M | All India Institute of Medical Sciences (AIIMS) New Delhi | 63 |
| 9 | Gupta R | Rajasthan University Health Science, Academic Res Dev Unit | 62 |
| 10 | Gupta A | Lady Hardinge Medical College & Associated Hospital | 61 |

The top 10 Indian author’s contribution to coronary artery disease research varied from 61 to 147 publications, and they together accounted for 17.33% (816) publication share output during 1990-2019. The table presents a scientometric profile of these 10 India authors. Two authors registered higher publications productivity than a group of ten authors: Kumar A (147 papers) from Sanjay Gandhi Postgraduate Institute of Medical Sciences and Kumar S (100 papers) from Indian Institute of Technology (IIT) - Guwahati, and other authors Prasad K (88 papers) from All India Institute of Medical Sciences (AIIMS) New Delhi, Sharma A (78 papers) from Zydus Hospital, Ahmadabad, Gujarat, Kaul S (75 papers) from Nizam's Institute of Medical Sciences, Mohan V (74 papers) from Madras Diabetes Research Foundation, Singh N (68 papers) from All India Institute of Medical Sciences (AIIMS) New Delhi, Singh M (63 papers) from All India Institute of Medical Sciences (AIIMS) New Delhi, Gupta R (62 papers) from Rajasthan University Health Science, Academic Res Dev Unit, and Gupta A (61 papers) from Lady Hardinge Medical College & Associated Hospital during 1990-2019.

It could be concluded from the above analysis, the authors “Kumar A”, “Kumar S”, and “Prasad K” have contributed the more number of publications and were identified the most productive authors on coronary artery disease research output from India.

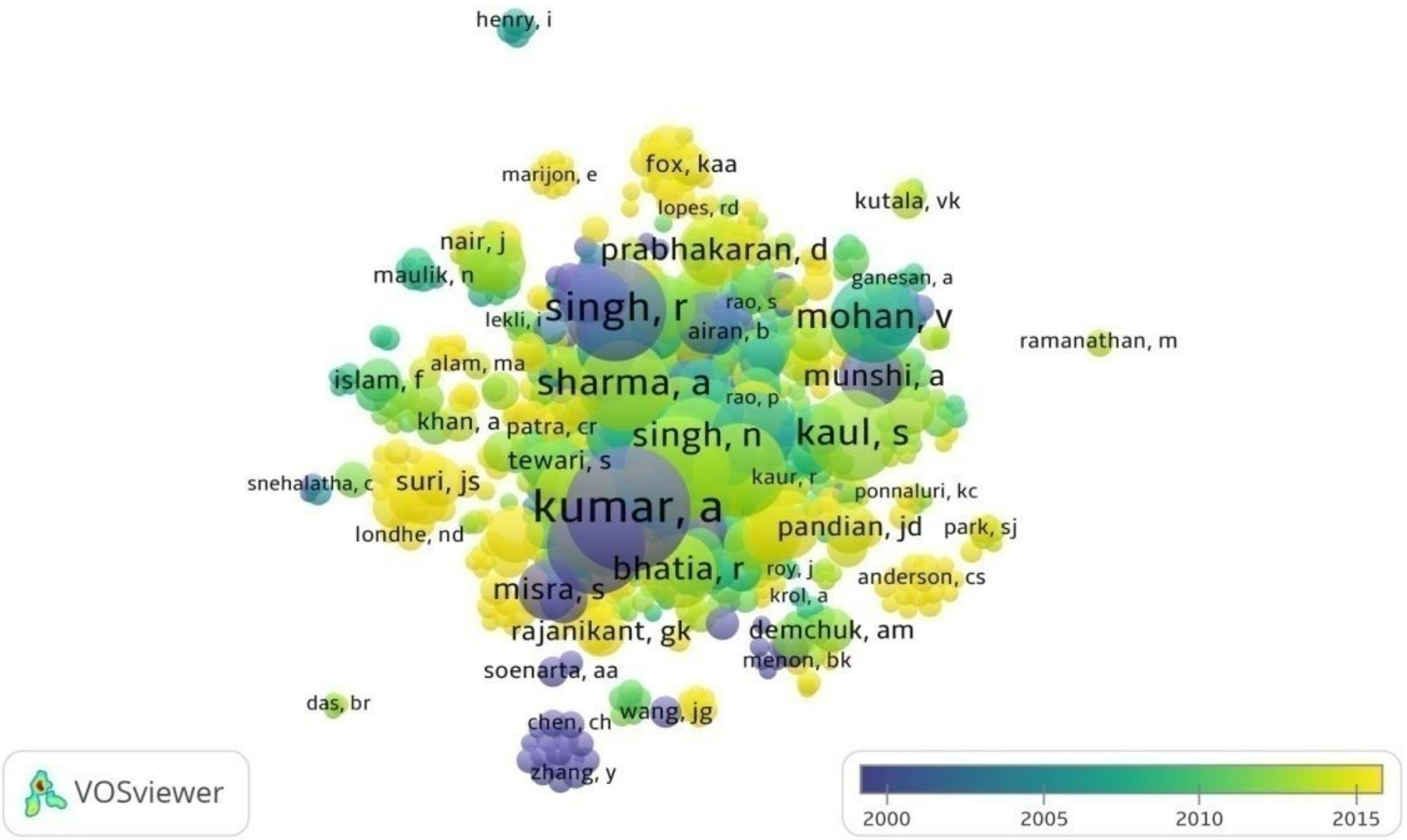

**Figure 15: Prolific Authors from India**

### 4.25.4 Prolific Authors from the Peoples Republic of China

| Table 25D: Prolific Authors on CAD from China | | | |
|---|---|---|---|
| S.No. | Authors | Affiliated Institutions | Publications |
| 1 | Zhang Y | University of Shanghai for Science & Technology | 1196 |
| 2 | Wang Y | Nanchang University | 1007 |
| 3 | Li Y | Hebei University | 796 |
| 4 | Liu Y | Hunan Polytechnic Environmental & Biology College | 774 |
| 5 | Wang J | Guangzhou Medical University | 708 |
| 6 | Zhang L | Anhui Medical University | 657 |
| 7 | Li J | Hubei Province Hospital Traditional Chinese Medical | 638 |
| 8 | Zhang J | Hangzhou Medical College | 615 |
| 9 | Wang L | Chinese Academy of Science | 580 |
| 10 | Chen J | Capital Medical University | 551 |

Among the 384758 authors, “Zhang Y” from the University of Shanghai for Science & Technology, School Medical Instrument & Food Engineering has published 1196 articles, contributed the highest number of publications in coronary artery disease research and occupied the first rank in the present context. Next to that, “Wang Y” from Nanchang University, Affiliated Hospital has published 1007 articles, and this author is also from the Peoples Republic of China, and the author occupied the second rank. “Li Y,” which is from Hebei University, College of Electronic and Information Engineering, has published 796 publications were measured and has occupied the third rank. Next in line is “Liu Y” affiliated to Hunan Polytechnic Environmental and Biology College, College of Medical Hengyang contributed 774 publications, Wang J from Guangzhou Medical University, Department Pathology & Pathophysiology contributed 708 publications, “Zhang L” from Anhui Medical University, Department Anesthesiology & Perioperative Medical contributed 657, “Li J” contributed 638 affiliated to Hubei Province Hospital Traditional Chinese Medical, Emergency Department, and “Zhang J” from Hangzhou Medical College, Zhejiang Province Peoples Hospital has contributed 615 publication on coronary artery disease and the author occupies the eighth rank among the authors contribution scientific literature on coronary artery disease in BRICS countries. The remaining authors produced less than 600 contributions to coronary artery disease literature.

It could be concluded from the above analysis, the authors “Zhang Y” from the University of Shanghai for Science & Technology, “Wang Y” from Nanchang University, and “Li Y” from Hebei University have contributed the more number of publications and were identified the most productive authors on coronary artery disease research output.

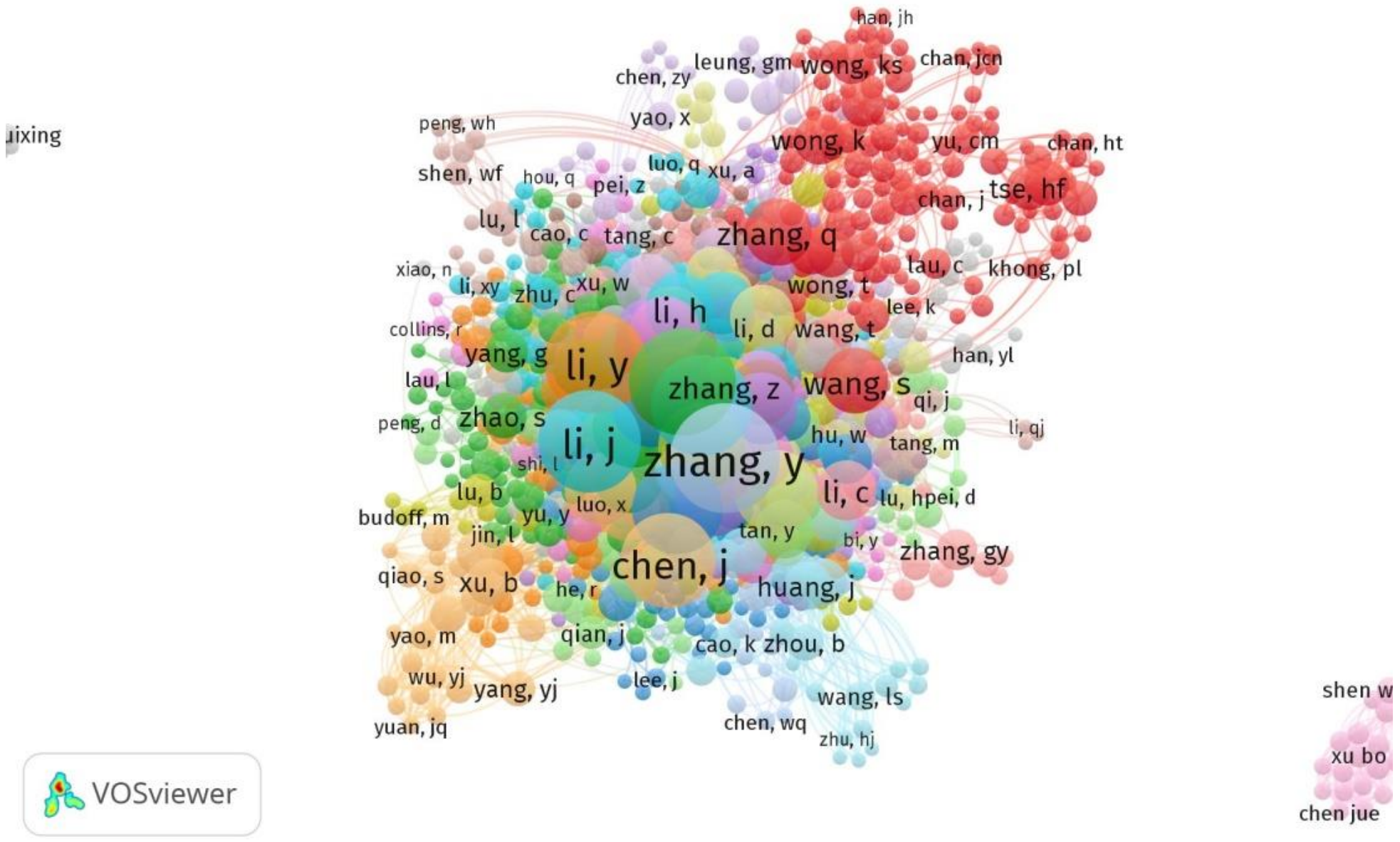

**Figure 16: Prolific Authors from China**

## 4.25.5 Prolific Authors from South Africa

| Table 25E: Prolific Authors on CAD from South Africa | | | |
|---|---|---|---|
| S.No. | Authors | Affiliated Institutions | Publications |
| 1 | Opie LH | University of Cape Town | 114 |
| 2 | Lochner A | Stellenbosch University | 68 |
| 3 | Lecour S | University of Cape Town | 53 |
| 4 | Sliwa K | University of Cape Town | 50 |
| 5 | Pretorius E | Stellenbosch University | 35 |
| 6 | Genade S | Stellenbosch University | 33 |
| 7 | Mayosi BM | University of Cape Town | 31 |
| 8 | Raal FJ | University of Witwatersrand | 29 |
| 9 | Marais AD | University Cape Town, Faculty of Health Science | 29 |
| 10 | Malan L | North West University - South Africa | 26 |

The list of ten top authors who produced the highest contribution to research output on CAD from South Africa is given in the table. In terms of a number of publications, Opie LH from the University of Cape Town is the most productive author with 114 publications followed by Lochner A 68 from Stellenbosch University, Lecour S 53 from University of Cape Town, and Sliwa K 50 publications from University of Cape Town. It is also noted that 1 out of 10 prolific authors contributed more than a hundred research publications while rest nine authors contributed more than 20 publications each.

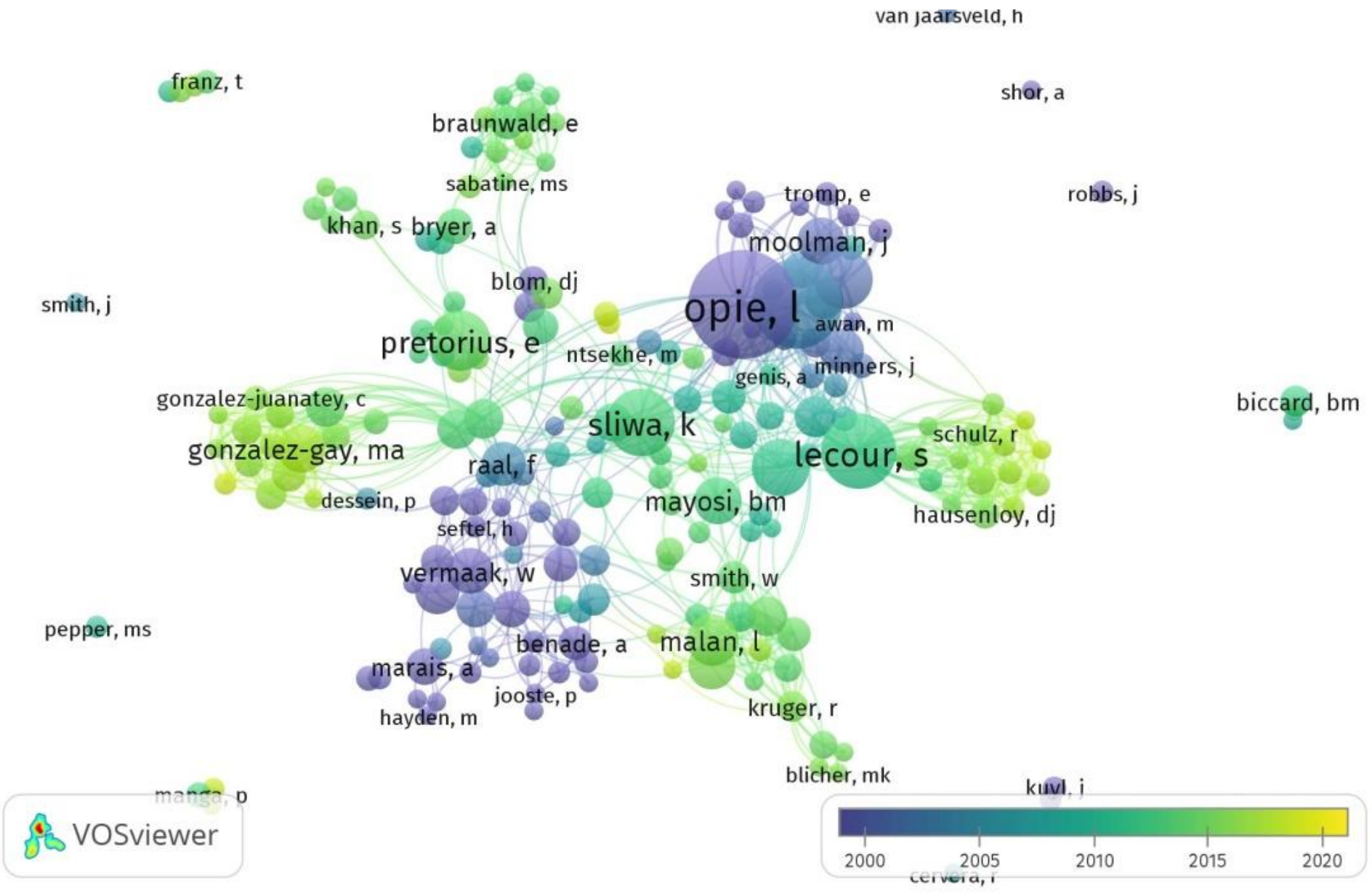

**Figure 17: Prolific Authors from South Africa**

It could be concluded from the above analysis, the authors “Opie LH” from the University of Cape Town, “Lochner A” from Stellenbosch University, and “Lecour S” from the University of Cape Town have contributed the more number of publications and were identified the most productive authors on coronary artery disease research output.

### 4.26 Analysis of Year Wise Top 20 Author Productivity from BRICS

Another descriptive analysis commonly used in scientometric studies concerns the most productive authors. This type of analysis draws up a list of the most productive authors and the total number of publications year wise. The percentage of publications by these authors concerning the total number of publications included in the study or the period in which they have been most productive can also be included in this descriptive analysis.

From among the prolific authors, twenty authors year-wise productivity has been given in the table given below. The author Zhang Y which is from the University of Shanghai for Science & Technology, School Medical Instrument & Food Engineering, has started publishing in the year 1999. The author produced two research papers in the year 1999 and has the production of 172 in the year 2019. Wang Y from Nanchang University, Affiliated Hospital has contributed 1007 research articles and has begun his work in the year 1993 and has produced one article in the same year and his production increased with the years and contributed 162 articles in the year 2019. The author Li Y from Hebei University has also begun his research contribution in the year 2003. Since then, the author continuously is contributing to the field of coronary artery disease research. The author has produced 796 articles during the study period of thirty years, and 121 articles have been contributed in the year 2019. Liu Y from Hunan Polytechnic Environmental and Biology College, College of Medical Hengyang, China, has also been very productive has contributed from the year 1991 and has been continuously contributing. The author has contributed most of his research published in the year 2019. It is observed that almost all the prolific author’s contribution is highest in the year 2019.

**Table 26: Year Wise Top 20Author Productivity from BRICS**

| S.No. | Authors | Year | | | | | | | | | | | | | | | | | | | | | | | | | | | | | | Total |
|---|---|---|---|---|---|---|---|---|---|---|---|---|---|---|---|---|---|---|---|---|---|---|---|---|---|---|---|---|---|---|---|---|
| | | 1990 | 1991 | 1992 | 1993 | 1994 | 1995 | 1996 | 1997 | 1998 | 1999 | 2000 | 2001 | 2002 | 2003 | 2004 | 2005 | 2006 | 2007 | 2008 | 2009 | 2010 | 2011 | 2012 | 2013 | 2014 | 2015 | 2016 | 2017 | 2018 | 2019 | |
| 1 | Zhang Y | 0 | 0 | 0 | 0 | 0 | 0 | 0 | 0 | 0 | 2 | 3 | 3 | 2 | 3 | 8 | 4 | 13 | 28 | 20 | 39 | 41 | 49 | 43 | 76 | 86 | 135 | 148 | 165 | 156 | 172 | 1196 |
| 2 | Wang Y | 0 | 0 | 0 | 1 | 0 | 0 | 0 | 0 | 1 | 0 | 0 | 1 | 1 | 3 | 2 | 4 | 10 | 19 | 12 | 28 | 26 | 45 | 65 | 71 | 97 | 82 | 115 | 114 | 148 | 162 | 1007 |
| 3 | Li Y | 0 | 0 | 0 | 1 | 1 | 0 | 0 | 0 | 0 | 0 | 0 | 1 | 1 | 0 | 4 | 8 | 12 | 7 | 15 | 21 | 29 | 35 | 53 | 58 | 70 | 73 | 91 | 98 | 97 | 121 | 796 |
| 4 | Liu Y | 0 | 1 | 0 | 0 | 0 | 0 | 0 | 0 | 1 | 1 | 0 | 0 | 2 | 3 | 1 | 2 | 9 | 9 | 15 | 31 | 30 | 38 | 47 | 74 | 77 | 72 | 80 | 76 | 87 | 118 | 774 |
| 5 | Wang J | 0 | 0 | 0 | 0 | 0 | 0 | 0 | 3 | 0 | 0 | 0 | 0 | 0 | 1 | 4 | 1 | 11 | 6 | 13 | 19 | 19 | 25 | 31 | 60 | 75 | 91 | 78 | 85 | 84 | 102 | 708 |
| 6 | Zhang L | 0 | 0 | 1 | 1 | 1 | 0 | 0 | 0 | 1 | 0 | 0 | 0 | 6 | 1 | 4 | 9 | 12 | 7 | 14 | 26 | 37 | 35 | 44 | 35 | 50 | 49 | 84 | 73 | 83 | 84 | 657 |
| 7 | Li J | 0 | 0 | 1 | 0 | 0 | 0 | 0 | 1 | 1 | 1 | 0 | 1 | 2 | 1 | 2 | 0 | 7 | 14 | 18 | 25 | 23 | 30 | 30 | 44 | 53 | 76 | 60 | 73 | 83 | 92 | 638 |
| 8 | Zhang J | 0 | 0 | 0 | 1 | 0 | 0 | 1 | 0 | 0 | 0 | 1 | 0 | 1 | 3 | 3 | 3 | 8 | 10 | 13 | 16 | 24 | 29 | 30 | 46 | 55 | 76 | 67 | 70 | 70 | 88 | 615 |
| 9 | Wang L | 0 | 0 | 0 | 0 | 0 | 0 | 0 | 0 | 0 | 0 | 0 | 1 | 0 | 4 | 3 | 1 | 6 | 2 | 10 | 13 | 15 | 24 | 32 | 46 | 54 | 70 | 68 | 81 | 61 | 89 | 580 |
| 10 | Chen J | 0 | 0 | 0 | 0 | 0 | 0 | 0 | 0 | 0 | 0 | 0 | 0 | 0 | 4 | 3 | 1 | 12 | 16 | 19 | 16 | 25 | 22 | 24 | 41 | 59 | 42 | 49 | 60 | 66 | 92 | 551 |
| 11 | Liu J | 0 | 0 | 0 | 0 | 0 | 0 | 4 | 0 | 0 | 0 | 0 | 1 | 0 | 0 | 4 | 1 | 7 | 8 | 7 | 12 | 14 | 13 | 33 | 46 | 57 | 54 | 60 | 63 | 66 | 89 | 539 |
| 12 | Li L | 0 | 1 | 0 | 0 | 0 | 0 | 2 | 0 | 0 | 0 | 1 | 1 | 1 | 1 | 5 | 2 | 3 | 5 | 8 | 11 | 26 | 26 | 39 | 39 | 49 | 57 | 54 | 52 | 62 | 66 | 511 |
| 13 | Wang YJ | 0 | 0 | 0 | 0 | 0 | 0 | 0 | 0 | 0 | 0 | 0 | 0 | 0 | 0 | 2 | 0 | 5 | 0 | 12 | 9 | 18 | 17 | 26 | 41 | 48 | 48 | 59 | 63 | 50 | 78 | 476 |
| 14 | Zhang H | 0 | 0 | 1 | 0 | 0 | 0 | 0 | 0 | 0 | 0 | 0 | 1 | 0 | 1 | 2 | 5 | 6 | 11 | 8 | 11 | 15 | 25 | 23 | 29 | 52 | 33 | 61 | 49 | 50 | 84 | 467 |
| 15 | Wang X | 0 | 0 | 0 | 0 | 0 | 1 | 0 | 0 | 0 | 0 | 0 | 0 | 0 | 1 | 5 | 4 | 4 | 11 | 16 | 8 | 17 | 16 | 28 | 18 | 37 | 43 | 48 | 57 | 69 | 73 | 456 |
| 16 | Wang YL | 0 | 1 | 0 | 0 | 1 | 0 | 1 | 0 | 0 | 0 | 0 | 0 | 0 | 1 | 3 | 0 | 7 | 2 | 6 | 6 | 7 | 13 | 20 | 33 | 44 | 39 | 61 | 58 | 59 | 83 | 445 |
| 17 | Yang Y | 0 | 0 | 0 | 0 | 0 | 0 | 0 | 0 | 0 | 0 | 0 | 0 | 0 | 1 | 1 | 1 | 3 | 2 | 12 | 8 | 15 | 15 | 23 | 39 | 31 | 43 | 50 | 64 | 62 | 71 | 441 |
| 18 | Li H | 0 | 0 | 1 | 0 | 0 | 0 | 0 | 2 | 0 | 0 | 1 | 0 | 1 | 3 | 1 | 3 | 5 | 5 | 8 | 3 | 10 | 22 | 17 | 31 | 44 | 35 | 53 | 53 | 59 | 81 | 438 |
| 19 | Wang W | 0 | 0 | 0 | 0 | 1 | 0 | 0 | 0 | 1 | 1 | 1 | 0 | 0 | 1 | 1 | 3 | 4 | 15 | 12 | 17 | 22 | 18 | 29 | 21 | 35 | 30 | 48 | 46 | 43 | 64 | 413 |
| 20 | Zhang Q | 0 | 0 | 0 | 0 | 0 | 0 | 0 | 0 | 1 | 0 | 1 | 0 | 0 | 3 | 3 | 5 | 8 | 9 | 9 | 21 | 15 | 16 | 26 | 30 | 35 | 39 | 48 | 60 | 40 | 38 | 407 |

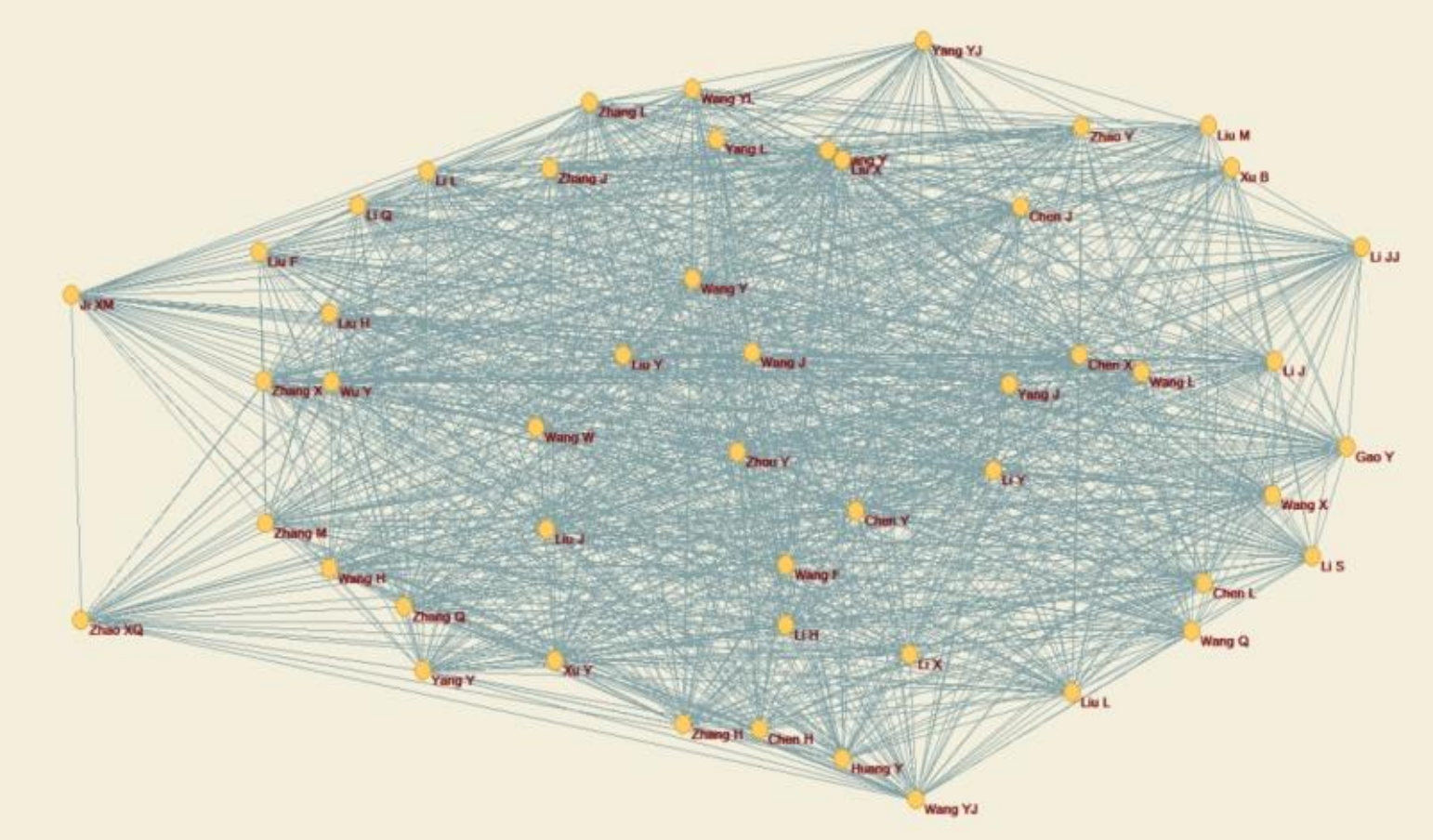

**Figure 18: Author Collaboration**

### 4.27 H-Index and Other Metrices of Most Prolific Authors on CAD from BRICS

The h-index is an author level metric that attempts to measure both the productivity and citation impact of a scientist or a scholar's publications. The index is based on the set of the scientist's most cited papers and the number of citations that they have received in other publications. The top 10 authors with the highest H-index have been analyzed to measure their publication output's quality impact. The complete publication, total citation, h core, and other indices like h, R, AR, a, M, Q2, e, and P were analyzed, and the values are shown in tables.

#### 4.27.1 Analysis of Top Ten Prolific Authors from Brazil with Metric Indices

The table highlights that Santos RD from the Hospital Israelita Albert Einstein has got the second highest publications of 146 with 5804 citations, an h-index of 32, R-index (70.150), AR-index (13.909), hnom-index (0.219), a-index (181.375), M-index (1.067), Q2-index (5.842), e-index (62.426) and P-index (61.334), which is followed by Ramires JAF from Universidade de Sao Paulo who has scored 3750 citations with 146 articles and has recorded 29 h-index. Among the top 10 authors, Santos RD has appended the highest h-index 32 with 5804 citations and is the second highest contributor in terms of output on CAD from Brazil, followed by Ramires JAF has second place in terms of h-index having 29 with 3750 citations and 146 publications.

To achieve the twelfth objective of evaluate the various indices in authors research output performance, the analysis of various indices in authors are carried out and it is expounded in tables 27A, 27B, 27C, 27D, and 27E.

| Table 27A: Prolific Authors from Brazil with Metric Indices | | | | | | | | | | | | | |
|---|---|---|---|---|---|---|---|---|---|---|---|---|---|
| S.No. | Authors | Citation Sum within h-core | All Citations | All Publications | h-index | R-Index | AR-Index | hnom Index | a- index | M-index | Q2 index | e-index | P-index |
| 1 | Santos RD | 4921 | 5804 | 146 | 32 | 70.150 | 13.909 | 0.219 | 181.375 | 1.067 | 5.842 | 62.426 | 61.334 |
| 2 | Ramires JAF | 2502 | 3750 | 231 | 29 | 50.020 | 11.180 | 0.126 | 129.310 | 0.967 | 5.295 | 40.755 | 39.338 |
| 3 | Avezum A | 8979 | 9097 | 45 | 26 | 94.758 | 17.414 | 0.578 | 349.885 | 0.867 | 4.747 | 91.121 | 122.517 |
| 4 | Abizaid A | 2408 | 2831 | 93 | 26 | 49.071 | 9.714 | 0.280 | 108.885 | 0.867 | 4.747 | 41.617 | 44.170 |
| 5 | Nicolau JC | 5583 | 5990 | 94 | 25 | 74.719 | 14.130 | 0.266 | 239.600 | 0.833 | 4.564 | 70.413 | 72.540 |
| 6 | Rochitte CE | 2937 | 3412 | 99 | 25 | 54.194 | 10.665 | 0.253 | 136.480 | 0.833 | 4.564 | 48.083 | 48.992 |
| 7 | Duncan BB | 5974 | 6145 | 38 | 24 | 77.292 | 14.312 | 0.632 | 256.042 | 0.800 | 4.382 | 73.471 | 99.790 |
| 8 | Lotufo PA | 8057 | 8517 | 90 | 23 | 89.761 | 16.849 | 0.256 | 370.304 | 0.767 | 4.199 | 86.764 | 93.063 |
| 9 | Netto CA | 1159 | 1370 | 43 | 22 | 34.044 | 6.758 | 0.512 | 62.273 | 0.733 | 4.017 | 25.981 | 35.209 |
| 10 | Lemos PA | 1354 | 1676 | 68 | 22 | 36.797 | 7.474 | 0.324 | 76.182 | 0.733 | 4.017 | 29.496 | 34.568 |

The study observed that highly citations and h-index by the authors; Santos RD from the Hospital Israelita Albert Einstein has the first position in the h-index having 32 h-index among 146 contributions. Followed by Ramires JAF from Universidade de Sao Paulo has the second position in the h-index (29), and other scientists' have h-index, publications and other indices are also shown in the table. It is evident from the above data that some authors appear among the list having significantly less contribution on CAD. However, their h-index is high, e.g., Duncan BB from Universidade Federal do Rio Grande do Sul Sch Med, the author has only 38 publications but is at the top level regarding the h-index.

### 4.27.2 Analysis of Top Ten Prolific Authors from Russia with Metric Indices

**Table 27B: Prolific Authors from Russia with Metric Indices**

| S.No. | Authors | Citation Sum within h-core | All Citations | All Publications | h-index | R-Index | AR-Index | hnom Index | a- index | M-index | Q2 index | e-index | P-index |
|---|---|---|---|---|---|---|---|---|---|---|---|---|---|
| 1 | Orekhov AN | 903 | 1396 | 84 | 22 | 30.050 | 6.822 | 0.262 | 63.455 | 0.733 | 4.017 | 20.469 | 28.521 |
| 2 | Hankey GJ | 8076 | 8113 | 22 | 19 | 89.867 | 16.445 | 0.864 | 427.000 | 0.633 | 3.469 | 87.835 | 144.094 |
| 3 | Sobenin IA | 538 | 710 | 43 | 17 | 23.195 | 4.865 | 0.395 | 41.765 | 0.567 | 3.104 | 15.780 | 22.717 |
| 4 | Bobryshev YV | 515 | 605 | 28 | 15 | 22.694 | 4.491 | 0.536 | 40.333 | 0.500 | 2.739 | 17.029 | 23.557 |
| 5 | Roth GA | 7962 | 7965 | 16 | 15 | 89.230 | 16.294 | 0.938 | 531.000 | 0.500 | 2.739 | 87.960 | 158.277 |
| 6 | Diaz R | 6630 | 6657 | 22 | 14 | 81.425 | 14.896 | 0.636 | 475.500 | 0.467 | 2.556 | 80.212 | 126.293 |
| 7 | Chistiakov DA | 490 | 564 | 25 | 14 | 22.136 | 4.336 | 0.560 | 40.286 | 0.467 | 2.556 | 17.146 | 23.346 |
| 8 | Tertov VV | 428 | 484 | 24 | 13 | 20.688 | 4.017 | 0.542 | 37.231 | 0.433 | 2.373 | 16.093 | 21.371 |
| 9 | Meretoja A | 7879 | 7879 | 13 | 13 | 88.764 | 16.206 | 1.000 | 606.077 | 0.433 | 2.373 | 87.807 | 168.396 |
| 10 | Zorov DB | 1917 | 1986 | 30 | 13 | 43.784 | 8.136 | 0.433 | 152.769 | 0.433 | 2.373 | 41.809 | 50.849 |

The list of ten top authors who produced the highest contribution to research output on CAD in Russia is given in the table. In terms of the number of publications,

Orekhov AN from Research Institute of Human Morphology is the most productive author with 84 publications, 1396 citations, 22 h-index, 30.050 R-index, 6.822 AR-index, 0.262 hnom-index, 63.455 a-index, 0.733 M-index, 4.017 Q2-index, 20.469 e-index and 28.521 as P-index. It is followed by Sobenin IA from Research Institute of Human Morphology contribute 43, Zorov DB from Lomonosov Moscow State University contributed 30, and Bobryshev YV from Russian Academy of Medical Sciences contributed 28 publications. It is also noted that 1 out of 10 prolific authors contributed more than eighty research publications, while nine authors contributed more than 12 articles each. The h index is highest for Orekhov AN (22), followed by Hankey GJ (19), Sobenin IA (17), followed by Bobryshev YV (15) and Roth GA (15. The data set puts forth the authors' Hankey GJ with 8113 citations, Roth GA with 7965 citations, Meretoja A with 7879 citations and Diaz R with 6657 citations.

#### 4.27.3 Analysis of Top Ten Prolific Authors from India with Metric Indices

The table reflects that Kumar A from Sanjay Gandhi Postgraduate Institute of Medical Sciences has got the highest publications of 147 with 2233 citations an h-index of 23, R-index 74.465, AR-index 14.249, hnom-index 0.500, a-index 164.622, M-index 1.233, Q2-index 6.755, e-index 64.622 and P-index as 79.442, which is followed by Kaul S from Nizam's Institute of Medical Sciences who has scored 958 citations with 75 articles and has recorded 19 h-index. Among the top 10 authors, Mohan V has appended the highest h-index 37 with 6091 citations and is the highest contributor in terms of output on CAD, followed by Gupta R has second place in terms of h-index having 32 with 20427 citations and 62 publications.

The study observed that highly citations and h-index by the authors; Mohan V from Madras Diabetes Research Foundation has the first position in the h-index having 37 h-index among 74 contributions. Followed by Gupta R from Eternal Heart Care Centre & Research Institute has the second position in the h-index (32), and other scientists' have h-index and publications also shown in the table. It is evident from the above data that some authors appear among the list having significantly less contribution on CAD from India. However, their h-index is high, e.g., Prabhakaran D from Public Health Foundation of India, Rollins School Public Health, the author has only 59 publications but is at the top level regarding the h-index.

**Table 27C: Prolific Authors from India with Metric Indices**

| S.No. | Authors | Citation Sum within h-core | All Citations | All Publications | h-index | R-Index | AR-Index | hnom Index | a- index | M-index | Q2 index | e-index | P-index |
|---|---|---|---|---|---|---|---|---|---|---|---|---|---|
| 1 | Mohan V | 5545 | 6091 | 74 | 37 | 74.465 | 14.249 | 0.500 | 164.622 | 1.233 | 6.755 | 64.622 | 79.442 |
| 2 | Gupta R | 20150 | 20427 | 62 | 32 | 141.951 | 26.094 | 0.516 | 638.344 | 1.067 | 5.842 | 138.297 | 188.802 |
| 3 | Prabhakaran D | 8010 | 8343 | 59 | 24 | 89.499 | 16.676 | 0.407 | 347.625 | 0.800 | 4.382 | 86.221 | 105.665 |
| 4 | Kumar A | 1599 | 2233 | 147 | 23 | 39.987 | 8.627 | 0.156 | 97.087 | 0.767 | 4.199 | 32.711 | 32.371 |
| 5 | Deepa R | 2064 | 2155 | 32 | 23 | 45.431 | 8.475 | 0.719 | 93.696 | 0.767 | 4.199 | 39.179 | 52.551 |
| 6 | Niaz MA | 1711 | 1849 | 36 | 23 | 41.364 | 7.851 | 0.639 | 80.391 | 0.767 | 4.199 | 34.380 | 45.624 |
| 7 | Xavier D | 9363 | 9570 | 43 | 23 | 96.763 | 17.861 | 0.535 | 416.087 | 0.767 | 4.199 | 93.989 | 128.662 |
| 8 | Singh M | 896 | 1182 | 63 | 20 | 29.933 | 6.277 | 0.317 | 59.100 | 0.667 | 3.651 | 22.271 | 28.095 |
| 9 | Singh N | 2896 | 3187 | 68 | 20 | 53.814 | 10.307 | 0.294 | 159.350 | 0.667 | 3.651 | 49.960 | 53.058 |
| 10 | Kaul S | 637 | 958 | 75 | 19 | 25.239 | 5.651 | 0.253 | 50.421 | 0.633 | 3.469 | 16.613 | 23.044 |

### 4.27.4 Analysis of Top Ten Prolific Authors from China with Metric Indices

The table describes that Zhang Y from the University of Shanghai for Science & Technology, School Medical Instrument & Food Engineering has got the highest publications of 1196 with 16315 citations an h-index of 51, R-index 76.831, AR-index 23.32, hnom-index 0.04, a-index 115.75, M-index 1.7, Q2-index 9.311, e-index 57.463 and P-index as 60.6, which is followed by Wang Y from Nanchang

University, Affiliated Hospital, who has scored 14254 citations with 1007 articles and has recorded 50 h-index. Among the top 10 authors, Zhang Y has appended the highest h-index 51 with 16315 citations and is the highest contributor in terms of output on CAD, followed by Wang Y has second place in terms of h-index having 50 with 14254 citations and 1007 publications.

**Table27D: Prolific Authors from China with Metric Indices**

| S.No. | Authors | Citation sum within h-core | All citations | All articles | h-index | R-Index | AR-Index | $h_{nom}$ Index | a- index | M-index | Q2 index | e-index | P-index |
|---|---|---|---|---|---|---|---|---|---|---|---|---|---|
| 1 | Zhang Y | 5903 | 16315 | 1196 | 51 | 76.831 | 23.32 | 0.04 | 115.75 | 1.7 | 9.311 | 57.463 | 60.6 |
| 2 | Wang Y | 5758 | 14254 | 1007 | 50 | 75.881 | 21.798 | 0.05 | 115.16 | 1.67 | 9.129 | 57.079 | 58.65 |
| 3 | Liu Y | 9056 | 15810 | 774 | 49 | 95.163 | 22.956 | 0.06 | 184.82 | 1.63 | 8.946 | 81.578 | 68.61 |
| 4 | Chen J | 7504 | 12200 | 551 | 49 | 86.626 | 20.166 | 0.09 | 153.14 | 1.63 | 8.946 | 71.435 | 64.64 |
| 5 | Li Y | 9904 | 16196 | 796 | 49 | 99.519 | 23.235 | 0.06 | 202.12 | 1.63 | 8.946 | 86.62 | 69.07 |
| 6 | Zhang L | 3612 | 9347 | 657 | 46 | 60.1 | 17.651 | 0.07 | 78.52 | 1.53 | 8.398 | 38.678 | 51.04 |
| 7 | Wang J | 6075 | 11637 | 708 | 42 | 77.942 | 19.695 | 0.06 | 144.64 | 1.4 | 7.668 | 65.658 | 57.62 |
| 8 | Li J | 2848 | 7936 | 638 | 41 | 53.367 | 16.264 | 0.06 | 69.46 | 1.37 | 7.486 | 34.161 | 46.22 |
| 9 | Wang YJ | 4209 | 7674 | 476 | 40 | 64.877 | 15.994 | 0.08 | 105.23 | 1.33 | 7.303 | 51.078 | 49.83 |
| 10 | Wang X | 3504 | 6415 | 456 | 40 | 59.195 | 14.623 | 0.09 | 87.6 | 1.33 | 7.303 | 43.635 | 44.85 |

The study observed that highly citations and h-index by the authors; Zhang Y from the University of Shanghai for Science & Technology, School Medical

Instrument & Food Engineering has the first position in the h-index having 51 h-index among 1196 contributions. Followed by Wang Y from Nanchang University, Affiliated Hospital has the second position in the h-index (50), and other scientists' have h-index and publications also shown in the table. It is evident from the above data that some authors appear among the list having significantly less contribution on CAD. However, their h-index is high, e.g., Chen J from Capital Medical University, Department Neurosurgery, the author has only 551 publications but is at the top level regarding the h-index.

### 4.27.5 Analysis of Top Ten Prolific Authors from South Africa with Metric Indices

**Table 27E: Prolific Authors from South Africa with Metric Indices**

| S.No. | Authors | Citation Sum within h-core | All Citations | All Publications | h-index | R-Index | AR-Index | hnom Index | a- index | M-index | Q2 index | e-index | P-index |
|---|---|---|---|---|---|---|---|---|---|---|---|---|---|
| 1 | Opie LH | 3854 | 4815 | 114 | 38 | 62.081 | 12.669 | 0.333 | 126.711 | 1.267 | 6.938 | 49.092 | 58.807 |
| 2 | Lochner A | 1582 | 1945 | 68 | 28 | 39.774 | 8.052 | 0.412 | 69.464 | 0.933 | 5.112 | 28.249 | 38.175 |
| 3 | Sliwa K | 15103 | 15289 | 50 | 27 | 122.894 | 22.575 | 0.540 | 566.259 | 0.900 | 4.930 | 119.892 | 167.210 |
| 4 | Lecour S | 2002 | 2183 | 53 | 24 | 44.744 | 8.530 | 0.453 | 90.958 | 0.800 | 4.382 | 37.762 | 44.800 |
| 5 | Genade S | 879 | 946 | 33 | 21 | 29.648 | 5.615 | 0.636 | 45.048 | 0.700 | 3.834 | 20.928 | 30.044 |
| 6 | Mayosi BM | 16140 | 16191 | 31 | 20 | 127.043 | 23.231 | 0.645 | 809.550 | 0.667 | 3.651 | 125.459 | 203.733 |
| 7 | Raal FJ | 5375 | 5438 | 29 | 18 | 73.314 | 13.464 | 0.621 | 302.111 | 0.600 | 3.286 | 71.070 | 100.653 |
| 8 | Marais AD | 1289 | 1372 | 29 | 18 | 35.903 | 6.763 | 0.621 | 76.222 | 0.600 | 3.286 | 31.064 | 40.189 |
| 9 | Mensah GA | 20553 | 20559 | 19 | 18 | 143.363 | 26.178 | 0.947 | 1142.167 | 0.600 | 3.286 | 142.229 | 281.244 |
| 10 | Dalby AJ | 2986 | 3036 | 25 | 15 | 54.644 | 10.060 | 0.600 | 202.400 | 0.500 | 2.739 | 52.545 | 71.706 |

The list of top ten authors who produced the highest contribution to research output on CAD from South Africa is given in the table. In terms of a number of publications, Opie LH is the most productive author with 114 publications having h-index 38, R-index 62.081, AR-index 12.669, hnom-index 0.333, a-index 126.711, M-index 1.267, Q2-index 6.938, e-index 49.092 and P-index as 58.807 followed by Lochner A 68, Lecour S 53, and Sliwa K 50 publications. It is also noted that 4 out of 10 prolific authors contributed more than fifty research publications each while the rest 6 authors contributed more than 18 publications each. The h index is highest for Opie LH (38) followed by Lochner A (28) followed by Sliwa K (27) and Lecour S (24). The data set puts forth that the authors Mensah GA with 20559 citations, Mayosi BM with 16191 citations, Sliwa K with 15289 citations and Raal FJ with 5375 citations.

**4.28 Analysis of Average Authors and Pages per Paper Year Wise**

Table 28 displays the Average Authors, which is calculated as the number of authors divided by the number of publications, and Average Pages, which is calculated as the number of publications divided by the number of pages. It is observed that the average number of authors is between 3.80 in the year 1990 to 10.99 in the year 2019, and the average number of pages is between 4.36 in the year 1990 and 9.11 in the year 2019. The average author shows a consistently increasing trend from 1990 up to 2019 and is high in the year 2019 (10.99). The Average Pages is also showing an increasing trend in its growth from 1990 to 2019, and the highest value is in the year 2018 (9.11), and it slightly decrease in the year 2019 (9.03).

This study explicates the author productivity of contributions shows an increasing trend in terms of authors as well as papers used to communicate the research. The study indicates that the majority of the articles are contributed by many authors. Especially in the year 2019, the contribution is the highest than the other year productivity. It indicates that the authors and pages are increasing in coronary artery disease literature.

To achieve the eighth objective of identify the most prolific authors and productive sources on Coronary Artery Disease research in BRICS countries, the analysis of most prolific authors and productive sources is carried out and it is demonstrated in tables 25A, 25B, 25C, 25D, 25E, 26 and 29.

| Table 28: Average Authors and Pages | | | | | | |
|---|---|---|---|---|---|---|
| S.No. | Year | Papers | Authors | No. of Pages | AA | AP |
| 1 | 1990 | 158 | 600 | 689 | 3.80 | 4.36 |
| 2 | 1991 | 266 | 1099 | 1388 | 4.13 | 5.22 |
| 3 | 1992 | 234 | 974 | 1361 | 4.16 | 5.82 |
| 4 | 1993 | 182 | 1051 | 1204 | 5.77 | 6.62 |
| 5 | 1994 | 196 | 879 | 1080 | 4.48 | 5.51 |
| 6 | 1995 | 197 | 1012 | 1194 | 5.14 | 6.06 |
| 7 | 1996 | 291 | 1342 | 1646 | 4.61 | 5.66 |
| 8 | 1997 | 297 | 1399 | 1693 | 4.71 | 5.70 |
| 9 | 1998 | 311 | 1683 | 1829 | 5.41 | 5.88 |
| 10 | 1999 | 343 | 1651 | 2007 | 4.81 | 5.85 |
| 11 | 2000 | 399 | 2021 | 2498 | 5.07 | 6.26 |
| 12 | 2001 | 360 | 1844 | 2242 | 5.12 | 6.23 |
| 13 | 2002 | 490 | 2530 | 2979 | 5.16 | 6.08 |
| 14 | 2003 | 559 | 2936 | 3361 | 5.25 | 6.01 |
| 15 | 2004 | 737 | 4114 | 4465 | 5.58 | 6.06 |
| 16 | 2005 | 794 | 4561 | 4880 | 5.74 | 6.15 |
| 17 | 2006 | 994 | 6048 | 6311 | 6.08 | 6.35 |
| 18 | 2007 | 1224 | 7578 | 8181 | 6.19 | 6.68 |
| 19 | 2008 | 1486 | 9047 | 9820 | 6.09 | 6.61 |
| 20 | 2009 | 1799 | 11382 | 12339 | 6.33 | 6.86 |
| 21 | 2010 | 2149 | 13600 | 14406 | 6.33 | 6.70 |
| 22 | 2011 | 2400 | 16233 | 16853 | 6.76 | 7.02 |
| 23 | 2012 | 2727 | 19110 | 20382 | 7.01 | 7.47 |
| 24 | 2013 | 3336 | 24577 | 24715 | 7.37 | 7.41 |
| 25 | 2014 | 3685 | 29500 | 28972 | 8.01 | 7.86 |
| 26 | 2015 | 4130 | 31775 | 34097 | 7.69 | 8.26 |
| 27 | 2016 | 4577 | 38678 | 38571 | 8.45 | 8.43 |
| 28 | 2017 | 4822 | 40623 | 43785 | 8.42 | 9.08 |
| 29 | 2018 | 5192 | 44281 | 47298 | 8.53 | 9.11 |
| 30 | 2019 | 5701 | 62630 | 51470 | 10.99 | 9.03 |
| Total | | 50036 | 384758 | 391716 | 7.69 | 7.83 |

AA – Average Authors, AP – Average Pages

### 4.29 Analysis of Average Authors and Pages per Block Year

Table 29 displays the average author and average pages in block years. There are six blocks in total, and the block 2015-2019 has shown the highest average authors and average pages of 8.93 and 8.81, respectively. It is analyzed that the average number of authors is between 4.44 and 8.93 from 1990-1994 to 2015-2019. It is also evident that average pages are between 5.52 and 8.81 from 1990-1994 to 2015-2019. The inference leads to demonstrate that there is an increasing trend in collaboration, as the average authors are increasing and the number of pages using by authors is also increasing. The argument is supported by the inference taken from the collaborative author's tables, where it shows that collaboration is increasing in coronary artery disease research among BRICS countries.

**Table 29: Average Authors & Pages in Block Years**

| S.No. | Year | Publications | No. of Authors | No. of Pages | AA | AP |
|---|---|---|---|---|---|---|
| 1 | 1990-1994 | 1036 | 4603 | 5722 | 4.44 | 5.52 |
| 2 | 1995-1999 | 1439 | 7087 | 8369 | 4.92 | 5.82 |
| 3 | 2000-2004 | 2545 | 13445 | 15545 | 5.28 | 6.11 |
| 4 | 2005-2009 | 6297 | 38616 | 41531 | 6.13 | 6.60 |
| 5 | 2010-2014 | 14297 | 103020 | 105328 | 7.21 | 7.37 |
| 6 | 2015-2019 | 24422 | 217987 | 215221 | 8.93 | 8.81 |
| Total | | 50036 | 384758 | 391716 | 7.69 | 7.83 |

### 4.30 Analysis of Lotka's Law on CAD

Although fundamental descriptive analyses regarding the most productive authors can be attempted to identify the most highly productive people in a given research area, the data can also be treated in another way. Indeed, author productivity is usually analyzed according to a widely used bibliometric law: Lotka's law. An analysis based on Lotka's law provides a more specific interpretation of author productivity.

(Lotka, 1926) studied author productivity patterns and developed one of the primary laws in bibliometrics. He observed that, in a given area of science, there are a lot of authors who publish only one study, while a small group of prolific authors contributes with a large number of publications. This premise is based on Lotka's law, also known as the *inverse square law* on author productivity. According to this inverse square law, the law takes the number of authors who have contributed with a single study and then predicts how many authors would have published x studies. In summary, the number of authors who produce $x$ studies is proportional to $1/x2$.

| Table 30: Lotka's Law on CAD | | | | | |
|---|---|---|---|---|---|
| x | y | X=Log x | Y=Log y | XY | $X^2$ |
| 1 | 60220 | 0.0000 | 4.7797 | 0.0000 | 0.0000 |
| 2 | 17056 | 0.3010 | 4.2319 | 1.2739 | 0.0906 |
| 3 | 7052 | 0.4771 | 3.8483 | 1.8361 | 0.2276 |
| 4 | 4273 | 0.6021 | 3.6307 | 2.1859 | 0.3625 |
| 5 | 2723 | 0.6990 | 3.4350 | 2.4010 | 0.4886 |
| 6 | 1872 | 0.7782 | 3.2723 | 2.5463 | 0.6055 |
| 7 | 2190 | 0.8451 | 3.3404 | 2.8230 | 0.7142 |
| 8 | 1178 | 0.9031 | 3.0711 | 2.7735 | 0.8156 |
| 9 | 1032 | 0.9542 | 3.0137 | 2.8758 | 0.9106 |
| 10 | 782 | 1.0000 | 2.8932 | 2.8932 | 1.0000 |
| 11 | 614 | 1.0414 | 2.7882 | 2.9036 | 1.0845 |
| 12 | 490 | 1.0792 | 2.6902 | 2.9032 | 1.1646 |
| 13 | 415 | 1.1139 | 2.6180 | 2.9164 | 1.2409 |
| 14 | 336 | 1.1461 | 2.5263 | 2.8955 | 1.3136 |
| 15 | 317 | 1.1761 | 2.5011 | 2.9415 | 1.3832 |
| 16 | 277 | 1.2041 | 2.4425 | 2.9410 | 1.4499 |
| 17 | 241 | 1.2304 | 2.3820 | 2.9310 | 1.5140 |
| 18 | 235 | 1.2553 | 2.3711 | 2.9763 | 1.5757 |
| 19 | 173 | 1.2788 | 2.2380 | 2.8619 | 1.6352 |
| 20 | 170 | 1.3010 | 2.2304 | 2.9019 | 1.6927 |
| 21 | 132 | 1.3222 | 2.1206 | 2.8039 | 1.7483 |
| 22 | 145 | 1.3424 | 2.1614 | 2.9015 | 1.8021 |
| 23 | 136 | 1.3617 | 2.1335 | 2.9053 | 1.8543 |
| 24 | 103 | 1.3802 | 2.0128 | 2.7781 | 1.9050 |
| 25 | 107 | 1.3979 | 2.0294 | 2.8370 | 1.9542 |
| 26 | 77 | 1.4150 | 1.8865 | 2.6693 | 2.0021 |
| 27 | 80 | 1.4314 | 1.9031 | 2.7240 | 2.0488 |
| 28 | 68 | 1.4472 | 1.8325 | 2.6519 | 2.0943 |
| 29 | 64 | 1.4624 | 1.8062 | 2.6414 | 2.1386 |
| 30 | 71 | 1.4771 | 1.8513 | 2.7345 | 2.1819 |
| 31 | 77 | 1.4914 | 1.8865 | 2.8134 | 2.2242 |
| 32 | 51 | 1.5051 | 1.7076 | 2.5701 | 2.2655 |
| 33 | 50 | 1.5185 | 1.6990 | 2.5799 | 2.3059 |
| 34 | 45 | 1.5315 | 1.6532 | 2.5319 | 2.3454 |
| 35 | 45 | 1.5441 | 1.6532 | 2.5527 | 2.3841 |
| 36 | 54 | 1.5563 | 1.7324 | 2.6961 | 2.4221 |
| 37 | 52 | 1.5682 | 1.7160 | 2.6910 | 2.4593 |
| 38 | 46 | 1.5798 | 1.6628 | 2.6268 | 2.4957 |
| 39 | 37 | 1.5911 | 1.5682 | 2.4951 | 2.5315 |
| 40 | 35 | 1.6021 | 1.5441 | 2.4737 | 2.5666 |
| 41 | 46 | 1.6128 | 1.6628 | 2.6817 | 2.6011 |
| 42 | 28 | 1.6232 | 1.4472 | 2.3491 | 2.6349 |
| 43 | 45 | 1.6335 | 1.6532 | 2.7005 | 2.6682 |
| 44 | 33 | 1.6435 | 1.5185 | 2.4956 | 2.7009 |
| 45 | 42 | 1.6532 | 1.6232 | 2.6836 | 2.7331 |
| 46 | 24 | 1.6628 | 1.3802 | 2.2950 | 2.7648 |
| 47 | 20 | 1.6721 | 1.3010 | 2.1754 | 2.7959 |
| 48 | 20 | 1.6812 | 1.3010 | 2.1873 | 2.8266 |
| 49 | 25 | 1.6902 | 1.3979 | 2.3628 | 2.8568 |
| 50 | 27 | 1.6990 | 1.4314 | 2.4318 | 2.8865 |
| 51 | 25 | 1.7076 | 1.3979 | 2.3871 | 2.9158 |
| 52 | 18 | 1.7160 | 1.2553 | 2.1541 | 2.9447 |
| 53 | 20 | 1.7243 | 1.3010 | 2.2433 | 2.9731 |
| 54 | 13 | 1.7324 | 1.1139 | 1.9298 | 3.0012 |
| 55 | 20 | 1.7404 | 1.3010 | 2.2643 | 3.0289 |
| 56 | 23 | 1.7482 | 1.3617 | 2.3806 | 3.0562 |
| 57 | 14 | 1.7559 | 1.1461 | 2.0125 | 3.0831 |
| 58 | 18 | 1.7634 | 1.2553 | 2.2136 | 3.1097 |
| 59 | 11 | 1.7709 | 1.0414 | 1.8442 | 3.1359 |
| 60 | 19 | 1.7782 | 1.2788 | 2.2738 | 3.1618 |
| 61 | 12 | 1.7853 | 1.0792 | 1.9267 | 3.1874 |
| 62 | 15 | 1.7924 | 1.1761 | 2.1080 | 3.2127 |
| 63 | 8 | 1.7993 | 0.9031 | 1.6250 | 3.2376 |
| 64 | 12 | 1.8062 | 1.0792 | 1.9492 | 3.2623 |
| 65 | 9 | 1.8129 | 0.9542 | 1.7300 | 3.2867 |
| 66 | 8 | 1.8195 | 0.9031 | 1.6432 | 3.3107 |
| 67 | 13 | 1.8261 | 1.1139 | 2.0341 | 3.3345 |
| 68 | 13 | 1.8325 | 1.1139 | 2.0413 | 3.3581 |
| 69 | 14 | 1.8388 | 1.1461 | 2.1076 | 3.3814 |
| 70 | 10 | 1.8451 | 1.0000 | 1.8451 | 3.4044 |
| 71 | 11 | 1.8513 | 1.0414 | 1.9279 | 3.4272 |
| 72 | 5 | 1.8573 | 0.6990 | 1.2982 | 3.4497 |
| 73 | 9 | 1.8633 | 0.9542 | 1.7781 | 3.4720 |
| 74 | 7 | 1.8692 | 0.8451 | 1.5797 | 3.4940 |
| 75 | 4 | 1.8751 | 0.6021 | 1.1289 | 3.5159 |
| 76 | 10 | 1.8808 | 1.0000 | 1.8808 | 3.5375 |
| 77 | 11 | 1.8865 | 1.0414 | 1.9646 | 3.5588 |
| 78 | 6 | 1.8921 | 0.7782 | 1.4723 | 3.5800 |
| 79 | 7 | 1.8976 | 0.8451 | 1.6037 | 3.6010 |
| 80 | 8 | 1.9031 | 0.9031 | 1.7187 | 3.6218 |
| 81 | 8 | 1.9085 | 0.9031 | 1.7235 | 3.6423 |
| 82 | 5 | 1.9138 | 0.6990 | 1.3377 | 3.6627 |
| 83 | 8 | 1.9191 | 0.9031 | 1.7331 | 3.6829 |
| 84 | 5 | 1.9243 | 0.6990 | 1.3450 | 3.7029 |
| 85 | 5 | 1.9294 | 0.6990 | 1.3486 | 3.7227 |
| 86 | 7 | 1.9345 | 0.8451 | 1.6348 | 3.7423 |
| 87 | 5 | 1.9395 | 0.6990 | 1.3557 | 3.7617 |
| 88 | 4 | 1.9445 | 0.6021 | 1.1707 | 3.7810 |
| 89 | 5 | 1.9494 | 0.6990 | 1.3626 | 3.8001 |
| 90 | 7 | 1.9542 | 0.8451 | 1.6515 | 3.8191 |
| 91 | 3 | 1.9590 | 0.4771 | 0.9347 | 3.8378 |
| 92 | 6 | 1.9638 | 0.7782 | 1.5281 | 3.8565 |
| 93 | 2 | 1.9685 | 0.3010 | 0.5926 | 3.8749 |
| 94 | 8 | 1.9731 | 0.9031 | 1.7819 | 3.8932 |
| 95 | 2 | 1.9777 | 0.3010 | 0.5954 | 3.9114 |
| 96 | 5 | 1.9823 | 0.6990 | 1.3855 | 3.9294 |
| 97 | 6 | 1.9868 | 0.7782 | 1.5460 | 3.9473 |
| 98 | 1 | 1.9912 | 0.0000 | 0.0000 | 3.9650 |
| 99 | 6 | 1.9956 | 0.7782 | 1.5529 | 3.9826 |
| 100 | 1 | 2.0000 | 0.0000 | 0.0000 | 4.0000 |
| 101 | 8 | 2.0043 | 0.9031 | 1.8101 | 4.0173 |
| 102 | 2 | 2.0086 | 0.3010 | 0.6046 | 4.0345 |
| 103 | 5 | 2.0128 | 0.6990 | 1.4069 | 4.0515 |
| 104 | 4 | 2.0170 | 0.6021 | 1.2144 | 4.0684 |
| 105 | 5 | 2.0212 | 0.6990 | 1.4128 | 4.0852 |
| 106 | 5 | 2.0253 | 0.6990 | 1.4156 | 4.1019 |
| 107 | 1 | 2.0294 | 0.0000 | 0.0000 | 4.1184 |
| 108 | 4 | 2.0334 | 0.6021 | 1.2242 | 4.1348 |
| 109 | 4 | 2.0374 | 0.6021 | 1.2267 | 4.1511 |
| 110 | 4 | 2.0414 | 0.6021 | 1.2290 | 4.1673 |
| 111 | 4 | 2.0453 | 0.6021 | 1.2314 | 4.1833 |
| 112 | 5 | 2.0492 | 0.6990 | 1.4323 | 4.1993 |
| 113 | 2 | 2.0531 | 0.3010 | 0.6180 | 4.2151 |
| 114 | 4 | 2.0569 | 0.6021 | 1.2384 | 4.2309 |
| 115 | 5 | 2.0607 | 0.6990 | 1.4404 | 4.2465 |
| 116 | 3 | 2.0645 | 0.4771 | 0.9850 | 4.2620 |
| 117 | 5 | 2.0682 | 0.6990 | 1.4456 | 4.2774 |
| 118 | 3 | 2.0719 | 0.4771 | 0.9885 | 4.2927 |
| 119 | 2 | 2.0755 | 0.3010 | 0.6248 | 4.3079 |
| 120 | 3 | 2.0792 | 0.4771 | 0.9920 | 4.3230 |
| 121 | 1 | 2.0828 | 0.0000 | 0.0000 | 4.3380 |
| 122 | 4 | 2.0864 | 0.6021 | 1.2561 | 4.3529 |
| 123 | 1 | 2.0899 | 0.0000 | 0.0000 | 4.3677 |
| 124 | 3 | 2.0934 | 0.4771 | 0.9988 | 4.3824 |
| 125 | 3 | 2.0969 | 0.4771 | 1.0005 | 4.3970 |
| 127 | 4 | 2.1038 | 0.6021 | 1.2666 | 4.4260 |
| 128 | 1 | 2.1072 | 0.0000 | 0.0000 | 4.4403 |
| 129 | 2 | 2.1106 | 0.3010 | 0.6354 | 4.4546 |
| 130 | 1 | 2.1139 | 0.0000 | 0.0000 | 4.4688 |
| 131 | 2 | 2.1173 | 0.3010 | 0.6374 | 4.4828 |
| 133 | 2 | 2.1239 | 0.3010 | 0.6393 | 4.5107 |
| 134 | 3 | 2.1271 | 0.4771 | 1.0149 | 4.5246 |
| 135 | 4 | 2.1303 | 0.6021 | 1.2826 | 4.5383 |
| 137 | 2 | 2.1367 | 0.3010 | 0.6432 | 4.5656 |
| 138 | 4 | 2.1399 | 0.6021 | 1.2883 | 4.5791 |
| 139 | 2 | 2.1430 | 0.3010 | 0.6451 | 4.5925 |
| 140 | 3 | 2.1461 | 0.4771 | 1.0240 | 4.6059 |
| 142 | 1 | 2.1523 | 0.0000 | 0.0000 | 4.6323 |
| 143 | 1 | 2.1553 | 0.0000 | 0.0000 | 4.6455 |
| 144 | 3 | 2.1584 | 0.4771 | 1.0298 | 4.6585 |
| 145 | 1 | 2.1614 | 0.0000 | 0.0000 | 4.6715 |
| 146 | 1 | 2.1644 | 0.0000 | 0.0000 | 4.6844 |
| 147 | 1 | 2.1673 | 0.0000 | 0.0000 | 4.6973 |
| 148 | 2 | 2.1703 | 0.3010 | 0.6533 | 4.7100 |
| 149 | 1 | 2.1732 | 0.0000 | 0.0000 | 4.7227 |
| 150 | 2 | 2.1761 | 0.3010 | 0.6551 | 4.7354 |
| 152 | 3 | 2.1818 | 0.4771 | 1.0410 | 4.7604 |
| 153 | 3 | 2.1847 | 0.4771 | 1.0424 | 4.7729 |
| 155 | 2 | 2.1903 | 0.3010 | 0.6594 | 4.7976 |
| 157 | 1 | 2.1959 | 0.0000 | 0.0000 | 4.8220 |
| 158 | 1 | 2.1987 | 0.0000 | 0.0000 | 4.8341 |
| 159 | 1 | 2.2014 | 0.0000 | 0.0000 | 4.8461 |
| 160 | 2 | 2.2041 | 0.3010 | 0.6635 | 4.8581 |
| 162 | 2 | 2.2095 | 0.3010 | 0.6651 | 4.8820 |
| 163 | 1 | 2.2122 | 0.0000 | 0.0000 | 4.8938 |
| 164 | 2 | 2.2148 | 0.3010 | 0.6667 | 4.9055 |
| 165 | 2 | 2.2175 | 0.3010 | 0.6675 | 4.9172 |
| 166 | 1 | 2.2201 | 0.0000 | 0.0000 | 4.9289 |
| 167 | 2 | 2.2227 | 0.3010 | 0.6691 | 4.9405 |
| 168 | 2 | 2.2253 | 0.3010 | 0.6699 | 4.9520 |
| 169 | 1 | 2.2279 | 0.0000 | 0.0000 | 4.9635 |
| 170 | 1 | 2.2304 | 0.0000 | 0.0000 | 4.9749 |
| 171 | 1 | 2.2330 | 0.0000 | 0.0000 | 4.9863 |
| 173 | 1 | 2.2380 | 0.0000 | 0.0000 | 5.0089 |
| 175 | 1 | 2.2430 | 0.0000 | 0.0000 | 5.0312 |
| 178 | 1 | 2.2504 | 0.0000 | 0.0000 | 5.0644 |
| 179 | 2 | 2.2529 | 0.3010 | 0.6782 | 5.0753 |
| 180 | 2 | 2.2553 | 0.3010 | 0.6789 | 5.0863 |
| 181 | 1 | 2.2577 | 0.0000 | 0.0000 | 5.0971 |
| 184 | 2 | 2.2648 | 0.3010 | 0.6818 | 5.1294 |
| 186 | 4 | 2.2695 | 0.6021 | 1.3664 | 5.1507 |
| 188 | 2 | 2.2742 | 0.3010 | 0.6846 | 5.1718 |
| 189 | 1 | 2.2765 | 0.0000 | 0.0000 | 5.1823 |
| 190 | 1 | 2.2788 | 0.0000 | 0.0000 | 5.1927 |
| 193 | 1 | 2.2856 | 0.0000 | 0.0000 | 5.2238 |
| 197 | 2 | 2.2945 | 0.3010 | 0.6907 | 5.2646 |
| 198 | 3 | 2.2967 | 0.4771 | 1.0958 | 5.2747 |
| 202 | 1 | 2.3054 | 0.0000 | 0.0000 | 5.3146 |
| 204 | 3 | 2.3096 | 0.4771 | 1.1020 | 5.3344 |
| 206 | 1 | 2.3139 | 0.0000 | 0.0000 | 5.3540 |
| 209 | 1 | 2.3201 | 0.0000 | 0.0000 | 5.3831 |
| 210 | 1 | 2.3222 | 0.0000 | 0.0000 | 5.3927 |
| 212 | 1 | 2.3263 | 0.0000 | 0.0000 | 5.4118 |
| 215 | 1 | 2.3324 | 0.0000 | 0.0000 | 5.4403 |
| 220 | 1 | 2.3424 | 0.0000 | 0.0000 | 5.4869 |
| 221 | 1 | 2.3444 | 0.0000 | 0.0000 | 5.4962 |
| 224 | 1 | 2.3502 | 0.0000 | 0.0000 | 5.5237 |
| 225 | 1 | 2.3522 | 0.0000 | 0.0000 | 5.5328 |
| 227 | 1 | 2.3560 | 0.0000 | 0.0000 | 5.5509 |
| 229 | 1 | 2.3598 | 0.0000 | 0.0000 | 5.5688 |
| 230 | 1 | 2.3617 | 0.0000 | 0.0000 | 5.5778 |
| 231 | 1 | 2.3636 | 0.0000 | 0.0000 | 5.5867 |
| 232 | 1 | 2.3655 | 0.0000 | 0.0000 | 5.5955 |
| 234 | 1 | 2.3692 | 0.0000 | 0.0000 | 5.6132 |
| 239 | 1 | 2.3784 | 0.0000 | 0.0000 | 5.6568 |
| 246 | 1 | 2.3909 | 0.0000 | 0.0000 | 5.7166 |
| 247 | 1 | 2.3927 | 0.0000 | 0.0000 | 5.7250 |
| 251 | 2 | 2.3997 | 0.3010 | 0.7224 | 5.7584 |
| 254 | 1 | 2.4048 | 0.0000 | 0.0000 | 5.7832 |
| 255 | 1 | 2.4065 | 0.0000 | 0.0000 | 5.7914 |
| 257 | 1 | 2.4099 | 0.0000 | 0.0000 | 5.8078 |
| 260 | 1 | 2.4150 | 0.0000 | 0.0000 | 5.8321 |
| 261 | 1 | 2.4166 | 0.0000 | 0.0000 | 5.8402 |
| 262 | 1 | 2.4183 | 0.0000 | 0.0000 | 5.8482 |
| 265 | 1 | 2.4232 | 0.0000 | 0.0000 | 5.8721 |
| 272 | 1 | 2.4346 | 0.0000 | 0.0000 | 5.9271 |
| 273 | 1 | 2.4362 | 0.0000 | 0.0000 | 5.9349 |
| 274 | 1 | 2.4378 | 0.0000 | 0.0000 | 5.9426 |
| 282 | 1 | 2.4502 | 0.0000 | 0.0000 | 6.0037 |
| 283 | 1 | 2.4518 | 0.0000 | 0.0000 | 6.0113 |
| 286 | 1 | 2.4564 | 0.0000 | 0.0000 | 6.0337 |
| 287 | 1 | 2.4579 | 0.0000 | 0.0000 | 6.0412 |
| 288 | 1 | 2.4594 | 0.0000 | 0.0000 | 6.0486 |
| 294 | 1 | 2.4683 | 0.0000 | 0.0000 | 6.0927 |
| 314 | 1 | 2.4969 | 0.0000 | 0.0000 | 6.2347 |
| 323 | 1 | 2.5092 | 0.0000 | 0.0000 | 6.2961 |
| 324 | 2 | 2.5105 | 0.3010 | 0.7557 | 6.3028 |
| 326 | 2 | 2.5132 | 0.3010 | 0.7566 | 6.3163 |
| 340 | 1 | 2.5315 | 0.0000 | 0.0000 | 6.4084 |
| 354 | 1 | 2.5490 | 0.0000 | 0.0000 | 6.4974 |
| 366 | 1 | 2.5635 | 0.0000 | 0.0000 | 6.5714 |
| 382 | 1 | 2.5821 | 0.0000 | 0.0000 | 6.6671 |
| 384 | 1 | 2.5843 | 0.0000 | 0.0000 | 6.6788 |
| 390 | 1 | 2.5911 | 0.0000 | 0.0000 | 6.7136 |
| 407 | 1 | 2.6096 | 0.0000 | 0.0000 | 6.8100 |
| 413 | 1 | 2.6160 | 0.0000 | 0.0000 | 6.8432 |
| 438 | 1 | 2.6415 | 0.0000 | 0.0000 | 6.9774 |
| 441 | 1 | 2.6444 | 0.0000 | 0.0000 | 6.9931 |
| 445 | 1 | 2.6484 | 0.0000 | 0.0000 | 7.0138 |
| 456 | 1 | 2.6590 | 0.0000 | 0.0000 | 7.0701 |
| 467 | 1 | 2.6693 | 0.0000 | 0.0000 | 7.1253 |
| 476 | 1 | 2.6776 | 0.0000 | 0.0000 | 7.1696 |
| 511 | 1 | 2.7084 | 0.0000 | 0.0000 | 7.3355 |
| 539 | 1 | 2.7316 | 0.0000 | 0.0000 | 7.4616 |
| 551 | 1 | 2.7412 | 0.0000 | 0.0000 | 7.5139 |
| 580 | 1 | 2.7634 | 0.0000 | 0.0000 | 7.6365 |
| 615 | 1 | 2.7889 | 0.0000 | 0.0000 | 7.7778 |
| 638 | 1 | 2.8048 | 0.0000 | 0.0000 | 7.8670 |
| 657 | 1 | 2.8176 | 0.0000 | 0.0000 | 7.9387 |
| 708 | 1 | 2.8500 | 0.0000 | 0.0000 | 8.1227 |
| 774 | 1 | 2.8887 | 0.0000 | 0.0000 | 8.3448 |
| 796 | 1 | 2.9009 | 0.0000 | 0.0000 | 8.4153 |
| 1007 | 1 | 3.0030 | 0.0000 | 0.0000 | 9.0182 |
| 1196 | 1 | 3.0777 | 0.0000 | 0.0000 | 9.4724 |
| | 104160 | 492.5385 | 180.7916 | 263.6368 | 1050.6941 |

To achieve the ninth objective of testing the applicability of Lotka's law of author production in Coronary Artery Disease research, the analysis of Lotka's law is carried out and it is simplified in tables 30 and 31.

Consequently, the first step is to calculate the exponent *n* using the least-squares method and according to the following formula:

$$n = \frac{N\sum XY - \sum X \sum Y}{N\sum X^2 - \left(\sum X\right)^2}$$

All the data needed for the *n* formula can be obtained from table 30. The only index that requires further work is *N*, which represents the number of pairs considered. In this example, those authors who have published between one and 1196 articles will be considered, representing 244 pairs of data (*N* = 244).

$$n = \frac{244(263.6368) - (492.5385)(180.7916)}{244(1050.6941) - (492.5385)^2}$$

$$n = \frac{24719.4443}{13775.186}$$

$$n = 1.79$$

Thus, the value of *n* (absolute value) is 1.79, which will then be the specific value of the coefficient in Lotka's formula that will explain author productivity in this particular case.

The theoretical value of '*n*' =1.79 is matched with the table value of R. Rosseau for getting C.S. value 0.5270

| Constant Value of Present Study | *n* Value |
|---|---|
| 0.5270 | 1.79 |
| Lotka's Constant Value | *n* Value |
| 0.6079 | 2 |

| D-Max Value Present Study | D-Max Value of Lotka's Study |
|---|---|
| 0.05115 | 0.01177 |

Finally, the Kolmogorov-Smirnov test is applied to verify whether the observed data fit the theoretical distribution according to Lotka's law. The highest value in column (*D*max) is taken as reference for comparison with the critical value (c.v.), whose general formulation is:

$D_{Max} = F(x) – En(x)$ á = 1.79

Theoretical Value of $C = 0.5270\,Fe+ = 0.5270\left(\frac{1}{x^{1.79}}\right)$

D-Max = 0.05115

Critical Value at 0.01 level of significance = $\frac{1.79}{\sqrt{104160}} = 0.0055$

## 4.31 Kolmogorov- Smirnov Test

**Table 31: Kolmogorov- Smirnov Test**

| x | y | Observed=y x/Σyx | Value=Σ(yx /Σyx) | Expected Frequency | Value of Frequency/ Cum. | Diff.(D) |
|---|---|---|---|---|---|---|
| 1 | 60220 | 0.57815 | 0.57815 | 0.52700 | 0.52700 | **0.05115** |
| 2 | 17056 | 0.16375 | 0.74190 | 0.15192 | 0.67892 | 0.01183 |
| 3 | 7052 | 0.06770 | 0.80960 | 0.07339 | 0.75231 | -0.00568 |
| 4 | 4273 | 0.04102 | 0.85063 | 0.04379 | 0.79610 | -0.00277 |
| 5 | 2723 | 0.02614 | 0.87677 | 0.02934 | 0.82545 | -0.00320 |
| 6 | 1872 | 0.01797 | 0.89474 | 0.02116 | 0.84660 | -0.00318 |
| 7 | 2190 | 0.02103 | 0.91577 | 0.01604 | 0.86265 | 0.00498 |
| 8 | 1178 | 0.01131 | 0.92707 | 0.01262 | 0.87527 | -0.00132 |
| 9 | 1032 | 0.00991 | 0.93698 | 0.01022 | 0.88549 | -0.00031 |
| 10 | 782 | 0.00751 | 0.94449 | 0.00846 | 0.89395 | -0.00095 |
| 11 | 614 | 0.00589 | 0.95039 | 0.00713 | 0.90108 | -0.00123 |
| 12 | 490 | 0.00470 | 0.95509 | 0.00610 | 0.90718 | -0.00139 |
| 13 | 415 | 0.00398 | 0.95907 | 0.00528 | 0.91246 | -0.00130 |
| 14 | 336 | 0.00323 | 0.96230 | 0.00462 | 0.91709 | -0.00140 |
| 15 | 317 | 0.00304 | 0.96534 | 0.00409 | 0.92117 | -0.00104 |
| 16 | 277 | 0.00266 | 0.96800 | 0.00364 | 0.92481 | -0.00098 |
| 17 | 241 | 0.00231 | 0.97032 | 0.00326 | 0.92808 | -0.00095 |
| 18 | 235 | 0.00226 | 0.97257 | 0.00295 | 0.93102 | -0.00069 |
| 19 | 173 | 0.00166 | 0.97423 | 0.00267 | 0.93370 | -0.00101 |
| 20 | 170 | 0.00163 | 0.97587 | 0.00244 | 0.93613 | -0.00081 |
| 21 | 132 | 0.00127 | 0.97713 | 0.00223 | 0.93837 | -0.00097 |
| 22 | 145 | 0.00139 | 0.97852 | 0.00206 | 0.94042 | -0.00066 |
| 23 | 136 | 0.00131 | 0.97983 | 0.00190 | 0.94232 | -0.00059 |
| 24 | 103 | 0.00099 | 0.98082 | 0.00176 | 0.94408 | -0.00077 |
| 25 | 107 | 0.00103 | 0.98185 | 0.00163 | 0.94571 | -0.00061 |
| 26 | 77 | 0.00074 | 0.98259 | 0.00152 | 0.94724 | -0.00078 |
| 27 | 80 | 0.00077 | 0.98335 | 0.00142 | 0.94866 | -0.00066 |
| 28 | 68 | 0.00065 | 0.98401 | 0.00133 | 0.94999 | -0.00068 |
| 29 | 64 | 0.00061 | 0.98462 | 0.00125 | 0.95124 | -0.00064 |
| 30 | 71 | 0.00068 | 0.98530 | 0.00118 | 0.95242 | -0.00050 |
| 31 | 77 | 0.00074 | 0.98604 | 0.00111 | 0.95353 | -0.00037 |
| 32 | 51 | 0.00049 | 0.98653 | 0.00105 | 0.95458 | -0.00056 |
| 33 | 50 | 0.00048 | 0.98701 | 0.00099 | 0.95557 | -0.00051 |
| 34 | 45 | 0.00043 | 0.98744 | 0.00094 | 0.95652 | -0.00051 |
| 35 | 45 | 0.00043 | 0.98788 | 0.00089 | 0.95741 | -0.00046 |
| 36 | 54 | 0.00052 | 0.98839 | 0.00085 | 0.95826 | -0.00033 |
| 37 | 52 | 0.00050 | 0.98889 | 0.00081 | 0.95907 | -0.00031 |
| 38 | 46 | 0.00044 | 0.98933 | 0.00077 | 0.95984 | -0.00033 |
| 39 | 37 | 0.00036 | 0.98969 | 0.00074 | 0.96057 | -0.00038 |
| 40 | 35 | 0.00034 | 0.99003 | 0.00070 | 0.96128 | -0.00037 |
| 41 | 46 | 0.00044 | 0.99047 | 0.00067 | 0.96195 | -0.00023 |
| 42 | 28 | 0.00027 | 0.99074 | 0.00064 | 0.96259 | -0.00038 |
| 43 | 45 | 0.00043 | 0.99117 | 0.00062 | 0.96321 | -0.00019 |
| 44 | 33 | 0.00032 | 0.99149 | 0.00059 | 0.96380 | -0.00028 |
| 45 | 42 | 0.00040 | 0.99189 | 0.00057 | 0.96437 | -0.00017 |
| 46 | 24 | 0.00023 | 0.99212 | 0.00055 | 0.96492 | -0.00032 |
| 47 | 20 | 0.00019 | 0.99231 | 0.00053 | 0.96545 | -0.00033 |
| 48 | 20 | 0.00019 | 0.99250 | 0.00051 | 0.96595 | -0.00031 |
| 49 | 25 | 0.00024 | 0.99274 | 0.00049 | 0.96644 | -0.00025 |
| 50 | 27 | 0.00026 | 0.99300 | 0.00047 | 0.96691 | -0.00021 |
| 51 | 25 | 0.00024 | 0.99324 | 0.00045 | 0.96737 | -0.00021 |
| 52 | 18 | 0.00017 | 0.99341 | 0.00044 | 0.96780 | -0.00027 |
| 53 | 20 | 0.00019 | 0.99361 | 0.00042 | 0.96823 | -0.00023 |
| 54 | 13 | 0.00012 | 0.99373 | 0.00041 | 0.96864 | -0.00029 |
| 55 | 20 | 0.00019 | 0.99392 | 0.00040 | 0.96904 | -0.00020 |
| 56 | 23 | 0.00022 | 0.99414 | 0.00038 | 0.96942 | -0.00016 |
| 57 | 14 | 0.00013 | 0.99428 | 0.00037 | 0.96979 | -0.00024 |
| 58 | 18 | 0.00017 | 0.99445 | 0.00036 | 0.97015 | -0.00019 |
| 59 | 11 | 0.00011 | 0.99456 | 0.00035 | 0.97050 | -0.00024 |
| 60 | 19 | 0.00018 | 0.99474 | 0.00034 | 0.97084 | -0.00016 |
| 61 | 12 | 0.00012 | 0.99486 | 0.00033 | 0.97117 | -0.00021 |
| 62 | 15 | 0.00014 | 0.99500 | 0.00032 | 0.97149 | -0.00018 |
| 63 | 8 | 0.00008 | 0.99508 | 0.00031 | 0.97180 | -0.00023 |
| 64 | 12 | 0.00012 | 0.99519 | 0.00030 | 0.97211 | -0.00019 |
| 65 | 9 | 0.00009 | 0.99528 | 0.00029 | 0.97240 | -0.00021 |
| 66 | 8 | 0.00008 | 0.99535 | 0.00029 | 0.97269 | -0.00021 |
| 67 | 13 | 0.00012 | 0.99548 | 0.00028 | 0.97297 | -0.00015 |
| 68 | 13 | 0.00012 | 0.99560 | 0.00027 | 0.97324 | -0.00015 |
| 69 | 14 | 0.00013 | 0.99574 | 0.00026 | 0.97350 | -0.00013 |
| 70 | 10 | 0.00010 | 0.99583 | 0.00026 | 0.97376 | -0.00016 |
| 71 | 11 | 0.00011 | 0.99594 | 0.00025 | 0.97401 | -0.00015 |
| 72 | 5 | 0.00005 | 0.99599 | 0.00024 | 0.97425 | -0.00020 |
| 73 | 9 | 0.00009 | 0.99607 | 0.00024 | 0.97449 | -0.00015 |
| 74 | 7 | 0.00007 | 0.99614 | 0.00023 | 0.97473 | -0.00017 |
| 75 | 4 | 0.00004 | 0.99618 | 0.00023 | 0.97495 | -0.00019 |
| 76 | 10 | 0.00010 | 0.99628 | 0.00022 | 0.97518 | -0.00013 |
| 77 | 11 | 0.00011 | 0.99638 | 0.00022 | 0.97539 | -0.00011 |
| 78 | 6 | 0.00006 | 0.99644 | 0.00021 | 0.97561 | -0.00015 |
| 79 | 7 | 0.00007 | 0.99651 | 0.00021 | 0.97581 | -0.00014 |
| 80 | 8 | 0.00008 | 0.99658 | 0.00020 | 0.97602 | -0.00013 |
| 81 | 8 | 0.00008 | 0.99666 | 0.00020 | 0.97621 | -0.00012 |
| 82 | 5 | 0.00005 | 0.99671 | 0.00019 | 0.97641 | -0.00015 |
| 83 | 8 | 0.00008 | 0.99678 | 0.00019 | 0.97660 | -0.00011 |
| 84 | 5 | 0.00005 | 0.99683 | 0.00019 | 0.97678 | -0.00014 |
| 85 | 5 | 0.00005 | 0.99688 | 0.00018 | 0.97696 | -0.00013 |
| 86 | 7 | 0.00007 | 0.99695 | 0.00018 | 0.97714 | -0.00011 |
| 87 | 5 | 0.00005 | 0.99700 | 0.00017 | 0.97732 | -0.00013 |
| 88 | 4 | 0.00004 | 0.99703 | 0.00017 | 0.97749 | -0.00013 |
| 89 | 5 | 0.00005 | 0.99708 | 0.00017 | 0.97765 | -0.00012 |
| 90 | 7 | 0.00007 | 0.99715 | 0.00016 | 0.97782 | -0.00010 |
| 91 | 3 | 0.00003 | 0.99718 | 0.00016 | 0.97798 | -0.00013 |
| 92 | 6 | 0.00006 | 0.99724 | 0.00016 | 0.97814 | -0.00010 |
| 93 | 2 | 0.00002 | 0.99726 | 0.00015 | 0.97829 | -0.00014 |
| 94 | 8 | 0.00008 | 0.99733 | 0.00015 | 0.97844 | -0.00007 |
| 95 | 2 | 0.00002 | 0.99735 | 0.00015 | 0.97859 | -0.00013 |
| 96 | 5 | 0.00005 | 0.99740 | 0.00015 | 0.97874 | -0.00010 |
| 97 | 6 | 0.00006 | 0.99746 | 0.00014 | 0.97888 | -0.00009 |
| 98 | 1 | 0.00001 | 0.99747 | 0.00014 | 0.97902 | -0.00013 |
| 99 | 6 | 0.00006 | 0.99752 | 0.00014 | 0.97916 | -0.00008 |
| 100 | 1 | 0.00001 | 0.99753 | 0.00014 | 0.97930 | -0.00013 |
| 101 | 8 | 0.00008 | 0.99761 | 0.00013 | 0.97943 | -0.00006 |
| 102 | 2 | 0.00002 | 0.99763 | 0.00013 | 0.97956 | -0.00011 |
| 103 | 5 | 0.00005 | 0.99768 | 0.00013 | 0.97969 | -0.00008 |
| 104 | 4 | 0.00004 | 0.99772 | 0.00013 | 0.97982 | -0.00009 |
| 105 | 5 | 0.00005 | 0.99776 | 0.00012 | 0.97994 | -0.00008 |
| 106 | 5 | 0.00005 | 0.99781 | 0.00012 | 0.98006 | -0.00007 |
| 107 | 1 | 0.00001 | 0.99782 | 0.00012 | 0.98018 | -0.00011 |
| 108 | 4 | 0.00004 | 0.99786 | 0.00012 | 0.98030 | -0.00008 |
| 109 | 4 | 0.00004 | 0.99790 | 0.00012 | 0.98042 | -0.00008 |
| 110 | 4 | 0.00004 | 0.99794 | 0.00011 | 0.98053 | -0.00008 |
| 111 | 4 | 0.00004 | 0.99798 | 0.00011 | 0.98065 | -0.00007 |
| 112 | 5 | 0.00005 | 0.99802 | 0.00011 | 0.98076 | -0.00006 |
| 113 | 2 | 0.00002 | 0.99804 | 0.00011 | 0.98087 | -0.00009 |
| 114 | 4 | 0.00004 | 0.99808 | 0.00011 | 0.98097 | -0.00007 |
| 115 | 5 | 0.00005 | 0.99813 | 0.00011 | 0.98108 | -0.00006 |
| 116 | 3 | 0.00003 | 0.99816 | 0.00010 | 0.98118 | -0.00008 |
| 117 | 5 | 0.00005 | 0.99821 | 0.00010 | 0.98128 | -0.00005 |
| 118 | 3 | 0.00003 | 0.99823 | 0.00010 | 0.98139 | -0.00007 |
| 119 | 2 | 0.00002 | 0.99825 | 0.00010 | 0.98149 | -0.00008 |
| 120 | 3 | 0.00003 | 0.99828 | 0.00010 | 0.98158 | -0.00007 |
| 121 | 1 | 0.00001 | 0.99829 | 0.00010 | 0.98168 | -0.00009 |
| 122 | 4 | 0.00004 | 0.99833 | 0.00010 | 0.98177 | -0.00006 |
| 123 | 1 | 0.00001 | 0.99834 | 0.00009 | 0.98187 | -0.00008 |
| 124 | 3 | 0.00003 | 0.99837 | 0.00009 | 0.98196 | -0.00006 |
| 125 | 3 | 0.00003 | 0.99840 | 0.00009 | 0.98205 | -0.00006 |
| 127 | 4 | 0.00004 | 0.99844 | 0.00009 | 0.98214 | -0.00005 |
| 128 | 1 | 0.00001 | 0.99845 | 0.00009 | 0.98223 | -0.00008 |
| 129 | 2 | 0.00002 | 0.99846 | 0.00009 | 0.98231 | -0.00007 |
| 130 | 1 | 0.00001 | 0.99847 | 0.00008 | 0.98240 | -0.00008 |
| 131 | 2 | 0.00002 | 0.99849 | 0.00008 | 0.98248 | -0.00006 |
| 133 | 2 | 0.00002 | 0.99851 | 0.00008 | 0.98256 | -0.00006 |
| 134 | 3 | 0.00003 | 0.99854 | 0.00008 | 0.98264 | -0.00005 |
| 135 | 4 | 0.00004 | 0.99858 | 0.00008 | 0.98272 | -0.00004 |
| 137 | 2 | 0.00002 | 0.99860 | 0.00008 | 0.98280 | -0.00006 |
| 138 | 4 | 0.00004 | 0.99864 | 0.00008 | 0.98288 | -0.00004 |
| 139 | 2 | 0.00002 | 0.99866 | 0.00008 | 0.98295 | -0.00006 |
| 140 | 3 | 0.00003 | 0.99869 | 0.00007 | 0.98303 | -0.00005 |
| 142 | 1 | 0.00001 | 0.99870 | 0.00007 | 0.98310 | -0.00006 |
| 143 | 1 | 0.00001 | 0.99870 | 0.00007 | 0.98317 | -0.00006 |
| 144 | 3 | 0.00003 | 0.99873 | 0.00007 | 0.98324 | -0.00004 |
| 145 | 1 | 0.00001 | 0.99874 | 0.00007 | 0.98331 | -0.00006 |
| 146 | 1 | 0.00001 | 0.99875 | 0.00007 | 0.98338 | -0.00006 |
| 147 | 1 | 0.00001 | 0.99876 | 0.00007 | 0.98345 | -0.00006 |
| 148 | 2 | 0.00002 | 0.99878 | 0.00007 | 0.98351 | -0.00005 |
| 149 | 1 | 0.00001 | 0.99879 | 0.00007 | 0.98358 | -0.00006 |
| 150 | 2 | 0.00002 | 0.99881 | 0.00007 | 0.98365 | -0.00005 |
| 152 | 3 | 0.00003 | 0.99884 | 0.00006 | 0.98371 | -0.00004 |
| 153 | 3 | 0.00003 | 0.99887 | 0.00006 | 0.98377 | -0.00003 |
| 155 | 2 | 0.00002 | 0.99889 | 0.00006 | 0.98383 | -0.00004 |
| 157 | 1 | 0.00001 | 0.99890 | 0.00006 | 0.98389 | -0.00005 |
| 158 | 1 | 0.00001 | 0.99891 | 0.00006 | 0.98395 | -0.00005 |
| 159 | 1 | 0.00001 | 0.99892 | 0.00006 | 0.98401 | -0.00005 |
| 160 | 2 | 0.00002 | 0.99894 | 0.00006 | 0.98407 | -0.00004 |
| 162 | 2 | 0.00002 | 0.99895 | 0.00006 | 0.98413 | -0.00004 |
| 163 | 1 | 0.00001 | 0.99896 | 0.00006 | 0.98419 | -0.00005 |
| 164 | 2 | 0.00002 | 0.99898 | 0.00006 | 0.98424 | -0.00004 |
| 165 | 2 | 0.00002 | 0.99900 | 0.00006 | 0.98430 | -0.00004 |
| 166 | 1 | 0.00001 | 0.99901 | 0.00005 | 0.98435 | -0.00005 |
| 167 | 2 | 0.00002 | 0.99903 | 0.00005 | 0.98441 | -0.00003 |
| 168 | 2 | 0.00002 | 0.99905 | 0.00005 | 0.98446 | -0.00003 |
| 169 | 1 | 0.00001 | 0.99906 | 0.00005 | 0.98451 | -0.00004 |
| 170 | 1 | 0.00001 | 0.99907 | 0.00005 | 0.98456 | -0.00004 |
| 171 | 1 | 0.00001 | 0.99908 | 0.00005 | 0.98462 | -0.00004 |
| 173 | 1 | 0.00001 | 0.99909 | 0.00005 | 0.98467 | -0.00004 |
| 175 | 1 | 0.00001 | 0.99910 | 0.00005 | 0.98472 | -0.00004 |
| 178 | 1 | 0.00001 | 0.99911 | 0.00005 | 0.98477 | -0.00004 |
| 179 | 2 | 0.00002 | 0.99913 | 0.00005 | 0.98481 | -0.00003 |
| 180 | 2 | 0.00002 | 0.99915 | 0.00005 | 0.98486 | -0.00003 |
| 181 | 1 | 0.00001 | 0.99916 | 0.00005 | 0.98491 | -0.00004 |
| 184 | 2 | 0.00002 | 0.99918 | 0.00005 | 0.98495 | -0.00003 |
| 186 | 4 | 0.00004 | 0.99921 | 0.00004 | 0.98500 | -0.00001 |
| 188 | 2 | 0.00002 | 0.99923 | 0.00004 | 0.98504 | -0.00002 |
| 189 | 1 | 0.00001 | 0.99924 | 0.00004 | 0.98508 | -0.00003 |
| 190 | 1 | 0.00001 | 0.99925 | 0.00004 | 0.98513 | -0.00003 |
| 193 | 1 | 0.00001 | 0.99926 | 0.00004 | 0.98517 | -0.00003 |
| 197 | 2 | 0.00002 | 0.99928 | 0.00004 | 0.98521 | -0.00002 |
| 198 | 3 | 0.00003 | 0.99931 | 0.00004 | 0.98525 | -0.00001 |
| 202 | 1 | 0.00001 | 0.99932 | 0.00004 | 0.98529 | -0.00003 |
| 204 | 3 | 0.00003 | 0.99935 | 0.00004 | 0.98533 | -0.00001 |
| 206 | 1 | 0.00001 | 0.99936 | 0.00004 | 0.98536 | -0.00003 |
| 209 | 1 | 0.00001 | 0.99937 | 0.00004 | 0.98540 | -0.00003 |
| 210 | 1 | 0.00001 | 0.99938 | 0.00004 | 0.98543 | -0.00003 |
| 212 | 1 | 0.00001 | 0.99939 | 0.00004 | 0.98547 | -0.00003 |
| 215 | 1 | 0.00001 | 0.99940 | 0.00003 | 0.98550 | -0.00002 |
| 220 | 1 | 0.00001 | 0.99941 | 0.00003 | 0.98554 | -0.00002 |
| 221 | 1 | 0.00001 | 0.99942 | 0.00003 | 0.98557 | -0.00002 |
| 224 | 1 | 0.00001 | 0.99942 | 0.00003 | 0.98560 | -0.00002 |
| 225 | 1 | 0.00001 | 0.99943 | 0.00003 | 0.98563 | -0.00002 |
| 227 | 1 | 0.00001 | 0.99944 | 0.00003 | 0.98566 | -0.00002 |
| 229 | 1 | 0.00001 | 0.99945 | 0.00003 | 0.98570 | -0.00002 |
| 230 | 1 | 0.00001 | 0.99946 | 0.00003 | 0.98573 | -0.00002 |
| 231 | 1 | 0.00001 | 0.99947 | 0.00003 | 0.98576 | -0.00002 |
| 232 | 1 | 0.00001 | 0.99948 | 0.00003 | 0.98579 | -0.00002 |
| 234 | 1 | 0.00001 | 0.99949 | 0.00003 | 0.98582 | -0.00002 |
| 239 | 1 | 0.00001 | 0.99950 | 0.00003 | 0.98584 | -0.00002 |
| 246 | 1 | 0.00001 | 0.99951 | 0.00003 | 0.98587 | -0.00002 |
| 247 | 1 | 0.00001 | 0.99952 | 0.00003 | 0.98590 | -0.00002 |
| 251 | 2 | 0.00002 | 0.99954 | 0.00003 | 0.98592 | -0.00001 |
| 254 | 1 | 0.00001 | 0.99955 | 0.00003 | 0.98595 | -0.00002 |
| 255 | 1 | 0.00001 | 0.99956 | 0.00003 | 0.98597 | -0.00002 |
| 257 | 1 | 0.00001 | 0.99957 | 0.00002 | 0.98600 | -0.00002 |
| 260 | 1 | 0.00001 | 0.99958 | 0.00002 | 0.98602 | -0.00001 |
| 261 | 1 | 0.00001 | 0.99959 | 0.00002 | 0.98605 | -0.00001 |
| 262 | 1 | 0.00001 | 0.99960 | 0.00002 | 0.98607 | -0.00001 |
| 265 | 1 | 0.00001 | 0.99961 | 0.00002 | 0.98610 | -0.00001 |
| 272 | 1 | 0.00001 | 0.99962 | 0.00002 | 0.98612 | -0.00001 |
| 273 | 1 | 0.00001 | 0.99963 | 0.00002 | 0.98614 | -0.00001 |
| 274 | 1 | 0.00001 | 0.99964 | 0.00002 | 0.98616 | -0.00001 |
| 282 | 1 | 0.00001 | 0.99965 | 0.00002 | 0.98618 | -0.00001 |
| 283 | 1 | 0.00001 | 0.99966 | 0.00002 | 0.98621 | -0.00001 |
| 286 | 1 | 0.00001 | 0.99966 | 0.00002 | 0.98623 | -0.00001 |
| 287 | 1 | 0.00001 | 0.99967 | 0.00002 | 0.98625 | -0.00001 |
| 288 | 1 | 0.00001 | 0.99968 | 0.00002 | 0.98627 | -0.00001 |
| 294 | 1 | 0.00001 | 0.99969 | 0.00002 | 0.98629 | -0.00001 |
| 314 | 1 | 0.00001 | 0.99970 | 0.00002 | 0.98630 | -0.00001 |
| 323 | 1 | 0.00001 | 0.99971 | 0.00002 | 0.98632 | -0.00001 |
| 324 | 2 | 0.00002 | 0.99973 | 0.00002 | 0.98634 | 0.00000 |
| 326 | 2 | 0.00002 | 0.99975 | 0.00002 | 0.98635 | 0.00000 |
| 340 | 1 | 0.00001 | 0.99976 | 0.00002 | 0.98637 | -0.00001 |
| 354 | 1 | 0.00001 | 0.99977 | 0.00001 | 0.98638 | 0.00000 |
| 366 | 1 | 0.00001 | 0.99978 | 0.00001 | 0.98640 | 0.00000 |
| 382 | 1 | 0.00001 | 0.99979 | 0.00001 | 0.98641 | 0.00000 |
| 384 | 1 | 0.00001 | 0.99980 | 0.00001 | 0.98642 | 0.00000 |
| 390 | 1 | 0.00001 | 0.99981 | 0.00001 | 0.98643 | 0.00000 |
| 407 | 1 | 0.00001 | 0.99982 | 0.00001 | 0.98644 | 0.00000 |
| 413 | 1 | 0.00001 | 0.99983 | 0.00001 | 0.98645 | 0.00000 |
| 438 | 1 | 0.00001 | 0.99984 | 0.00001 | 0.98646 | 0.00000 |
| 441 | 1 | 0.00001 | 0.99985 | 0.00001 | 0.98647 | 0.00000 |
| 445 | 1 | 0.00001 | 0.99986 | 0.00001 | 0.98648 | 0.00000 |
| 456 | 1 | 0.00001 | 0.99987 | 0.00001 | 0.98649 | 0.00000 |
| 467 | 1 | 0.00001 | 0.99988 | 0.00001 | 0.98650 | 0.00000 |
| 476 | 1 | 0.00001 | 0.99989 | 0.00001 | 0.98651 | 0.00000 |
| 511 | 1 | 0.00001 | 0.99990 | 0.00001 | 0.98651 | 0.00000 |
| 539 | 1 | 0.00001 | 0.99990 | 0.00001 | 0.98652 | 0.00000 |
| 551 | 1 | 0.00001 | 0.99991 | 0.00001 | 0.98653 | 0.00000 |
| 580 | 1 | 0.00001 | 0.99992 | 0.00001 | 0.98653 | 0.00000 |
| 615 | 1 | 0.00001 | 0.99993 | 0.00001 | 0.98654 | 0.00000 |
| 638 | 1 | 0.00001 | 0.99994 | 0.00000 | 0.98654 | 0.00000 |
| 657 | 1 | 0.00001 | 0.99995 | 0.00000 | 0.98655 | 0.00000 |
| 708 | 1 | 0.00001 | 0.99996 | 0.00000 | 0.98655 | 0.00001 |
| 774 | 1 | 0.00001 | 0.99997 | 0.00000 | 0.98656 | 0.00001 |
| 796 | 1 | 0.00001 | 0.99998 | 0.00000 | 0.98656 | 0.00001 |
| 1007 | 1 | 0.00001 | 0.99999 | 0.00000 | 0.98656 | 0.00001 |
| 1196 | 1 | 0.00001 | 1.00000 | 0.00000 | 0.98656 | 0.00001 |
| Total | 104160 | Present Study D.Max = 0.05115 | | | | |

The theoretical value of C as 0.5270 for *n* = 1.79 is taken from the book 'Power Laws in the Information Production Process: Lotkaian Informetrics' by Egghe (2005).

Kolmogorov Simonov test is applied for the wellness of the Lotka's law for the estimations of Lotka's types acquired from least square methods. The outcomes tabulated in the above table show that the estimation of D-max, i.e., 0.05115 decided with Lotka's type, i.e., *n*=1.79. The critical value decided at the 0.005 level of significance is 0.0055, which is less noteworthy than the D-max value and henceforth, the watched authorship information distribution holds good for the Lotka's law, and consequently, the Lotka's law for the coronary artery disease literature research from BRICS acknowledge for the authorship distributions. ***The hypothesis has to be accepted. We can, therefore, conclude that author productivity in CAD research from the BRICS area fits Lotka's law. Hence, the Fourth Hypothesis has been proved.***

## 4.32 Price Square Root Law of CAD

**Table 32: Price Square Root Law of CAD**

| No. of Contributions (A) | No. of Contributors (B) | % | AXB | % | Cum. (AXB) | Cum.% |
|---|---|---|---|---|---|---|
| 98 | 1 | 0.00 | 98 | 0.03 | 98 | 0.03 |
| 100 | 1 | 0.00 | 100 | 0.03 | 198 | 0.05 |
| 107 | 1 | 0.00 | 107 | 0.03 | 305 | 0.08 |
| 121 | 1 | 0.00 | 121 | 0.03 | 426 | 0.11 |
| 123 | 1 | 0.00 | 123 | 0.03 | 549 | 0.14 |
| 128 | 1 | 0.00 | 128 | 0.03 | 677 | 0.18 |
| 130 | 1 | 0.00 | 130 | 0.03 | 807 | 0.21 |
| 142 | 1 | 0.00 | 142 | 0.04 | 949 | 0.25 |
| 143 | 1 | 0.00 | 143 | 0.04 | 1092 | 0.28 |
| 145 | 1 | 0.00 | 145 | 0.04 | 1237 | 0.32 |
| 146 | 1 | 0.00 | 146 | 0.04 | 1383 | 0.36 |
| 147 | 1 | 0.00 | 147 | 0.04 | 1530 | 0.40 |
| 149 | 1 | 0.00 | 149 | 0.04 | 1679 | 0.44 |
| 157 | 1 | 0.00 | 157 | 0.04 | 1836 | 0.48 |
| 158 | 1 | 0.00 | 158 | 0.04 | 1994 | 0.52 |
| 159 | 1 | 0.00 | 159 | 0.04 | 2153 | 0.56 |
| 163 | 1 | 0.00 | 163 | 0.04 | 2316 | 0.60 |
| 166 | 1 | 0.00 | 166 | 0.04 | 2482 | 0.65 |
| 169 | 1 | 0.00 | 169 | 0.04 | 2651 | 0.69 |
| 170 | 1 | 0.00 | 170 | 0.04 | 2821 | 0.73 |
| 171 | 1 | 0.00 | 171 | 0.04 | 2992 | 0.78 |
| 173 | 1 | 0.00 | 173 | 0.04 | 3165 | 0.82 |
| 175 | 1 | 0.00 | 175 | 0.05 | 3340 | 0.87 |
| 178 | 1 | 0.00 | 178 | 0.05 | 3518 | 0.91 |
| 181 | 1 | 0.00 | 181 | 0.05 | 3699 | 0.96 |
| 189 | 1 | 0.00 | 189 | 0.05 | 3888 | 1.01 |
| 190 | 1 | 0.00 | 190 | 0.05 | 4078 | 1.06 |
| 193 | 1 | 0.00 | 193 | 0.05 | 4271 | 1.11 |
| 202 | 1 | 0.00 | 202 | 0.05 | 4473 | 1.16 |
| 206 | 1 | 0.00 | 206 | 0.05 | 4679 | 1.22 |
| 209 | 1 | 0.00 | 209 | 0.05 | 4888 | 1.27 |
| 210 | 1 | 0.00 | 210 | 0.05 | 5098 | 1.32 |
| 212 | 1 | 0.00 | 212 | 0.06 | 5310 | 1.38 |
| 215 | 1 | 0.00 | 215 | 0.06 | 5525 | 1.44 |
| 220 | 1 | 0.00 | 220 | 0.06 | 5745 | 1.49 |
| 221 | 1 | 0.00 | 221 | 0.06 | 5966 | 1.55 |
| 224 | 1 | 0.00 | 224 | 0.06 | 6190 | 1.61 |
| 225 | 1 | 0.00 | 225 | 0.06 | 6415 | 1.67 |
| 227 | 1 | 0.00 | 227 | 0.06 | 6642 | 1.73 |
| 229 | 1 | 0.00 | 229 | 0.06 | 6871 | 1.79 |
| 230 | 1 | 0.00 | 230 | 0.06 | 7101 | 1.85 |
| 231 | 1 | 0.00 | 231 | 0.06 | 7332 | 1.91 |
| 232 | 1 | 0.00 | 232 | 0.06 | 7564 | 1.97 |
| 234 | 1 | 0.00 | 234 | 0.06 | 7798 | 2.03 |
| 239 | 1 | 0.00 | 239 | 0.06 | 8037 | 2.09 |
| 246 | 1 | 0.00 | 246 | 0.06 | 8283 | 2.15 |
| 247 | 1 | 0.00 | 247 | 0.06 | 8530 | 2.22 |
| 254 | 1 | 0.00 | 254 | 0.07 | 8784 | 2.28 |
| 255 | 1 | 0.00 | 255 | 0.07 | 9039 | 2.35 |
| 257 | 1 | 0.00 | 257 | 0.07 | 9296 | 2.42 |
| 260 | 1 | 0.00 | 260 | 0.07 | 9556 | 2.48 |
| 261 | 1 | 0.00 | 261 | 0.07 | 9817 | 2.55 |
| 262 | 1 | 0.00 | 262 | 0.07 | 10079 | 2.62 |
| 265 | 1 | 0.00 | 265 | 0.07 | 10344 | 2.69 |
| 272 | 1 | 0.00 | 272 | 0.07 | 10616 | 2.76 |
| 273 | 1 | 0.00 | 273 | 0.07 | 10889 | 2.83 |
| 274 | 1 | 0.00 | 274 | 0.07 | 11163 | 2.90 |
| 282 | 1 | 0.00 | 282 | 0.07 | 11445 | 2.97 |
| 283 | 1 | 0.00 | 283 | 0.07 | 11728 | 3.05 |
| 286 | 1 | 0.00 | 286 | 0.07 | 12014 | 3.12 |
| 287 | 1 | 0.00 | 287 | 0.07 | 12301 | 3.20 |
| 288 | 1 | 0.00 | 288 | 0.07 | 12589 | 3.27 |
| 294 | 1 | 0.00 | 294 | 0.08 | 12883 | 3.35 |
| 314 | 1 | 0.00 | 314 | 0.08 | 13197 | 3.43 |
| 323 | 1 | 0.00 | 323 | 0.08 | 13520 | 3.51 |
| 340 | 1 | 0.00 | 340 | 0.09 | 13860 | 3.60 |
| 354 | 1 | 0.00 | 354 | 0.09 | 14214 | 3.69 |
| 366 | 1 | 0.00 | 366 | 0.10 | 14580 | 3.79 |
| 382 | 1 | 0.00 | 382 | 0.10 | 14962 | 3.89 |
| 384 | 1 | 0.00 | 384 | 0.10 | 15346 | 3.99 |
| 390 | 1 | 0.00 | 390 | 0.10 | 15736 | 4.09 |
| 407 | 1 | 0.00 | 407 | 0.11 | 16143 | 4.20 |
| 413 | 1 | 0.00 | 413 | 0.11 | 16556 | 4.30 |
| 438 | 1 | 0.00 | 438 | 0.11 | 16994 | 4.42 |
| 441 | 1 | 0.00 | 441 | 0.11 | 17435 | 4.53 |
| 445 | 1 | 0.00 | 445 | 0.12 | 17880 | 4.65 |
| 456 | 1 | 0.00 | 456 | 0.12 | 18336 | 4.77 |
| 467 | 1 | 0.00 | 467 | 0.12 | 18803 | 4.89 |
| 476 | 1 | 0.00 | 476 | 0.12 | 19279 | 5.01 |
| 511 | 1 | 0.00 | 511 | 0.13 | 19790 | 5.14 |
| 539 | 1 | 0.00 | 539 | 0.14 | 20329 | 5.28 |
| 551 | 1 | 0.00 | 551 | 0.14 | 20880 | 5.43 |
| 580 | 1 | 0.00 | 580 | 0.15 | 21460 | 5.58 |
| 615 | 1 | 0.00 | 615 | 0.16 | 22075 | 5.74 |
| 638 | 1 | 0.00 | 638 | 0.17 | 22713 | 5.90 |
| 657 | 1 | 0.00 | 657 | 0.17 | 23370 | 6.07 |
| 708 | 1 | 0.00 | 708 | 0.18 | 24078 | 6.26 |
| 774 | 1 | 0.00 | 774 | 0.20 | 24852 | 6.46 |
| 796 | 1 | 0.00 | 796 | 0.21 | 25648 | 6.67 |
| 1007 | 1 | 0.00 | 1007 | 0.26 | 26655 | 6.93 |
| 1196 | 1 | 0.00 | 1196 | 0.31 | 27851 | 7.24 |
| 93 | 2 | 0.00 | 186 | 0.05 | 28037 | 7.29 |
| 95 | 2 | 0.00 | 190 | 0.05 | 28227 | 7.34 |
| 102 | 2 | 0.00 | 204 | 0.05 | 28431 | 7.39 |
| 113 | 2 | 0.00 | 226 | 0.06 | 28657 | 7.45 |
| 119 | 2 | 0.00 | 238 | 0.06 | 28895 | 7.51 |
| 129 | 2 | 0.00 | 258 | 0.07 | 29153 | 7.58 |
| 131 | 2 | 0.00 | 262 | 0.07 | 29415 | 7.64 |
| 133 | 2 | 0.00 | 266 | 0.07 | 29681 | 7.71 |
| 137 | 2 | 0.00 | 274 | 0.07 | 29955 | 7.79 |
| 139 | 2 | 0.00 | 278 | 0.07 | 30233 | 7.86 |
| 148 | 2 | 0.00 | 296 | 0.08 | 30529 | 7.93 |
| 150 | 2 | 0.00 | 300 | 0.08 | 30829 | 8.01 |
| 155 | 2 | 0.00 | 310 | 0.08 | 31139 | 8.09 |
| 160 | 2 | 0.00 | 320 | 0.08 | 31459 | 8.18 |
| 162 | 2 | 0.00 | 324 | 0.08 | 31783 | 8.26 |
| 164 | 2 | 0.00 | 328 | 0.09 | 32111 | 8.35 |
| 165 | 2 | 0.00 | 330 | 0.09 | 32441 | 8.43 |
| 167 | 2 | 0.00 | 334 | 0.09 | 32775 | 8.52 |
| 168 | 2 | 0.00 | 336 | 0.09 | 33111 | 8.61 |
| 179 | 2 | 0.00 | 358 | 0.09 | 33469 | 8.70 |
| 180 | 2 | 0.00 | 360 | 0.09 | 33829 | 8.79 |
| 184 | 2 | 0.00 | 368 | 0.10 | 34197 | 8.89 |
| 188 | 2 | 0.00 | 376 | 0.10 | 34573 | 8.99 |
| 197 | 2 | 0.00 | 394 | 0.10 | 34967 | 9.09 |
| 251 | 2 | 0.00 | 502 | 0.13 | 35469 | 9.22 |
| 324 | 2 | 0.00 | 648 | 0.17 | 36117 | 9.39 |
| 326 | 2 | 0.00 | 652 | 0.17 | 36769 | 9.56 |
| 91 | 3 | 0.00 | 273 | 0.07 | 37042 | 9.63 |
| 116 | 3 | 0.00 | 348 | 0.09 | 37390 | 9.72 |
| 118 | 3 | 0.00 | 354 | 0.09 | 37744 | 9.81 |
| 120 | 3 | 0.00 | 360 | 0.09 | 38104 | 9.90 |
| 124 | 3 | 0.00 | 372 | 0.10 | 38476 | 10.00 |
| 125 | 3 | 0.00 | 375 | 0.10 | 38851 | 10.10 |
| 134 | 3 | 0.00 | 402 | 0.10 | 39253 | 10.20 |
| 140 | 3 | 0.00 | 420 | 0.11 | 39673 | 10.31 |
| 144 | 3 | 0.00 | 432 | 0.11 | 40105 | 10.42 |
| 152 | 3 | 0.00 | 456 | 0.12 | 40561 | 10.54 |
| 153 | 3 | 0.00 | 459 | 0.12 | 41020 | 10.66 |
| 198 | 3 | 0.00 | 594 | 0.15 | 41614 | 10.82 |
| 204 | 3 | 0.00 | 612 | 0.16 | 42226 | 10.97 |
| 75 | 4 | 0.00 | 300 | 0.08 | 42526 | 11.05 |
| 88 | 4 | 0.00 | 352 | 0.09 | 42878 | 11.14 |
| 104 | 4 | 0.00 | 416 | 0.11 | 43294 | 11.25 |
| 108 | 4 | 0.00 | 432 | 0.11 | 43726 | 11.36 |
| 109 | 4 | 0.00 | 436 | 0.11 | 44162 | 11.48 |
| 110 | 4 | 0.00 | 440 | 0.11 | 44602 | 11.59 |
| 111 | 4 | 0.00 | 444 | 0.12 | 45046 | 11.71 |
| 114 | 4 | 0.00 | 456 | 0.12 | 45502 | 11.83 |
| 122 | 4 | 0.00 | 488 | 0.13 | 45990 | 11.95 |
| 127 | 4 | 0.00 | 508 | 0.13 | 46498 | 12.08 |
| 135 | 4 | 0.00 | 540 | 0.14 | 47038 | 12.22 |
| 138 | 4 | 0.00 | 552 | 0.14 | 47590 | 12.37 |
| 186 | 4 | 0.00 | 744 | 0.19 | 48334 | 12.56 |
| 72 | 5 | 0.00 | 360 | 0.09 | 48694 | 12.66 |
| 82 | 5 | 0.00 | 410 | 0.11 | 49104 | 12.76 |
| 84 | 5 | 0.00 | 420 | 0.11 | 49524 | 12.87 |
| 85 | 5 | 0.00 | 425 | 0.11 | 49949 | 12.98 |
| 87 | 5 | 0.00 | 435 | 0.11 | 50384 | 13.09 |
| 89 | 5 | 0.00 | 445 | 0.12 | 50829 | 13.21 |
| 96 | 5 | 0.00 | 480 | 0.12 | 51309 | 13.33 |
| 103 | 5 | 0.00 | 515 | 0.13 | 51824 | 13.47 |
| 105 | 5 | 0.00 | 525 | 0.14 | 52349 | 13.60 |
| 106 | 5 | 0.00 | 530 | 0.14 | 52879 | 13.74 |
| 112 | 5 | 0.00 | 560 | 0.15 | 53439 | 13.89 |
| 115 | 5 | 0.00 | 575 | 0.15 | 54014 | 14.04 |
| 117 | 5 | 0.00 | 585 | 0.15 | 54599 | 14.19 |
| 78 | 6 | 0.01 | 468 | 0.12 | 55067 | 14.31 |
| 92 | 6 | 0.01 | 552 | 0.14 | 55619 | 14.45 |
| 97 | 6 | 0.01 | 582 | 0.15 | 56201 | 14.61 |
| **99 (37275)** | **6(325)** | **0.01** | **594 (56795)** | **0.15** | **56795** | **14.76** |
| 74 | 7 | 0.01 | 518 | 0.13 | 57313 | 14.90 |
| 79 | 7 | 0.01 | 553 | 0.14 | 57866 | 15.04 |
| 86 | 7 | 0.01 | 602 | 0.16 | 58468 | 15.20 |
| 90 | 7 | 0.01 | 630 | 0.16 | 59098 | 15.36 |
| 63 | 8 | 0.01 | 504 | 0.13 | 59602 | 15.49 |
| 66 | 8 | 0.01 | 528 | 0.14 | 60130 | 15.63 |
| 80 | 8 | 0.01 | 640 | 0.17 | 60770 | 15.79 |
| 81 | 8 | 0.01 | 648 | 0.17 | 61418 | 15.96 |
| 83 | 8 | 0.01 | 664 | 0.17 | 62082 | 16.13 |
| 94 | 8 | 0.01 | 752 | 0.20 | 62834 | 16.33 |
| 101 | 8 | 0.01 | 808 | 0.21 | 63642 | 16.54 |
| 65 | 9 | 0.01 | 585 | 0.15 | 64227 | 16.69 |
| 73 | 9 | 0.01 | 657 | 0.17 | 64884 | 16.86 |
| 70 | 10 | 0.01 | 700 | 0.18 | 65584 | 17.04 |
| 76 | 10 | 0.01 | 760 | 0.20 | 66344 | 17.24 |
| 59 | 11 | 0.01 | 649 | 0.17 | 66993 | 17.41 |
| 71 | 11 | 0.01 | 781 | 0.20 | 67774 | 17.61 |
| 77 | 11 | 0.01 | 847 | 0.22 | 68621 | 17.83 |
| 61 | 12 | 0.01 | 732 | 0.19 | 69353 | 18.02 |
| 64 | 12 | 0.01 | 768 | 0.20 | 70121 | 18.22 |
| 54 | 13 | 0.01 | 702 | 0.18 | 70823 | 18.41 |
| 67 | 13 | 0.01 | 871 | 0.23 | 71694 | 18.63 |
| 68 | 13 | 0.01 | 884 | 0.23 | 72578 | 18.86 |
| 57 | 14 | 0.01 | 798 | 0.21 | 73376 | 19.07 |
| 69 | 14 | 0.01 | 966 | 0.25 | 74342 | 19.32 |
| 62 | 15 | 0.01 | 930 | 0.24 | 75272 | 19.56 |
| 52 | 18 | 0.02 | 936 | 0.24 | 76208 | 19.81 |
| 58 | 18 | 0.02 | 1044 | 0.27 | 77252 | 20.08 |
| 60 | 19 | 0.02 | 1140 | 0.30 | 78392 | 20.37 |
| 47 | 20 | 0.02 | 940 | 0.24 | 79332 | 20.62 |
| 48 | 20 | 0.02 | 960 | 0.25 | 80292 | 20.87 |
| 53 | 20 | 0.02 | 1060 | 0.28 | 81352 | 21.14 |
| 55 | 20 | 0.02 | 1100 | 0.29 | 82452 | 21.43 |
| 56 | 23 | 0.02 | 1288 | 0.33 | 83740 | 21.76 |
| 46 | 24 | 0.02 | 1104 | 0.29 | 84844 | 22.05 |
| 49 | 25 | 0.02 | 1225 | 0.32 | 86069 | 22.37 |
| 51 | 25 | 0.02 | 1275 | 0.33 | 87344 | 22.70 |
| 50 | 27 | 0.03 | 1350 | 0.35 | 88694 | 23.05 |
| 42 | 28 | 0.03 | 1176 | 0.31 | 89870 | 23.36 |
| 44 | 33 | 0.03 | 1452 | 0.38 | 91322 | 23.73 |
| 40 | 35 | 0.03 | 1400 | 0.36 | 92722 | 24.10 |
| 39 | 37 | 0.04 | 1443 | 0.38 | 94165 | 24.47 |
| 45 | 42 | 0.04 | 1890 | 0.49 | 96055 | 24.96 |
| 34 | 45 | 0.04 | 1530 | 0.40 | 97585 | 25.36 |
| 35 | 45 | 0.04 | 1575 | 0.41 | 99160 | 25.77 |
| 43 | 45 | 0.04 | 1935 | 0.50 | 101095 | 26.27 |
| 38 | 46 | 0.04 | 1748 | 0.45 | 102843 | 26.73 |
| 41 | 46 | 0.04 | 1886 | 0.49 | 104729 | 27.22 |
| 33 | 50 | 0.05 | 1650 | 0.43 | 106379 | 27.65 |
| 32 | 51 | 0.05 | 1632 | 0.42 | 108011 | 28.07 |
| 37 | 52 | 0.05 | 1924 | 0.50 | 109935 | 28.57 |
| 36 | 54 | 0.05 | 1944 | 0.51 | 111879 | 29.08 |
| 29 | 64 | 0.06 | 1856 | 0.48 | 113735 | 29.56 |
| 28 | 68 | 0.07 | 1904 | 0.49 | 115639 | 30.05 |
| 30 | 71 | 0.07 | 2130 | 0.55 | 117769 | 30.61 |
| 26 | 77 | 0.07 | 2002 | 0.52 | 119771 | 31.13 |
| 31 | 77 | 0.07 | 2387 | 0.62 | 122158 | 31.75 |
| 27 | 80 | 0.08 | 2160 | 0.56 | 124318 | 32.31 |
| 24 | 103 | 0.10 | 2472 | 0.64 | 126790 | 32.95 |
| 25 | 107 | 0.10 | 2675 | 0.70 | 129465 | 33.65 |
| 21 | 132 | 0.13 | 2772 | 0.72 | 132237 | 34.37 |
| 23 | 136 | 0.13 | 3128 | 0.81 | 135365 | 35.18 |
| 22 | 145 | 0.14 | 3190 | 0.83 | 138555 | 36.01 |
| 20 | 170 | 0.16 | 3400 | 0.88 | 141955 | 36.89 |
| 19 | 173 | 0.17 | 3287 | 0.85 | 145242 | 37.75 |
| 18 | 235 | 0.23 | 4230 | 1.10 | 149472 | 38.85 |
| 17 | 241 | 0.23 | 4097 | 1.06 | 153569 | 39.91 |
| 16 | 277 | 0.27 | 4432 | 1.15 | 158001 | 41.06 |
| 15 | 317 | 0.30 | 4755 | 1.24 | 162756 | 42.30 |
| 14 | 336 | 0.32 | 4704 | 1.22 | 167460 | 43.52 |
| 13 | 415 | 0.40 | 5395 | 1.40 | 172855 | 44.92 |
| 12 | 490 | 0.47 | 5880 | 1.53 | 178735 | 46.45 |
| 11 | 614 | 0.59 | 6754 | 1.76 | 185489 | 48.21 |
| 10 | 782 | 0.75 | 7820 | 2.03 | 193309 | 50.24 |
| 9 | 1032 | 0.99 | 9288 | 2.41 | 202597 | 52.65 |
| 8 | 1178 | 1.13 | 9424 | 2.45 | 212021 | 55.10 |
| 6 | 1872 | 1.80 | 11232 | 2.92 | 223253 | 58.02 |
| 7 | 2190 | 2.10 | 15330 | 3.98 | 238583 | 62.01 |
| 5 | 2723 | 2.61 | 13615 | 3.54 | 252198 | 65.54 |
| 4 | 4273 | 4.10 | 17092 | 4.44 | 269290 | 69.99 |
| 3 | 7052 | 6.77 | 21156 | 5.50 | 290446 | 75.48 |
| 2 | 17056 | 16.37 | 34112 | 8.87 | 324558 | 84.35 |
| 1 | 60220 | 57.81 | 60220 | 15.65 | 384778 | 100.00 |
| 40825 | 104160 | 100.00 | 384778 | 100.00 | | |

From the table, it is seen that all records from above 104160 and also seen that all scientific papers from above are 384778. As indicated by value square root law $\sqrt{104160}$ , contributors ought to contribute 384778/2 = 192389 papers.

In this way $\sqrt{104160}$ = 322.74 (323) authors, ½ of 384778 = 192389 papers.

We can see from the table that 323 authors contribute just 56795 papers

The worth is too far away from 50 % (half of the writing regarding a matter); hence it does not fulfil the value square root law.

**4.32.1 Pareto Principle (80 × 20 Rule)**

The researcher has obtained the analysis from the table 32 to validate the Pareto Principle and test whether 80 per cent of contributions have come from 20 per cent of contributors. Since the total authors' number is 104160, that means the 20 per cent total author number is 77252. The total number of publications is 40825, and 80 per cent of publications value is 324558.

Based on the analysis, the value of Accumulated % of A*B is 20.08 per cent of contributed more than twenty per cent of contributions, once the contributors are 77252. In the 80 × 20 rule view, the value should be very close to 80 per cent. The remaining 80 (79.92) per cent of the author's publications are 324558.

## 4.33 Analysis of Year Wise Total Citation of Publications

**Table 33: Year Wise Total Citations**

| S.No. | Year | Publications | All citations | Citation % | RG | CPP |
|---|---|---|---|---|---|---|
| 1 | 1990 | 158 | 866 | 0.11 | 3.02 | 5.48 |
| 2 | 1991 | 266 | 2617 | 0.34 | 1.29 | 9.84 |
| 3 | 1992 | 234 | 3381 | 0.45 | 1.57 | 14.45 |
| 4 | 1993 | 182 | 5319 | 0.70 | 0.53 | 29.23 |
| 5 | 1994 | 196 | 2834 | 0.37 | 1.51 | 14.46 |
| 6 | 1995 | 197 | 4287 | 0.57 | 1.25 | 21.76 |
| 7 | 1996 | 291 | 5378 | 0.71 | 1.02 | 18.48 |
| 8 | 1997 | 297 | 5486 | 0.72 | 1.63 | 18.47 |
| 9 | 1998 | 311 | 8919 | 1.18 | 0.82 | 28.68 |
| 10 | 1999 | 343 | 7338 | 0.97 | 1.43 | 21.39 |
| 11 | 2000 | 399 | 10500 | 1.38 | 0.99 | 26.32 |
| 12 | 2001 | 360 | 10423 | 1.37 | 1.55 | 28.95 |
| 13 | 2002 | 490 | 16125 | 2.13 | 1.05 | 32.91 |
| 14 | 2003 | 559 | 17003 | 2.24 | 1.25 | 30.42 |
| 15 | 2004 | 737 | 21184 | 2.79 | 1.05 | 28.74 |
| 16 | 2005 | 794 | 22326 | 2.94 | 1.28 | 28.12 |
| 17 | 2006 | 994 | 28485 | 3.76 | 1.23 | 28.66 |
| 18 | 2007 | 1224 | 35040 | 4.62 | 1.21 | 28.63 |
| 19 | 2008 | 1486 | 42332 | 5.58 | 1.03 | 28.49 |
| 20 | 2009 | 1799 | 43617 | 5.75 | 1.12 | 24.25 |
| 21 | 2010 | 2149 | 49059 | 6.47 | 1.11 | 22.83 |
| 22 | 2011 | 2400 | 54593 | 7.20 | 1.18 | 22.75 |
| 23 | 2012 | 2727 | 64572 | 8.51 | 1.01 | 23.68 |
| 24 | 2013 | 3336 | 65501 | 8.63 | 0.90 | 19.63 |
| 25 | 2014 | 3685 | 58726 | 7.74 | 0.93 | 15.94 |
| 26 | 2015 | 4130 | 54829 | 7.23 | 0.89 | 13.28 |
| 27 | 2016 | 4577 | 48588 | 6.41 | 0.88 | 10.62 |
| 28 | 2017 | 4822 | 42790 | 5.64 | 0.48 | 8.87 |
| 29 | 2018 | 5192 | 20748 | 2.74 | 0.28 | 4.00 |
| 30 | 2019 | 5701 | 5707 | 0.75 | 132.92 | 1.00 |
| Total | | 50036 | 758573 | 100.00 | | 15.16 |

RG- Ratio of Growth, CPP-Citation per Paper

The distribution of year-wise citations has been indicated in table 33. Citations per paper were additionally determined utilizing the formula as all out of the number of citations of a year separated by the total number of cited papers in that year and multiplied by 100 and furthermore citation percentage and Ratio of Growth has also been calculated.

In the table above, it is observed that the distribution of citation shows an increasing trend from 1990 to 2008, which is very typical, and from 2009 to 2019, it goes decreasing (24.25 to 1.00) continuously. The ratio of growth rate values is high in the years 2019 (132.92). It is also analyzed that the average citation per paper is increasing from 1990 to 2008 with the values from 5.48 to 32.91, and the same is decreasing from 2009 to 2019 with the values from 24.25 to 1.00. The overall citation per paper value is 15.16.

### 4.34 Analysis of Block Year Wise Citations

**Table 34: Block Year Wise Citations**

| S.No. | Year | Publications | Citations | Citation % | RG | CPP |
|---|---|---|---|---|---|---|
| 1 | 1990-1994 | 1036 | 15017 | 1.98 | 2.09 | 14.50 |
| 2 | 1995-1999 | 1439 | 31408 | 4.14 | 2.40 | 21.83 |
| 3 | 2000-2004 | 2545 | 75235 | 9.92 | 2.28 | 29.56 |
| 4 | 2005-2009 | 6297 | 171800 | 22.65 | 1.70 | 27.28 |
| 5 | 2010-2014 | 14297 | 292451 | 38.55 | 0.59 | 20.46 |
| 6 | 2015-2019 | 24422 | 172662 | 22.76 | 4.39 | 7.07 |
| Total | | 50036 | 758573 | 100.00 | | 15.16 |

RG- Ratio of Growth, CPP-Citation per Paper

It is highlighted that CPP in the third block is highest (29.56), and the sixth block has the lowest (7.07), and the overall performance of CPP is 15.16. The citation percentage is increasing, but in the last block, there is a slight decrease in the percentage of citation percentage (22.76%). The citations of the first block have 15017 citations, and it continuously increases up to the last block having 172662 citations.

### 4. 35 Citation Range of Publications

| Table 35: Citation Range of Publications | | | |
|---|---|---|---|
| S.No. | Range of Citations | Count (Records) | % |
| 1 | Zero | 11714 | 23.41 |
| 2 | 1-100 | 37456 | 74.86 |
| 3 | 101-200 | 584 | 1.17 |
| 4 | 201-300 | 117 | 0.23 |
| 5 | 301-400 | 45 | 0.09 |
| 6 | 401-500 | 31 | 0.06 |
| 7 | 501-600 | 25 | 0.05 |
| 8 | 601-700 | 13 | 0.03 |
| 9 | 701-800 | 9 | 0.02 |
| 10 | 801-900 | 6 | 0.01 |
| 11 | 901-1000 | 9 | 0.02 |
| 12 | 1001-6324 | 27 | 0.05 |
| Total Records | | 50036 | 100 |

Citation analysis is an increasingly common way to evaluate the research impact. However, there seems to be a general lack of understanding of how different data sources and citation metrics might impact comparisons between disciplines. The extent to which other articles cite a scientific article is often seen as one indicator of its importance, since the more critical an article, the more likely it is that others will refer to it.

As is readily apparent from table 35, the pattern of reduced citation scores for the ranged set of citations can be seen. Out of the 50036 publications understudy, only 27 (0.05%) records have citations in the range 1001-6324. It depicts that the publication which is having the highest citation record accounts for 6324 citations. Similarly, the publications having citations in the range 901-1000 are 9 (0.02%) records. The publications which have zero citations consist of 11714 records having 23.41%. The highest number of records having citations between 1 and 100 is 37456 consisting of 74.86 percentages. The publications having citations 101-200 are 584 (1.17%). The inference from the table above can be that a big chunk of publications 37456, which consists of 74.86% of records, have a citation range 1-100.

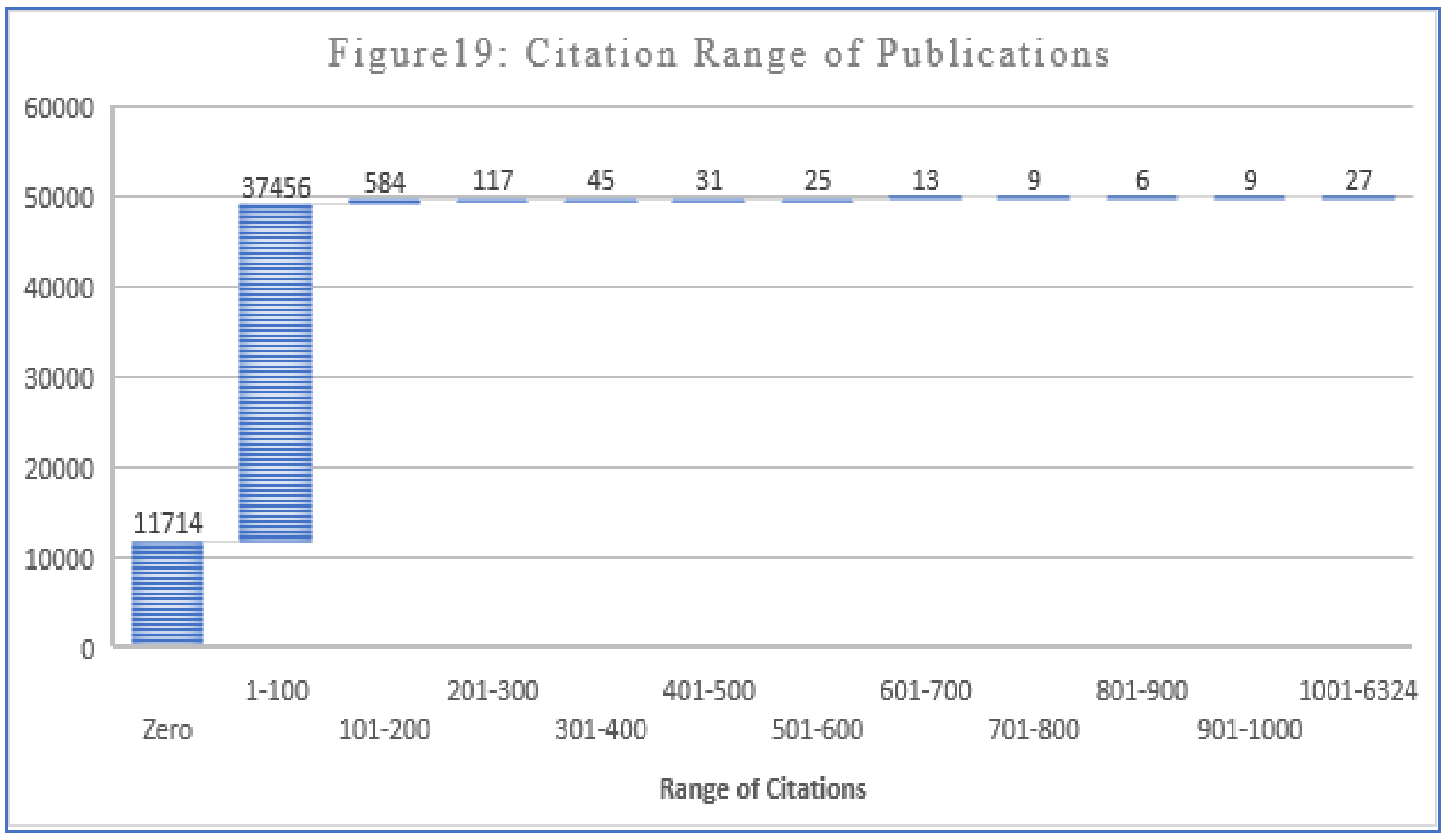

### 4.36 Analysis of Bradford's Law of Scattering on CAD

Another standard unit of analysis in a scientometric study concerns the journals in which the articles gathered are published. As in the case of author productivity, descriptive analyses about the most productive journals can be carried out. However, a more detailed analysis of the scatter of journal productivity may be interesting for research purposes. This part of the analysis considers the primary law applied to journal productivity, namely Bradford's law, and provides a detailed explanation of its application.

Bradford's law can be used as a tool for collection management in libraries by identifying core journals in subject areas, thereby providing evidence for journal subscription decision-making (Wolfram, 2003). By applying this law to a given set of data, it is possible to identify those journals that will account for most of the studies published in a given area.

#### 4.36.1 Distribution of Journals in various Zones in the research output of CAD

Bradford's law has been applied and examined in the publications in coronary artery disease literature. It has listed the journals containing that field in descending order of productivity and then divided the list into three zones. The distribution of journals in various zones is as follows:

To achieve the tenth objective of testing the applicability of Bradford's Law of scattering in Coronary Artery Disease research, the analysis of Bradford's Law of scattering is carried out and it is delineated in tables 36 and 37.

| Table 36A: Three Zones | | | |
|---|---|---|---|
| Zone | No. of Journals | No. of Articles | Multiplier Factors |
| Zone 1 | 41 (1.48) | 13259 (33.25) | - |
| Zone 2 | 202 (7.31) | 13125 (32.92) | 4.93 |
| Zone 3 | 2519 (91.20) | 13489 (33.83) | 12.47 |
| | 2762 | 39873 | 17.4 (8.7) |

The above table determines the analysis of a small group of forty-three journals identified to the nuclear or core zone representing 1.48% of journals covered 13259 (33.25%) of articles. The second more extensive group of 202 (7.31%) journals provides 13125 (32.92%) articles, and the third-largest zone, 2519 (91.20%) of journals yield the next 13489 (33.83%) articles. The Bradford multiplier between the number of references in zone 1 and zone 2 is 4.93, while it is 12.47 between zone 2 and zone 3. The average multiplier value is 8.7.

| Table 36 B: Bradford's Law of Scattering on CAD | | | |
|---|---|---|---|
| **No. of Journals** | **No. of Articles** | **Total Articles** | **Cum. Articles** |
| 1 | 1247 | 1247 | |
| 1 | 965 | 965 | 2212 |
| 1 | 699 | 699 | 2911 |
| 1 | 571 | 571 | 3482 |
| 1 | 490 | 490 | 3972 |
| 1 | 483 | 483 | 4455 |
| 1 | 474 | 474 | 4929 |
| 1 | 412 | 412 | 5341 |
| 1 | 410 | 410 | 5751 |
| 1 | 406 | 406 | 6157 |
| 1 | 397 | 397 | 6554 |
| 1 | 396 | 396 | 6950 |
| 1 | 376 | 376 | 7326 |
| 1 | 360 | 360 | 7686 |
| 1 | 358 | 358 | 8044 |
| 1 | 302 | 302 | 8346 |
| 1 | 263 | 263 | 8609 |
| 1 | 246 | 246 | 8855 |
| 1 | 230 | 230 | 9085 |
| 1 | 227 | 227 | 9312 |
| 1 | 225 | 225 | 9537 |
| 1 | 218 | 218 | 9755 |
| 1 | 216 | 216 | 9971 |
| 1 | 215 | 215 | 10186 |
| 1 | 208 | 208 | 10394 |
| 2 | 205 | 410 | 10804 |
| 1 | 201 | 201 | 11005 |
| 1 | 191 | 191 | 11196 |
| 1 | 188 | 188 | 11384 |
| 1 | 185 | 185 | 11569 |
| 1 | 182 | 182 | 11751 |
| 1 | 179 | 179 | 11930 |
| 1 | 176 | 176 | 12106 |
| 1 | 173 | 173 | 12279 |
| 1 | 172 | 172 | 12451 |
| 1 | 167 | 167 | 12618 |
| 1 | 164 | 164 | 12782 |
| 1 | 162 | 162 | 12944 |
| 1 | 161 | 161 | 13105 |
| **1 (41)** | **154** | **154** | **13259** |
| 2 | 148 | 296 | 13555 |
| 1 | 139 | 139 | 13694 |
| 1 | 138 | 138 | 13832 |
| 1 | 137 | 137 | 13969 |
| 1 | 136 | 136 | 14105 |
| 1 | 131 | 131 | 14236 |
| 2 | 130 | 260 | 14496 |
| 1 | 129 | 129 | 14625 |
| 1 | 128 | 128 | 14753 |
| 2 | 126 | 252 | 15005 |
| 2 | 123 | 246 | 15251 |
| 1 | 121 | 121 | 15372 |
| 1 | 120 | 120 | 15492 |
| 1 | 118 | 118 | 15610 |
| 1 | 117 | 117 | 15727 |
| 2 | 116 | 232 | 15959 |
| 1 | 115 | 115 | 16074 |
| 2 | 112 | 224 | 16298 |
| 1 | 111 | 111 | 16409 |
| 3 | 109 | 327 | 16736 |
| 1 | 108 | 108 | 16844 |
| 1 | 107 | 107 | 16951 |
| 1 | 106 | 106 | 17057 |
| 2 | 103 | 206 | 17263 |
| 1 | 101 | 101 | 17364 |
| 3 | 99 | 297 | 17661 |
| 1 | 98 | 98 | 17759 |
| 2 | 97 | 194 | 17953 |
| 1 | 95 | 95 | 18048 |
| 1 | 94 | 94 | 18142 |
| 1 | 93 | 93 | 18235 |
| 2 | 92 | 184 | 18419 |
| 1 | 91 | 91 | 18510 |
| 1 | 90 | 90 | 18600 |
| 2 | 88 | 176 | 18776 |
| 1 | 84 | 84 | 18860 |
| 1 | 83 | 83 | 18943 |
| 2 | 81 | 162 | 19105 |
| 1 | 79 | 79 | 19184 |
| 1 | 77 | 77 | 19261 |
| 1 | 76 | 76 | 19337 |
| 2 | 75 | 150 | 19487 |
| 4 | 73 | 292 | 19779 |
| 1 | 72 | 72 | 19851 |
| 2 | 71 | 142 | 19993 |
| 4 | 70 | 280 | 20273 |
| 2 | 69 | 138 | 20411 |
| 2 | 68 | 136 | 20547 |
| 5 | 67 | 335 | 20882 |
| 1 | 65 | 65 | 20947 |
| 3 | 64 | 192 | 21139 |
| 2 | 63 | 126 | 21265 |
| 2 | 62 | 124 | 21389 |
| 2 | 61 | 122 | 21511 |
| 5 | 59 | 295 | 21806 |
| 1 | 58 | 58 | 21864 |
| 2 | 57 | 114 | 21978 |
| 5 | 56 | 280 | 22258 |
| 2 | 55 | 110 | 22368 |
| 2 | 54 | 108 | 22476 |
| 3 | 53 | 159 | 22635 |
| 5 | 52 | 260 | 22895 |
| 1 | 51 | 51 | 22946 |
| 2 | 50 | 100 | 23046 |
| 2 | 49 | 98 | 23144 |
| 3 | 47 | 141 | 23285 |
| 4 | 46 | 184 | 23469 |
| 4 | 45 | 180 | 23649 |
| 7 | 44 | 308 | 23957 |
| 8 | 43 | 344 | 24301 |
| 2 | 42 | 84 | 24385 |
| 3 | 41 | 123 | 24508 |
| 6 | 40 | 240 | 24748 |
| 8 | 39 | 312 | 25060 |
| 2 | 38 | 76 | 25136 |
| 7 | 37 | 259 | 25395 |
| 6 | 36 | 216 | 25611 |
| 7 | 35 | 245 | 25856 |
| 8 | 34 | 272 | 26128 |
| 4 | 33 | 132 | 26260 |
| **5 (202)** | **32** | **160** | **26420** |
| 6 | 31 | 186 | 26606 |
| 8 | 30 | 240 | 26846 |
| 7 | 29 | 203 | 27049 |
| 4 | 28 | 112 | 27161 |
| 6 | 27 | 162 | 27323 |
| 17 | 26 | 442 | 27765 |
| 13 | 25 | 325 | 28090 |
| 22 | 24 | 528 | 28618 |
| 15 | 23 | 345 | 28963 |
| 13 | 22 | 286 | 29249 |
| 19 | 21 | 399 | 29648 |
| 17 | 20 | 340 | 29988 |
| 8 | 19 | 152 | 30140 |
| 18 | 18 | 324 | 30464 |
| 23 | 17 | 391 | 30855 |
| 23 | 16 | 368 | 31223 |
| 27 | 15 | 405 | 31628 |
| 39 | 14 | 546 | 32174 |
| 42 | 13 | 546 | 32720 |
| 36 | 12 | 432 | 33152 |
| 33 | 11 | 363 | 33515 |
| 48 | 10 | 480 | 33995 |
| 76 | 9 | 684 | 34679 |
| 66 | 8 | 528 | 35207 |
| 68 | 7 | 476 | 35683 |
| 93 | 6 | 558 | 36241 |
| 136 | 5 | 680 | 36921 |
| 151 | 4 | 604 | 37525 |
| 245 | 3 | 735 | 38260 |
| 373 | 2 | 746 | 39006 |
| **867 (2519)** | **1** | **867** | **39873** |
| 2762 | | 39873 | |

$$\frac{39873}{3} = 13291$$

$$1 : n : n^2$$

$$= \frac{41}{41} : \frac{202}{41} : \frac{2519}{41}$$

$$= \quad 1: (4.93): (61.44)$$

$$= \quad 1: (4.93): (4.93)^2$$

$$= \quad 1:4.93:\frac{61.44}{24.30} = 1:4.93:2.53$$

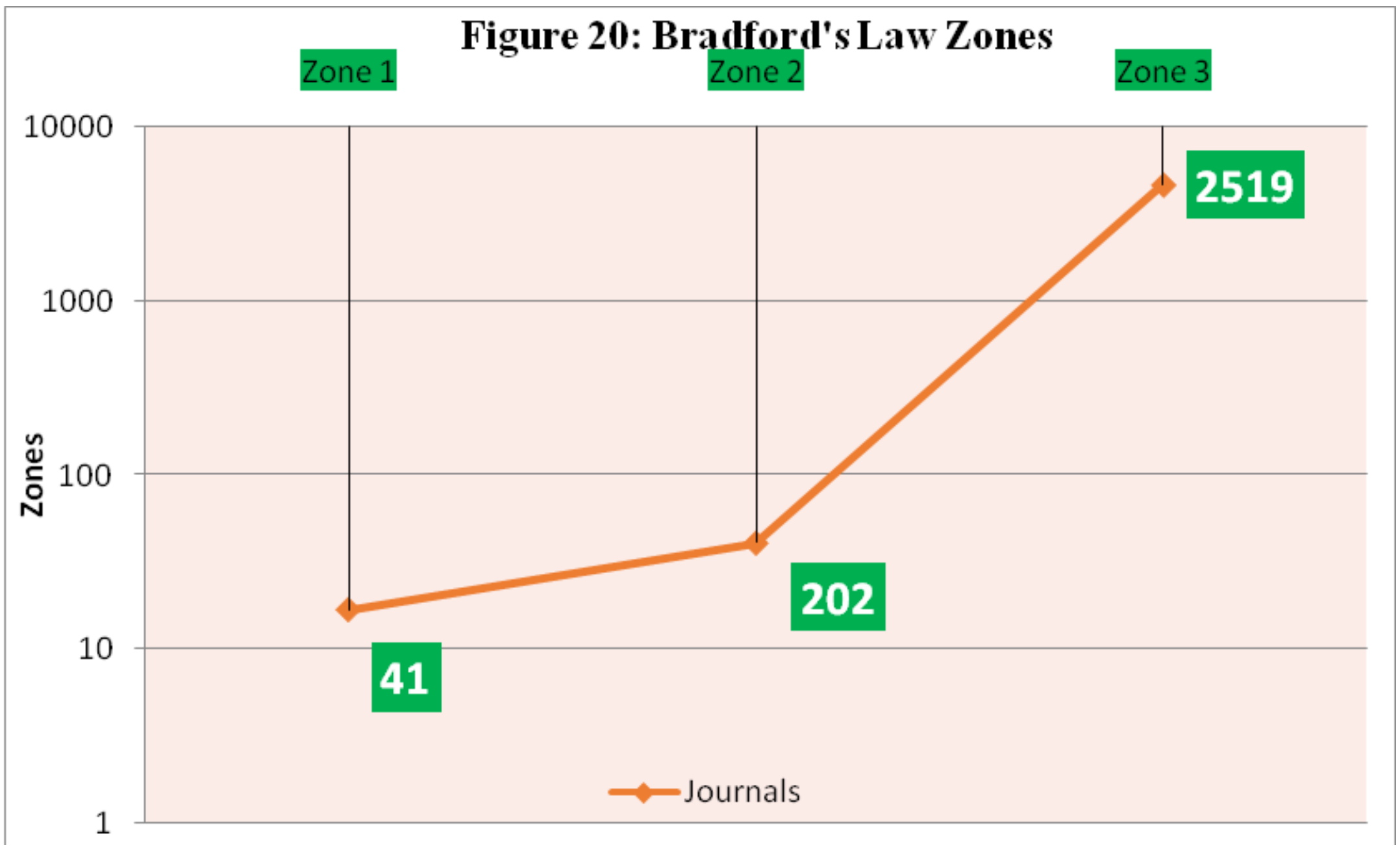

According to Bradford's distribution, the relationship between the zone is 1: $n$: $n^2$. In contrast, the relationship in each of the present study is 41:202:2519. This shows that 41 journals give core contributions. The second zone consists of more than double the number of the first core. i.e., 202. The number of journals that fall in the third zone is 2519. Accordingly, to Bradford's distribution, it should be 41:1681:40804, and in the present study, the second zone was increased as 202 instead of 1: *n*, and on the other hand, the third core supports Bradford's formula and brought the result of 2519 journals. This is a clear indication that the core zone is concentrated and the second zone is much extended, showing the scattering of journals on coronary artery disease research. When this analysis is done for a broader range of periods, the extend of scattering can increase. The distribution of coronary artery disease research output journals and articles relatively confirms implications of Bradford's law. It observed from the above analysis each zone, core, zone 2 = z2, and zone 3 = z3, consists of approximately 39873 records. The documents are scattered over 2759 journals; the highest concentration is in the core with 41 journals z2 consists of 202 and 13489 articles in z3 are scattered across 2519 journals. ***The distribution of coronary artery disease research output Journals and articles relatively confirm Bradford's law implications. Hence, Fifth Hypothesis has been proved.***

### 4.37 Analysis of Core-I Journal Wise List on CAD

Another formal analysis included in scientometric research and studies journal productivity involves ranking journals according to the frequency of the documents they publish. Obtaining a list of journals in order of decreasing productivity is comfortable and provides essential information. Furthermore, it is a preliminary step in applying Bradford's law, so it is worth performing. It is to be stated, identifying those journals that publish most articles about a given subject has practical implications.

By ranking the journals according to the number of documents published, it is possible to identify the most productive ones. Table 37 shows the forty one most productive journals in coronary artery disease research from 1990 to 2019. As can be seen in the table, there is one journal that stands out in this research field: the journal *Kardiologiya* from Moscow, Russia accounts for 1247 of total production in the coronary artery disease research area.

| Table 37: Core-I Journals Wise List on CAD | | | |
|---|---|---|---|
| S.No. | Core Journals | Country | Publications |
| 1 | Kardiologiya | Moscow, Russia | 1247 |
| 2 | Plos One | San Francisco, USA | 965 |
| 3 | Chinese Medical Journal | Beijing, China | 699 |
| 4 | International Journal of Cardiology | Clare, Ireland | 571 |
| 5 | Stroke | Philadelphia, USA | 490 |
| 6 | International Journal of Clinical and Experimental Medicine | Madison, USA | 483 |
| 7 | Journal of The American College of Cardiology | New York, USA | 474 |
| 8 | Terapevticheskii Arkhiv | Moscow, Russia | 412 |
| 9 | Scientific Reports | Berlin, Germany | 410 |
| 10 | Atherosclerosis | Clare, Ireland | 406 |
| 11 | Circulation | Philadelphia, USA | 397 |
| 12 | Medicine | Philadelphia, USA | 396 |
| 13 | Arquivos Brasileiros De Cardiologia | Rio De Janeiro, Brazil | 376 |
| 14 | European Heart Journal | Oxford, England | 360 |
| 15 | Experimental and Therapeutic Medicine | Athens, Greece | 358 |
| 16 | Neural Regeneration Research | Mumbai, India | 302 |
| 17 | Journal of Stroke & Cerebrovascular Diseases | Amsterdam, Netherlands | 263 |
| 18 | Brain Research | Amsterdam, Netherlands | 246 |
| 19 | Molecular Medicine Reports | Athens, Greece | 230 |
| 20 | International Journal of Stroke | London, England | 227 |
| 21 | Cerebrovascular Diseases | Basel, Switzerland | 225 |
| 22 | Heart | London, England | 218 |
| 23 | Journal of the Neurological Sciences | Amsterdam, Netherlands | 216 |
| 24 | Neurological Research | Abingdon, England | 215 |
| 25 | Acta Pharmacologica Sinica | London, England | 208 |
| 26 | Bulletin of Experimental Biology and Medicine | New York, United States | 205 |
| 27 | Arquivos De Neuro-Psiquiatria | Sao Paulo SP, Brazil | 205 |
| 28 | American Journal of Cardiology | Bridgewater, USA | 201 |
| 29 | Zhurnal Nevrologii I Psikhiatrii Imeni S S Korsakova | Moscow, Russia | 191 |
| 30 | Cellular Physiology and Biochemistry | Hoboken, USA | 188 |
| 31 | Neuroscience Letters | Clare, Ireland | 185 |
| 32 | Lipids in Health and Disease | London, England | 182 |
| 33 | Biochemical and Biophysical Research Communications | San Diego, USA | 179 |
| 34 | International Journal of Clinical and Experimental Pathology | Madison, USA | 176 |
| 35 | Clinica Chimica Acta | Amsterdam, Netherlands | 173 |
| 36 | Medical Science Monitor | Melville, USA | 172 |
| 37 | Cardiology | Basel, Switzerland | 167 |
| 38 | Biomed Research International | London, England | 164 |
| 39 | European Review for Medical and Pharmacological Sciences | Rome, Italy | 162 |
| 40 | Oncotarget and Therapy | Auckland, New Zealand | 161 |
| 41 | Life Sciences | Oxford, England | 154 |

The journals were ranked based on their published papers on coronary artery disease research output. Table 37 suggests the journals in which coronary artery disease scientists of BRICS countries preferred to publish their research articles. Further, it was analyzed to find out the critical journals in coronary artery disease, which have brought out the most number of publications made by coronary artery disease scientists.

There were 3066 journals in which coronary artery disease scientists have published their articles throughout the study. There have been 1247 contributors published by a single journal *Kardiologiya* from Russia, and it is ranked in the first position. The second position is taken by *Plos One* journal from San Francisco, USA, which has accounted for 965 publications, and the third position has *Chinese Medical Journal* from Beijing, Peoples Republic of China accumulated 699 publications on CAD. The other journals, namely: *International Journal of Cardiology, Stroke, International Journal of Clinical and Experimental Medicine, Journal of the American College of Cardiology, Terapevticheskii Arkhiv, Scientific Reports* stands at the next six ranks in terms of publishing having 571 articles, 490 articles, 483 articles, 474 articles, 412 articles, and 410 articles respectively. The other journals that published articles on coronary artery disease research include *Atherosclerosis* (406 articles), *Circulation* (397 articles), *Medicine* (396 articles), *Arquivos Brasileiros De Cardiologia* (376 articles), *European Heart Journal* accounts for 360 articles, *Experimental and Therapeutic Medicine* accounts 358 articles, *Neural Regeneration Research* accumulated 302 articles and *Journal of Stroke & Cerebrovascular Diseases* contributed 263 articles.  A detailed list of journals, along with their related ranks, has been provided in table 37.

It could be observed clearly from the above discussion that the journals are ranked on the basis of their maximum number of productivity papers. It is established that the first position was recorded by *Kardiologiya* journal, which is from Russia have contributed 1247 of total publications. There is a large gap between the top-ranking journal and other ranked journals on the list. It is supported by the fact that the *Kardiologiya* journal, which is from Russia, contributed 2.49%. In comparison, the remaining journals contribute less than 2% of publications individually to the CAD literature over three decades of research.

### 4.38 Ranking of Word Occurrence in Zipf's Law

Zipf's law was proposed to be applied to recorded discourse; it is used in social sciences disciplines such as linguistics and other fields as well. Zipf stated that if one takes the words making up a vast body of text and ranks them by frequency of occurrence, then the rank of words multiplied by their frequency of occurrence will be approximately constant. Zipf's law takes two other variables: the number of words in a text and their occurrence frequency.

Word frequency analysis, counting the number of times each word in a document is used, and correcting any excess. It says that the most frequent word will occur approximately twice as often as the second most frequent word, which will occur approximately twice as often as the fourth most frequent word.

**Table 38: Ranking of Word Occurrence in Zipf's Law**

| S.No. | Words | Frequency | Rank | Log F | R | C |
|---|---|---|---|---|---|---|
| 1 | Risk | 4005 | 1 | 3.6026 | 0.0000 | 3.6026 |
| 2 | Expression | 3877 | 2 | 3.5885 | 0.3010 | 3.8895 |
| 3 | Disease | 3779 | 3 | 3.5774 | 0.4771 | 4.0545 |
| 4 | Atherosclerosis | 3358 | 4 | 3.5261 | 0.6021 | 4.1281 |
| 5 | Coronary-Artery-Disease | 3267 | 5 | 3.5141 | 0.6990 | 4.2131 |
| 6 | Myocardial-Infarction | 3076 | 6 | 3.4880 | 0.7782 | 4.2661 |
| 7 | Association | 2696 | 7 | 3.4307 | 0.8451 | 4.2758 |
| 8 | Stroke | 2546 | 8 | 3.4059 | 0.9031 | 4.3089 |
| 9 | Activation | 2529 | 9 | 3.4029 | 0.9542 | 4.3572 |
| 10 | Mortality | 2377 | 10 | 3.3760 | 1.0000 | 4.3760 |
| 11 | Ischemic-Stroke | 2310 | 11 | 3.3636 | 1.0414 | 4.4050 |
| 12 | Cardiovascular-Disease | 2262 | 12 | 3.3545 | 1.0792 | 4.4337 |
| 13 | Injury | 2151 | 13 | 3.3326 | 1.1139 | 4.4466 |
| 14 | Oxidative Stress | 2130 | 14 | 3.3284 | 1.1461 | 4.4745 |
| 15 | Heart-Disease | 2075 | 15 | 3.3170 | 1.1761 | 4.4931 |
| 16 | Risk-Factors | 2061 | 16 | 3.3141 | 1.2041 | 4.5182 |
| 17 | Inflammation | 2057 | 17 | 3.3132 | 1.2304 | 4.5437 |
| 18 | Mechanisms | 1790 | 18 | 3.2529 | 1.2553 | 4.5081 |
| 19 | Artery-Disease | 1700 | 19 | 3.2304 | 1.2788 | 4.5092 |
| 20 | Therapy | 1699 | 20 | 3.2302 | 1.3010 | 4.5312 |
| 21 | Brain | 1674 | 21 | 3.2238 | 1.3222 | 4.5460 |
| 22 | Apoptosis | 1659 | 22 | 3.2198 | 1.3424 | 4.5623 |
| 23 | Cells | 1472 | 23 | 3.1679 | 1.3617 | 4.5296 |
| 24 | Coronary-Heart-Disease | 1470 | 24 | 3.1673 | 1.3802 | 4.5475 |
| 25 | Inhibition | 1339 | 25 | 3.1268 | 1.3979 | 4.5247 |
| 26 | Outcomes | 1321 | 26 | 3.1209 | 1.4150 | 4.5359 |
| 27 | Management | 1315 | 27 | 3.1189 | 1.4314 | 4.5503 |
| 28 | Acute Ischemic-Stroke | 1251 | 28 | 3.0973 | 1.4472 | 4.5444 |
| 29 | Model | 1215 | 29 | 3.0846 | 1.4624 | 4.5470 |
| 30 | Meta Analysis | 1202 | 30 | 3.0799 | 1.4771 | 4.5570 |
| 31 | Prevalence | 1200 | 31 | 3.0792 | 1.4914 | 4.5705 |
| 32 | Heart | 1187 | 32 | 3.0745 | 1.5051 | 4.5796 |
| 33 | Acute Myocardial-Infarction | 1187 | 33 | 3.0745 | 1.5185 | 4.5930 |
| 34 | Cerebral-Ischemia | 1141 | 34 | 3.0573 | 1.5315 | 4.5888 |
| 35 | Reperfusion Injury | 1109 | 35 | 3.0449 | 1.5441 | 4.5890 |
| 36 | Ischemia | 1101 | 36 | 3.0418 | 1.5563 | 4.5981 |
| 37 | Population | 1089 | 37 | 3.0370 | 1.5682 | 4.6052 |
| 38 | Dysfunction | 1078 | 38 | 3.0326 | 1.5798 | 4.6124 |
| 39 | Mice | 1043 | 39 | 3.0183 | 1.5911 | 4.6093 |
| 40 | Infarction | 1040 | 40 | 3.0170 | 1.6021 | 4.6191 |
| 41 | Rats | 1015 | 41 | 3.0065 | 1.6128 | 4.6192 |
| 42 | In-Vitro | 998 | 42 | 2.9991 | 1.6232 | 4.6224 |
| 43 | C-Reactive Protein | 982 | 43 | 2.9921 | 1.6335 | 4.6256 |
| 44 | Blood-Pressure | 969 | 44 | 2.9863 | 1.6435 | 4.6298 |
| 45 | In-Vivo | 960 | 45 | 2.9823 | 1.6532 | 4.6355 |
| 46 | Trial | 910 | 46 | 2.9590 | 1.6628 | 4.6218 |
| 47 | Heart-Failure | 904 | 47 | 2.9562 | 1.6721 | 4.6283 |
| 48 | Focal Cerebral-Ischemia | 902 | 48 | 2.9552 | 1.6812 | 4.6364 |
| 49 | Angiogenesis | 894 | 49 | 2.9513 | 1.6902 | 4.6415 |
| 50 | Reperfusion | 887 | 50 | 2.9479 | 1.6990 | 4.6469 |

It could be seen from the table and figure that the word 'Risk' has repeatedly been used 4005 times by coronary artery disease scientists, and it is dominated in the first rank with 3.6026 constant value. The word 'Expression' has been used 3877 times, which stood in the second rank in the repeated words frequency list with a 'C' value as 3.8895. The word 'Disease' is occupied in the third rank with used constant frequently 3779 times with c value as 4.0545, and it is calculated and occupied at the third position of the frequent occurrence in the sample data. The word 'Atherosclerosis' has 3358 frequencies with a 'C' value 4.1281, followed by "Coronary-Artery-Disease" 2262 with C value 4.4337 and "Myocardial-Infarction" frequency of 3076 and C value as 4.2661. Moreover, the following eleven words have used frequency at above two thousand times. The following twenty-four words are frequently used by above thousand, and the remaining words have above eight hundread times frequently used in the table.

It is observed that the top 50 words that have been used more than 800 times have been taken for study. The above 4000 times frequently used the word “Risk” (4005) in the present study. Zipf’s law's applicability is tested to which the constant the equal value ranging from 3.6026 to 4.6469. Thus, it is proved that Zipf’s law is valid in the present study.

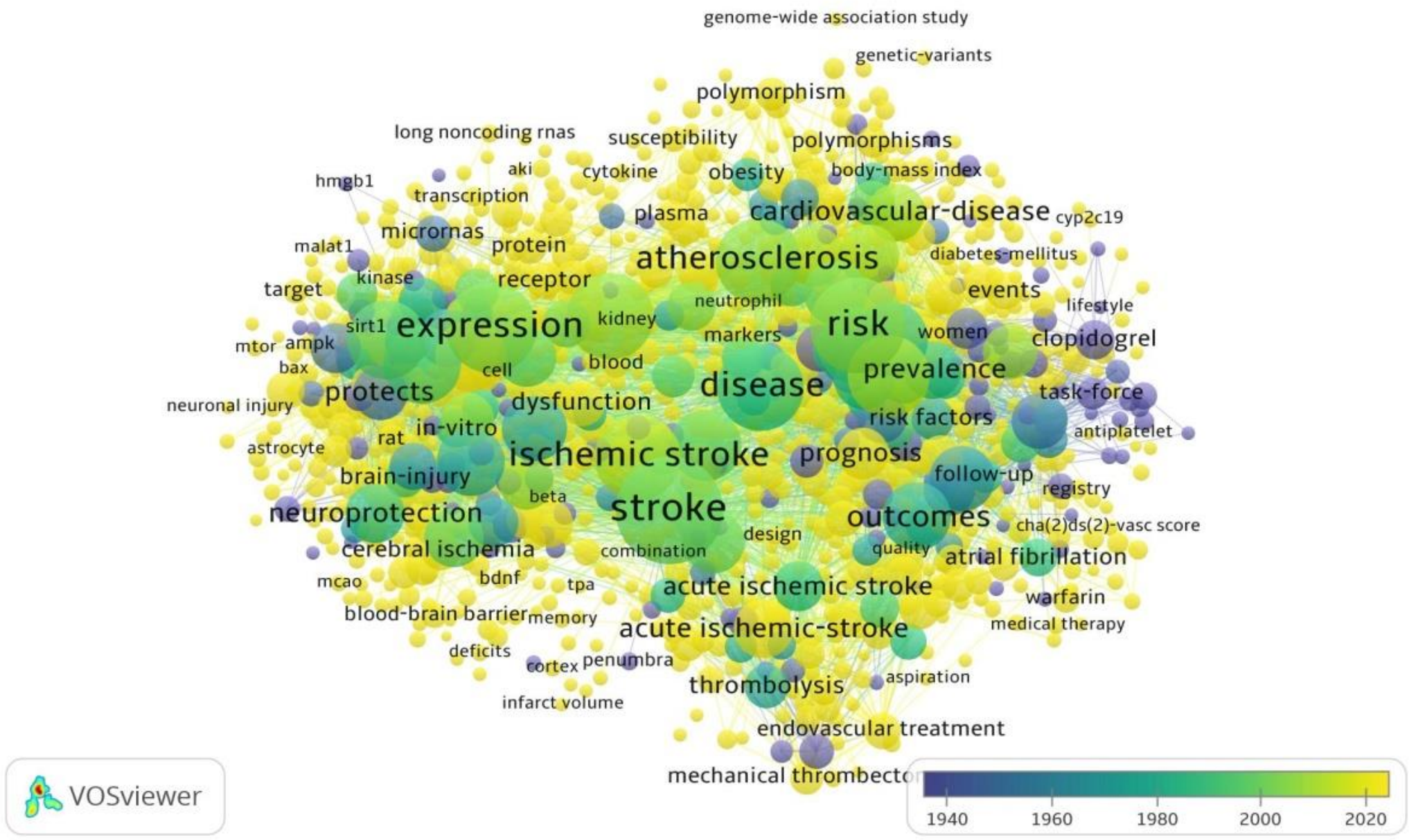

**Figure 21: Word Frequency**

To achieve the eleventh objective of testing the applicability of Zipf’s law of word frequencies on the Coronary Artery Disease literature, the analysis of Zipf’s law is carried out and it is described in table 38.

# CHAPTER V

## DISCUSSION ON FINDINGS, SUGGESTIONS AND CONCLUSION

The present study is a scientometric analysis of coronary artery disease research literature output with particular reference to BRICS countries output, based on publications as found recorded and extracted from Web of Science database, based on the controlled vocabulary of Medical Subject Heading (MeSH). This chapter highlights the findings, suggestions, and conclusions to the research work carried out. A total of 50036 records spanning from 1990 through 2019, covering a period of thirty calendar years, were obtained, organized, sorted out by chosen specific fields in records, and analyzed.

### 5.1 Findings

In this concluding chapter, the significant findings based on the analysis and interpretation presented in Chapter 4 are listed under appropriate headings:

#### 5.1.1 Year-wise Analysis

The researcher has chosen the data for analysis from 1990 to 2019 (three decades) periods. The research output cumulated to 50036 records downloaded from the Web of Science database to analyze the subject of coronary artery disease research productivity in BRICS countries. The study reveals that the year wise growth trend is gradually increasing.

It is observed that the year 2019 occupies first place with 5701 (11.39%) publications, 2018 obtained second place with 5192 (10.38%) publications, 2017 settled third place with 4822 (9.64%) publications, 2016 got fourth place with 4577 (9.15%) publications, 2015 is at fifth place with 4130 (8.25%) publications.

The year 1990 has the minimum number of publications, 158 (0.32%) as a comparative study reveals. The last ten years show a remarkable growth in coronary artery disease research, i.e., 2010 to 2019, which is appreciable by the scientific community.

#### 5.1.2 Exponential Growth Rate

The exponential growth rate is very high in the third block period 2000-2004 (0.198641), so the study indicates that the exponential growth rate is gradually increasing and then a slight fluctuation in growth is observed in the fourth block (2005-2009) and again decreased in growth is noted in the fifth block (2010-2014) but the next block period, i.e. (2015-2019) have not shown remarkable growth.

It is observed from the study that the block year 2015-2019 has more publications compared to the other five-block years. It shows that in recent years, the growth of literature in coronary artery disease increases. It indicates that coronary artery disease has a very high exponential growth rate in the second block.

### 5.1.3 Relative Growth Rate and Doubling Time

The study observed that the relative growth rate has progressively increased from 0.4662 in 1990 to 2.1721 in the year 2019. The whole study period 'mean relative growth rate' is 1.7803. The doubling time for publications of all sources in coronary artery disease research output has decreased from 1.4864 in 1990 to 0.3190 in 2019. During the study period, the 'doubling time' value is 11.9507. The whole study period 'mean doubling time' has been calculated as 0.3984. The relative growth rate has shown an increasing trend, which means the rate of increase is high in terms of segment, and this has been highlighted by doubling time for publications, which reflects a decreasing trend.

### 5.1.4 Time Series Analysis of Research Productivity on CAD Literature

On the application of the formula of time series analysis and subsequently, from the research obtained separately for the year 2022 and 2037, it is found that the future growth trend in coronary artery disease output is assumed to be slow during the years to follow. It will be 5137 in the year 2022, and it is estimated to be 8028 in the year 2037. The inference is that there will be slow growth assumed at the BRICS level literature research output on coronary artery disease.

### 5.1.5 Page Wise Distribution

During the study period, the number of articles published was 50036. They were represented in 391716 pages. The total number of authors calculated in the present study was 384758. The highest number of pages contributed by authors through their research articles was found in 2019, having 51470 pages, followed by the year 2018 contributed in 47298 pages. The year 1990 recorded least number of pages contributed by the authors. The year 2019 seems the most productive in terms of length of pages as it has also shown a higher number of pages, i.e., more than ten pages.

It is observed from the analysis that in terms of the length of pages contributed the growth is observed uneven from 1990 to 1993. It was gradually increasing from 1080 in 1994 to 2498 in 2000, and it falls to 2242 in the year 2001. The gradual increase is noted from 2002 onwards.

### 5.1.6 Country Wise Collaboration of Publications on CAD

Peoples Republic China collaborated 32770 (65.49%) publications, followed by Brazil 6218 (12.43%), Russia 5058 (10.11%), India 4706 (9.41%), and South Africa 1284 (2.57%).

The Peoples Republic of China emerged as the top contributing country in coronary artery disease research, followed by Brazil ranked second, Russia, which ranked third, and India is ranked fourth. South Africa is ranked in fifth place.

### 5.1.7 Collaborating Countries with H-Index and CPP

The study interprets citation per paper of BRICS countries and citation per paper are in the range between 12.97 of Russia to 55.33 of South Africa. Peoples Republic of China, Brazil, India, and South Africa is ranked at the top places in terms of h-index. Peoples Republic of China, Brazil and India have high citations as compared to South Africa and Russia, but South Africa is ranked at the top place according to citation per paper.

### 5.1.8 Activity Index of BRICS Countries

Activity Index for BRICS countries has been calculated. It is to analyze how the BRICS country’s research performance change over different years. The comparison of the BRICS country’s research performance with the world’s research performance has been made using Activity Index calculation. The activity index of BRICS countries from the year 2010 onwards has shown a growth except for South Africa. It remains low from the year 1990 to 2009 of the cumulated output of all the countries together. Brazil shows an increasing trend from the year 2007 as from 1991 to 2006 its activity index is under activity.

### 5.1.9 Document Type

The findings clearly indicate that the 'article' type of document has shown a predominant contribution (79.689%). It occupies the first position concerning the total number of publications reported during the study period. The majority of the work on coronary artery disease by scientists preferred to publish their research papers in journal articles. The other preferred forms of publications among the researchers are 'reviews,' 'meeting abstract,' and 'letter'.

### 5.1.10 Language Wise Publication

The English language has been used as a significant communication language for coronary artery disease publications. Nearly 93.532% of publications appear in the English language and dominates in the first place out of ten languages, followed by Russian (5.744%), Portuguese (0.751%), Chinese (0.140%), and Spanish (0.088%). The remaining language's contributions are less than six articles each in coronary artery disease.

### 5.1.11 Authorship Pattern on CAD

The findings reveal regarding the authorship pattern that six-authorship pattern published 6642 papers on coronary artery disease research, constituting 13.27% of the total publications. It is observed that five authorship pattern of authors published 6146 papers on coronary artery disease research, constituting 12.28% of the total publications. It is observed that seven authors published 5588 papers on coronary artery disease research, consisting of 11.17% of the total publications. It is noted that four authors published 5387 papers in coronary artery disease research, constituting 10.77% of the total publications, followed by eight authors pattern published 4719 papers on coronary artery disease research, constituting 9.43% of the total publications. The remaining authors published less than 9 per cent of publications on CAD.

It is also evident that nearly 97.91% of research output published collaboratively either by double authors and more than two authors in the case of BRICS countries publications on CAD. Single authors' contribution has been noted in about 2.09% of publications.

### 5.1.12 Degree of Collaboration on CAD

The degree of collaboration on coronary artery disease research output from BRICS countries shows an increasing trend from 1990 to 2019. The degree of collaboration ranges between 0.91 and 0.99. The overall degree of collaboration is 0.98 during the study period. Out of the total 50036 publications, 97.91% are published under a collaborative venture of publication in coronary artery disease research. Based on this study, the result of the degree of collaboration DC = 0.98, which clearly supports that 98% of publications are brought out with an effort of collaborative authors rather than the individual.

### 5.1.13 Collaborative Indices

The collaborative Index (CI) ranges between 3.80 and 10.99 during the research period of 1990 to 2018. CI is determined minimum during the year 1990. It is highest in the year 2019. Therefore, it can be found that the collaborative Index is improving from 1990 onwards.

The collaboration coefficient (CC) ranges between 0.62786 and 0.81674 during the study period of 1990 to 2019. CC is found minimum in the year 1990, and it is increasing in the year to follow up to 2019 and is highest in the year 2017. Therefore, it can be understood that the collaborative coefficient is also showing an increasing trend from 1990 onwards.

Modified Collaborative Coefficient (MCC) is calculated to overcome the collaborative coefficient limitation, which ranges from 0.63186 to 0.81691. The value of MCC is lowest during the year 1990, and it is highest in the year 2017.

### 5.1.14 Co-Authorship Index

It is revealed from the study that the co-authorship index for single-author papers was 98.90 in the year 1990, which increased to 100.89 in the year 2019. Subsequently, it shows the inclining trend wherein it was 99.66 in the year 2012. It is noted that publications follow the same pattern of increasing CAI from two to four authorship patterns except for five and above five authors, which is decreasing from the year 1990 to 2019.

The inference depicts that the CAI pattern increases regarding single, two, three, and four authorships, but five and above five authorship pattern has a decreasing trend in terms of CAI.

### 5.1.15 Prolific Authors on CAD from BRICS Countries

Among the 384758 authors, "Zhang Y" from the University of Shanghai for Science & Technology, School Medical Instrument & Food Engineering has published 1196 articles, and the author who is from the Peoples Republic of China contributed the highest number of publications in coronary artery disease research and occupied the first rank in the present context. Next to that, "Wang Y" from Nanchang University, Affiliated Hospital has published 1007 articles, and this author is also from the Peoples Republic of China, and the author occupied the second rank. "Li Y," which is from Hebei University, College of Electronic and Information Engineering, has published 796 publications and has occupied the third rank. It is to be noted that the most productive authors listed in the study are from the Peoples Republic of China, and none among the other BRICS countries authors have spotted in the list of most prolific authors.

### 5.1.16 H-Index of Most Prolific Authors on CAD

Zhang Y from the University of Shanghai for Science & Technology, School Medical Instrument & Food Engineering has got the highest publications of 1196 with 16315 citations and an h-index of 51, which is followed by Wang Y from Nanchang University, Affiliated Hospital who has scored 14254 citations with 1007 articles and has recorded 50 h-index. Among the top 50 authors, Zhang Y has appended the highest h-index 51 with 16315 citations and is the highest contributor in terms of output on CAD. Wang Y has second place in terms of h-index, having 50 with 14254 citations and 1007 publications.

### 5.1.17 Average Authors and Pages per Paper

It is observed that the average number of authors is between 3.80 in the year 1990 to 10.99 in the year 2019, and the average number of pages is between 4.36 in the year 1990 and 9.11 in the year 2019. The average author can be seen showing a consistently increasing trend from 1990 up to 2019 and is high in the year 2019 (10.99). The Average Pages are also showing an increasing trend in its growth from 1990 to 2019, and the highest value is in the year 2018 (9.11), and it slightly decreased in the year 2019 (9.03).

The inference highlighted an increasing trend in collaboration as the average authors are increasing, and the number of pages using by authors is also increasing. The argument is supported by the inference taken from the collaborative authors' study done already. It studied that collaboration is increasing in coronary artery disease research among BRICS countries.

### 5.1.18 Lotka's Law on CAD literature

In order to test the Lotka's law, the Kolmogorov Simonov test was applied for the wellness of the Lotka's law for the estimations of Lotka's types acquired from least square methods. The outcomes tabulated show that the estimation of D-max, i.e., 0.05115 decided with Lotka's type, i.e., n=1.79. The critical value decided at the 0.005 level of significance is 0.0055, which is less noteworthy than the D-max value. Henceforth, the watched authorship information distribution hold suitable for Lotka's law. Consequently, Lotka's law for coronary artery disease literature research from BRICS acknowledges the authorship distributions.

### 5.1.19 Price Square Root Law of CAD

In this study, Price Square Root Law was applied, and square root was applied to the number of contributors, i.e., 104160 and the result was thus matched with the formula applied, and it is seen that all records 104160 and also seen that all scientific papers from above are 384778. As indicated by value square root law $\sqrt{104160}$, contributors ought to contribute 384778/2 = 192389 papers.

In this way $\sqrt{104160}$ = 322.74 (323) authors, ½ of 384778 = 192389 papers.

It can be observed that 323 authors contribute just 56795 papers. The worth is too far away from 50 % (half of the writing regarding a matter); hence it does not fulfil the value square root law.

### 5.1.20 Pareto Principle (80 × 20 Rule)

The researcher has used this analysis with the same values from Price Square Root Law Data to validate the Pareto Principle and test whether 80 per cent of contributions have come from 20 per cent of contributors. Since the total authors' number is 104160, that means the 20 per cent total author number is 77252. The total number of publications is 40825, and 80 per cent of publications value is 324558.

Based on the analysis, the value of Accumulated % of A*B is 20.08 per cent of contributed more than twenty per cent of contributions, once the contributors are 77252. In the 80 × 20 rule view, the value should be very close to 80 per cent. The remaining 80 (79.92) per cent of the author's publications are 324558.

### 5.1.21 Year Wise Total Citation of Publications

The study observed that the citation distribution shows an increasing trend from 1990 to 2008, which is very obvious as the quality papers get more citations over time. From 2009 to 2019, it decreases (24.25 to 1.00) continuously. The ratio of growth rate values is high in the years 2019 (132.92). It is also analyzed that the average citation per paper is increasing from 1990 to 2008 with the values from 5.48 to 32.91, and the same is decreasing from 2009 to 2019 with the values from 24.25 to 1.00. The overall citation per paper value is 15.16.

### 5.1.22 Bradford's Law of Scattering on CAD

According to Bradford's distribution, the relationship between the zone is 1: $n$: $n^2$. In contrast, the relationship in each of the present study is 41:202:2519. This shows that 41 journals give core contributions. The second zone consists of more than double the number of the first core. i.e., 202. The number of journals that fall in the third zone is 2519. Accordingly, to Bradford's distribution, it should be 41:1681:40804, and in the present study, the second zone was increased 202 instead of 1: $n$, and on the other hand, the third core supports Bradford's formula and brought the result of 2519 journals. This is a clear indication that the core zone is concentrated and the second zone is much extended, showing the scattering of journals on coronary artery disease research. When this analysis is done for a broader range of periods, the extend of scattering can increase. The distribution of coronary artery disease research output journals and articles relatively confirms implications of Bradford's law. It observed from the above analysis each zone, core, zone 2 = z2, and zone 3 =z3, consists of approximately 39873 records. The documents are scattered over 2759 journals; the highest concentration is in the core with 41 journals, z2 consists of 202 and 13489 articles in z3 are scattered across 2519 journals. The distribution of coronary artery disease research output Journals and articles confirm the implications of Bradford's law.

#### 5.1.23 Core-I Journals Wise Publications

There were 3066 journals in which coronary artery disease scientists have published their articles throughout the study. There have been 1247 contributions published by a single journal *Kardiologiya* from Russia, and it is ranked in the first position. The second position is taken by *Plos One* journal from the USA, which has accounted for 965 publications, and at the third position, *Chinese Medical Journal* from the Peoples Republic of China accumulated 699 publications on CAD.

#### 5.1.24 Ranking of Word Occurrence in Zipf's Law

It could be seen that the word 'Risk' has repeatedly been used 4005 times by coronary artery disease scientists, and it is dominated in the first rank with 3.6026 constant value. The word 'Expression' has been used 3877 times, which stood in the second rank in the repeated words frequency list with a 'C' value of 3.8895. The word 'Disease' is occupied in the third rank with used constant frequently 3779 times with 'C' value as 4.0545, and it is calculated and occupied at the third position of the frequent occurrence in the sample data.

It is observed that the top 50 words that have been used more than 800 times have been taken for study. The above 4000 times frequently used the word "Risk" (4005) in the present study. The applicability of Zipf's law is tested to which the constant the equal value ranging from 3.6026 to 4.6469. Thus, it is proved that Zipf's law is valid in the present study.

### 5.2 Discussion on Findings

#### 5.2.1 Discussion on Year-wise Productivity and Page Length

The study reveals that the year wise growth trend is gradually increasing. The study indicates that the exponential growth rate is gradually increasing and then a slight fluctuation in growth is also observed. In recent years, the growth of literature in coronary artery disease increases. It indicates that coronary artery disease has a very high exponential growth rate in the second block. The relative growth rate has shown an increasing trend, which means the rate of increase is high in terms of segment, and this has been highlighted by doubling time for publications, which reflects a decreasing trend. The inference is that there will be slow growth assumed at the BRICS level literature research output on coronary artery disease. It is observed from the analysis that in terms of the length of pages contributed the growth is observed uneven from 1990 to 1993. It was gradually increasing from 1080 in 1994 to 2498 in 2000, and it falls to 2242 in the year 2001. The gradual increase is noted from 2002 onwards.

### 5.2.2 Country-wise Collaboration, Document Type and Language

The Peoples Republic of China emerged as the top contributing country in coronary artery disease research, followed by Brazil ranked second, Russia, which ranked third, and India is ranked fourth. South Africa is ranked in fifth place. Peoples Republic of China, Brazil and India have high citations as compared to South Africa and Russia, but South Africa is ranked at the top place according to citation per paper. The activity index of BRICS countries from the year 2010 onwards has shown a growth except for South Africa. The findings clearly indicate that the 'article' type of document has shown a predominant contribution (79.689%). The English language has been used as a significant communication language for coronary artery disease publications.

### 5.2.3 Discussion on Authorship and Degree of Collaboration

The findings reveal regarding the authorship pattern that six-authorship pattern published 6642 papers on coronary artery disease research, constituting 13.27% of the total publications. It is also evident that nearly 97.91% of research output published collaboratively either by double authors and more than two authors in the case of BRICS countries publications on CAD. Single authors' contribution has been noted in about 2.09% of publications. The overall degree of collaboration is 0.98 during the study period. Out of the total 50036 publications, 97.91% are published under a collaborative venture of publications on coronary artery disease research, which are brought out with an effort of collaborative authors rather than the individual. The inference depicts that the CAI pattern increases regarding single, two, three, and four authorships, but five and above five authorship pattern has a decreasing trend in terms of CAI. The most productive authors listed in the study are from the Peoples Republic of China, and none among the other BRICS countries authors have spotted in the list of most prolific authors. Among the top 50 authors, Zhang Y has appended the highest h-index 51 with 16315 citations and is the highest contributor in terms of output on CAD. Wang Y has second place in terms of h-index, having 50 with 14254 citations and 1007 publications. The inference highlighted an increasing trend in collaboration as the average authors are increasing, and the number of pages using by authors is also increasing. The argument is supported by the inference taken from the collaborative authors' study done already. It studied that collaboration is increasing in coronary artery disease research among BRICS countries.

### 5.2.4 Discussion on the Inferences Drawn from Lotka's Law and Price Square Law

Lotka's law for coronary artery disease literature research from BRICS acknowledges the authorship distributions. Price Square Root Law was also applied, and square root was applied to the number of contributors and among them 323 authors contributes just 56795 papers. The worth is too far away from 50 % (half of the writing regarding a matter); hence it does not fulfill the value square root law. Based on the analysis, the value of Accumulated % of A*B is 20.08 per cent of contributed more than twenty per cent of contributions, once the contributors are 77252. In the 80 × 20 rule view, the value should be very close to 80 per cent. The remaining 80 (79.92) per cent of the author's publications are 324558.

### 5.2.5 Discussion on Bradford's Law of Scattering and Zipf's Law

The average citation per paper is increasing from 1990 to 2008 with the values from 5.48 to 32.91, and the same is decreasing from 2009 to 2019 with the values from 24.25 to 1.00. The overall citation per paper value is 15.16. The distribution of coronary artery disease research output journals and articles relatively confirms implications of Bradford's law. It observed from the above analysis each zone, core, zone 2 = z2, and zone 3 =z3, consists of approximately 39873 records. There have been 1247 contributions published by a single journal Kardiologiya from Russia, and it is ranked in the first position. The second position is taken by Plos One journal from the USA and at the third position, Chinese Medical Journal from the Peoples Republic of China accumulated 699 publications. The top 50 words that have been used more than 800 times have been taken for study. The above 4000 times frequently used the word "Risk" (4005) in the present study. The applicability of Zipf's law is tested to which the constant the equal value ranging from 3.6026 to 4.6469 and it is proved that Zipf's law is valid in the present study.

## 5.3 Testing of Hypotheses

**5.3.1 Hypothesis I:** *There is an expanding trend in the relative growth rate and correspondingly a decreasing pattern in the doubling time in Coronary Artery Disease research.*

There is a progressive trend in the number of coronary artery disease literature publications in the present study. Subsequently, it is found that there is expanding in the relative growth and correspondingly a decreasing pattern in the doubling time in coronary artery disease research literature. The hypothesis relevant to this was tested, and results show that the first formulated hypothesis has been proved. Hence relative growth rate on coronary artery disease in BRICS countries level publication shows an increasing trend, and doubling time shows a decreasing pattern.

**5.3.2 Hypothesis II:** *The Journal source of distribution of Coronary Artery Disease output involves a predominant spot compared to other sources of productions.*

The study results reveal that among the source wise distribution on coronary artery disease, journal articles output involves a high publication output compared with other sources of publications. The result shows that the formulated second hypothesis has been proved. Hence the source wise distribution reflects that journal article publication has secured a predominant spot in coronary artery disease research output.

**5.3.3 Hypothesis III:** *There has been an increasing trend in collaborative research in coronary artery disease in recent years.*

The authorship distribution shows that among the total publications of coronary artery disease, multiple-authored papers dominate with a high percentage of 97.91%. The single-authored papers are less (2.09%); the distribution demonstrates that collaboration may have some advantages over individual researchers' research. Thus, the result shows that the formulated third hypothesis has been proved, and hence it is evident that collaborative research dominates coronary artery disease literature.

**5.3.4 Hypothesis IV:** *There has been a logical efficiency of the authors contributing to the Coronary Artery Disease in conformity to Lotka's Law.*

The productivity of authors based on Lotka's law equation $x^n y$= constant, for n = 1.79, the value of $x^n y$ is not constant. Since the scientific productivity of authors in coronary artery disease research literature conforms to Lotka's law of scientific productivity; hence the fourth formulated hypothesis is significantly proved.

**5.3.5 Hypothesis V:** *There has been a delivery of Coronary Artery Disease research productivity in journals and articles that comply with the implication of Bradford's law.*

Bradford's Law of Scattering analysis in the field of coronary artery disease research shows that the journal in the three zones is in the ratio of 41:202:2519, which is in the ratio 1: $n$: $n^2$ proving its validity. This is a clear indication that the core zone is concentrated and the second zone is much extended, showing the scattering of journals on coronary artery disease research. When this analysis is done for a broader range of periods, the extend of scattering can increase. Hence, the present study does agree with Bradford's law. Hence the fifth hypothesis is significantly proved.

**5.3.6 Hypothesis VI:** *The distribution of Coronary Artery Disease literature on output articles relatively confirms the implications of Zipf's law.*

The present study shows that the word frequency of Zipf's law in coronary artery disease research is found applicable. When it was tested, the constant value was ranging from 3.6026 to 4.6469. Hence the sixth formulated hypothesis has been proved.

## 5.4 Area for Further Research

As this study is carried out only on some limited parameters of the BRICS countries' research publication data for a window period of thirty years only, there left some scope for further research as follows:

1. A study can be conducted by taking research contributions of other nations with some other metrics. More particularly, the scope for Altmetrics has been inviting the attention of the Scientometricians. Further studies may be carried out to visualize research which otherwise beyond the scope of the database sources adopted for this work. This study is based on the Web of Science database. Further research can be conducted by taking bibliographic data from other bibliographic databases like Scopus, PubMed, Chemical Abstract, INSPEC, BIOSIS, and Dimensions to minimize the duplications for further analysis.

2. Scientometrics being a subject of rapid development, new metrics are proliferating every passing year. Adopting a more robust set of metrics may reveal some more fascinating facts on various dimensions of research.
3. Other countries from Asia and other nations from developed nations from the west can also be chosen for such studies. Cross comparison of research productivity among scientists may be expected to reveal fascinating facts.
4. Studies similar to this may be conducted by taking the research contributions for a wider time window.

**5.5 Suggestions**

As per the observations of the analysis and results of the present study, the following suggestions are given:

The scientists who work on coronary artery disease should focus on the new area to carry out more research activities in the subfield of CAD research. It is evident from the analysis of the present study, the productivity of the authors as individual authors contribution is very less. Therefore, the individual scientist may be inspired to distribute more number of contributions. There is a need to motivate and encourage researchers and scientists in the field of coronary artery disease research to identify the impact of research output. Provide strategic oversight for CAD research, identifying the gaps in the coronary artery disease research portfolio in the nationally, BRICS level and globally highlighting new scientific opportunities. It is required to initiate specific institutes to support research in the area of coronary artery literature. There are other international languages in the web of science, and it should include Indian languages and also cover journals BRICS countries and Indian as well. The present study investigates on the productivity of CAD research; further, it can be expanded in other forms of future research in the following areas:

1. Extension and collaborative research model in the subfield research on CAD.
2. Comparative study on CAD research in different countries apart from BRICS.
3. Study on productive institutions and research centres in different countries.

4. This study is based on the Web of Science database. Further research can be conducted collaborating all the databases like Scopus, PubMed, Chemical Abstract and BIOSIS with minimizing the duplications.

**5.6 Conclusion**

This scientometrics study investigates that coronary artery disease research among BRICS countries has revealed that the continuous rapid growth was found from 1990 to 2019. Also, the collaboration of country-wise and year-wise collaborative researches is in an increasing trend. The majority of the work on coronary artery disease by scientists preferred to publish their research papers in journal articles. From the study, it has been revealed that Peoples Republic China contributed the highest productivity of articles and the lowest publications contributed by South Africa among the BRICS Countries for three decades. English is by, and large the medium of research communication, for it is widely recognized worldwide. In the study, multiple authorship patterns are dominated compared to single author productivity in the BRICS Countries on coronary artery disease research publications. The research on coronary artery disease is an essential aspect in terms of its utility. This type of research could be increased by organizing seminars and conferences, and also more importance should be given in the field research and development. The funding agencies and governments should encourage coronary artery disease researchers to carry out more researches among the other four countries in BRICS apart from the Peoples Republic of China, which has multifold research on CAD compared to other countries. The Government and other agencies should prepare policies to promote research and development in this area from these countries, particularly South Africa, which lags in CAD research. The country needs to intensify the quality and quantity of coronary artery disease researches carried out by the Research and Development Organizations and Institutions with BRICS country collaboration. This shows the need for a high quality of research and improved scientific research in coronary artery disease.

The bibliometric/scientometric studies are frequently used to assess the research publications and to generate information that could be used by policymakers and experts. This study could be proven to be a useful tool in the assessment of research publications of scientists on coronary artery disease research. The present study illustrates the facts and figures on scientists' scientific publications in the field of coronary artery disease during the study period. Moreover, the present study mirrors the actual published results of the work of scientists on CAD.

Collaboration studies can be instrumental in developing research policy, as they provide an overview of the scientific communication pattern. Developing countries might use these communication patterns and output trends to help identify the strategies propelling Peoples Republic China and other top-publishing BRICS countries. There are a few potential areas, including the following, which may be considered to improve the scientific research outcome in coronary artery disease.

Developing countries would need to address various issues relating to research, skills development, technology development, regulations, and governance to improve their competitive position in the coronary artery disease research field. This study's findings will help understand the behaviour and the impact of coronary artery disease literature. They may assist policymakers as well as the academic community in determining gaps to be addressed.

# Coronary Artery Disease Research in India: A Scientometric Assessment of Publication during 1990-2019

Muneer Ahmad
muneerbangroo@gmail.com

Dr. M.Sadik Batcha
*Annamalai University*

# Coronary Artery Disease Research in India: A Scientometric Assessment of Publication during 1990-2019

**Muneer Ahmad[1] Dr. M Sadik Batcha[2]**

[1]*Research Scholar, Department of Library and Information Science, Annamalai University, Annamalai nagar, muneerbangroo@gmail.com*
[2]*Research Supervisor & Mentor, Professor and University Librarian, Annamalai University, Annamalai nagar, msbau@rediffmail.com*

**Abstract**

The present study examined 4698 Indian Coronary Artery Disease research publications, as indexed in Web of Science database during 1990-2019, with a view to understand their growth rate, global share, citation impact, international collaborative papers, distribution of publications by broad subjects, productivity and citation profile of top organizations and authors, and preferred media of communication. The Indian publications registered an annual average growth rate of 11.47%, global share of 1.14%, international collaborative publications share of 38.89% and its citation impact averaged to 25.58 citations per paper. Among broad subjects, Cardiovascular System & Cardiology contributed the largest publications share of 19.14% in Indian coronary artery disease output, followed by Neurosciences & Neurology (14.94%), Pharmacology & Pharmacy (8.51%), etc. during 1990-2019. Among various organizations and authors contributing to Indian coronary artery disease research, the top 20 organizations and top 30 authors together contributed 40.70% and 37.29% respectively as their share of Indian publication output and 38.36% and 33.13% respectively as their share of Indian citation output during 1990-2019. Among 1222 contributing journals in Indian coronary artery disease research, the top 30 journals registered 30.80% share during 1990-2019. There is an urgent need to increase the publication output, improve research quality and improve international collaboration. Indian government also needs to come up with a policy for identification, screening, diagnosis and treatment of coronary artery disease patients, besides curriculum reform in teaching, capacity building, patient education and political support are badly needed.

## Introduction

*Coronary artery disease (CAD)*-often called coronary heart disease or CHD, is generally used to refer to the pathologic process affecting the coronary arteries (usually atherosclerosis). CAD includes the diagnoses of angina pectoris, myocardial infarction (MI), silent myocardial ischemia, and CAD mortality that result from CAD. Hard CAD endpoints generally include MI and CAD death. The term CHD is often used interchangeably with CAD. *CAD death*—Includes sudden cardiac death (SCD) for circumstances when the death has occurred within 24 hours of the abrupt onset of symptoms, and the term non- SCD applies when the time course from the clinical presentation until the time of death exceeds 24 hours or has not been specifically identified. *Atherosclerotic cardiovascular disease (ASCVD, often shortened to CVD)*-the pathologic process affecting the entire arterial circulation, not just the coronary arteries. Stroke, transient ischemic attacks, angina, MI, CAD death, claudication, and critical limb ischemia are manifestations of ASCVD (Lemos & Omland, 2018).

Coronary artery disease (CAD) is a major cause of death and disability in developed countries. Although CAD mortality rates worldwide have declined over the past 4 decades, CAD remains responsible for approximately one-third or more of all deaths in individuals over age 35, and it has been estimated that nearly half of all middle-aged men and one-third of middle aged women in the United States will develop clinical CAD (Mozaffarian et al., 2016). A Global Burden of Disease Study Group report from 2013 estimated that 17.3 million deaths worldwide in 2013 were related to ASCVD, a 41% increase since 1990 (GBD: 2013 Mortality and Causes of Death Collaborators, 2015). Although the absolute numbers of ASCVD deaths had increased significantly since 1990, the age-standardized death rate decreased by 22% in the same period, primarily due to shifting age demographics and causes of death worldwide (Towfighi, Zheng, & Ovbiagele, 2009).

Heart disease mortality has declined since the 1970s in the United States and in regions where economies and healthcare systems are relatively advanced. Ischemic heart disease remains the number one cause of death in adults on a worldwide basis (GBD: 2013 Mortality and Causes of Death Collaborators, 2015). In a 2014 study using World Health Organization data from 49 countries in Europe and northern Asia, over 4 million annual deaths were attributable to ASCVD (Nichols, Townsend, Scarborough, & Rayner, 2014). Current worldwide estimates for heart disease mortality show Eastern European countries have the highest ASCVD death rates (> 200 per 100,000/year), followed by an intermediate group that includes most countries with modern economies (100–200 per 100,000/ year), and the lowest levels (0–100 per 100,000/year) are largely observed in European countries and a few non- European countries with advanced healthcare systems. A detailed analysis of European country specific data showed that CHD mortality rates dropped by more than 50% over the 1980–2009 interval, and the decline was observed across virtually all European countries for both sexes. The authors of the report concluded that the downward trends did not appear to show a plateau. Rather, CHD mortality was stable or continuing to decline across Europe (Nichols, Townsend, Scarborough, & Rayner, 2013). Complementary analyses have been undertaken in the United States, and CHD mortality has been demonstrated to have peaked in the 1970s and declined since that date (Mozaffarian et al., 2016).

**Indian Perspective**

The office of the RGI has periodically reported data on cardiovascular mortality rates in India (Registrar General of India, 2013). These data have been summarized as circulatory system deaths in the Medical Certification of Cause of Deaths reports, and in 1980s and 1990s it was reported that CVD led to 15%-20% of deaths in the country (Gupta, Misra, Pais, Rastogi, & Gupta, 2006). An increasing trend in proportionate CVD mortality has been reported, with 20.6% deaths in 1990, 21.4% in 1995, 24.3% in 2000, 27.5% in 2005, and 29.0% in 2013 (Registrar General of India, 2013).

However, these reports were based on incomplete data (mainly rural health surveys) from which national data were extrapolated. The Million Death Study Group in collaboration with RGI reported deaths for the year 2001-2003 using a validated verbal autopsy instrument (Registrar General of India, 2013). This study used the existing sample registration surveys of the Indian government and evaluated more than 120,000 death reports obtained from 661 districts of the country using a nationally representative sample of more than 6 million participants. CVD emerged as the most important cause of death in men and women, in urban and rural populations, and in developed and developing states of the country (Registrar General of India, 2013). In India, more than 10.5 million deaths occur annually, and it was reported that CVD led to 20.3% of these deaths in men and 16.9% of all deaths in women (Registrar General of India, 2013)**.** According to 2010-2013 RGI data, (Registrar General of India, 2011) proportionate mortality from CVD increased to 23% of total and 32% of adult deaths in years 2010-2013. The mortality varies from <10% in rural locations in less developed states to >35% in more developed urban locations(Institute for Health Metrics and Evaluation (IHME), 2014). Geographic distribution of CVD mortality in India indicates that in less developed regions, such as the eastern and northeastern states with low Human Development indices, there is lower proportionate mortality compared with better developed states in southern and western regions. There is a linear relationship of increasing proportionate CVD mortality with regional Human Development Index, which confirms the presence of the epidemiological transition introduced earlier (Gaziano & Gaziano, 2008; Kuate Defo, 2014).

**Literature Review**

The review, in general, provides an overview of the theory and the research literature, with a special emphasis on the literature specific to the topic of investigation. It provides support to the proposition of one's research, with ample evidences drawn from subject experts and authorities in the concerned field. The sources consulted for the review of literature here includes Scientometric studies related materials drawn from Primary periodicals.

(Batcha & Ahmad, 2017) obtained the analysis of two journals Indian Journal of Information Sources and Services (IJSS) which is of Indian origin and Pakistan Journal of Library and Information Science (PJLIS) from Pakistan origin and studied them comparatively with scientometric indicators like year wise distribution of articles, pattern of authorship and productivity, degree of collaboration, pattern of co-authorship, average length of papers, average keywords, etc and found 138 (94.52%) of contributions from IJISS were made by Indian authors and similarly 94 (77.05) of contributions from PJLIS were done by Pakistani authors. The collaboration with foreign authors of both the countries is negligible (1.37% of articles) from India and (4.10% of articles) from Pakistan.

(Ahmad, Batcha, Wani, Khan, & Jahina, 2018) studied Webology journal one of the reputed journals from Iran through scientometric analysis. The study aims to provide a comprehensive analysis regarding the journal like year wise growth of research articles, authorship pattern, author productivity, and subjects taken by the authors over the period of 5 years from 2013 to 2017. The findings indicate that 62 papers were published in the journal during the study period. The articles having collaborative nature were high in number. Regarding the subject concentration of papers of the journal, Social Networking, Web 2.0, Library 2.0 and Scientometrics or Bibliometrics were highly noted. The results were formulated through standard formulas and statistical tools.

(Batcha, Jahina, & Ahmad, 2018) has examined the DESIDOC Journal by means of various scientometric indicators like year wise growth of research papers , authorship pattern, subjects and themes of the articles over the period of five years from 2013 to 2017. The study reveals that 227 articles were published over the five years from 2013 to 2017. The authorship pattern was highly collaborative in nature. The maximum numbers of articles (65 %) have ranged their thought contents between 6 and 10 pages.

(Ahmad & Batcha, 2019) analyzed research productivity in Journal of Documentation (JDoc) for a period of 30 years between 1989 and 2018. Web of Science a service from Clarivate Analytics has been consulted to obtain bibliographical data and it has been analysed through Bibexcel and Histcite tools to present the datasets. Analysis part deals with local and global citation level impact, highly prolific authors and their research output, ranking of prominent institution and countries. In addition to this scientographical mapping of bibliographical data is obtainable through VOSviewer, which is open source mapping software.

(Ahmad & Batcha, 2019) studied the scholarly communication of Bharathiar University which is one of the vibrant universities in Tamil Nadu. The study find out the impact of research produced, year-wise research output, citation impact at local and global level, prominent authors and their total output, top journals of publications, top collaborating countries which collaborate with the university authors, highly industrious departments and trends in publication of the university during 2009 through 2018. During the 10 years of study under consideration it indicates that a total of 3440 research articles have been published receiving 38104 citations having h-index as 68. In addition the study used scientographical mapping of data and presented it through graphs using VOSviewer software mapping technique.

(Ahmad, Batcha, & Jahina, 2019) quantitatively measured the research productivity in the area of artificial intelligence at global level over the study period of ten years (2008-2017). The study acknowledged the trends and features of growth and collaboration pattern of artificial intelligence research output. Average growth rate of artificial intelligence per year increases at the rate of 0.862. The multi-authorship pattern in the study is found high and the average number of authors per paper is 3.31. Collaborative Index is noted to be the highest range in the year 2014 with 3.50. Mean CI during the period of study is 3.24. This is also supported by the mean degree of collaboration at the percentage of 0.83 .The mean CC observed is 0.4635. Regarding the application of Lotka’s Law of authorship productivity in the artificial intelligence literature it proved to be fit for the study. The distribution frequency of the authorship follows the exact Lotka’s Inverse Law with the exponent á = 2. The modified form of the inverse square law, i.e.,

Inverse Power Law with á and C parameters as 2.84 and 0.8083 for artificial intelligence literature is applicable and appears to provide a good fit. Relative Growth Rate [Rt(P)] of an article gradually increases from -0.0002 to 1.5405, correspondingly the value of doubling time of the articles Dt(P) decreases from 1.0998 to 0.4499 (2008-2017). At the outset the study reveals the fact that the artificial intelligence literature research study is one of the emerging and blooming fields in the domain of information sciences.

(Batcha, Dar, & Ahmad, 2019) presented a scientometric analysis of the journal titled "Cognition" for a period of 20 years from 1999 to 2018. The study was conducted with an aim to provide a summary of research activity in the journal and characterize its most aspects. The research coverage includes the year wise distribution of articles, authors, institutions, countries and citation analysis of the journal. The analysis showed that 2870 papers were published in journal of Cognition from 1999 to 2018. The study identified top 20 prolific authors, institutions and countries of the journal. Researchers from USA have made the most percentage of contributions.

**Objectives**

The present manuscript aims to study the various dimensions of Indian coronary artery disease research output in terms of various bibliometric indicators, based on publications and citation data, derived from Web of Science database during 1990-2020. In particular, the study analyzed overall annual and cumulative growth of Indian publications, its global share among top 6 most productive countries, its citation impact, its international collaborative papers share, publication output distribution by broad sub-fields, productivity and citation impact of most productive organizations and authors, and leading media of communications.

**Methodology**

For the present study, the publication data was retrieved and downloaded from the Web of Science database (http://apps.webofknowledge.com/) on coronary artery disease research during 1990-2020. A main search strategy for global output was formulated, where the keyword such as ("coronary artery disease'') and mesh terms ("coronary arteriosclerosis" OR "coronary atherosclerosis" OR "coronary ischemic" OR ''arterial sclerosis'' AND CU="India") were searched together in the "Topic tag" and further limited the search output to period '1990-2019' within "date range tag". This search strategy generated 4698 Indian publications on coronary artery disease from the Web of Science database. Detailed analysis was carried out on 4698 Indian publications using the Histcite and Bibexcel tools to get data distribution by subject, collaborating countries, author-wise, organization-wise and journal-wise, etc. Further, mapping tool such as VOSviewer was used to study the collaboration behavior and citation network.

**Analysis**

The global and Indian research output in coronary artery disease research cumulated to 411668 and 4698 publications in 30 years during 1990-2019 and they increased from 1801 and 15 in the year 1990 to 22483 and 390 publications in the year 2019, registering 8.77% and 11.47% growth per annum. Their ten-year cumulative output increased from 69679 and 331 to 133574 and 1136 to 208415 and 3231 publications from 1990-1999 to 2000-2009 to 2010-2019, registering 6.72%, 13.12%, 4.55% and 11.02% growth respectively. The share of Indian publications in global output was 1.14% during 1990-2019, which increased from 0.48% to 0.85% to 1.55% from 1990-1999 to 2000-2009 to 2010-2019 respectively. Amongst Indian publications on coronary artery disease, 69.9% (3286) was published as articles, 11.7% (551) as meeting abstract, 11.4% (535) as review, 2.8% (133) as letter, 2.0% (94) as editorial material, 1.3% (59) as article; proceedings paper, 0.2%(11) as article; early access, 0.2% (8) as note, 0.1% (5) review; early access, 0.1% (4) Article; retracted publication and correction, 0.1% (3) Review; book chapter, 0.0% (2) article; book chapter and reprint and 0.0% (1) biographical-item . The research impact as measured by citations per paper registered by Indian publications in coronary artery disease averaged to 25.58 citations per publication (CPP) during 1990-2019; ten-yearly impact averaged to 21.46 CPP for the period 1990-1999 which increased to 30.38 CPP in the succeeding ten-year 2000-2009 and then declined to 24.32 CPP for the period 2010-2019 (Table 1).

***Table 1:*** *World and India's Output in Coronary Artery Disease Research, 1990-2019.*

| **Publication Period** | **World** | **India** | | | |
|---|---|---|---|---|---|
| | **TP** | **TP** | **TGCS** | **CPP** | **%TP** |
| 1990 | 1801 | 15 | 80 | 5.33 | 0.83 |
| 1991 | 5434 | 24 | 303 | 12.63 | 0.44 |
| 1992 | 5929 | 21 | 270 | 12.86 | 0.35 |
| 1993 | 6236 | 25 | 361 | 14.44 | 0.40 |
| 1994 | 6756 | 26 | 484 | 18.62 | 0.38 |
| 1995 | 7255 | 24 | 821 | 34.21 | 0.33 |
| 1996 | 7804 | 45 | 1137 | 25.27 | 0.58 |
| 1997 | 9154 | 49 | 1421 | 29.00 | 0.54 |
| 1998 | 9384 | 53 | 1393 | 26.28 | 0.56 |
| 1999 | 9926 | 49 | 834 | 17.02 | 0.49 |
| 2000 | 10656 | 55 | 1543 | 28.05 | 0.52 |
| 2001 | 10564 | 57 | 1792 | 31.44 | 0.54 |
| 2002 | 10771 | 60 | 2080 | 34.67 | 0.56 |
| 2003 | 11837 | 72 | 1299 | 18.04 | 0.61 |
| 2004 | 12977 | 101 | 2580 | 25.54 | 0.78 |
| 2005 | 13662 | 110 | 2260 | 20.55 | 0.81 |
| 2006 | 14362 | 103 | 4441 | 43.12 | 0.72 |
| 2007 | 15130 | 143 | 5202 | 36.38 | 0.95 |
| 2008 | 16241 | 212 | 7381 | 34.82 | 1.31 |
| 2009 | 17374 | 223 | 5932 | 26.60 | 1.28 |
| 2010 | 17695 | 255 | 6281 | 24.63 | 1.44 |
| 2011 | 18664 | 265 | 11131 | 42.00 | 1.42 |
| 2012 | 19071 | 307 | 17519 | 57.07 | 1.61 |
| 2013 | 21172 | 314 | 14773 | 47.05 | 1.48 |
| 2014 | 20683 | 330 | 5355 | 16.23 | 1.60 |
| 2015 | 21689 | 272 | 5199 | 19.11 | 1.25 |
| 2016 | 22363 | 421 | 7456 | 17.71 | 1.88 |
| 2017 | 22420 | 312 | 7797 | 24.99 | 1.39 |
| 2018 | 22175 | 365 | 2462 | 6.75 | 1.65 |
| 2019 | 22483 | 390 | 611 | 1.57 | 1.73 |
| 1990-1999 | 69679 | 331 | 7104 | 21.46 | 0.48 |
| 2000-2009 | 133574 | 1136 | 34510 | 30.38 | 0.85 |
| 2010-2019 | 208415 | 3231 | 78584 | 24.32 | 1.55 |
| 1990-2019 | 411668 | 4698 | 120198 | 25.58 | 1.14 |

*TP: Total Papers; TC: Total Citations; CPP: Citations Per Paper; ICP: International Collaborative Papers*

**Publication Profile of Top 6 Most Productive Countries**

More than 140 countries of the world participated in global research in coronary artery disease research during 1990-2019. Between 4698 and 139222 publications were contributed by top 6 most productive countries in coronary artery disease research and they together accounted for 65.69% of global publication share during 1990-2019. Their ten-year publications output decreased from 65.34% to 63.27% from 1990-1999 to 2000-2009 and then increased 67.35% in 2010-2019. Each of top 6 countries had global publication share between 1.14% and 33.82% during 1990-2019. USA accounted for the highest publication share (33.82%), followed by Germany (8.12%), Republic of China (8.02%), Japan (7.61%), England (6.97%) and India (1.14%) during 1990-2019. Their ten-year global publication share have increased by 2.45% in Republic of China, followed by India (0.38%), Germany (0.28%), and England (0.04%), as against decline by 4.83% in USA and 0.37% in Japan from 1990-1999 to 2000-2009 and then again ten-year global share have increased by 10.61% in Republic of China, followed by India (0.70%) and England (0.01%) as against decline by 4.27% in USA, 1.54% in Japan and 1.44% in Germany from 2000-2009 to 2010-2019 (Table 2).

***Table 2:*** *Global Publication Output and Share of Top 6 Countries in Coronary Artery Disease Research during 1990-2019*

| S.No. | Country Name | TP | | | | %TP | | | |
|---|---|---|---|---|---|---|---|---|---|
| | | 1990-1999 | 2000-2009 | 2010-2019 | 1990-2019 | 1990-1999 | 2000-2009 | 2010-2019 | 1990-2019 |
| 1 | USA | 27869 | 46971 | 64382 | 139222 | 40.00 | 35.16 | 30.89 | 33.82 |
| 2 | Germany | 6008 | 11886 | 15553 | 33447 | 8.62 | 8.90 | 7.46 | 8.12 |
| 3 | Republic of China | 430 | 4093 | 28506 | 33029 | 0.62 | 3.06 | 13.68 | 8.02 |
| 4 | Japan | 6063 | 11126 | 14158 | 31347 | 8.70 | 8.33 | 6.79 | 7.61 |
| 5 | England | 4825 | 9305 | 14547 | 28677 | 6.92 | 6.97 | 6.98 | 6.97 |
| 6 | India | 331 | 1136 | 3231 | 4698 | 0.48 | 0.85 | 1.55 | 1.14 |
| | Total of 6 Countries | 45526 | 84517 | 140377 | 270420 | 65.34 | 63.27 | 67.35 | 65.69 |
| | World Output | 69679 | 133574 | 208415 | 411668 | 100.00 | 100.00 | 100.00 | 100.00 |
| | Share of 6 in World Output | 65.34 | 63.27 | 67.35 | | 65.34 | 63.27 | 67.35 | |

**India's International Collaboration**

The share of India's international collaborative publications (ICP) in its national output in coronary artery disease research was 38.88% during 1990-2019, which increased from 0.89% during 1990-1999 to 6.00% during 2000-2009 and then again increased to 31.99% during 2010-2019. About 139 foreign countries collaborated with India in 1827 coronary artery disease research papers during 1990-2019. These 1827 papers together registered 285,336 citations, with 156 citations per paper. USA, among foreign countries, contributed the largest share (40.45%) to India's international collaborative papers in coronary artery disease research, followed by England (14.72%), Canada (14.07%), Peoples Republic of China (10.73%), Australia (10.24%), and Germany (9.80%) during 1990-2019. The share of ICP increased by 7.40% in Canada, followed by 5.57% in USA, 4.91% in England, as against decrease by 10.03% Republic of China, 4.46% in Australia and 3.39% in Germany from 1990-1999 to 2000-2009 and then again share of ICP increased by 7.59% in Peoples Republic of China, followed by 3.38% in England, 3.27% in Australia and 1.47% in Germany, as against decrease by 15.33% USA and 0.37% in Canada from 2000-2009 to 2010-2019 (Table 3).

***Table 3:*** *The Share of Top 6 Foreign Countries in India's International Collaborative Papers in India's Coronary Artery Disease Research during 1990-2019.*

| S.No. | Collaborative Country | Number of International Collaborative Papers | | | | Share of International Collaborative Papers | | | |
|---|---|---|---|---|---|---|---|---|---|
| | | 1990-1999 | 2000-2009 | 2010-2019 | 1990-2019 | 1990-1999 | 2000-2009 | 2010-2019 | 1990-2019 |
| 1 | USA | 20 | 150 | 569 | 739 | 47.62 | 53.19 | 37.86 | 40.45 |
| 2 | England | 3 | 34 | 232 | 269 | 7.14 | 12.06 | 15.44 | 14.72 |
| 3 | Canada | 3 | 41 | 213 | 257 | 7.14 | 14.54 | 14.17 | 14.07 |
| 4 | Peoples Republic of China | 6 | 12 | 178 | 196 | 14.29 | 4.26 | 11.84 | 10.73 |
| 5 | Australia | 5 | 21 | 161 | 187 | 11.90 | 7.45 | 10.71 | 10.24 |
| 6 | Germany | 5 | 24 | 150 | 179 | 11.90 | 8.51 | 9.98 | 9.80 |
| Total | | 42 | 282 | 1503 | 1827 | 100.00 | 100.00 | 100.00 | 100.00 |

**Subject-Wise Distribution of Indian Research Output**

As per the Web of Science database classification, India's coronary artery disease research output is distributed across 88 subjects during 1990-2019. Among subjects, cardiovascular system and cardiology registered the highest publications share (19.14%), followed by neurosciences and neurology (14.94%), pharmacology and pharmacy (8.51%), general and internal medicine (4.40%), biochemistry and molecular biology (4.22%), research and experimental medicine (3.81%), surgery (3.23%), cell biology (2.96%), endocrinology and metabolism (2.76%), pediatrics (2.12%) and other subjects respectively during 1990-2019 (Table 4).

*Table 4: Subject-Wise Breakup of Indian Publications in Coronary artery Disease Research during 1990-2019*

| S.No. | *Subject wise | TP | % |
|---|---|---|---|
| 1 | Cardiovascular System & Cardiology | 1298 | 19.14 |
| 2 | Neurosciences & Neurology | 1013 | 14.94 |
| 3 | Pharmacology & Pharmacy | 577 | 8.51 |
| 4 | General & Internal Medicine | 298 | 4.40 |
| 5 | Biochemistry & Molecular Biology | 286 | 4.22 |
| 6 | Research & Experimental Medicine | 258 | 3.81 |
| 7 | Surgery | 219 | 3.23 |
| 8 | Cell Biology | 201 | 2.96 |
| 9 | Endocrinology & Metabolism | 187 | 2.76 |
| 10 | Pediatrics | 144 | 2.12 |
| 11 | Science & Technology - Other Topics | 125 | 1.84 |
| 12 | Engineering | 121 | 1.78 |
| 13 | Immunology | 120 | 1.77 |
| 14 | Hematology | 119 | 1.76 |
| 15 | Genetics & Heredity | 111 | 1.64 |
| 16 | Radiology, Nuclear Medicine & Medical Imaging | 98 | 1.45 |
| 17 | Nutrition & Dietetics | 86 | 1.27 |
| 18 | Respiratory System | 85 | 1.25 |
| 19 | Chemistry | 71 | 1.05 |
| 20 | Ophthalmology | 69 | 1.02 |

**There is overlapping of literature covered under various subjects*

**Significant Keywords**

Around 7357 significant keywords have been identified from the literature, which highlight possible research trends in Indian coronary artery disease research. The 40 keywords are listed in table 5 in the decreasing order of their frequency of occurrence in 30 years during 1990-2019.

*Table 5: List of Significant Keywords in Literature on Indian Coronary Artery Disease Research during 1990-2019.*

| S.No. | Name of Key Words | Frequency | S.No. | Name of Key Words | Frequency |
|---|---|---|---|---|---|
| 1 | Coronary | 981 | 21 | Cerebral | 214 |
| 2 | Disease | 956 | 22 | Population | 214 |
| 3 | Ischemic | 936 | 23 | Cardiovascular | 203 |
| 4 | Stroke | 857 | 24 | Rats | 191 |
| 5 | Artery | 850 | 25 | Factors | 176 |
| 6 | Patients | 719 | 26 | Reperfusion | 173 |
| 7 | Acute | 437 | 27 | Clinical | 168 |
| 8 | Risk | 409 | 28 | Diabetes | 167 |
| 9 | Indian | 359 | 29 | Infarction | 163 |
| 10 | Effect | 288 | 30 | Case | 147 |
| 11 | India | 277 | 31 | North | 142 |
| 12 | Myocardial | 276 | 32 | Polymorphism | 142 |
| 13 | Heart | 263 | 33 | Using | 142 |
| 14 | Association | 249 | 34 | Based | 141 |
| 15 | Gene | 236 | 35 | South | 138 |
| 16 | Ischemia | 228 | 36 | Cardiac | 137 |
| 17 | Induced | 224 | 37 | Stress | 137 |
| 18 | Role | 223 | 38 | Therapy | 135 |
| 19 | Injury | 215 | 39 | Trial | 135 |
| 20 | Analysis | 214 | 40 | Rat | 132 |

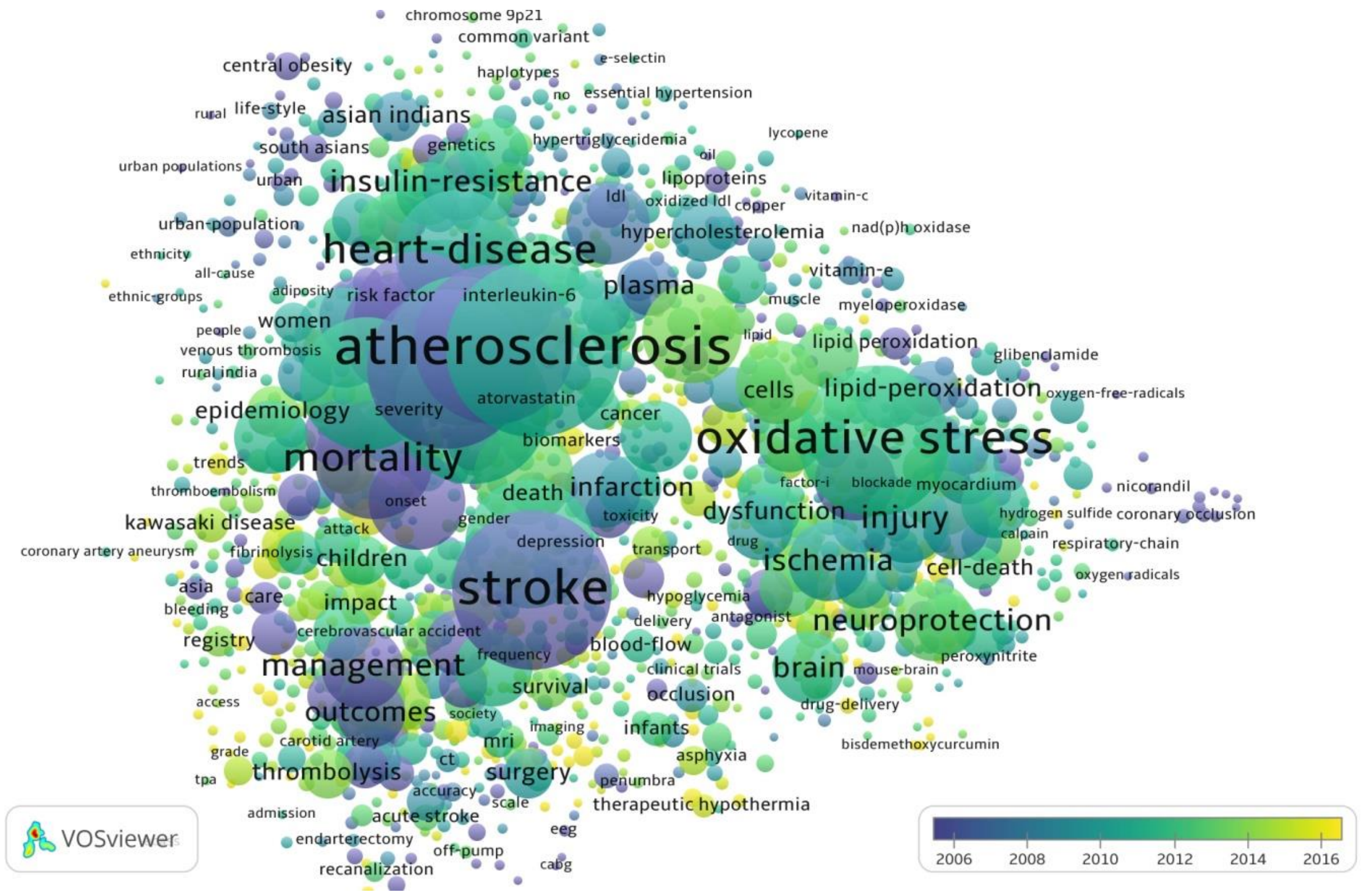

*Cluster of Keywords on Indian Coronary Artery Disease Research*

**Profile of Top 20 Most Productive Indian Organizations**

The top 20 Indian organizations contribution to coronary artery disease research varied from 45 to 496 publications and they together accounted for 40.10% (1887) publication share and 82.20% (98807) citation share to its cumulative publications output during 1990-2019. Table 6 presents a scientometric profile of these 20 India organizations.

***Table 6:*** *Scientometric Profile of Top 20 Most Productive Indian Organizations in Coronary Artery Disease Research during 1990-2019*

| S.No. | Name of Organization | TP | % | TGCS | CPP |
|---|---|---|---|---|---|
| 1 | All India Institute of Medical Sciences (AIIMS), New Delhi | 496 | 10.60 | 29360 | 59.19 |
| 2 | Postgraduate Institute of Medical Education & Research (PGIMER), Chandigarh | 212 | 4.50 | 8165 | 38.51 |
| 3 | Christian Medical College & Hospital, Vellore | 121 | 2.60 | 8814 | 72.84 |
| 4 | Sree Chitra Tirunal Institute Medical Science & Technology | 105 | 2.20 | 6108 | 58.17 |
| 5 | Nizams Institute of Medical Science | 104 | 2.20 | 1727 | 16.61 |
| 6 | Sanjay Gandhi Postgraduate Institute Medical Science | 100 | 2.10 | 18847 | 188.47 |
| 7 | Osmania University | 67 | 1.40 | 866 | 12.93 |
| 8 | Natl Institution Mental Health & NeuroScience | 62 | 1.30 | 1046 | 16.87 |
| 9 | Madras Diabet Research Foundation | 57 | 1.20 | 5028 | 88.21 |
| 10 | Banaras Hindu University | 56 | 1.20 | 3866 | 69.04 |
| 11 | Panjab University | 56 | 1.20 | 1478 | 26.39 |
| 12 | University Delhi | 53 | 1.10 | 4382 | 82.68 |
| 13 | GB Pant Hospital | 52 | 1.10 | 628 | 12.08 |
| 14 | Manipal University | 52 | 1.10 | 902 | 17.35 |
| 15 | Govt Medical College | 51 | 1.10 | 1514 | 29.69 |
| 16 | Post Graduate Institution of Medical Education & Research | 51 | 1.10 | 3057 | 59.94 |
| 17 | Punjabi University | 50 | 1.10 | 629 | 12.58 |
| 18 | Sir Ganga Ram Hospital | 49 | 1.00 | 694 | 14.16 |
| 19 | Apollo Hospital | 48 | 1.00 | 813 | 16.94 |
| 20 | Maulana Azad Medical College | 45 | 1.00 | 883 | 19.62 |
| | Total of 20 Organizations | 1887 | 40.10 | 98807 | 52.36 |
| | Total of India | 4698 | 100 | 270420 | 57.56 |
| | Share of 20 Organizations in Indian total output | 40.17 | | 36.54 | |

*TP: Total Papers; TGCS: Total Global Citations Score; CPP: Citations Per Paper*

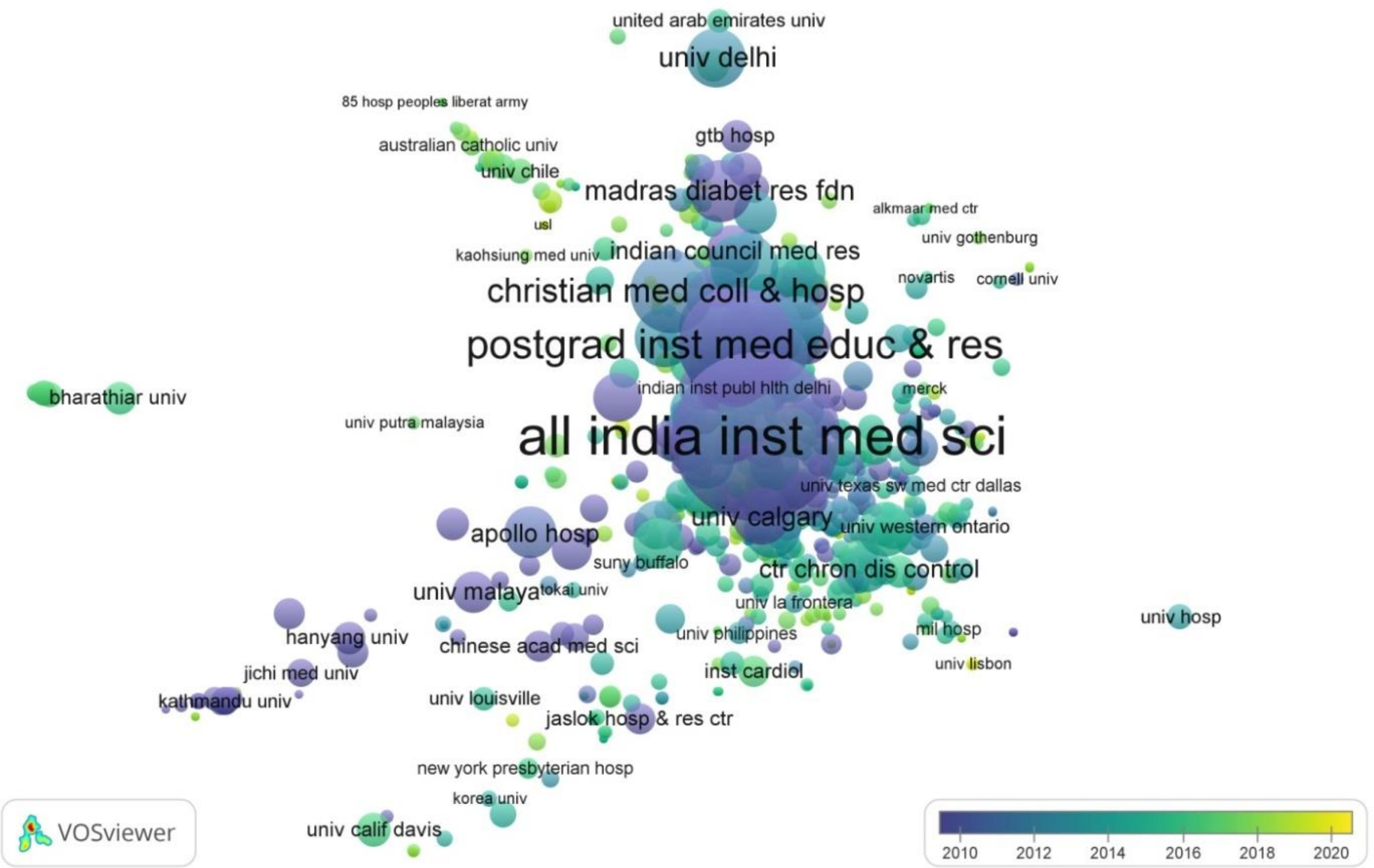

*Cluster of Most Productive Indian Organizations in Coronary Artery Disease Research*

Six organizations registered higher productivity than the group average of 94.35: All India Institute of Medical Sciences (AIIMS), New Delhi (496 papers), Postgraduate Institute of Medical Education & Research (PGIMER), Chandigarh (212 papers), Christian Medical College & Hospital, Vellore (121 papers), Sree Chitra Tirunal Institute Medical Science & Technology (105 papers), Nizams Institute of Medical Science (104 papers), Sanjay Gandhi Postgraduate Institute Medical Science (100) during 1990-2019.

Eight organizations registered higher citation impact than the group average of 52.36: All India Institute of Medical Sciences (AIIMS), New Delhi (59.19), Christian Medical College & Hospital, Vellore (72.84), Sree Chitra Tirunal Institute Medical Science & Technology (58.17), Sanjay Gandhi Postgraduate Institute Medical Science (188.47), Madras Diabet Research Foundation (88.21), Banaras Hindu University (69.04), University Delhi (82.68) and Post Graduate Institution of Medical Education & Research (59.94) during 1990-2019.

**Profile of Top 30 Most Productive Authors**

The top 30 Indian author's contribution to coronary artery disease research varied from 35 to 146 publications and they together accounted for 37.29% (1752) publication share and 38.36% (103744) citation share to its cumulative publications output during 1990-2019. Table 7 presents a scientometric profile of these 20 India authors.

*Table 7: Scientometric Profile of Top 20 Most Productive Authors in Coronary Artery Disease Research during 1990-2019*

| S.No. | Authors | TP | % | TGCS | CPP |
|---|---|---|---|---|---|
| 1 | Kumar A | 146 | 3.1 | 2233 | 15.29 |
| 2 | Kumar S | 100 | 2.1 | 732 | 7.32 |
| 3 | Prasad K | 87 | 1.9 | 1090 | 12.53 |
| 4 | Singh S | 80 | 1.7 | 1197 | 14.96 |
| 5 | Sharma A | 78 | 1.7 | 590 | 7.56 |
| 6 | Kaul S | 74 | 1.6 | 958 | 12.95 |
| 7 | Mohan V | 74 | 1.6 | 6091 | 82.31 |
| 8 | Kumar P | 70 | 1.5 | 2881 | 41.16 |
| 9 | Singh N | 68 | 1.4 | 3187 | 46.87 |
| 10 | Singh RB | 68 | 1.4 | 2611 | 38.40 |
| 11 | Singh M | 63 | 1.3 | 1182 | 18.76 |
| 12 | Gupta R | 62 | 1.3 | 20427 | 329.47 |
| 13 | Gupta A | 61 | 1.3 | 713 | 11.69 |
| 14 | Prabhakaran D | 59 | 1.3 | 8343 | 141.41 |
| 15 | Sylaja PN | 53 | 1.1 | 1390 | 26.23 |
| 16 | Bhatia R | 51 | 1.1 | 875 | 17.16 |
| 17 | Das S | 50 | 1.1 | 993 | 19.86 |
| 18 | Sharma S | 47 | 1 | 282 | 6.00 |
| 19 | Pandian JD | 44 | 0.9 | 16111 | 366.16 |
| 20 | Xavier D | 43 | 0.9 | 9570 | 222.56 |
| 21 | Munshi A | 40 | 0.9 | 670 | 16.75 |
| 22 | Karthikeyan G | 39 | 0.8 | 14634 | 375.23 |
| 23 | Gupta S | 38 | 0.8 | 801 | 21.08 |
| 24 | Jaggi AS | 38 | 0.8 | 592 | 15.58 |
| 25 | Kapoor A | 38 | 0.8 | 564 | 14.84 |
| 26 | Trehan N | 38 | 0.8 | 901 | 23.71 |
| 27 | Khurana D | 36 | 0.8 | 131 | 3.64 |
| 28 | Kumar R | 36 | 0.8 | 989 | 27.47 |
| 29 | Niaz MA | 36 | 0.8 | 1849 | 51.36 |
| 30 | Ghosh S | 35 | 0.7 | 1157 | 33.06 |
| Total of 30 authors | | 1752 | 37.29 | 103744 | 59.21 |
| Total of India | | 4698 | | 270420 | 57.56 |
| Share of 30 authors in India's output | | 37.29 | | 38.36 | |

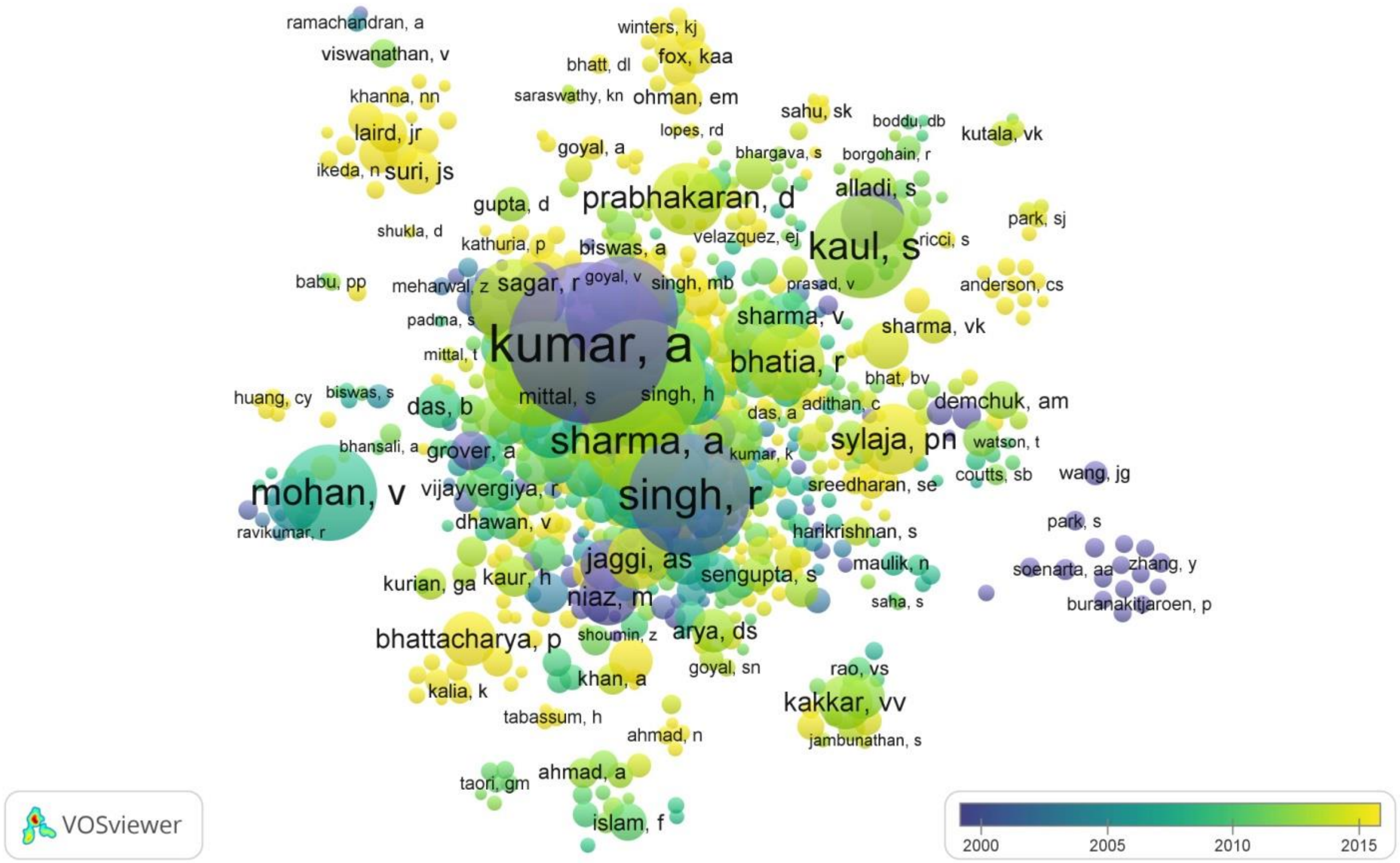

*Cluster of Most Productive Authors in Coronary Artery Disease Research*

Fourteen authors registered higher publications productivity than group average of 58.4: Singh S (80 papers), Sharma A (78 papers), Kaul S and Mohan V (74 papers each), Kumar P (70 papers), Singh N and Singh RB (68 papers each), Singh M (63 papers), Gupta R (62 papers), Gupta A (61 papers), and Prabhakaran D (59 papers) during 1990-2019.

Six authors registered higher citation impact than the group average of 59.21 citations per publication: Karthikeyan G (375.23), Pandian D (366.16), Gupta R (329.47), Xavier D (222.56), Prabhakaran D (141.41) and Mohan V (82.31) during 1990-2019.

**Medium of Communication**

Among India's coronary artery disease output, Indian publications on coronary artery disease, 69.9% (3286) was published as articles, 11.7% (551) as meeting abstract, 11.4% (535) as review, 2.8% (133) as letter, 2.0% (94) as editorial material, 1.3% (59) as article; proceedings paper, and other forms are less than one percent. The top 30 most productive journals accounted for 22 to 143 papers. The top 30 journals publishing Indian papers in coronary artery disease together accounted for 30.80% share (1447 papers) of total Indian journal publication output during 1990-2019. Neurology India *was* the most productive journals with 143 papers, followed by International Journal of Cardiology (139 papers), Annals of Indian Academy of Neurology (99 papers), Journal of the Neurological Sciences (77 papers), Indian Journal of Medical Research (69 papers), Molecular And Cellular Biochemistry (56 papers) Stroke (55 papers), Journal of the American College of Cardiology and PLOS One (52 papers each) etc. during 1990-2019 (Table 8).

***Table 8:*** *Productivity of Top 30 Most Productive Journals in Indian Coronary Artery Disease Research during 1990-2019*

| S.No. | Name of the Journals | Number of Papers | % | TLCS | TGCS |
|---|---|---|---|---|---|
| 1 | Neurology India | 143 | 3.00 | 135 | 945 |
| 2 | International Journal of Stroke | 139 | 3.00 | 52 | 682 |
| 3 | International Journal of Cardiology | 108 | 2.30 | 131 | 2110 |
| 4 | Annals of Indian Academy of Neurology | 99 | 2.10 | 40 | 559 |
| 5 | Journal of The Neurological Sciences | 77 | 1.60 | 105 | 573 |
| 6 | Indian Journal of Medical Research | 69 | 1.50 | 87 | 1393 |
| 7 | Molecular And Cellular Biochemistry | 56 | 1.20 | 71 | 1201 |
| 8 | Stroke | 55 | 1.20 | 56 | 1564 |
| 9 | Journal of The American College of Cardiology | 52 | 1.10 | 82 | 2702 |
| 10 | PLOS One | 52 | 1.10 | 0 | 964 |
| 11 | Indian Journal of Pharmacology | 43 | 0.90 | 11 | 130 |
| 12 | Circulation | 39 | 0.80 | 51 | 1388 |
| 13 | Journal of Stroke & Cerebrovascular Diseases | 36 | 0.80 | 43 | 313 |
| 14 | American Journal of Cardiology | 33 | 0.70 | 34 | 567 |
| 15 | Annals of Thoracic Surgery | 33 | 0.70 | 13 | 666 |
| 16 | Indian Journal of Ophthalmology | 33 | 0.70 | 5 | 105 |
| 17 | Atherosclerosis | 32 | 0.70 | 35 | 666 |
| 18 | European Heart Journal | 32 | 0.70 | 27 | 551 |
| 19 | Lancet | 31 | 0.70 | 160 | 24289 |
| 20 | Catheterization And Cardiovascular Interventions | 30 | 0.60 | 11 | 468 |
| 21 | Indian Journal of Pediatrics | 30 | 0.60 | 10 | 240 |
| 22 | Gene | 29 | 0.60 | 28 | 324 |
| 23 | Cerebrovascular Diseases | 28 | 0.60 | 16 | 73 |
| 24 | Biomedicine & Pharmacotherapy | 27 | 0.60 | 10 | 370 |
| 25 | Current Science | 25 | 0.50 | 34 | 477 |
| 26 | European Journal of Pharmacology | 24 | 0.50 | 72 | 648 |
| 27 | Life Sciences | 24 | 0.50 | 52 | 552 |
| 28 | Atherosclerosis Supplements | 23 | 0.50 | 0 | 3 |
| 29 | Clinica Chimica Acta | 23 | 0.50 | 56 | 562 |
| 30 | Biomedical Research-India | 22 | 0.50 | 5 | 58 |
| | Total of 30 Journals | 1447 | 30.80 | 1432 | 45143 |
| | Total Indian Journal Output | 4698 | 100.00 | 4038 | 120198 |
| | Share of 30 journals in Indian journal output | 30.80 | 30.80 | 35.46 | 37.56 |

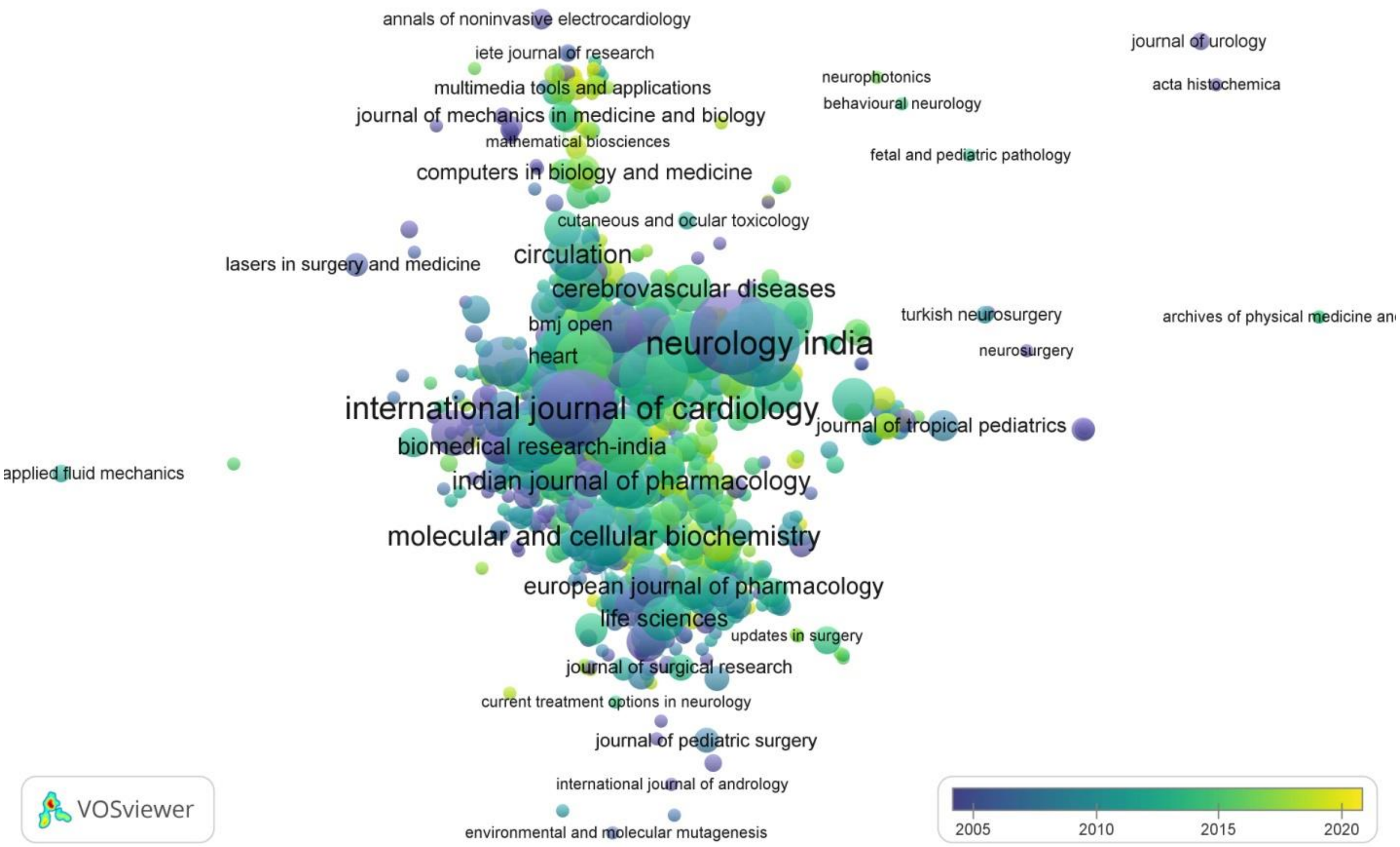

*Cluster of Most Productive Journals in Indian Coronary Artery Disease Research*

**Summary and Conclusion**

1168 Indian publications in coronary artery disease research as indexed in Web of Science database, was published during 1990-2019 and they increased from 15 to 390 in the year 1990 to the year 2019, registering 11.47% growth per annum. Their cumulative Indian output increased from 331 to 1136 to 3231, witnessing 13.12% and 11.02% growth respectively from 1990-1999 to 2000-2009 to 2010-2019. India's global publications share in coronary artery disease research was only 1.14% during 1990-2019, witnessing increase from 0.48% to 0.85% to 1.55% from 1990-1999 to 2000-2009 to 2010-2019. The citation impact per paper of Indian publications on coronary artery disease research was averaged to 25.58 citations, however, increasing from 21.46 during 1990-1999 to 30.38 during 2000-2009 and then decreasing from 30.38 during 2000-2009 to 24.32 during 2010-2019.

The share of India's international collaborative publications in coronary artery disease research was 38.88% during 1990-2019, showing increase from 0.89% during 1990-1999 to 6.00% during 2000-2009 and then again increased to 31.99% during 2010-2019. USA in India's international collaborative papers, contributed the largest publications share of 40.45%, followed by England (14.72%), Canada (14.07%), Peoples Republic of China (10.73%), Australia (10.24%), and Germany (9.80%) during 1990-2019.

Cardiovascular system and cardiology, among main subjects contributed the highest publications share (19.14%), followed by neurosciences and neurology (14.94%), pharmacology and pharmacy (8.51%), general and internal medicine (4.40%), biochemistry and molecular biology (4.22%), research and experimental medicine (3.81%), surgery (3.23%), cell biology (2.96%), endocrinology and metabolism (2.76%), pediatrics (2.12%) and other subjects respectively during 1990-2019.

Among leading organizations and authors participating in India's coronary artery disease research, the top 20 organizations and top 30 authors together contributed 40.17% and 37.29% respectively as their share of Indian publication output and 36.54% and 38.36% respectively as their share of Indian citation output during 1990-2019. The leading organizations in research productivity were: All India Institute of Medical Sciences (AIIMS), New Delhi (496 papers), Postgraduate Institute of Medical Education & Research (PGIMER), Chandigarh (212 papers), Christian Medical College & Hospital, Vellore (121 papers), Sree Chitra Tirunal Institute Medical Science & Technology (105 papers), Nizams Institute of Medical Science (104 papers), Sanjay Gandhi Postgraduate Institute Medical Science (100) during 1990-2019. The leading authors in publication productivity were Singh S (80 papers), Sharma A (78 papers), Kaul S and Mohan V (74 papers each), Kumar P (70 papers), Singh N and Singh RB (68 papers each), Singh M (63 papers), Gupta R (62 papers), Gupta A (61 papers), and Prabhakaran D (59 papers) during 1990-2019.

Among the total journal output of 4698 papers, the top 30 journals publishing Indian papers in coronary artery disease together accounted for 30.80% share of total Indian journal publication output during 1990-2019. Among journals contributing to Indian coronary artery disease research, *Neurology India* was the most productive journal with 143 papers, followed by *International Journal of Cardiology* (139 papers), *Annals of Indian Academy of Neurology* (99 papers), *Journal of the Neurological Sciences* (77 papers), *Indian Journal of Medical Research* (69 papers), *Molecular And Cellular Biochemistry* (56 papers) Stroke (55 papers), *Journal of the American College of Cardiology* and *PLOS One* (52 papers each) etc. during 1990-2019.

Concludes that coronary artery disease research have been a neglected subspecialty in India, both in teaching and research. There is an urgent need to increase the publication output, improve research quality and improve international collaboration. Review coronary artery disease studies in India indicate that this has become an important public health problem in India. CAD is one of the most important causes of mortality and morbidity in the country. With higher patient coming for treatment and shortage of trained cardiologist specialists are some of the challenges that confront coronary artery disease research at the national level. To address the problems with coronary artery disease research in India, Indian government needs to come up with a policy for identification, screening, diagnosis and treatment of coronary artery disease patients, besides curriculum reform in teaching, capacity building, patient education and political support are badly needed. There is an urgent need to promote primordial, primary, and secondary prevention strategies. Primordial strategies such as promotion of smoking/tobacco cessation, physical activity, and healthy dietary habits should prevent risk factors from occurring in the first place. Primary prevention should focus on screening and better control of risk factors (hypertension, hypercholesterolemia, and diabetes) to prevent incidence of overt CAD. Good quality secondary prevention and better management of acute and chronic events will prevent premature mortality and morbidity.

2020

# Measuring Research Productivity and Performance of Medical Scientists on Coronary Artery Disease in Brazil: A Metric Study

Muneer Ahmad
muneerbangroo@gmail.com

Dr. M.Sadik Batcha
*Annamalai University*

# Measuring Research Productivity and Performance of Medical Scientists on Coronary Artery Disease in Brazil: A Metric Study

**Muneer Ahmad[1] Dr. M Sadik Batcha[2]**

[1]*Research Scholar, Department of Library and Information Science, Annamalai University, Annamalai nagar, muneerbangroo@gmail.com*
[2]*Research Supervisor & Mentor, Professor and University Librarian, Annamalai University, Annamalai nagar, msbau@rediffmail.com*

**Abstract**

The aim of this study was to analyze Brazil research performance on Coronary Artery Disease as reflected in indexed publications in Web of Science with a view to understand their distribution of research output, top journals for publications, most prolific authors, authorship pattern, and citations pattern on CAD. The results indicate that highest growth rate of publications occurred between the years 1995-1999. University Sao Paulo topped the scene among all institutes. The maximum publications were more than ten authored publications. Ramires JAF and Santos RD were found to be the most prolific authors. It is also found that most of the prolific authors (by number of publications) do not emerge in highly cited publications' list. CAD researchers mostly preferred using article publications to communicate their findings.

## 1. Introduction

Coronary artery disease (CAD) refers to the build-up of atherosclerotic plaque in the blood vessels that supply oxygen and nutrients to the heart (Braunwald & Bonow, 2012). The complex process of atherosclerosis begins early in life and is thought to initiate with dysfunction of endothelial cells that line the coronary arteries; these cells are no longer able to appropriately regulate vascular tone (narrowing or constriction of the vessels) with nitric oxide signaling. Progressive infiltration of the vessel wall by lipoprotein particles carrying cholesterol propagates an inflammatory response by cholesterol-loaded macrophage 'foam cells'. Smooth muscle cells underlying the vessel wall proliferate and lead to remodeling of the vessel that can ultimately lead to a narrowing of the vessel that obstructs blood flow. A myocardial infarction (heart attack) is typically caused when a blood clot is incited by a rupture in the surface of the plaque; this process deprives the heart muscle downstream of the blood clot of adequate blood flow and leads to cell death (Khera & Kathiresan, 2017).

The prevalence of CAD, also known as coronary heart disease (CHD), has been observed to vary greatly according to the geographical locations, ethnicity, and gender (Go et al., 2014). Epidemiological studies on such cardiovascular diseases have provided information which could guide the strategies of prevention and eradication of these diseases both at the individual and population levels (Wong, 2014). Even before the field of cardiovascular epidemiology existed, in Minnesota (United States) the first prospective studies of CAD prevalence in population was conducted in 1946 (Keys et al., 1963). In the seven countries study, the relationships between lifestyle, diet, CAD, and stroke were elucidated (Keys, 1980). This study also indicated that the rates of heart attack and stroke were directly related to the levels of total cholesterol and this remained constant across different countries and cultures (Epstein, Blackburn, & Gutzwiller, 1996).

## 2. Review of Literature

Numerous studies have been conducted in the areas of Scientometrics, Bibliometrics and related to it, Webometrics (Ahmad, Batcha, Rashid, & Hafiz, 2018). The discipline has been widely spread through different journals, conference articles, monographs, textbooks, etc, especially in the recent decades. In view of the huge amount of literature available in the field, an attempt has been made to review only significant and recent literature on the various aspects of scientometrics research. (Batcha & Ahmad, 2017) obtained the analysis of two journals Indian Journal of Information Sources and Services (IJSS) which is of Indian origin and Pakistan Journal of Library and Information Science (PJLIS) from Pakistan origin and studied them comparatively with scientometric indicators like year wise distribution of articles, pattern of authorship and productivity, degree of collaboration, pattern of co-authorship, average length of papers, average keywords. The collaboration with foreign authors of both the countries is negligible (1.37% of articles) from India and (4.10% of articles) from Pakistan.

(Ahmad, Batcha, Wani, Khan, & Jahina, 2018) studied Webology journal one of the reputed journals from Iran was explored through scientometric analysis. The study aims to provide a comprehensive analysis regarding the journal like year wise growth of research articles, authorship pattern, author productivity, and subjects taken by the authors over the period of 5 years from 2013 to 2017. The findings indicate that 62 papers were published in the journal during the study period. The articles having collaborative nature were high in number. Regarding the subject concentration of papers of the journal, Social Networking, Web 2.0, Library 2.0 and Scientometrics or Bibliometrics were highly noted.

(Batcha, Jahina, & Ahmad, 2018) has examined the DESIDOC Journal by means of various scientometric indicators like year wise growth of research papers , authorship pattern, subjects and themes of the articles over the period of five years from 2013 to 2017. The study reveals that 227 articles were published over the five years from 2013 to 2017. The authorship pattern was highly collaborative in nature. The maximum numbers of articles (65 %) have ranged their thought contents between 6 and 10 pages.

(Ahmad & Batcha, 2019) analyzed research productivity in Journal of Documentation (JDoc) for a period of 30 years between 1989 and 2018. Web of Science a service from Clarivate Analytics has been consulted to obtain bibliographical data and it has been analysed through Bibexcel and Histcite tools to present the datasets. Analysis part deals with local and global citation level impact, highly prolific authors and their research output, ranking of prominent institution and countries. In addition to this scientographical mapping of bibliographical data is obtainable through VOSviewer, which is open source mapping software.

(Ahmad & Batcha, in 2019) studied the scholarly communication of Bharathiar University which is one of the vibrant universities in Tamil Nadu. The study find out the impact of research produced, year-wise research output, citation impact at local and global level, prominent authors and their total output, top journals of publications, top collaborating countries which collaborate with the university authors, highly industrious departments and trends in publication of the university during 2009 through 2018. In addition the study used scientographical mapping of data and presented it through graphs using VOSviewer software mapping technique.

(Ahmad, Batcha, & Jahina, 2019) quantitatively measured the research productivity in the area of artificial intelligence at global level over the study period of ten years (2008-2017). The study acknowledged the trends and features of growth and collaboration pattern of artificial intelligence research output. Average growth rate of artificial intelligence per year increases at the rate of 0.862. The multi-authorship pattern in the study is found high and the average number of authors per paper is 3.31. Collaborative Index is noted to be the highest range in the year 2014 with 3.50. Mean CI during the period of study is 3.24. This is also supported by the mean degree of collaboration at the percentage of 0.83 .The mean CC observed is 0.4635. Regarding the application of Lotka’s Law of authorship productivity in the artificial intelligence literature it proved to be fit for the study.

(Batcha, Dar, & Ahmad, 2019) presented a scientometric analysis of the journal titled “Cognition” for a period of 20 years from 1999 to 2018. The present study was conducted with an aim to provide a summary of research activity in current journal and characterize its most aspects. The research coverage includes the year wise distribution of articles, authors, institutions, countries and citation analysis of the journal. The analysis showed that 2870 papers were published in journal of Cognition from 1999 to 2018. The study identified top 20 prolific authors, institutions and countries of the journal. Researchers from USA have made the most percentage of contributions.

(Batcha, Dar, & Ahmad, 2020) conducts a scientometric study of the *Modern Language Journal* literature from 1999 to 2018. A total of 2564 items resulted from the publication name using “Modern Language Journal” as the search term was retrieved from the Web of Science Database. Based on the number of publications during the study period, no consistent growth was observed in the research activities pertaining to the journal. The annual distribution of publications, number of authors, institution productivity, country wise publications and Citations are analyzed. Highly productive authors, institutions, and countries are identified. The results reveal that the maximum number of papers 179 is published in the year 1999. It was also observed that Byrnes H is the most productive, contributed 51 publications and Kramsch C is most cited author in the field having 543 global citations. The highest number (38.26%) of publications, contributed from USA and the foremost productive establishment was University of Iowa.

(Ahmad, Batcha, & Dar, 2020) studied the Brain and Language journal which is an interdisciplinary journal, publishes articles that explicate the complex relationships among language, brain, and behavior and is one such journal which is concerned with investigating the neural correlates of Language. The study aims at mapping the structure of the *Brain and Language* journal. The journal looks into the intrinsic relationship between language and brain. The study demonstrates and elaborates on the various aspects of the Journal, such as its chronology wise total papers, most productive authors, citations, average citation per paper, institution and country wise distribution of publications for a period of 20 years.

(Ahmad & Batcha, 2020) explores and analyses the trend of world literature on “Coronavirus Disease” in terms of the output of research publications as indexed in the Science Citation Index

Expanded (SCI-E) of Web of Science during the period from 2011 to 2020. The study found that 6071 research records have been published on Coronavirus Disease. The various scientometric components of the research records published in the study period were studied. The study reveals the various aspects of Coronavirus Disease literature such as year wise distribution, relative growth rate, doubling time of literature, geographical wise, organization wise, language wise, form wise , most prolific authors, and source wise.

(Ahmad & Batcha, 2020) analyzed the application of Lotka's law to the research publication, in the field of Dyslexia disease. The data related to Dyslexia were extracted from web of science database, which is a scientific, citation and indexing service, maintained by Clarivate Analytics. A total of 5182 research publications were published by the researchers, in the field of Dyslexia. The study found out that, the Lotka's inverse square law is not fit for this data. The study also analyzed the authorship pattern, Collaborative Index (CI), Degree of Collaboration (DC), Co-authorship Index (CAI), Collaborative Co-efficient (CC), Modified Collaborative Co-efficient (MCC), Lotka's Exponent value, Kolmogorov-Smirnov Test (K-S Test), Relative Growth Rate and Doubling Time.

(Umar, Ahmad, & Batcha, 2020) studied and focused on the growth and development of Library and Culture research in forms of publications reflected in Web of Science database, during the span of 2010-2019. A total 890 publications were found and the highest 124 (13.93%) publications published in 2019.The analysis maps comprehensively the parameters of total output, growth of output, authorship, institution wise and country-level collaboration patterns, major contributors (individuals, top publication sources, institutions, and countries).

(Ahmad & Batcha, 2020) studied and examined 4698 Indian Coronary Artery Disease research publications, as indexed in Web of Science database during 1990-2019, with a view to understand their growth rate, global share, citation impact, international collaborative papers, distribution of publications by broad subjects, productivity and citation profile of top organizations and authors, and preferred media of communication.

(Jahina, Batcha, & Ahmad, 2020) study deals a scientometric analysis of 8486 bibliometric publications retrieved from the Web of Science database during the period 2008 to 2017. Data is collected and analyzed using Bibexcel software. The study focuses on various aspect of the quantitative research such as growth of papers (year wise), Collaborative Index (CI), Degree of Collaboration (DC), Co-authorship Index (CAI), Collaborative Co-efficient (CC), Modified

Collaborative Co-Efficient (MCC), Lotka's Exponent value, Kolmogorov-Smirnov test (K-S Test).

**3. Objectives**

The main objective of the present study is to study the growth of research output in Coronary Artery Disease from Brazil. Moreover, the study has been performed:

- To find out the type of documents containing Coronary Artery Disease research output in Brazil during 1990-2019;
- To analyse the year wise distribution and growth of literature on Coronary Artery Disease in Brazil during 1990-2019;
- To identify the top institutions conducting research on Coronary Artery Disease;
- To identify the most prolific authors conducting research on Coronary Artery Disease;
- To study the authorship pattern in Coronary Artery Disease research;
- To study the top sources preferred by authors for publishing Coronary Artery Disease research.

**4. Methodology**

The present study is a scientometric analysis of Coronary Artery Disease research publications. A total of 6211 records have been extracted from the Web of Science database in the '.txt' format covering the period (1990-2019). The search string used for data extraction is:

"TS=(Artery Disease, Coronary OR Artery Diseases, Coronary OR Coronary Artery Diseases OR Disease, Coronary Artery OR Diseases, Coronary Artery OR Coronary Arteriosclerosis OR Arterioscleroses, Coronary OR Coronary Arterioscleroses OR Atherosclerosis, Coronary OR Atheroscleroses, Coronary OR Coronary Atheroscleroses OR Coronary Atherosclerosis OR Arteriosclerosis, Coronary OR Ischaemic OR Ischemic OR hardening of the Arteries OR Induration of the Arteries OR Arterial Sclerosis ) AND CU=(Brazil)"

This search has been refined to limit the period from 1990 to 2019. Data filtering has been performed manually to remove irrelevant record entries. Bibliometrix Package in RStudio has been used for analyzing the data and it has also been used for tabulation and visualization of Results.

Calculations and statistical techniques were applied in the excel sheet to draw specific results. Total Publications (TP), Total Citations (TC), Average Citations per Paper (ACPP) h-index, g-index and m-index was calculated during analysis. ACPP is calculated by dividing the total citations received by the number of papers. The h-index was suggested by Jorge H. Hirch in 2005 (Hirsch, 2010). A scientist/ journal/ institution has index h if its h papers have atleast h citations each. Egghe defines g-index as “the highest rank such that the top g papers have, together, at least g2 citations. This also means that the top g + 1 have less than (g + 1)2 papers”. The g-index is always higher or equal to h-index, as has been also stated by (Egghe, 2006). m-index is another variant of the h-index that displays h-index per year since first publication. The h-index tends to increase with career length, and m-index can be used in situations where this is a shortcoming, such as comparing researchers within a field but with very different career lengths. The m-index inherently assumes unbroken research activity since the first publication

## 5. Data Analysis and Findings

### 5.1. Type of Publications

Different kind of publications in which research work on Coronary Artery Disease from Brazil is contributed during last 30 years is listed in Table 1. Out of total publications 4668 (75.16 %) are research articles, 565 (9.10 %) are meeting abstracts, 527 (8.48 %) are reviews, 157 (2.53 %) are editorial material, 136 (2.19 %) are article; proceedings paper, 117 (1.88 %) are letter, 13 (0.21 %) are article; early access, 6 (0.10 %) are article; book chapter, 6 (0.10 %) is note, 6 (0.10%) are review & book chapter, 4 (0.06%) are correction, 3 (0.05%) are new item and 1 (0.02%) are article; retracted publication, editorial material; early access, and review; retracted publication . It is apparent that more research output was produced in the form of articles and is having highest ACPP (30.49) than other forms of publications. It is also evident that in spite of more research output was produced in articles but ACPP of research output published as reviews and article; proceedings paper was also fair amount (64.20) compared to articles (30.49). ACPP of review; book chapter having (25.17), note (13.50), editorial Material (9.41), letter (4.04). Article; retracted publication published on CAD also received 3.00 ACPP. Other type of documents had ACPP less than 3. Thus; it was observed that articles, reviews and article; proceedings paper received more citations than other forms of documents.

**Table 1: Publication Type**

| S.No. | Document Type | Publications | % | TC | ACPP |
|---|---|---|---|---|---|
| 1 | Article | 4668 | 75.16 | 142311 | 30.49 |
| 2 | Meeting Abstract | 565 | 9.10 | 81 | 0.14 |
| 3 | Review | 527 | 8.48 | 16363 | 31.05 |
| 4 | Editorial Material | 157 | 2.53 | 1477 | 9.41 |
| 5 | Article; Proceedings Paper | 136 | 2.19 | 4508 | 33.15 |
| 6 | Letter | 117 | 1.88 | 473 | 4.04 |
| 7 | Article; Early Access | 13 | 0.21 | 9 | 0.69 |
| 8 | Article; Book Chapter | 6 | 0.10 | 17 | 2.83 |
| 9 | Note | 6 | 0.10 | 81 | 13.50 |
| 10 | Review; Book Chapter | 6 | 0.10 | 151 | 25.17 |
| 11 | Correction | 4 | 0.06 | 2 | 0.50 |
| 12 | News Item | 3 | 0.05 | 3 | 1.00 |
| 13 | Article; Retracted Publication | 1 | 0.02 | 3 | 3.00 |
| 14 | Editorial Material; Early Access | 1 | 0.02 | 0 | 0.00 |
| 15 | Review; Retracted Publication | 1 | 0.02 | 17 | 17.00 |

TC= "Total Citations", ACPP= "Average Citations per Paper"

### 5.2. Distribution of Research Publications

There has been a continuous increase in publications from the first decade (1990-1999) to the latest decade (2010-2019). During last 30 years, about 68 per centre research output on CAD was contributed in decade third (2010-2019). Table 2 shows the distribution of research output in five blocks of five years each. It is very apparent that highest growth rate occurs in the block year 1995-1999 (70.40 %) followed by 2005-2009 (62.26 %). Almost one-third (35.36 %) research output on CAD was contributed during 2010-2014. In first block year, research output was (1.06%) and in second block it increased (3.59%) and in third block it again increased (7.63%) and afterwards increased continuously by every block year. Highest number of research was contributed in the block 2015-2019 (30.21%).

**Table 2 : Distribution of Papers during 1990-2019**

| **Year** | Articles | % of TP | CO | % of Growth |
|---|---|---|---|---|
| 1990-1994 | 66 | 1.06 | 66 | -- |
| 1995-1999 | 223 | 3.59 | 289 | 70.40 |
| 2000-2004 | 474 | 7.63 | 763 | 52.95 |
| 2005-2009 | 1256 | 20.22 | 2019 | 62.26 |
| 2010-2014 | 1943 | 31.28 | 3962 | 35.36 |
| 2015-2019 | 2249 | 36.21 | 6211 | 13.61 |
| Total | 6211 | 100.00 | | |

TP= "Total Publications", CO= "Cumulative Output", Formula of Growth= "Final Value-Start Value/Start Value X100"

### 5.3. Institution-wise Research Share

The top 20 institutions that produced highest research outputs on CAD during the period under study are listed in Table 3. Table 3 summarizes total articles, the total citation score, and average citation per paper of the publications of these institutions. In total, 11033 institutions, including 20765 subdivisions published 6211 research papers during 1990 – 2019. The topmost twenty institutions involved in this research have published 83 and more research articles. The mean average is 0.56 research articles per Institution. Out of 11033 institutions, top 20 institutions published 8020 collaboratively research papers. It is also observed that among twenty top Institutions which contributed highest research output on CAD, University Sao Paulo took the lead by producing research output of 2211 publications followed by University Fed Sao Paulo with 502 research publications followed by University Fed Rio Grande do Sul with 402 research publications followed by University Fed Minas Gerais with 303 research publications. Eleven institutions produced 100 or more than 100 research publications on CAD. In terms of citations, University Sao Paulo received highest citations i.e. 66817 for 2211 total research publications. It is also noticed that Harvard University, had highest ACPP (194.74).

**Table 3: Top Institutions Research Output**

| S.No. | Institution | Publications | % | TC | ACPP |
|---|---|---|---|---|---|
| 1 | University Sao Paulo | 2211 | 39.47 | 66817 | 30.22 |
| 2 | University Fed Sao Paulo | 502 | 8.96 | 27489 | 54.76 |
| 3 | University Fed Rio Grande do Sul | 402 | 7.18 | 15158 | 37.71 |
| 4 | University Fed Minas Gerais | 303 | 5.41 | 11352 | 37.47 |
| 5 | University Fed Rio de Janeiro | 280 | 5.00 | 6105 | 21.80 |
| 6 | University Estadual Campinas | 225 | 4.02 | 4694 | 20.86 |
| 7 | Inst Dante Pazzanese Cardiol | 176 | 3.14 | 12231 | 69.49 |
| 8 | Hospital Clin Porto Alegre | 152 | 2.71 | 7011 | 46.13 |
| 9 | Hospital Israelita Albert Einstein | 150 | 2.68 | 1904 | 12.69 |
| 10 | Harvard University | 137 | 2.45 | 26680 | 194.74 |
| 11 | Brigham & Womens Hospital | 135 | 2.41 | 13704 | 101.51 |
| 12 | Johns Hopkins University | 125 | 2.23 | 18350 | 146.80 |
| 13 | University Fed Fluminense | 117 | 2.09 | 1476 | 12.62 |
| 14 | University Toronto | 110 | 1.96 | 12589 | 114.45 |
| 15 | University Estado Rio De Janeiro | 107 | 1.91 | 1736 | 16.22 |
| 16 | University Fed Parana | 99 | 1.77 | 910 | 9.19 |
| 17 | Harvard Medical School | 97 | 1.73 | 8235 | 84.90 |
| 18 | Columbia University | 96 | 1.71 | 17383 | 181.07 |
| 19 | Duke University | 95 | 1.70 | 16103 | 169.51 |
| 20 | McMaster University | 83 | 1.48 | 16118 | 194.19 |

TP= "Total Publications", TC= "Total Citations", ACPP= "Average Citations per Paper".

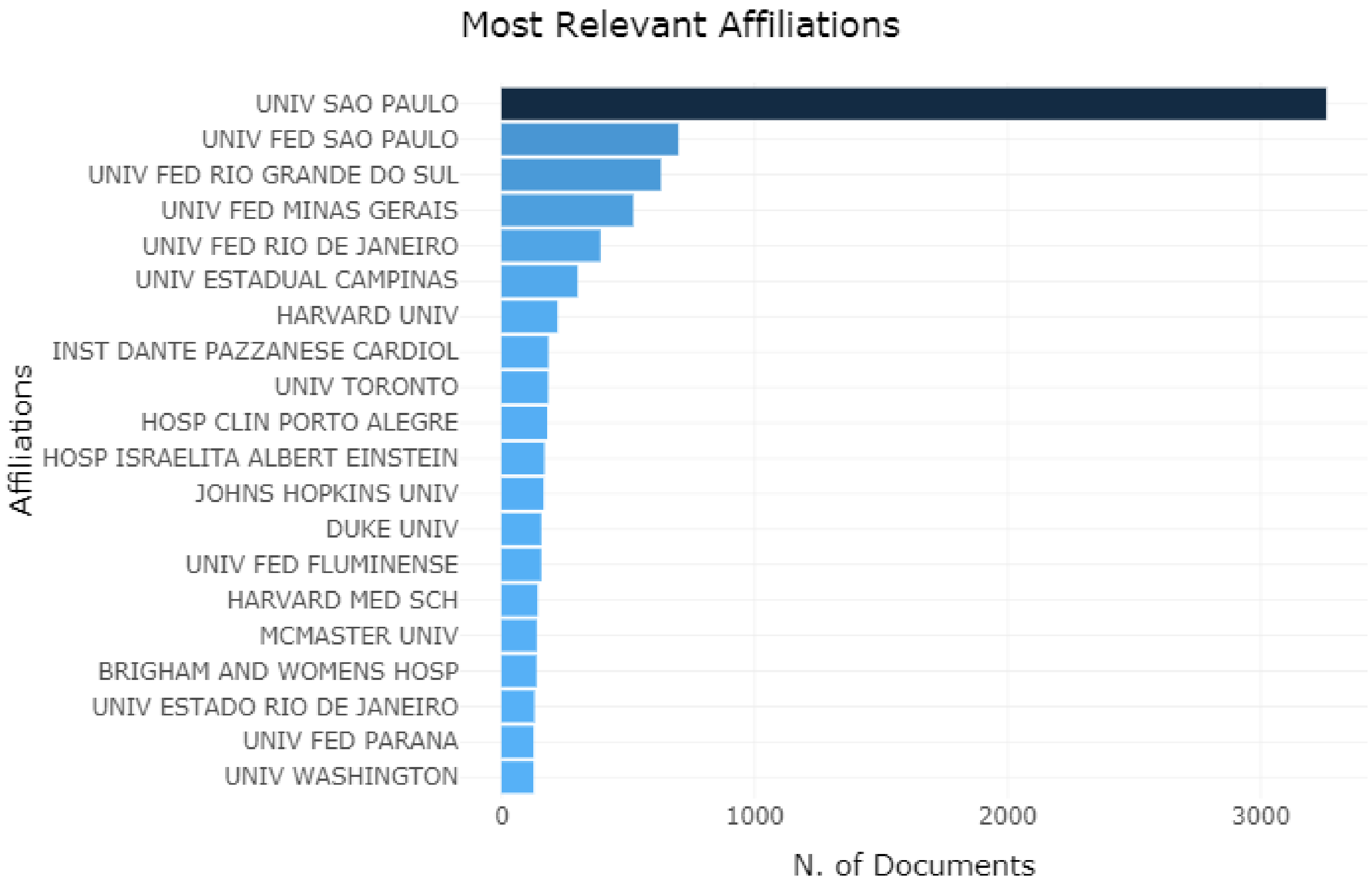

### 5.4. Most Prolific Authors

The list of twenty top authors who produced highest contribution to research output on CAD in Brazil is given in Table 4. In terms of number of publications, Ramires JAF is the most productive author with 231 publications followed by Santos RD 146, Hueb W 137, and Kalil R 117 publications. It is also noted that 5 out of 20 prolific authors contributed more than hundred research publications each while rest 15 authors contributed more than 60 publications each. The ACPP on research output contributed by Serruys PW (101.89) was recorded highest that was distantly followed by Lotufo PA (94.63). The h index is highest for Santos RD (32) followed by Ramires JAF (29) followed by Serruys PW (28) and Rochitte CE, Nicolau JC & Abizaid A (25). The data set puts forth that the authors Lotufo PA with 90 g index, Nicolau JC with 77 g index, Santos RD with 75 g index, Bensenor IM with 74 g index and Serruys PW with 73 g index. Rochitte CE (1.39), Abizaid A (1.25), Bittencourt MS (1.15) are having the highest m index respectively.

**Table 4: Most Prolific Authors**

| Author | NP | TC | ACPP | h-index | g-index | m-index |
|---|---|---|---|---|---|---|
| Ramires JAF | 231 | 3750 | 16.23 | 29 | 55 | 0.92 |
| Santos RD | 146 | 5804 | 39.75 | 32 | 75 | 0.57 |
| Hueb W | 137 | 3025 | 22.08 | 22 | 54 | 0.81 |
| Kalil R | 117 | 1008 | 8.62 | 15 | 30 | 0.87 |
| Pereira AC | 110 | 1324 | 12.04 | 19 | 31 | 0.95 |
| Rochitte CE | 99 | 3412 | 34.46 | 25 | 57 | 1.39 |
| Cesar LAM | 96 | 1121 | 11.68 | 17 | 31 | 0.57 |
| Nicolau JC | 94 | 5990 | 63.72 | 25 | 77 | 0.93 |
| Abizaid A | 88 | 2764 | 31.41 | 25 | 51 | 1.25 |
| Maranhao RC | 93 | 1234 | 13.27 | 21 | 30 | 0.56 |
| Lotufo PA | 90 | 8517 | 94.63 | 23 | 90 | 0.88 |
| Bittencourt MS | 78 | 1084 | 13.90 | 15 | 31 | 1.15 |
| Krieger JE | 77 | 974 | 12.65 | 17 | 27 | 0.85 |
| Bensenor IM | 74 | 6881 | 92.99 | 19 | 74 | 1.12 |
| Serruys PW | 73 | 7438 | 101.89 | 28 | 73 | 0.77 |
| Lemos PA | 68 | 1676 | 24.65 | 22 | 40 | 1.10 |
| Lima EG | 67 | 280 | 4.18 | 9 | 16 | 0.75 |
| Favarato D | 64 | 981 | 15.33 | 18 | 30 | 0.82 |
| Rezende PC | 64 | 224 | 3.50 | 9 | 14 | 0.90 |
| Serrano CV | 60 | 730 | 12.17 | 12 | 26 | 0.40 |

TP= “Total Publications”, TC= “Total Citations”, ACPP= “Average Citations per Paper”.

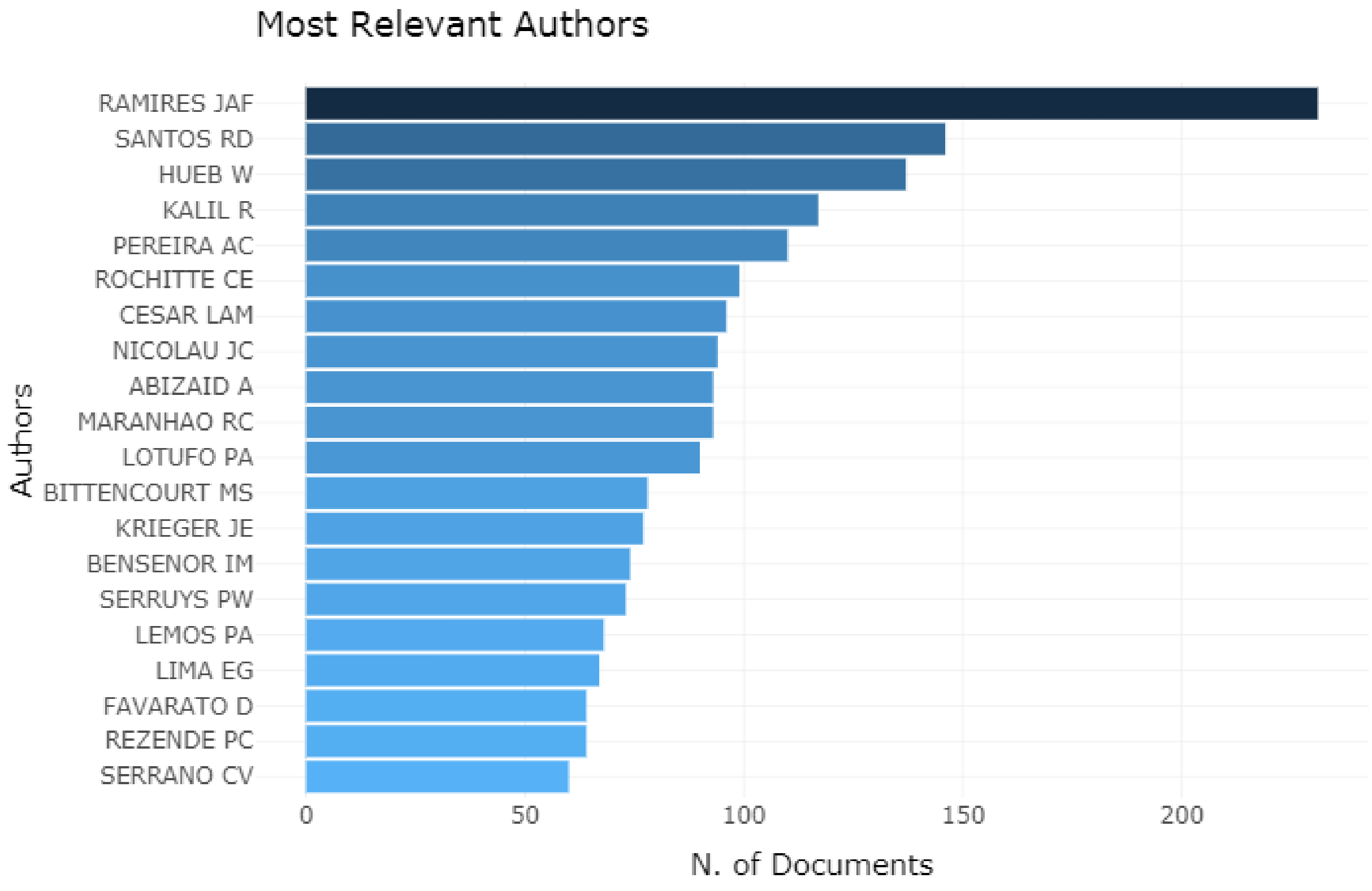

### 5.5. Authorship Pattern

Table 5 illustrates the overall and five year wise distribution of authorship trend. It is evident from the Table 5 that only 2.03 per cent publications were single authored publications while rest of 97.97 had two or more authors. The maximum number of publications were more than ten authored publications (16.99 %) nearly followed by six authored publications (12.78 %), five authored (11.30 %), seven authored (10.56 %) and eight authored publications (9.90 %). Two to nine authored publications accounted for 73.56 per cent while more than 10 authored publications accounted for 16.99 per cent.

**Table 5: Authorship Pattern**

| Author(s) | Total Research Output (5 Yearly) | | | | | | Total Research Output | |
|---|---|---|---|---|---|---|---|---|
| | 1990-1994 | 1995-1999 | 2000-2004 | 2005-2009 | 2010-2014 | 2015-2019 | Total | % |
| Single | 3 | 10 | 9 | 31 | 33 | 40 | 126 | 2.03 |
| Two | 5 | 15 | 34 | 54 | 99 | 116 | 323 | 5.20 |
| Three | 7 | 22 | 34 | 117 | 142 | 132 | 454 | 7.31 |
| Four | 11 | 36 | 55 | 122 | 192 | 171 | 587 | 9.45 |
| Five | 10 | 27 | 72 | 173 | 227 | 193 | 702 | 11.30 |
| Six | 8 | 26 | 81 | 186 | 230 | 263 | 794 | 12.78 |
| Seven | 7 | 36 | 59 | 148 | 219 | 187 | 656 | 10.56 |
| Eight | 5 | 18 | 42 | 138 | 207 | 205 | 615 | 9.90 |
| Nine | 4 | 11 | 27 | 85 | 138 | 173 | 438 | 7.05 |
| Ten | 1 | 9 | 17 | 80 | 173 | 181 | 461 | 7.42 |
| More than 10 | 5 | 13 | 44 | 122 | 283 | 588 | 1055 | 16.99 |
| Total | **66** | **223** | **474** | **1256** | **1943** | **2249** | **6211** | **100.00** |
| % | 1.06 | 3.59 | 7.63 | 20.22 | 31.28 | 36.21 | 100.00 | |

### 5.6. Top Journals Preferred for Publication

The total number of 6211 publications on CAD from 1990 to 2019 appeared in 1224 different sources. The top 20 journals preferred for publications on CAD are listed in Table 6 which accounted for 36.32 per cent of total research publications during the period under study. Circulation has published highest (162) publications on CAD followed by Journal of the American College of Cardiology (144). According to the journals preferred for publication output from the table 6 the journal wise distribution of research documents, Circulation has the highest number of research documents 162 with 9316 of total citation score and 42, 96 and .4 h index, g index and m index respectively and being prominent among the 20 journals and it stood in first rank position. Journal of the American College of Cardiology has 144 research documents and it stood in second position with 9113 of total citation score and 39, 95, 1.34 h index, g index and m index score were scaled. It is followed by the Lancet with 33 of records and it stood in third rank position along with 19436 of total citation score and 29, 33, and 1.12 h, g, and m index score measured.

**Table 6: Top 20 Sources for Publications**

| S.No. | Source of Publication | NP | TC | h-index | g-index | m-index |
|---|---|---|---|---|---|---|
| 1 | Circulation | 162 | 9316 | 42 | 96 | 1.4 |
| 2 | Journal of the American College of Cardiology | 144 | 9113 | 39 | 95 | 1.34 |
| 3 | Lancet | 33 | 19436 | 29 | 33 | 1.12 |
| 4 | Stroke | 80 | 2040 | 28 | 45 | 1.04 |
| 5 | European Heart Journal | 156 | 4477 | 26 | 66 | 0.96 |
| 6 | Atherosclerosis | 106 | 2108 | 25 | 42 | 0.96 |
| 7 | Arquivos Brasileiros De Cardiologia | 456 | 3088 | 22 | 35 | 1.57 |
| 8 | American Journal of Cardiology | 60 | 1455 | 22 | 37 | 0.76 |
| 9 | International Journal of Cardiology | 111 | 1503 | 21 | 32 | 0.72 |
| 10 | American Heart Journal | 46 | 1289 | 20 | 35 | 0.65 |
| 11 | Brazilian Journal of Medical and Biological Research | 133 | 1485 | 19 | 29 | 0.63 |
| 12 | Arquivos De Neuro-Psiquiatria | 240 | 1477 | 18 | 23 | 0.69 |
| 13 | Clinics | 84 | 1038 | 18 | 27 | 1.29 |
| 14 | Plos One | 62 | 841 | 18 | 26 | 1.5 |
| 15 | Revista Brasileira De Cirurgia Cardiovascular | 113 | 843 | 16 | 19 | 1.23 |
| 16 | Transplantation Proceedings | 59 | 440 | 13 | 16 | 0.72 |
| 17 | Catheterization and Cardiovascular Interventions | 46 | 875 | 12 | 29 | 0.7 |
| 18 | Acta Cirurgica Brasileira | 77 | 428 | 11 | 14 | 0.79 |
| 19 | Journal of Stroke & Cerebrovascular Diseases | 40 | 204 | 9 | 12 | 0.9 |
| 20 | Revista Da Associacao Medica Brasileira | 48 | 250 | 8 | 13 | 0.57 |

NP= “Number of Publications”, TC= “Total Citations”

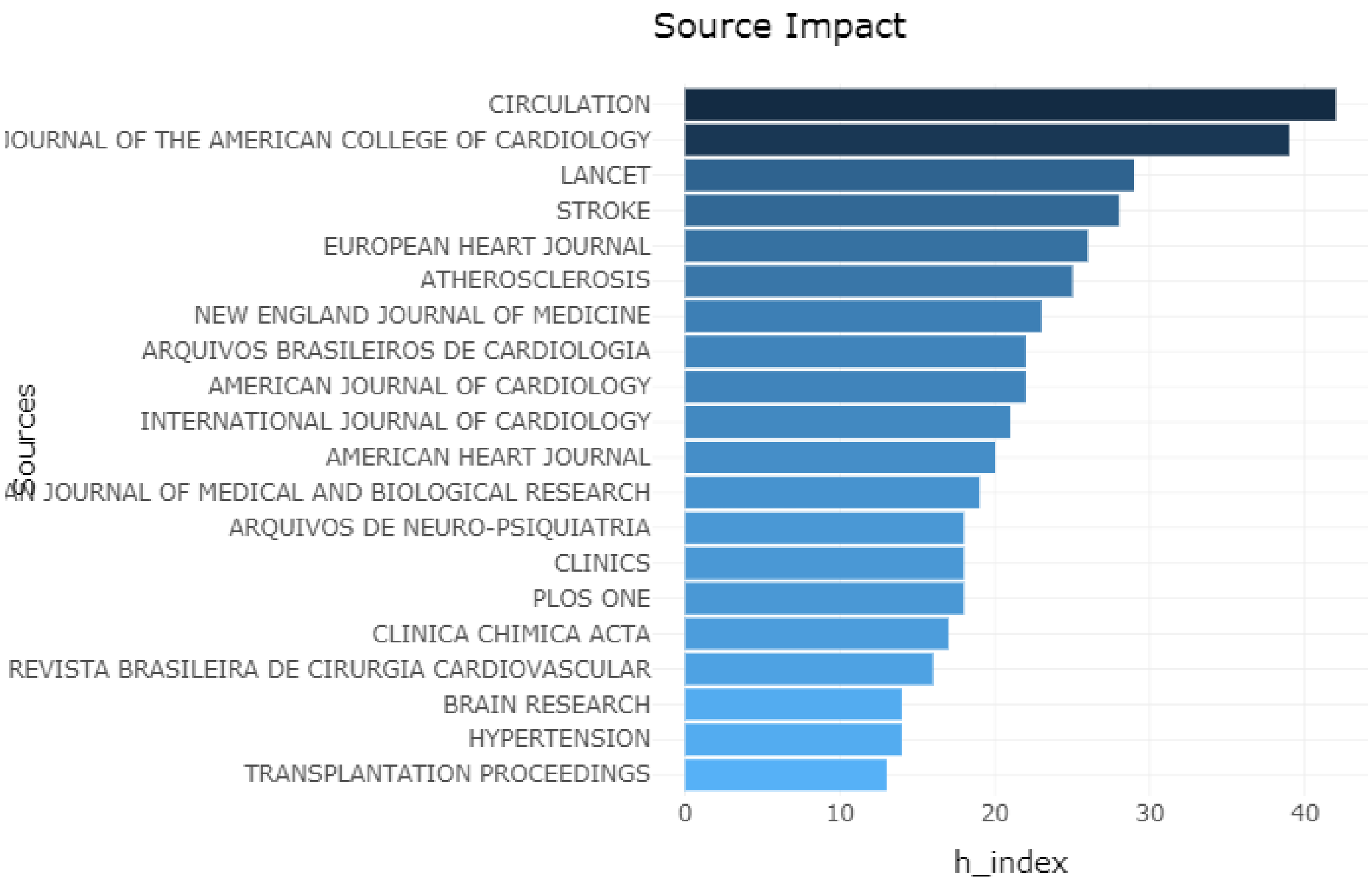

**6. Conclusion**

The study explores the 30 years research output on CAD in Brazil level. It was found that a total number of 6211 papers on CAD were published during 1990-2019 which received 160218 citations with ACPP of 25.80. Growth rate was highest (70.40%) in the block year 1995-1999. ACCP of Harvard University was highest with 195.74 average citations per paper. Nearly 36.32 per cent of research on CAD was published in 20 journals among which Circulation produced highest research output on CAD. Ramires JAF and Santos RD were the front runners in terms of number of publications but in terms of citations and ACPP Lotufo PA and Serruys PW remained at top. Only 2.03 per cent publications were single authored publications while rest of 97.97 had two or more authors. Among all type of publications, articles and reviews received more citations. The study depicts that research work on CAD was very less in earlier years or decades but increased during the later decades. Major research output was produced near 21st century especially during the last decade.